%% file: paper_v31h_final_eprint.tex
\documentclass[twocolumn,aps,prc,showpacs,superscriptaddress,nofootinbib,floatfix]{revtex4}
\usepackage{graphicx}
\usepackage{dcolumn} 
\usepackage{bm} 
\usepackage{longtable}

\def 	\snn 	{\sqrt{s_{_{NN}}}}
\def	\Nch 	{N_{\rm ch}}
\def	\dNdeta	{dN_{\rm ch}/d\eta}
\def	\dNchdy	{dN_{\rm ch}/dy}
\def	\dNraw	{dN_{\rm ch}^{\rm raw}/d\eta}
\def	\pt	{p_{\perp}}
\def	\mt	{m_{\perp}}
\def	\Et		{E_{\perp}}
\def	\meanpt {\langle p_{\perp}\rangle}
\def	\kt	{k_{\perp}}
\def	\Npart 	{N_{\rm part}}
\def	\Ncoll 	{N_{\rm coll}}
\def	\Sovlp 	{S_{\perp}}
\def	\dNdyS 	{\frac{dN_{\pi}/dy}{\Sovlp}}
\def	\dedx 	{dE/dx}
\def	\mdedx 	{\left\langle dE/dx\right\rangle}
\def	\Tch 	{T_{\rm chem}}
\def	\Tkin 	{T_{\rm kin}}
\def	\pbar	{\overline{p}}
\def	\sigmapp	{\sigma_{pp}}
\def	\sigmaAA	{\sigma_{\rm AA}}
\def	\ebj		{\epsilon_{Bj}}
\def	\npp		{n_{pp}}
\def	\dca	{d_{ca}}

\def	\jour#1#2#3#4	{{#1}~{\bf #2}, #3 (#4). }
\def	\reftit#1	{}
\def	\eprint#1	{}
\def	\Eprint#1	{arXiv:#1. }
\def	\collab#1	{(#1 Collaboration), }
\def	\etal		{{\it et al. }}

\begin{document}
\title{Systematic Measurements of Identified Particle Spectra in $pp$, d+Au and Au+Au Collisions from STAR}

\input authors_20090210

\date{\today, version 30h}

\begin{abstract}
Identified charged particle spectra of $\pi^{\pm}$, $K^{\pm}$, $p$ and $\pbar$ at mid-rapidity ($|y|<0.1$) measured by the $\dedx$ method in the STAR-TPC are reported for $pp$ and d+Au collisions at $\snn = 200$~GeV and for Au+Au collisions at 62.4~GeV, 130~GeV, and 200~GeV. Average transverse momenta, total particle production, particle yield ratios, strangeness and baryon production rates are investigated as a function of the collision system and centrality. The transverse momentum spectra are found to be flatter for heavy particles than for light particles in all collision systems; the effect is more prominent for more central collisions. The extracted average transverse momentum of each particle species follows a trend determined by the total charged particle multiplicity density. The Bjorken energy density estimate is at least several GeV/fm$^3$ for a formation time less than 1~fm/$c$. A significantly larger net-baryon density and a stronger increase of the net-baryon density with centrality are found in Au+Au collisions at 62.4~GeV than at the two higher energies. Antibaryon production relative to total particle multiplicity is found to be constant over centrality, but increases with the collision energy. Strangeness production relative to total particle multiplicity is similar at the three measured RHIC energies. Relative strangeness production increases quickly with centrality in peripheral Au+Au collisions, to a value about 50\% above the $pp$ value, and remains rather constant in more central collisions. Bulk freeze-out properties are extracted from thermal equilibrium model and hydrodynamics-motivated blast-wave model fits to the data. Resonance decays are found to have little effect on the extracted kinetic freeze-out parameters due to the transverse momentum range of our measurements. The extracted chemical freeze-out temperature is constant, independent of collision system or centrality; its value is close to the predicted phase-transition temperature, suggesting that chemical freeze-out happens in the vicinity of hadronization and the chemical freeze-out temperature is universal despite the vastly different initial conditions in the collision systems. The extracted kinetic freeze-out temperature, while similar to the chemical freeze-out temperature in $pp$, d+Au, and peripheral Au+Au collisions, drops significantly with centrality in Au+Au collisions, whereas the extracted transverse radial flow velocity increases rapidly with centrality. There appears to be a prolonged period of particle elastic scatterings from chemical to kinetic freeze-out in central Au+Au collisions. The bulk properties extracted at chemical and kinetic freeze-out are observed to evolve smoothly over the measured energy range, collision systems, and collision centralities.
\end{abstract}

\maketitle
\tableofcontents

\section{Introduction}
Quantum Chromodynamics (QCD) predicts a phase transition at sufficiently high energy density from normal hadronic matter to a deconfined state of quarks and gluons, the Quark-Gluon Plasma (QGP)~\cite{Shuryak,Bj,Karsch02}. Such a phase transition may be achievable in ultra-relativistic heavy-ion collisions. Many QGP signatures have been proposed which include rare probes (e.g. direct photon and dilepton production, jet modification) as well as bulk probes (e.g.~enhanced strangeness and antibaryon production, strong collective flow)~\cite{lastCall99}. While rare probes are more robust, they are relatively difficult to measure. On the other hand, signals of QGP that are related to the bulk of the collision are most probably disguised or diluted by other processes like final state interaction. Simultaneous observations and systematic studies of multiple QGP signals in the bulk would, however, serve as strong evidence for QGP formation. These bulk properties include strangeness and baryon production rates and collective radial  flow. These bulk observables can be studied via transverse momentum ($\pt$) spectra of identified particles in heavy-ion collisions in comparison to nucleon-nucleon and nucleon-nucleus reference systems. 

This paper reports results on identified charged pions ($\pi^{\pm}$), charged kaons ($K^{\pm}$), protons ($p$) and antiprotons ($\pbar$) at low $\pt$ at mid-rapidity~\cite{Molnar}. The results are measured by the STAR experiment in $pp$ and d+Au collisions at a nucleon-nucleon center-of-mass energy of $\snn=200$~GeV and in Au+Au collisions at $\snn=62.4$~GeV, 130~GeV and 200~GeV. The particles are identified by their specific ionization energy loss in the detector material -- the $\dedx$ method. Transverse momentum spectra, average transverse momenta, total particle production, particle yield ratios, antibaryon and strangeness production rates are presented as a function of the event multiplicity for $pp$, d+Au and Au+Au collisions. The paper also presents freeze-out parameters extracted from thermal equilibrium model and hydrodynamics-motivated blast-wave model fits to the data. The paper summarizes low $\pt$ results from STAR with $\dedx$ particle identification, including the previously published data~\cite{whitepaper}. 

The paper is organized as follows: Section~\ref{sec:Detector} describes the STAR detector, followed by descriptions of event selection, track quality cuts, and centrality definitions. Section~\ref{sec:dEdx} presents the $\dedx$ method for particle identification at low $\pt$. Section~\ref{sec:Corr} discusses the backgrounds and corrections applied at the event and track levels. Section~\ref{sec:systuncer} summarizes the systematic uncertainties of the measurements. Section~\ref{sec:Results} presents results on identified particle $\pt$ spectra, average $\meanpt$, particle yields and ratios. Section~\ref{sec:Model} discusses the systematics of bulk properties extracted from a statistical model and the hydrodynamics-motivated blast-wave model. Section~\ref{sec:Summary} summarizes the paper. Appendix~\ref{app:Glauber} describes the details of the Glauber model calculations used in this paper. Appendix~\ref{app:Resonance} discusses in detail the effect of resonance decays on the extracted kinetic freeze-out parameters. Appendix~\ref{app:spectra} lists tabulated data of transverse momentum spectra.

\section{Detector Setup and Data Samples\label{sec:Detector}}

\subsection{Detector Setup and Track Reconstruction}

Details of the STAR experiment can be found in Ref.~\cite{STAR}. The main detector of the STAR experiment is the Time Projection Chamber (TPC)~\cite{TPC1,TPC2}. The cylindrical axis of the TPC is aligned to the beam direction and is referred to as the $z$-direction. The TPC provides the full azimuthal coverage ($0\leq \phi \leq 2\pi$) and a pseudorapidity coverage of $-1.8 < \eta < 1.8$. 

Trigger selection of the experiment is obtained from the Zero Degree Calorimeters (ZDC)~\cite{ZDC}, the Beam-Beam Counters (BBC)~\cite{BBC} and the Central Trigger Barrel (CTB)~\cite{CTB}. The ZDC's are located at $\pm 18$~m along the $z$-direction from the TPC center and measure neutral energy. The scintillator-based BBC's provide the principal relative luminosity measurement in $pp$ data taking. The scintillator CTB surrounds the TPC and measures the charged particle multiplicity within $|\eta|<1$. The coincidence of the signals from the ZDC's and the BBC's selects minimum bias (MB) events in $pp$ and d+Au collisions. Our minimum bias $pp$ events correspond to non-singly diffractive (NSD) $pp$ collisions, whose cross-section is measured to be $30.0\pm3.5$~mb~\cite{highpt200}. The combination of the CTB and ZDC information provides the minimum bias trigger for Au+Au collisions. In addition, a central trigger is constructed by imposing an upper cut on the ZDCs' signal with a modest minimum CTB cut to exclude contamination from very peripheral events; the central trigger corresponds to approximately 12\% of the total cross-section. The trigger efficiencies are found to be approximately 86\% and 95\% in $pp$ and d+Au, respectively, and essentially 100\% in Au+Au collisions.

The TPC is filled with P-10 gas (90\% Argon and 10\% Methane). Charged particles interact with the gas atoms while traversing the TPC gas volume and ionize the electrons out of the gas atoms. Drift electric field is provided along the $z$-direction between the TPC central membrane and both ends of the TPC by a negative high voltage on the central membrane. Ionization electrons drift in the electric field towards the TPC ends. The TPC ends are divided into twelve equal-size bisectors, and are equipped with read-out pads and front-end electronics. Multi-Wire Proportional Chambers (MWPC) are installed close to the end pads inside the TPC. 
The drifting electrons avalanche in the high fields at the MWPC anode wires. The positive ions created in the avalanche induce a temporary image charge on the pads measured by a preamplifier/shaper/waveform digitizer system~\cite{TPC2,TPC3}. The original track positions (hits) are formed from the signals on each padrow (a row of read-out pads) by the hit reconstruction algorithm. Hits can be reconstructed to a small fraction of a pad width because the induced charge from an avalanche is shared over several adjacent pads.

The TPC is located inside a magnet which provides a magnetic field along the $z$-direction for particle momentum measurements. Data are taken at a maximum magnetic field of 0.5 Tesla. Inhomogeneities are on the level of $5\times10^{-3}$ Tesla and are incorporated in track reconstruction~\cite{MagnetNIM}. The direction of the magnetic field can be reversed to study systematic effects, which are found to be negligible for the bulk particles presented in this paper. 

Track reconstruction starts from the outermost hits in the TPC, projecting inward assuming an initial primary vertex position at the center of the TPC. Hits on the padrows are searched about the projected positions, and track segments are formed. Particle track momentum is estimated from the curvature of the track segments and the magnetic field strength. The momentum information is in turn used to refine further track projections. Track segments can be connected over short gaps from missed padrow signals. Tracks are formed from track segments and are allowed to cross the TPC sector boundaries. The reconstructed tracks are called global tracks. 

The primary interaction vertex is fit from the global tracks with at least 10 hits. The distance of closest approach ($\dca$) to the fit primary vertex is calculated for each global track. Iterations are made such that global tracks with $\dca>3$~cm are excluded from subsequent primary vertex fitting. Tracks with $\dca < 3$~cm (from the final fit primary vertex position) and at least 10 hits are called primary tracks. The primary tracks are refit including the primary vertex to improve particle track momentum determination. The reconstructed transverse momentum resolution is measured to be $\sigma_{\delta p_\perp}=0.01+\pt/(200$~GeV/$c$)~\cite{rdEdx}. The effect of the momentum resolution is negligible on the measured low $\pt$ particle spectra reported here, and is thus not corrected for. Only primary tracks are used in this analysis.

\subsection{Event Selection and Track Quality Cuts}

Data sets used in this paper are from $pp$ collisions at 200~GeV from Run II, d+Au collisions at 200~GeV from Run III, and Au+Au collisions at 62.4~GeV from Run IV, at 130~GeV from Run I, and at 200~GeV from Run II. The $pp$ and Au+Au data at 200~GeV from Run II have been published in Ref.~\cite{spec200}, and the Au+Au data for $K^{\pm}$, $p$ and $\pbar$ at 130~GeV from Run I have been published in Refs.~\cite{kaon130,pbar130,p130}. These data are incorporated in this paper to provide a systematic overview. The pion spectra from the 130~GeV Au+Au data are analyzed in this work. The data sets are summarized in Table~\ref{tab:datasets}.

\begin{table*}[htbp]
\caption{Summary of data sets, primary vertex cuts, and the numbers of good events (after cuts) used in the analysis.}
\label{tab:datasets}
\begin{ruledtabular}
\begin{tabular}{cccccccc}
Run 	& Data set & $\snn$ (GeV)	& Year	& Trigger & Max.~$|z_{\rm vtx}|$ & No.~of events\\ \hline
I	& Au+Au & 130 & 2000	& min.~bias	& 25 cm & 2.0 million\\
I	& Au+Au & 130 & 2000	& central	& 25 cm & 2.0 million\\
II	& Au+Au & 200 & 2001	& min.~bias	& 30 cm & 2.0 million\\
II	& $pp$ & 200 	& 2002	& min.~bias	& 30 cm & 3.9 million\\
III	& d+Au & 200 	& 2003	& min.~bias	& 50 cm & 8.8 million\\
IV	& Au+Au & 62.4 	& 2004 & min.~bias	& 30 cm & 6.4 million\\
\end{tabular}
\end{ruledtabular}
\end{table*}

The longitudinal, $z$ position of the interaction point is determined on-line by the measured signal time difference in the two ZDC's. A cut of the order of 50~cm on the $z$ position of the interaction point from the TPC center is applied on-line for all data sets (except $pp$) in order to maximize the amount of useful data for physics analysis, since events with primary vertex far away from the TPC center have a significantly non-uniform acceptance. In off-line data analyses further cuts are applied on the $z$ position of the reconstructed primary vertex, $z_{\rm vtx}$, to ensure nearly uniform detector acceptance and avoid multiplicity biases near the edges of the on-line cuts. These off-line cuts are listed in Table~\ref{tab:datasets}. In addition, the $x$ and $y$ position of the primary vertex are required to be within $\pm 3.5$ cm of the beam since the beam pipe diameter is 3~inches. 

The use of primary tracks significantly reduces contributions from background processes and pileup events in $pp$ data. Tracks can have a maximum of 45 hits. In the analysis at least 25 hits are required for each track to avoid track splitting effects. Singly charged particles must have a minimum $\pt$ of 0.15~GeV/$c$~to exit the TPC in the 0.5 Tesla magnetic field. In this analysis tracks are required to have $\pt>0.2$~GeV/$c$. For the identified particle results in this paper, the rapidity region is restricted within $|y| < 0.1$ (i.e. mid-rapidity). The full $2\pi$ azimuthal coverage of the TPC is utilized.

\subsection{Centrality Measures}

\subsubsection{Centrality Definitions}

\begin{figure}[htbp]
\centering
\includegraphics[width=0.48\textwidth]{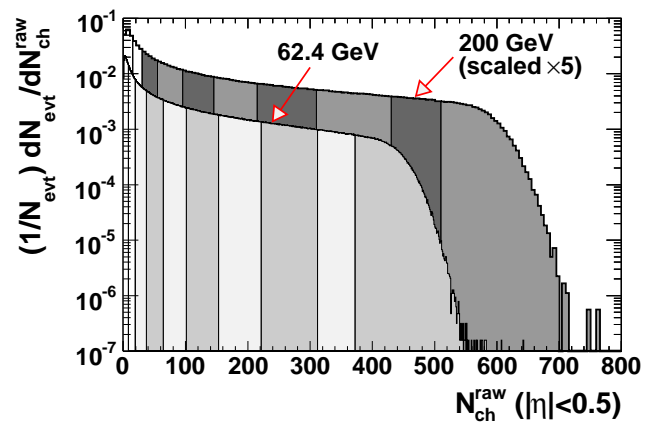}
\caption{Uncorrected charged particle multiplicity distribution measured in the TPC in $\left|\eta\right|<0.5$ for Au+Au collisions at 62.4~GeV and 200~GeV. The shaded regions indicate the centrality bins used in the analysis. The 200~GeV data are scaled by a factor 5 for clarity.}
\label{fig:auaumult}
\end{figure}

In Au+Au collisions, the measured (uncorrected) charged particle multiplicity density in the TPC within $|\eta| < 0.5$, $\dNraw$, is used for centrality selection. The primary tracks to be counted in the charged particle multiplicity are required to have at least 10 fit points (good primary tracks). The multiplicity distributions in Au+Au collisions at 62.4~GeV and 200~GeV are shown in Fig.~\ref{fig:auaumult} . Nine centrality bins are chosen, the same as in Ref.~\cite{spec200}; they correspond to the fraction of the measured total cross section from central to peripheral collisions of 0-5\%, 5-10\%, 10-20\%, 20-30\%, 30-40\%, 40-50\%, 50-60\%, 60-70\%, and 70-80\%. The 80-100\% centrality is not used in our analysis because of its significant trigger bias due to vertex inefficiency at low multiplicities and the contamination from electromagnetic interactions.

\begin{figure}[htbp]
\centering
\includegraphics[width=0.48\textwidth]{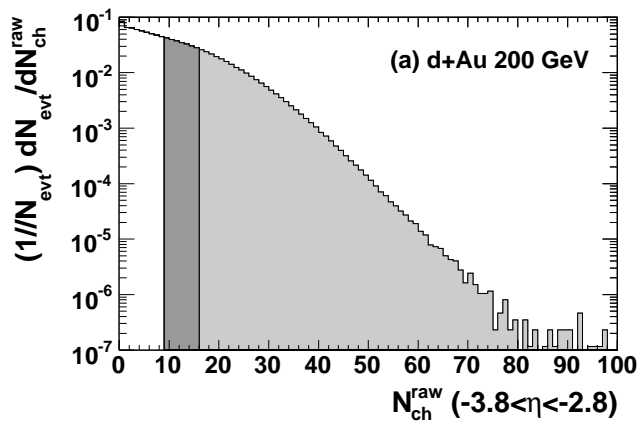}
\includegraphics[width=0.48\textwidth]{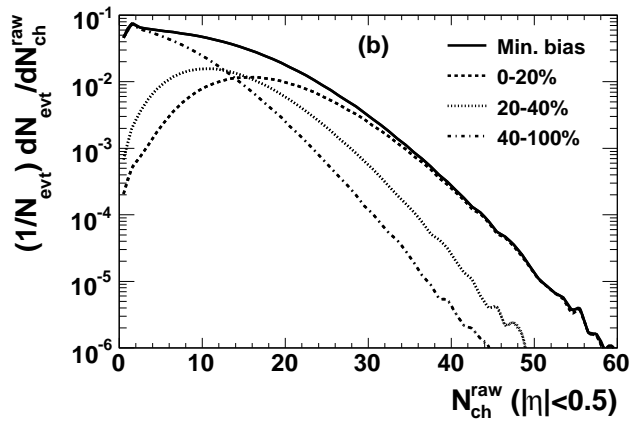}
\caption{(a) Uncorrected charged particle multiplicity distribution measured in the E-FTPC (Au-direction) within $-3.8<\eta<-2.8$ in d+Au collisions at 200~GeV. The shaded regions indicate the centrality bins used in the analysis. (b) The TPC mid-rapidity multiplicity distributions ($|\eta|<0.5$) for the corresponding E-FTPC selected centrality bins.}
\label{fig:pp_dau_mult}
\end{figure}

\begin{figure}[htbp]
\centering
\includegraphics[width=0.48\textwidth]{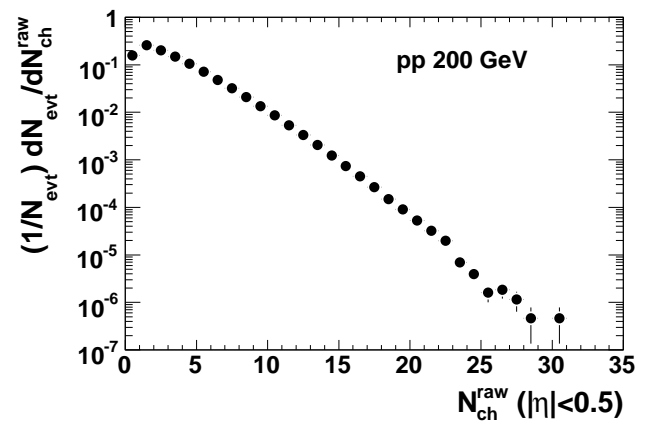}
\caption{Uncorrected charged particle multiplicity distribution measured in the TPC within $|\eta|<0.5$ in $pp$ collisions at 200~GeV.}
\label{fig:pp_multiplicity3}
\end{figure}

In d+Au collisions centralities are selected based on the charged particle multiplicity measured in the East (Au-direction) Forward Time Projection Chamber (E-FTPC)~\cite{ftpc} within the pseudo-rapidity range of $-3.8 < \eta < -2.8$. To be counted, tracks are required to have at least 6 hits out of 11 maximum and a $\dca < 3$~cm. Additionally the transverse momentum is required not to exceed 3~GeV/$c$ because of the reduced momentum resolution and a significant background contamination at high $\pt$~\cite{ftpc}. Figure~\ref{fig:pp_dau_mult}(a) shows the measured (uncorrected) E-FTPC charged particle multiplicity. Three centrality classes are defined, as indicated by the shaded regions, representing 40-100\%, 20-40\% and 0-20\% of the measured total cross section~\cite{dAuNch200}. The mid-rapidity multiplicities measured in the TPC for the selected centrality bins are shown in Fig.~\ref{fig:pp_dau_mult}(b). Positive correlation between the TPC multiplicity and the E-FTPC multiplicity is evident, although the correlation is not very strong due to the low multiplicities of d+Au collisions.

The reason to use the FTPC multiplicity instead of the TPC mid-rapidity multiplicity for centrality selection is to avoid auto-correlation between centrality and the measurements of charged particles which are made within $|y|<0.1$ in the TPC. The auto-correlation is not significant for Au+Au collisions due to their large multiplicities. The auto-correlation is significant for $pp$, and since the FTPC was not ready for data taking in the $pp$ run only minimum bias data are presented for $pp$. For completeness, the uncorrected multiplicity distribution in minimum bias $pp$ collisions is shown in Fig.~\ref{fig:pp_multiplicity3}.

Table~\ref{tab:collision_prop} summarizes the centralities for $pp$, d+Au, and Au+Au collisions.

\begin{table*}[htbp]
\caption{Summary of centralities in $pp$ and d+Au collisions at 200~GeV and in Au+Au collisions at 62.4~GeV, 130~GeV, and 200~GeV. Our minimum bias $pp$ data correspond to NSD events with total cross-section of $30.0\pm3.5$~mb.~\cite{highpt200}. The centrality percentages in other systems are in terms of the measured total cross-sections. The uncorrected charged particle multiplicity $\dNraw$ for d+Au is measured in the E-FTPC within $-3.8<\eta<-2.8$, and for all other systems in the TPC within $|\eta|<0.5$. The corrected charged particle multiplicity $\dNdeta$ (and the corrected negatively charged particle multiplicity $dN_{h^-}/d\eta$ for the 130~GeV Au+Au data) are from the TPC within $|\eta|<0.5$. The multiplicity rapidity density $dN/dy$ are from the rapidity slice of $|y|<0.1$. The 200~GeV $pp$ and Au+Au data are from Ref.~\cite{spec200}; the 130~GeV data are from Ref.~\cite{hminus130} and this work; and the 200~GeV d+Au and 62.4~GeV Au+Au data are from this work. The {\it Monte-Carlo} Glauber model is used in the calculation of the impact parameter ($b$), the number of participant nucleons ($\Npart$), the number of binary nucleon-nucleon collisions ($\Ncoll$), and the overlap area between the colliding nuclei in the transverse plane ($\Sovlp$). The nucleon-nucleon cross-sections used in the calculations are $36\pm2$~mb, $39\pm2$~mb, $41\pm2$~mb for 62.4~GeV, 130~GeV, and 200~GeV, respectively. The Glauber model results for d+Au are from Ref.~\cite{dAuPLB}, and for all other systems from this work. The quoted errors are total statistical and systematic uncertainties added in quadrature.} 
\label{tab:collision_prop}
\begin{ruledtabular}
\begin{tabular}{cccccc|ccccc}
Centrality & $\dNraw$ & $\dNraw$ & $\dNdeta$ & $dN_{h^-}/d\eta$ & $\dNchdy$ & $b$ (fm) & $b$ (fm) & $\Npart$ & $\Ncoll$ & $\Sovlp$ (fm$^2$) \\ 
 & range & mean & & & & range & mean & & & \\ 
\hline 
\multicolumn{6}{c|}{$pp$ 200~GeV} & \multicolumn{5}{c}{$pp$} \\ \hline
min.~bias & -- 	& 2.4 & $2.98\pm0.34$ && $3.40\pm0.23$ &--&--& 2 & 1 & $4.1\pm0.7$ \\ 
\hline 
\multicolumn{6}{c|}{d+Au 200~GeV} & \multicolumn{5}{c}{d+Au~\cite{dAuPLB}} \\ \hline
min.~bias & -- 	& 10.2 & $10.2\pm0.68$ && $11.3\pm0.7$ &&& $8.31\pm0.37$ & $7.51\pm0.39$ & \\ 
40-100\% & 0-9 & 6.2 & $6.23\pm0.34$ && $6.98\pm0.44$ &&& $5.14\pm0.44$  & $4.21\pm0.49$ & \\ 
20-40\% & 10-16 & 12.6 & $14.1\pm1.0$ && $14.9\pm0.9$ &&& $11.2\pm1.1$ & $10.6\pm0.8$ & \\ 
0-20\%  & $\geq$17 & 17.6 & $19.9\pm1.6$ && $20.9\pm1.3$ &&& $15.7\pm1.2$ & $15.1\pm1.3$ & \\ 
\hline 
\multicolumn{6}{c|}{Au+Au 200~GeV} & \multicolumn{5}{c}{Au+Au ($\sigmapp=41$ mb)} \\ \hline
70-80\% & 14-29  	& 22.5 & $22\pm2$ && $26.5\pm1.8$ & 12.3-13.2 & $12.8\pm0.3$ & $15.7\pm2.6$ & $15.0\pm3.2$ & $17.8\pm2.2$ \\
60-70\% & 30-55  	& 43.1 & $45\pm3$ && $52.1\pm3.5$ & 11.4-12.3 & $11.9\pm0.3$ & $28.8\pm3.7$ & $32.4\pm5.5$ & $27.2\pm2.5$ \\
50-60\% & 56-93  	& 74.8 & $78\pm6$ && $90.2\pm6.0$ & 10.5-11.4 & $11.0\pm0.3$ & $49.3\pm4.7$ & $66.8\pm9.0$ & $38.8\pm2.7$ \\
40-50\% & 94-145 	& 120 & $126\pm9$ && $146\pm10$ & 9.33-10.5 & $9.90\pm0.23$ & $78.3\pm5.3$ & $127\pm13$ & $52.1\pm2.7$ \\
30-40\% & 146-216 	& 181 & $195\pm14$ && $222\pm15$ & 8.10-9.33 & $8.73\pm0.19$ & $117.1\pm5.2$ & $221\pm17$ & $67.5\pm2.9$ \\
20-30\% & 217-311 	& 264 & $287\pm20$ && $337\pm23$ & 6.61-8.10 & $7.37\pm0.16$ & $167.6\pm5.4$ & $365\pm24$ & $86.1\pm3.1$ \\
10-20\% & 312-430 	& 370 & $421\pm30$ && $484\pm33$ & 4.66-6.61 & $5.70\pm0.14$ & $234.3\pm4.6$ & $577\pm36$ & $109.8\pm3.4$ \\
 5-10\% & 431-509 	& 470 & $558\pm40$ && $648\pm44$ & 3.31-4.66 & $4.03\pm0.13$ & $298.6\pm4.1$ & $805\pm50$ & $133.0\pm3.5$ \\
 0-5\%  & $\geq$510	& 559 & $691\pm49$ && $811\pm56$ & 0  -3.31 & $2.21\pm0.07$ & $350.6\pm2.4$ & $1012\pm59$ & $153.9\pm4.3$ \\
\hline 
\multicolumn{6}{c|}{Au+Au 130~GeV} & \multicolumn{5}{c}{Au+Au ($\sigmapp=39$ mb)} \\ \hline
58-85\% & 11-57  	& 30.8&& $17.9\pm1.3$ & $39.5\pm4.0$ & 11.1-13.4 & $12.3\pm0.4$ & $22.6\pm5.0$ & $24.4\pm7.0$ & $21.9\pm3.6$ \\
45-58\% & 57-105  	& 80.3&& $47.3\pm3.3$ & $105\pm8$  & 9.77-11.1 & $10.5\pm0.3$ & $61.0\pm7.8$ & $88\pm16$  & $43.4\pm3.8$ \\
34-45\% & 105-163 	& 133 && $78.9\pm5.5$ & $177\pm11$  & 8.50-9.77 & $9.15\pm0.28$ & $100.9\pm8.4$ & $175\pm22$  & $60.2\pm3.7$ \\
26-34\% & 163-217 	& 190 && $115\pm8$  & $257\pm18$  & 7.43-8.50 & $7.99\pm0.25$ & $141.9\pm8.4$ & $280\pm26$  & $75.6\pm3.9$ \\
18-26\% & 217-286 	& 251 && $154\pm11$  & $348\pm24$  & 6.19-7.43 & $6.82\pm0.21$ & $187.7\pm7.5$ & $411\pm31$  & $91.9\pm3.8$ \\
11-18\% & 286-368 	& 327 && $196\pm14$  & $460\pm34$  & 4.83-6.19 & $5.55\pm0.18$ & $237.8\pm6.8$ & $568\pm39$  & $109.7\pm3.7$ \\
 6-11\% & 368-417 	& 392 && $236\pm17$  & $562\pm35$  & 3.58-4.83 & $4.23\pm0.16$ & $289.0\pm5.4$ & $739\pm49$  & $127.8\pm3.7$ \\
 0-6\%  & $\geq$417	& 462 && $290\pm20$  & $695\pm45$  & 0  -3.58 & $2.39\pm0.09$ & $344.3\pm3.1$ & $945\pm58$  & $149.5\pm4.3$ \\
\hline
\multicolumn{6}{c|}{Au+Au 62.4~GeV} & \multicolumn{5}{c}{Au+Au ($\sigmapp=36$ mb)} \\ \hline
70-80\% & 9-19  	& 12.4 & $13.9\pm1.1$ && $17.7\pm1.3$ & 12.3-13.1 & $12.7\pm0.3$ & $15.3\pm2.4$ & $14.1\pm2.8$ & $16.1\pm2.0$ \\
60-70\% & 20-37  	& 26.8 & $29.1\pm2.2$ && $35.8\pm2.8$ & 11.4-12.3 & $11.8\pm0.3$ & $27.8\pm3.7$ & $30.0\pm5.2$ & $24.8\pm2.4$ \\
50-60\% & 38-64  	& 49.1 & $53.1\pm4.2$ && $65.0\pm5.0$ & 10.4-11.4 & $10.9\pm0.2$ & $47.9\pm4.7$ & $61.2\pm8.2$ & $35.8\pm2.6$ \\
40-50\% & 65-101  	& 81.0 & $87.2\pm7.1$ && $107\pm8$  & 9.27-10.4 & $9.83\pm0.23$ & $76.3\pm5.2$ & $115\pm12$  & $48.7\pm2.7$ \\
30-40\% & 102-153 	& 125.2 & $135\pm11$  && $166\pm11$  & 8.05-9.27 & $8.67\pm0.19$ & $114.3\pm5.1$ & $199\pm16$  & $63.6\pm2.8$ \\
20-30\% & 154-221 	& 184.8 & $202\pm17$  && $249\pm16$  & 6.56-8.05 & $7.32\pm0.17$ & $164.1\pm5.4$ & $325\pm23$  & $81.6\pm3.1$ \\
10-20\% & 222-312 	& 263.6 & $292\pm25$  && $359\pm24$  & 4.63-6.56 & $5.67\pm0.14$ & $229.8\pm4.6$ & $511\pm34$  & $104.6\pm3.3$ \\
 5-10\% & 313-372 	& 340.5 & $385\pm33$  && $476\pm30$  & 3.29-4.63 & $4.00\pm0.13$ & $293.9\pm4.2$ & $710\pm47$  & $127.2\pm3.6$ \\
 0-5\% & $\geq$373	& 411.8 & $472\pm41$  && $582\pm38$  & 0  -3.29 & $2.20\pm0.07$ & $346.5\pm2.8$ & $891\pm57$  & $147.5\pm4.3$ \\
\end{tabular} 
\end{ruledtabular}
\end{table*} 

\subsubsection{Corrected Charged Particle Multiplicity}

The results in this paper are presented as a function of centrality. As an experimental measure of centrality, the {\it corrected} charged particle rapidity density ($\dNchdy$) is used. It is obtained from the identified charged particle spectra ($\pi^{\pm}$, $K^{\pm}$, $\overline{p}$ and $p$) as the sum of the individual rapidity densities. The identified charged particle spectra are either from prior STAR publications~\cite{spec200,kaon130,pbar130,p130} or obtained by this work~\cite{Molnar}. The charged particle rapidity densities are listed in Table~\ref{tab:collision_prop} for various systems and centralities. The systematic uncertainties on $\dNchdy$ are discussed in Section~\ref{sec:dNdySystErr}.

Another commonly used centrality measure is the charged particle pseudo-rapidity density, either uncorrected ($\dNraw$) or corrected ($\dNdeta$) for detector losses and tracking efficiency. These quantities are also listed in Table~\ref{tab:collision_prop} for reference. The correction is done using reconstruction efficiency of pions obtained from embedding {\it Monte Carlo} (see Section~\ref{sec:efficiency}). This is because the efficiencies at high $\pt$ are the same for different particle species and at low $\pt$ charged particles are dominated by pions. The pseudorapidity multiplicity density data for 130~GeV are from Ref.~\cite{Manuel}, for $pp$ and Au+Au at 200~GeV from Ref.~\cite{spec200}, and for d+Au at 200~GeV and Au+Au at 62.4~GeV from this work.

\subsubsection{Glauber Model Calculations}

While the charged hadron multiplicity is a viable experimental way to characterize centrality, it is sometimes desirable to use other variables directly connected to the collision geometry. Those variables include the number of participant nucleons ($\Npart$), the number of nucleon-nucleon binary collisions ($\Ncoll$), and the ratio of the charged pion rapidity density to the transverse overlap area of the colliding nuclei ($\dNdyS$). Many models have studied particle production mechanisms based on these centrality variables. For example, the two-component model~\cite{Hwa88,WangMiniJet,ToporPopMiniJet}, characterizing particle production by a linear combination of $\Npart$ and $\Ncoll$, can describe the multiplicity density well, allowing the extraction of the relative fractions of the two components. The gluon saturation model~\cite{dima01,dima02,Schaffner-Bielich1,Schaffner-Bielich2,WangGluonShadowing} predicts a suppressed multiplicity in heavy-ion collisions relative to the $\Ncoll$-scaled $pp$ collision multiplicity, with an increased $\meanpt$ for the produced particles. The relevant, and perhaps only scale in such a gluon-saturation picture is $\dNdyS$~\cite{dima01}.

Unfortunately, $\Npart$, $\Ncoll$ and the transverse overlap area $\Sovlp$ cannot be measured directly from collider experiments, so they have to be extracted from the measured multiplicity distributions via models, such as the Glauber model~\cite{glauber,glauber2}. The essential ingredient is to match the calculated cross-section versus impact parameter ($d\sigma/db$) to the measured cross-section versus charged multiplicity ($d\sigma/d\Nch$), exploiting the fact that the average multiplicity should monotonically increase with decreasing impact parameter, $b$. The matching relates the measured $\Nch$ to $b$ (and thus $\Npart$ and $\Ncoll$). 

Two different schemes are used to implement the Glauber model, the {\it optical} calculation and the {\it Monte-Carlo} (MC) calculation. The details of the optical and MC Glauber calculations are described in Appendix~\ref{app:Glauber}. In this work the MC Glauber calculation is used except when otherwise noted. The $\Npart$, $\Ncoll$, and $\Sovlp$ for Au+Au collisions from the MC Glauber model calculation are listed in Table~\ref{tab:collision_prop}. For $pp$ collisions the overlap area $\Sovlp$ is simply taken as the $pp$ cross-section, $\sigmapp$. For d+Au collisions the $\Npart$ and $\Ncoll$ are calculated using the realistic wavefunction for the deuteron in Ref.~\cite{dAuPLB}.

\begin{figure}[hbtp]
\centering
\includegraphics[width=0.48\textwidth]{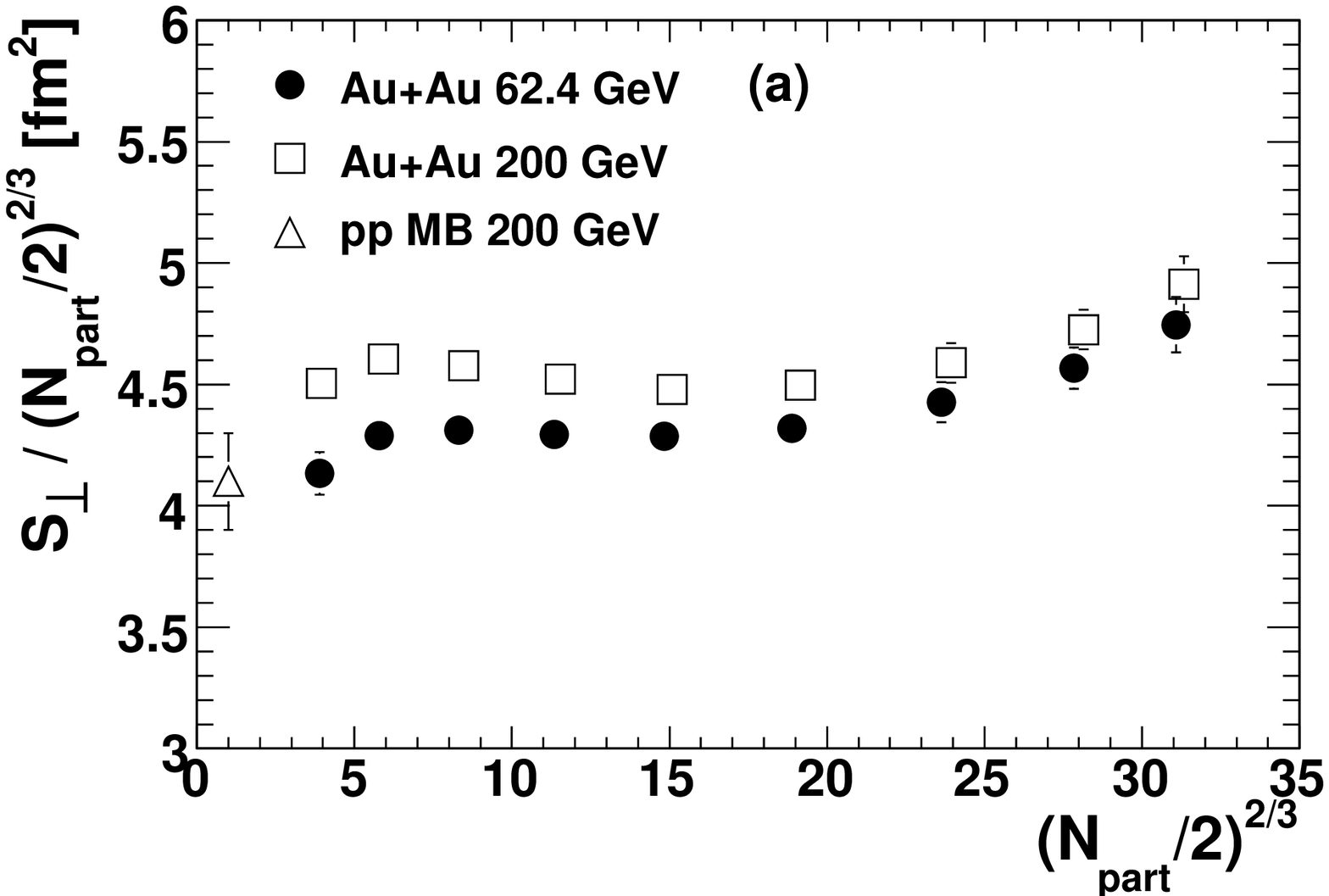}
\includegraphics[width=0.48\textwidth]{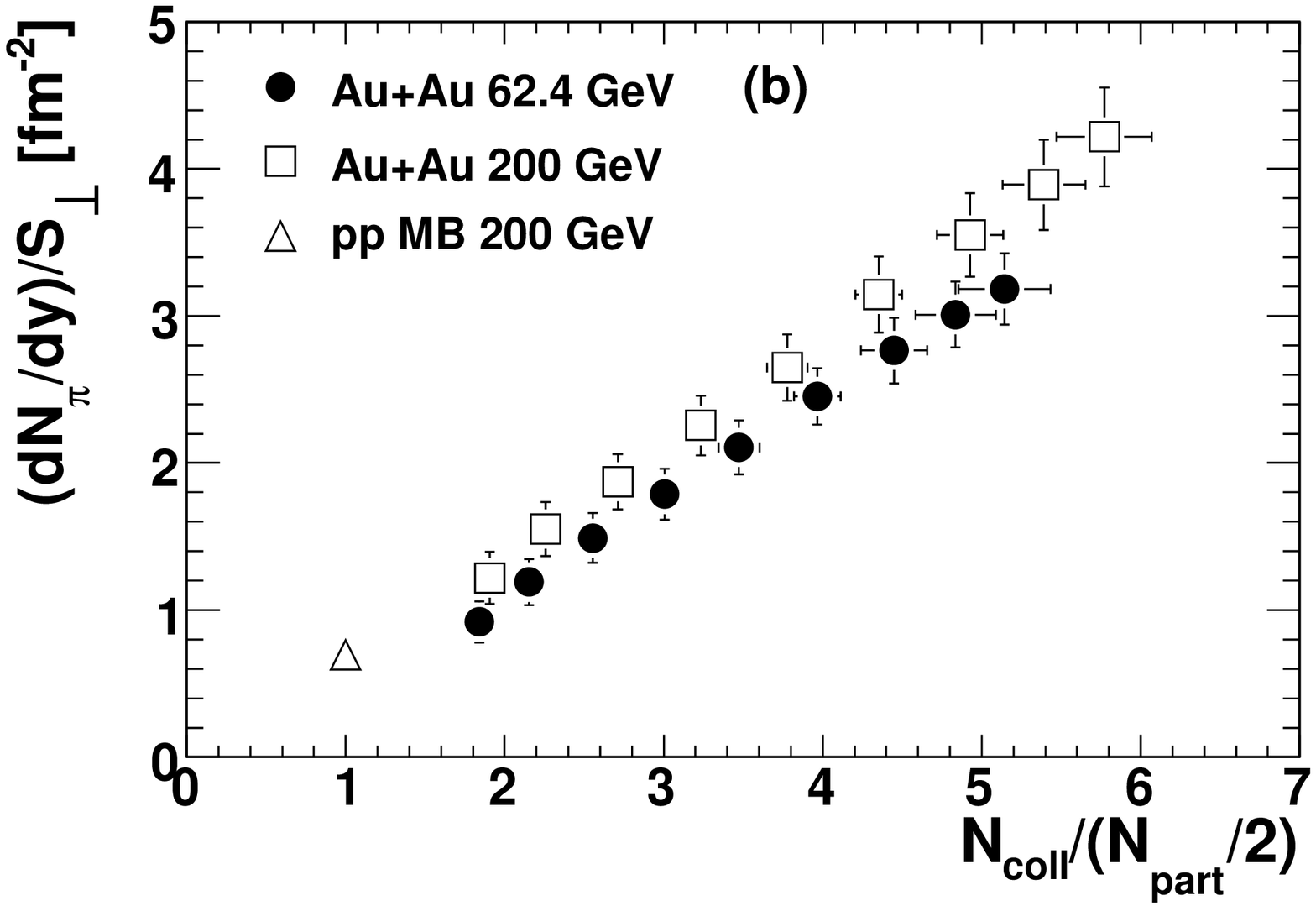}
\caption{(a) The ratio of the transverse overlap area $\Sovlp$ to $\left(\Npart/2\right)^{2/3}$ versus $\left(\Npart/2\right)^{2/3}$. (b) The ratio of the charged pion multiplicity to the transverse overlap area $\dNdyS$ versus $\Ncoll/\Npart$. Errors shown are total errors, dominated by systematic uncertainties. The systematic uncertainties are correlated between $\Npart$, $\Ncoll$, and $\Sovlp$, and are largely canceled in the plotted ratio quantities.}
\label{fig:S_Npart}
\end{figure}

Figure~\ref{fig:S_Npart}(a) shows the ratio of $\Sovlp$ to $\left(\Npart/2\right)^{2/3}$ as a function of $\left(\Npart/2\right)^{2/3}$. The overlap area $\Sovlp$ scales with $\left(\Npart/2\right)^{2/3}$ to a good approximation, and the scaling factor is the proton-proton cross-section used in the Glauber calculation, $\sigmapp=36$~mb and 41~mb for 62.4~GeV and 200~GeV, respectively. Figure~\ref{fig:S_Npart}(b) shows the ratio of the charged pion multiplicity to the transverse overlap area $\dNdyS$ as a function of $\frac{\Ncoll}{\Npart/2}$, the average number of binary collisions per participant nucleon pair. As seen from the figure, the two quantities have monotonical correspondence and have little dependence on beam energy (i.e. on the value of $\sigmapp$).

\section{Particle Identification by $\dedx$\label{sec:dEdx}}

Charged particles, while traversing the TPC gas volume, interact with the gas atoms and lose energy by ionizing electrons out of the gas atoms. This specific ionization energy loss, called the $\dedx$, is a function of the particle momentum magnitude. This property is used for particle identification. This paper focuses on particle identification in the low $\pt$ region. This section describes the low $\pt$ $\dedx$ particle identification method in detail. Extension of particle identification to high $\pt$ is possible by the Time of Flight (TOF) patch~\cite{TOFNIM1,TOFNIM2} and by using the relativistic rise of the specific ionization energy loss (r$\dedx$)~\cite{rdEdx}. The details of the TOF and r$\dedx$ methods are out of the scope of this paper.

The electron ionization process has large fluctuations; the measured $\dedx$ sample for a given track length follows the Landau distribution. The Landau tail results in a large fluctuation in the average $\dedx$. To reduce fluctuation, a truncated mean, $\mdedx$, is used to characterize the ionization energy loss of charged particles. In this analysis, the truncated mean $\mdedx$ is calculated from the lowest 70\% of the measured $\dedx$ values of the hits for each track. The resolution of the obtained $\mdedx$ depends on the track length and particle momentum. For a minimum ionizing pion at momentum $p = 0.5$~GeV/$c$ with long track length (45 hits), the resolution is measured to be 8-9\% in central Au+Au collisions. The resolution is better in $pp$, d+Au, and peripheral Au+Au collisions due to less cluster overlapping.

\begin{figure}[hbtp]
\centering
\includegraphics[width=0.48\textwidth]{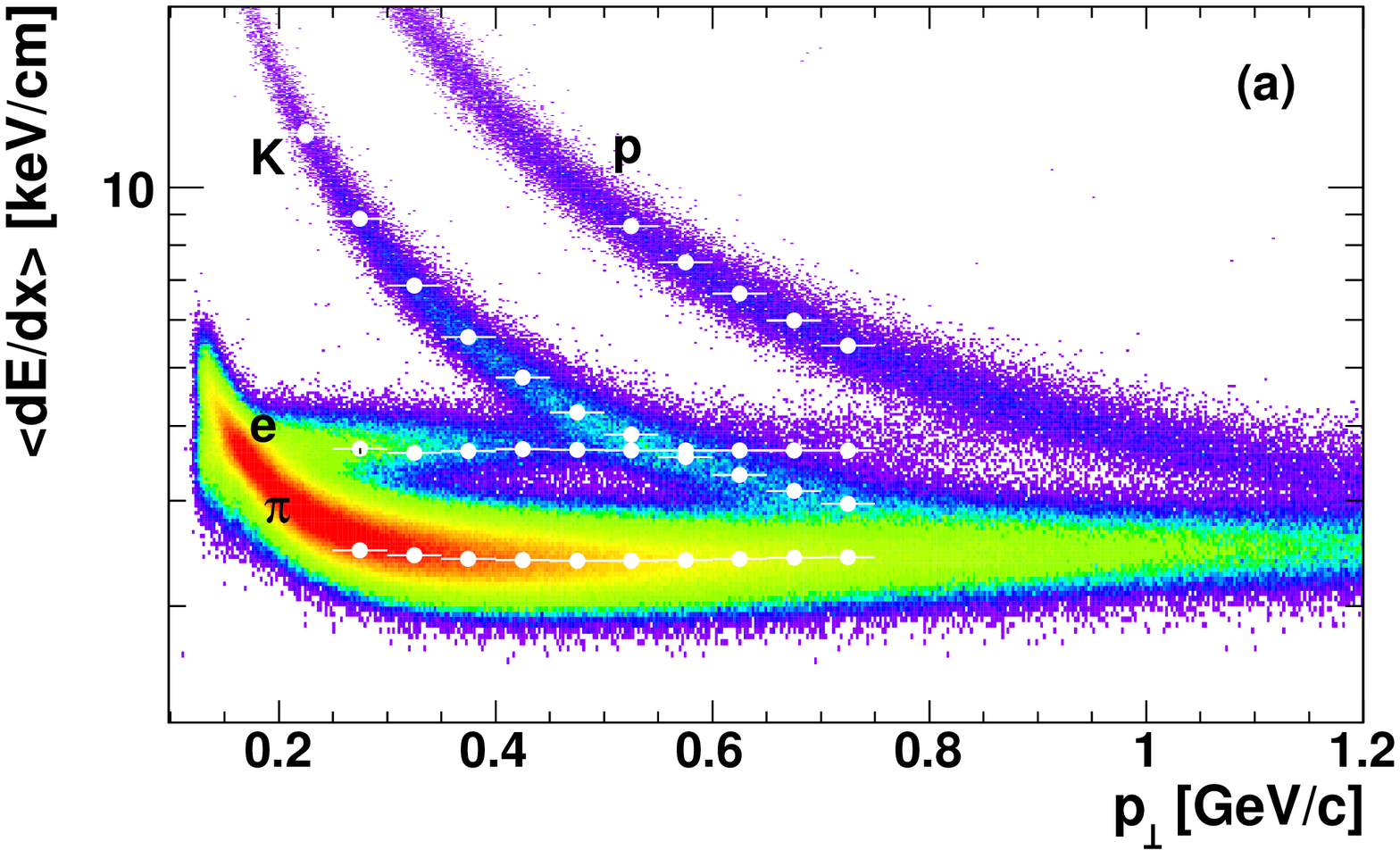}
\includegraphics[width=0.48\textwidth]{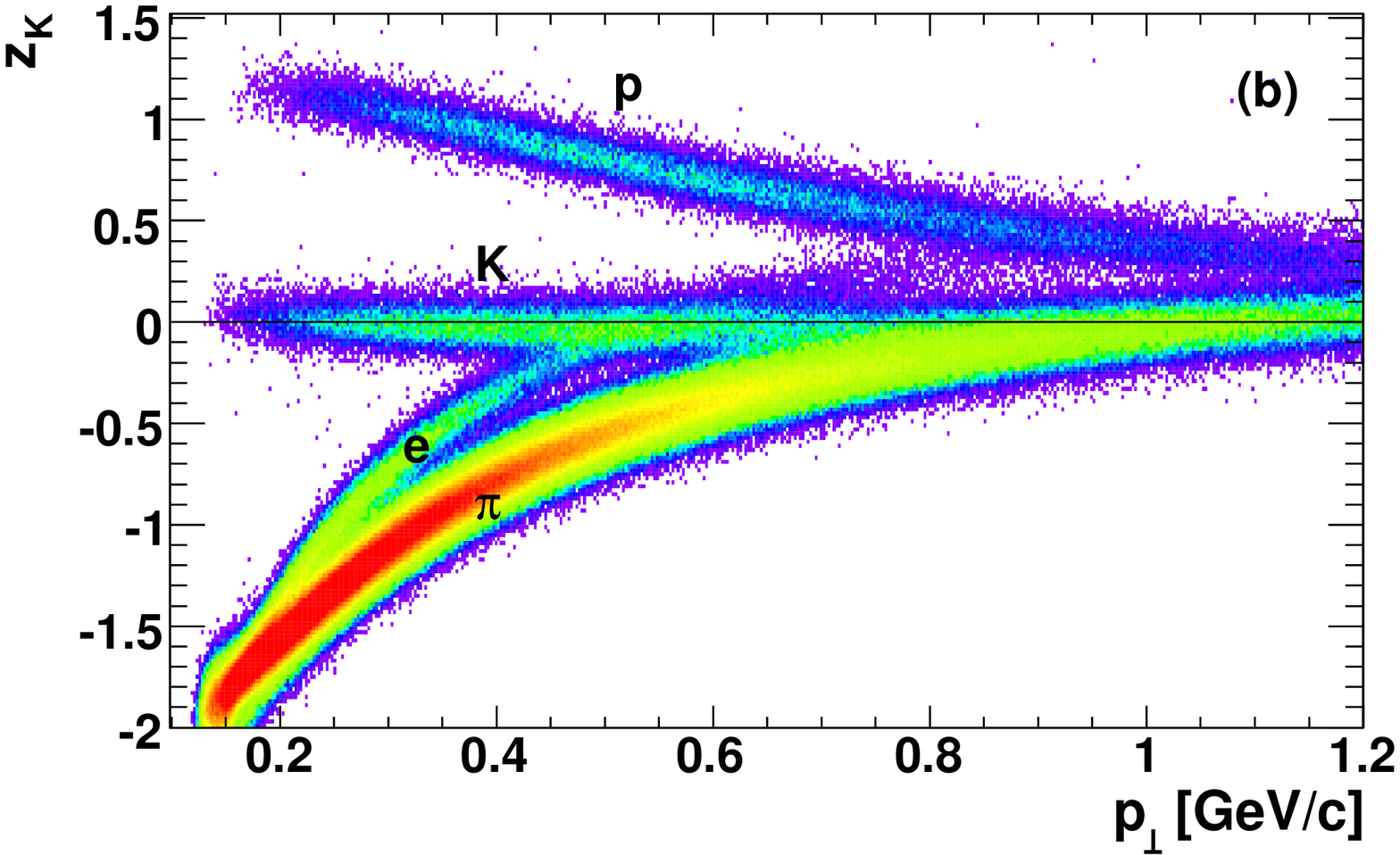}
\caption{(color online) (a) Truncated $\mdedx$ of specific ionization energy loss of $\pi^{-}, e^{-}, K^{-}, \pbar$ as a function of $\pt$ for particles in $|\eta|<0.1$ measured in 200~GeV minimum bias $pp$ collisions by the STAR-TPC. The Gaussian centroids for $\pi^{-}$, $e^{-}$, $K^{-}$ and $\overline{p}$ fit to the kaon $z_K$ distributions are shown with circles. (b) The $z_K$ variable for kaon versus $\pt$ in 200~GeV minimum bias $pp$ collisions. Particles are restricted in $|y_K|<0.1$ where the kaon mass is used in the rapidity calculation. In this narrow rapidity (or pseudo-rapidity) slice, $\pt$ is approximately equal to $p_{\rm mag}$.}
\label{fig:dedxPlotAllBands}
\end{figure}

The ionization energy loss by charged particles in material is given by the Bethe-Bloch formula~\cite{PDG} and for thin material by the more precise Bichsel formula~\cite{Bichsel}. At low momentum, ionization energy loss is approximately inversely proportional to the particle velocity squared. With the measured particle momentum and $\mdedx$, the particle type can be determined by comparing the measurements against the Bethe-Bloch expectation. Figure~\ref{fig:dedxPlotAllBands}(a) shows the measured $\mdedx$ versus momentum magnitude for particles in $|\eta|<0.1$. Various bands, corresponding to different mass particles, are clearly separated at low $\pt$. At modest $\pt$, the bands start to overlap: $e^{\pm}$ and $K^{\pm}$ merge at $\sim$ 0.5~GeV/$c$, $K^{\pm}$ and $\pi^{\pm}$ merge at $\sim$ 0.75~GeV/$c$, and $p$ ($\pbar$) and $\pi^{\pm}$ merge at $\sim$ 1.2~GeV/$c$. However, particles can still be {\em statistically} identified by a fitting procedure to deconvolute the overlapped distribution into several components. The separation of the $\dedx$ bands depends on the pseudorapidity region and decreases toward higher rapidities. To obtain maximal separation we only concentrate on the mid-rapidity region of $\left|y\right|<$ 0.1. 

Since the $\mdedx$ distribution for a fixed particle type is not Gaussian~\cite{z}, a new variable is useful in order to have a proper deconvolution into Gaussians. It is shown~\cite{z} that a better Gaussian variable, for a given particle type, is the $z$-variable, defined as 
\begin{equation}
z_i = \ln\left(\frac{\mdedx}{\mdedx_i^{BB}}\right),\label{eq:z}
\end{equation}
where $\mdedx_i^{BB}$ is the Bethe-Bloch (Bichsel~\cite{Bichsel}) expectation of $\mdedx$ for the given particle type $i$ ($i=\pi,K,p$). In this analysis, $\mdedx_i^{BB}$ is parameterized as
\begin{equation}
\mdedx_i^{BB}=A_i\left(1\ +\ \frac{m_i^{2}}{p_{\rm mag}^{2}}\right),\label{eq:dedx_BB_fit}
\end{equation}
where $m_i$ is the particle's rest mass and $p_{\rm mag}$ is the particle momentum magnitude. This parameterization is found to describe the data well, with the normalization factor $A_i$ determined from data. The expected value of $z_i$ for the particle in study is around 0. The $z_K$ variable is shown for $K^-$ in Fig.~\ref{fig:dedxPlotAllBands}(b), where the kaon band is situated around 0. 

\begin{figure*}[hbtp]
\centering
\includegraphics[width=0.48\textwidth]{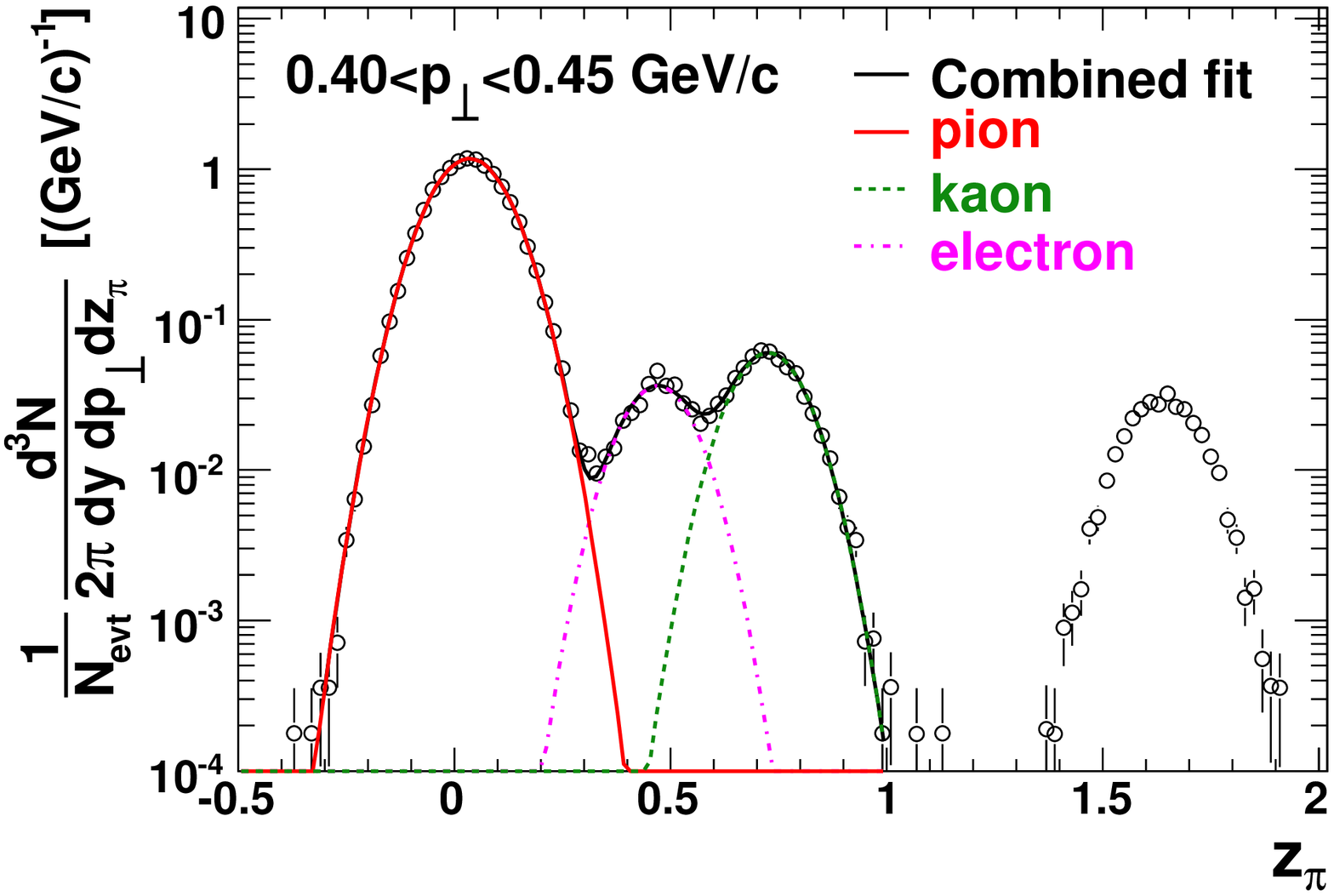}
\includegraphics[width=0.48\textwidth]{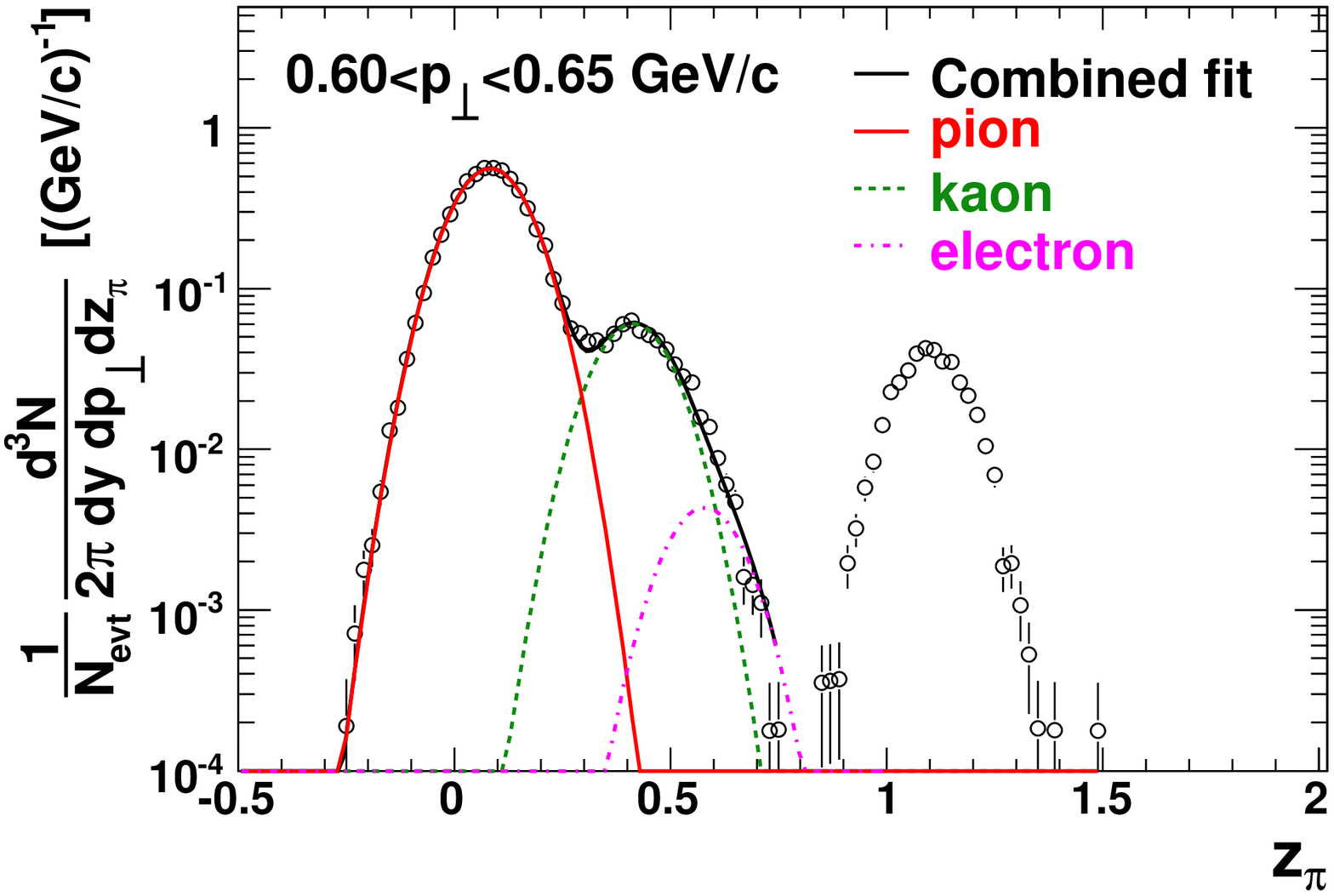}
\includegraphics[width=0.48\textwidth]{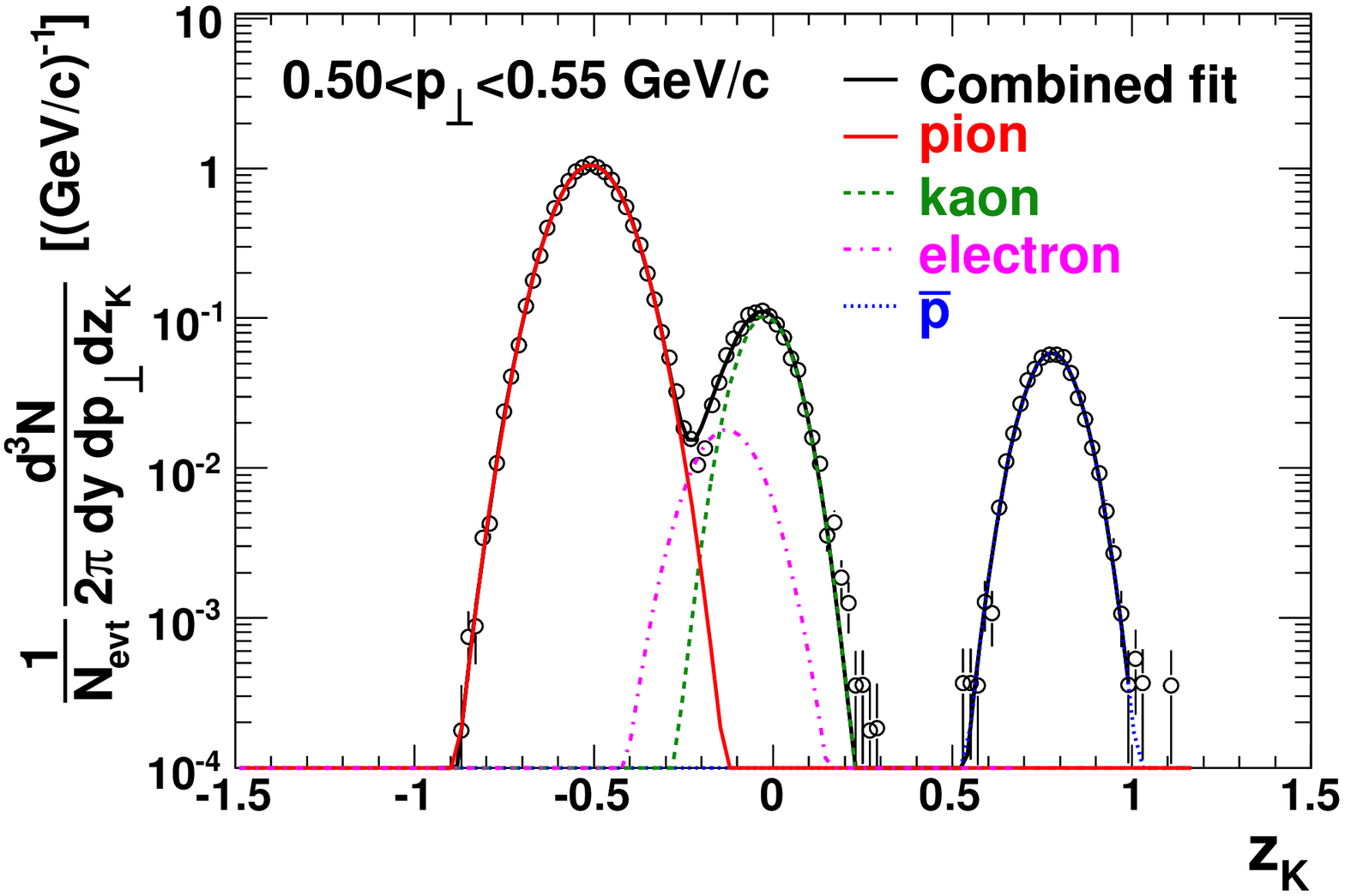}
\includegraphics[width=0.48\textwidth]{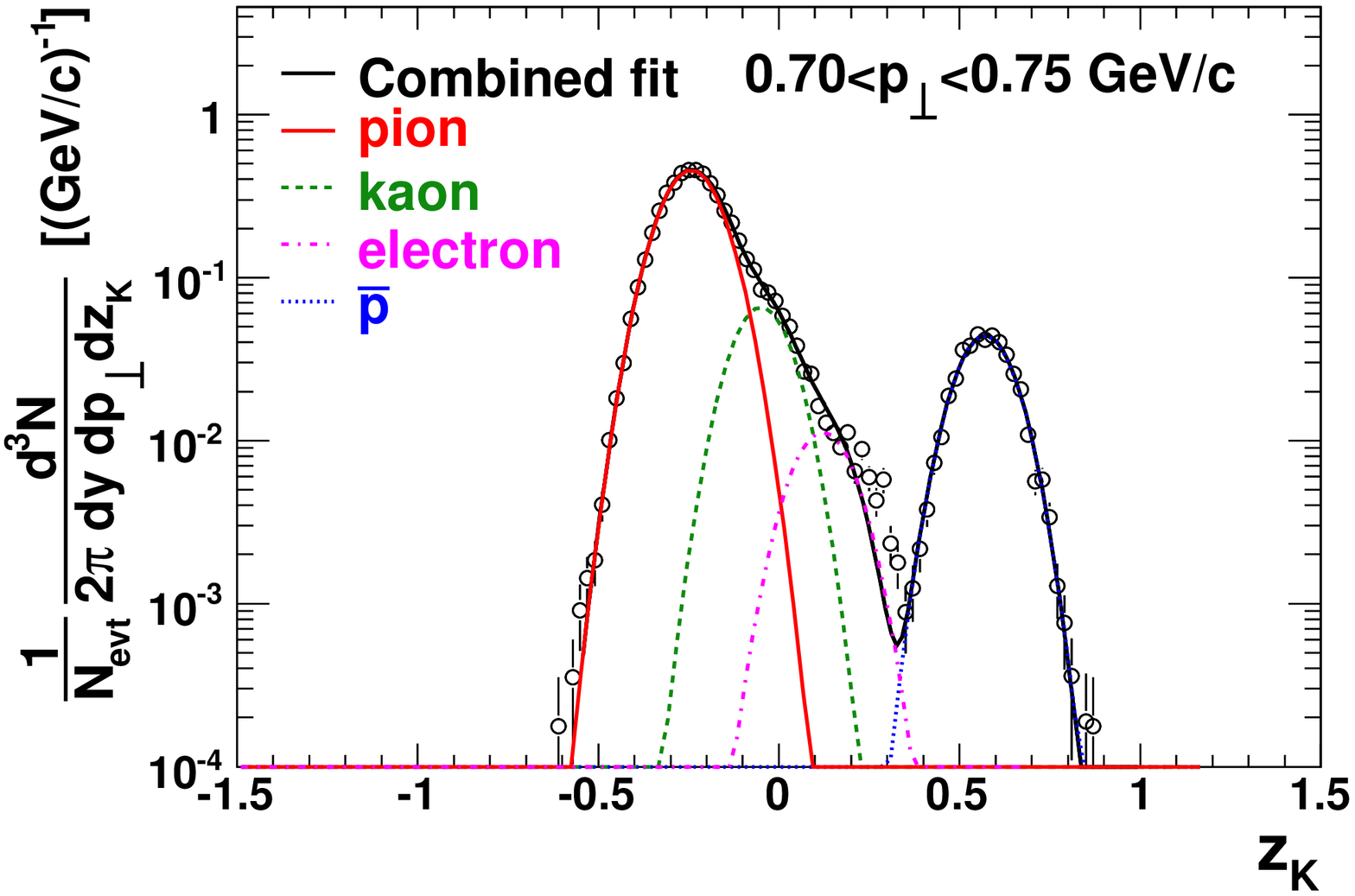}
\caption{(color online) Distributions of $z_{\pi}$ for $\pi^{-}$ (upper panels) and $z_K$ for $K^{-}$ (lower panels) in 200~GeV minimum bias $pp$ collisions. Four $\pt$ bins are shown. Errors shown are statistical only. The curves represent the Gaussian fits to the $z_{\pi}$ and $z_K$ distributions, with individual particle peaks plotted separately.}
\label{fig:kaondedx}
\end{figure*}

The $z_i$ distribution is constructed for a given particle type in a given $\pt$ bin within $|y| < 0.1$. Figure~\ref{fig:kaondedx} shows the $z_\pi$ and $z_K$ distributions, each for two $\pt$ bins. The distributions show a multi-Gaussian structure. To extract the raw particle yield for a given particle type, a multi-Gaussian fit is applied to the $z_i$ distribution as superimposed in Fig.~\ref{fig:kaondedx}. The parameters of the multi-Gaussian fit are the centroids, widths, and amplitudes for $\pi^\pm$, $e^\pm$, $K^\pm$, $\overline{p}$ and $p$. The positive and negative particle $z_i$-distributions are fit simultaneously. The particle and antiparticle centroids and widths are kept the same. The centroid of the particle type in study is not fixed at zero, but treated as a free parameter because the parameterization by Eq.~(\ref{eq:dedx_BB_fit}) is only approximate. For the large $\pt$ bins where the $\mdedx$ bands merge, the Gaussian widths for all three particle species are kept the same. The fit centroids of $\pi^-$, $e^{-}$, $K^-$, and $\overline{p}$, where $K^-$ is the particle type in study, are superimposed on Fig.~\ref{fig:dedxPlotAllBands}(a) as a function of $\pt$. The kaon $z_K$ fit centroid is very close to zero, affirming the good description of $\mdedx^{BB}_K$ by Eq.~(\ref{eq:dedx_BB_fit}) at low $\pt$. 

The particle yield extracted from the fit to the corresponding $z$ distribution is the raw yield. The fit yields for the other particle peaks cannot be used, because the rapidity calculation is incorrect for those particle types. Thus, the same procedure is repeated for each particle type separately. 

As shown in Fig.~\ref{fig:dedxPlotAllBands}, particle identification as a function of momentum magnitude is limited due to the merging of the $\dedx$ bands at large $\pt$. Pions can be identified in the momentum range of 0.2-0.7~GeV/$c$, kaons 0.2-0.7~GeV/$c$ and (anti)protons 0.35-1.2~GeV/$c$. Kaon identification is particularly difficult because electrons are merged into the kaon band above $\pt>0.5$~GeV/$c$. In order to extract the kaon yield at relatively large $\pt$, electron contributions are interpolated to the $\dedx$ overlapping $\pt$ range and are then fixed. The uncertainties in the estimation of electron contaminations are the main source of systematic uncertainties on the extracted kaon yields at large $\pt$, as discussed in Section~\ref{sec:systuncer}.

\section{Corrections and Backgrounds\label{sec:Corr}}

\subsection{Monte-Carlo Embedding Technique}

The correction factors are obtained by the multi-step embedding MC technique. First, simulated tracks are blended into real events at the raw data level. Real data events to be used in the embedding are sampled over the entire data-taking period in order to have proper representation of the whole data set used in the analysis. MC tracks are simulated with primary vertex position taken from the real events. The MC track kinematics are taken from flat distributions in $\eta$ and $\pt$. The flat $\pt$ distribution is used in order to have similar statistics in different $\pt$ bins. The number of embedded MC tracks is of the order of 5\% of the measured multiplicity in real events. The tracks are propagated through the full simulation of the STAR detector and geometry using GEANT with a realistic simulation of the STAR-TPC response. The simulation starts with the initial ionization of the TPC gas by charged particles, followed by electron transport and multiplication in the drift field, and finally the induced signal on the TPC read-out pads and the response of the read-out electronics. All physical processes (hadronic interaction, decay, multiple scattering, etc.) are turned on in the GEANT simulation. The obtained raw data pixel information for the simulated particles are added on to the existing information of the real data. Detector effects such as the saturation of ADC channels are taken into account. The format of the resulting combined events is identical to that of the real raw data events.

Second, the mixed events are treated just as real data and are processed through the full reconstruction chain. Clusters and hits are formed from the pixel information; tracks are reconstructed from the hits.

Third, an association map is created between the input MC tracks and the reconstructed tracks of the mixed event. The association is made by matching hits by proximity~\footnote{Another possible matching algorithm is the identity truth method, where the track identity information is propagated to the reconstructed hits.}. For each MC hit from GEANT, a search for reconstructed hits from the embedded event is performed with a window of $\pm 6$~mm in $x$, $y$, and $z$~\cite{Manuel}. The window size is chosen based on the hit resolution and the typical occupancy of the TPC in central Au+Au collisions. If a reconstructed hit is found in the search window, the MC hit is marked as matched. The MC track is considered to be reconstructed if more than 10 of its hits are matched to a single reconstructed track in the embedded event. Multiple associations are allowed, but the probability is small to have a single MC track matched with two or more reconstructed tracks or vice versa. From the multiple associations, the effects of track splitting (two reconstructed tracks matched to one MC track) and track merging (two MC tracks matched to a single reconstructed track) can be studied. The reconstruction efficiency is obtained by the ratio of the number of matched MC tracks to the number of input MC tracks. The reconstruction efficiency contains the net effect of tracking efficiency, detector acceptance, decays, and interaction losses.

\begin{figure}[htbp]
\centering
\includegraphics[width=0.52\textwidth]{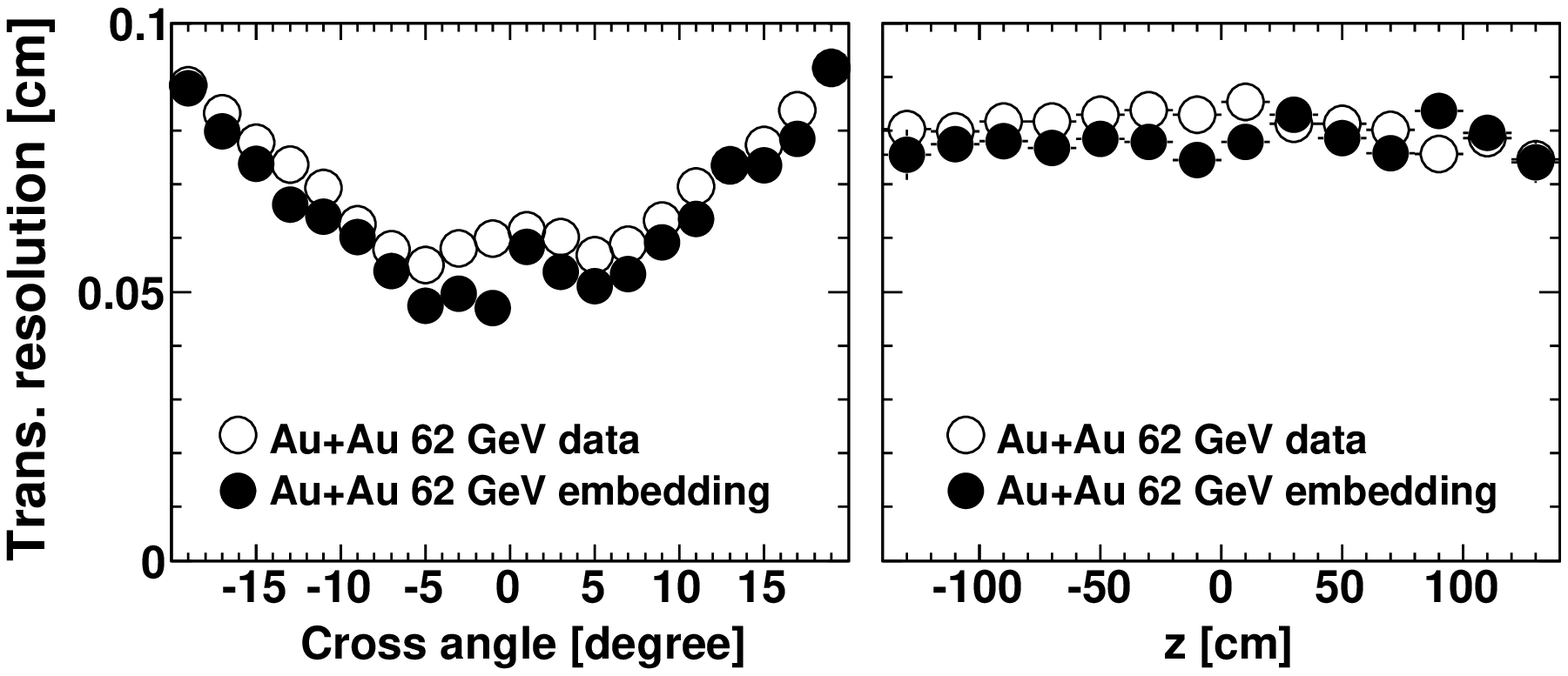}
\includegraphics[width=0.52\textwidth]{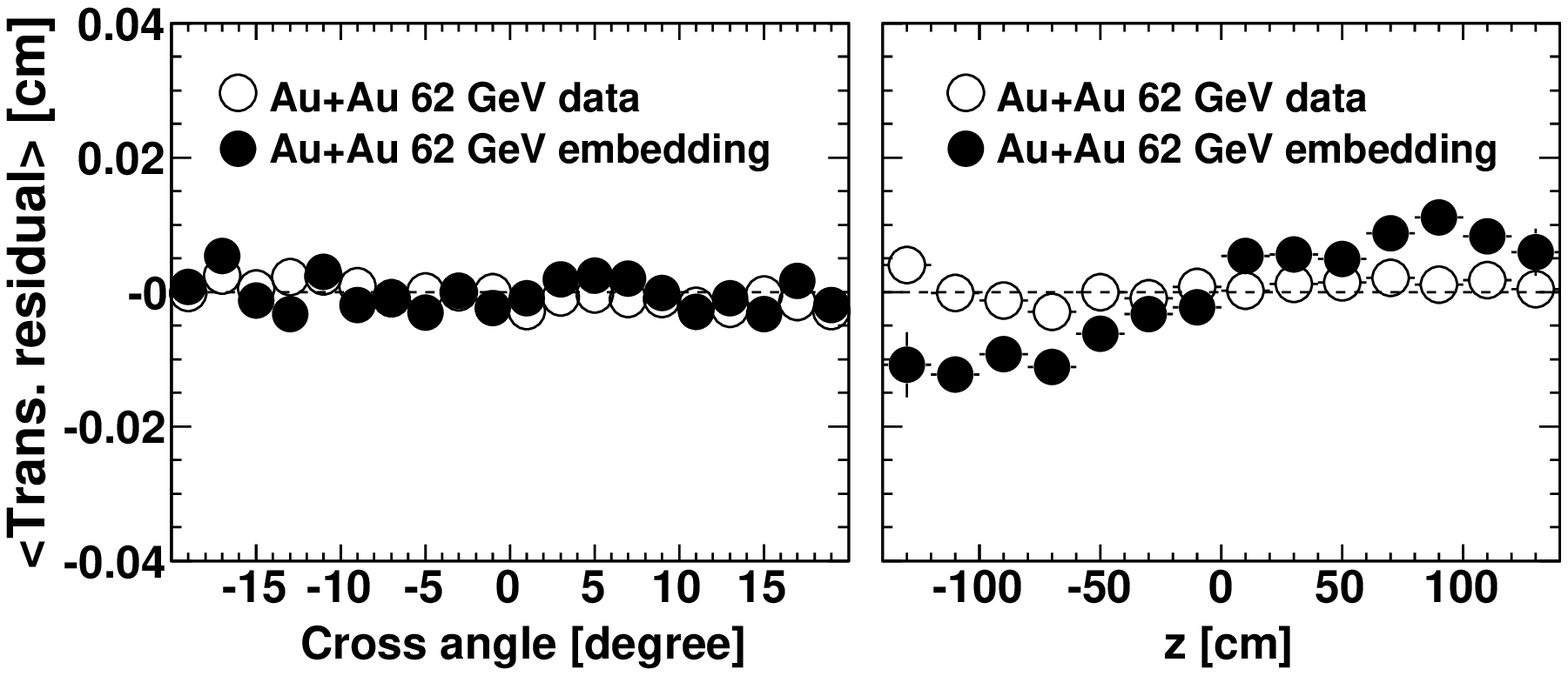}
\includegraphics[width=0.52\textwidth]{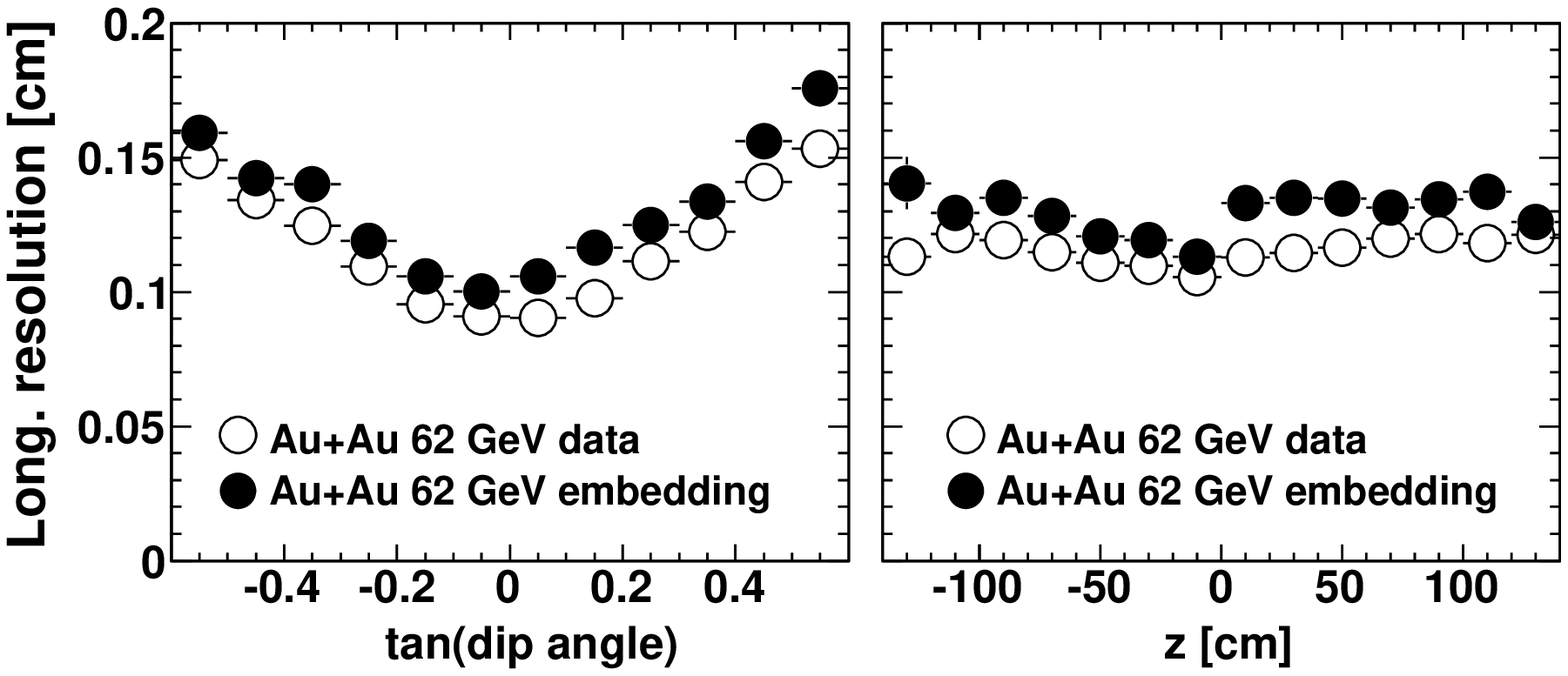}
\includegraphics[width=0.52\textwidth]{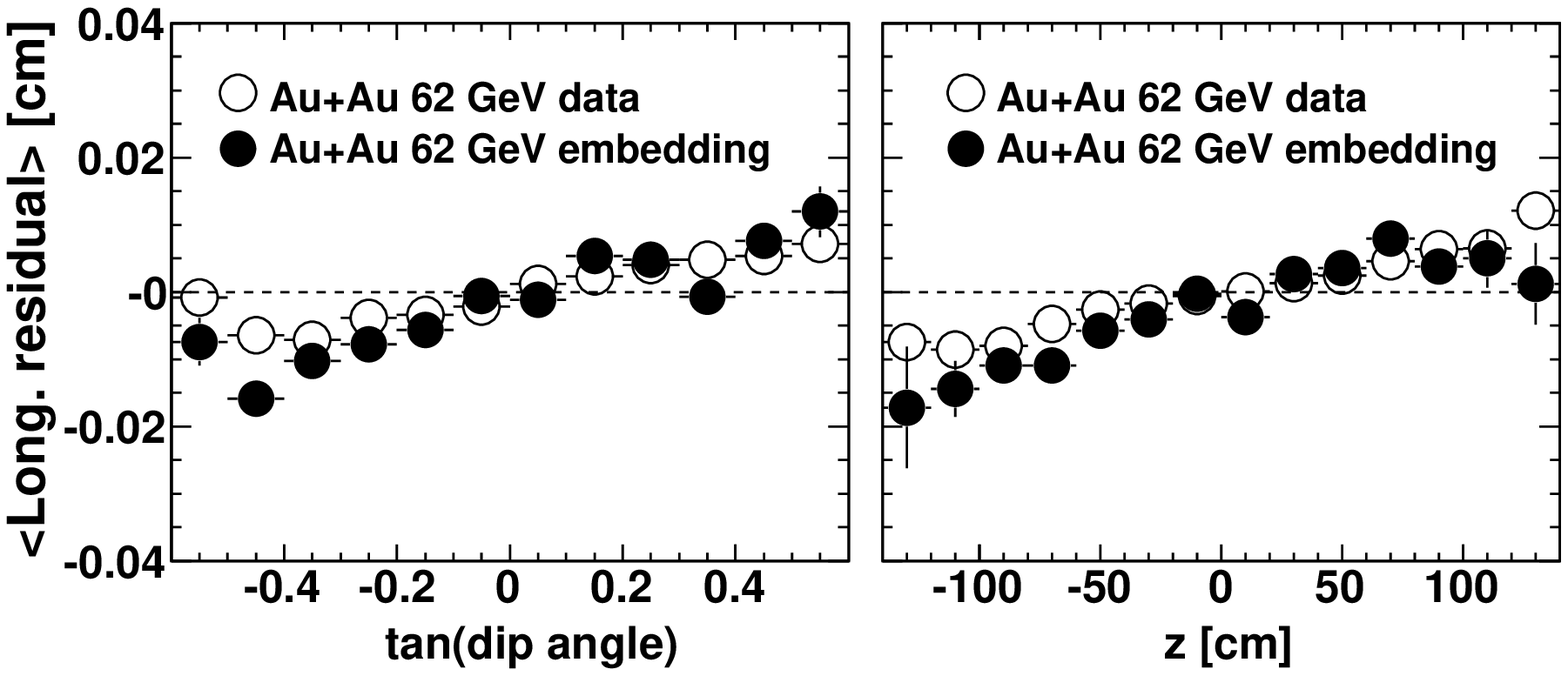}
\caption{Hit resolution and mean hit residual as a function of the track crossing angle at the hit position, the track dip angle, and the hit $z$ coordinate. The data are an enriched $K^+$ sample (via $\dedx$ cut) within $|y|<1$ and $0.4<\pt<0.5$~GeV/$c$ in 62.4~GeV Au+Au collisions. Errors shown are statistical only.}
\label{fig:trackresid}
\end{figure}

\begin{figure}[hbt]
\begin{center}	
\includegraphics[width=0.238\textwidth]{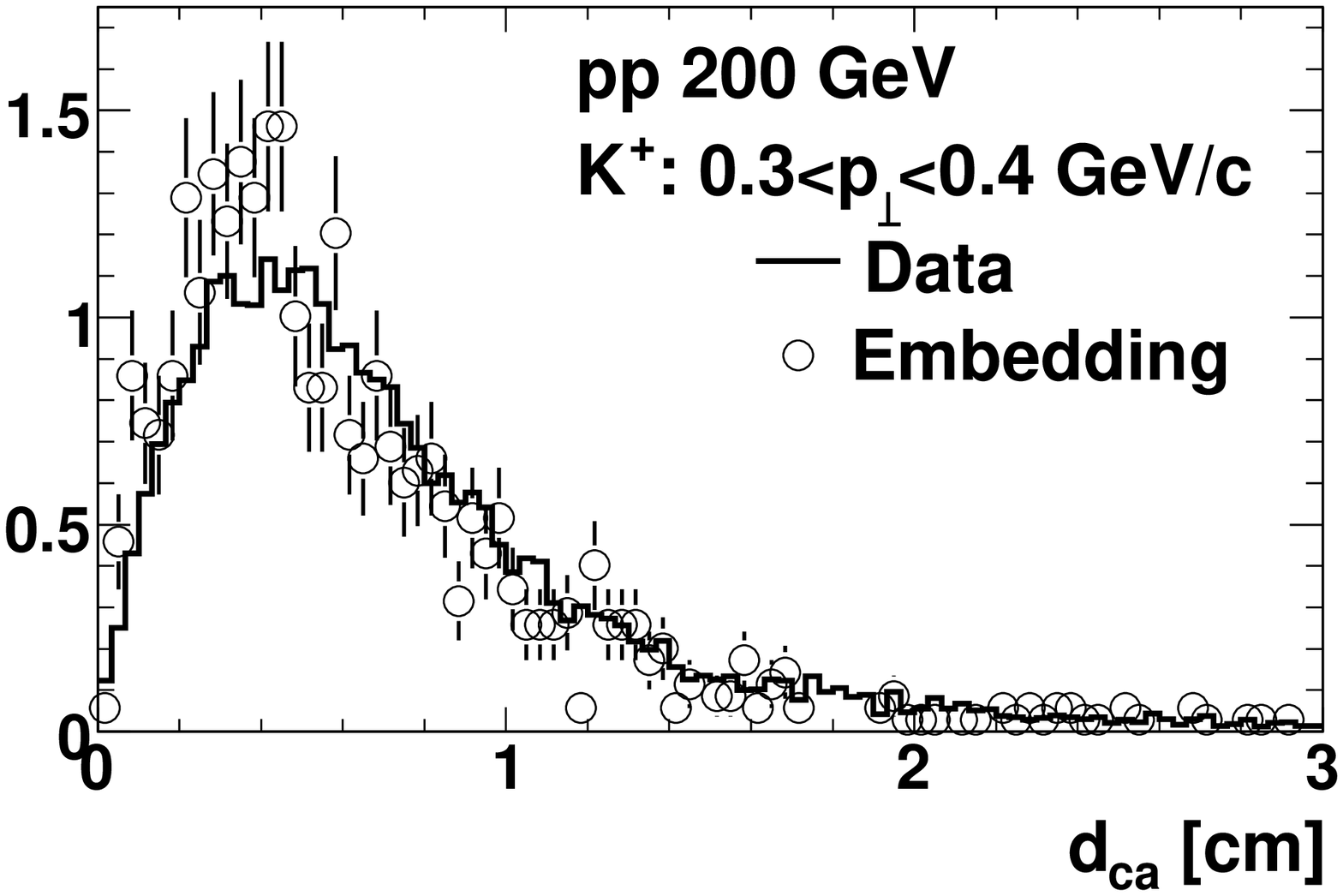}
\includegraphics[width=0.238\textwidth]{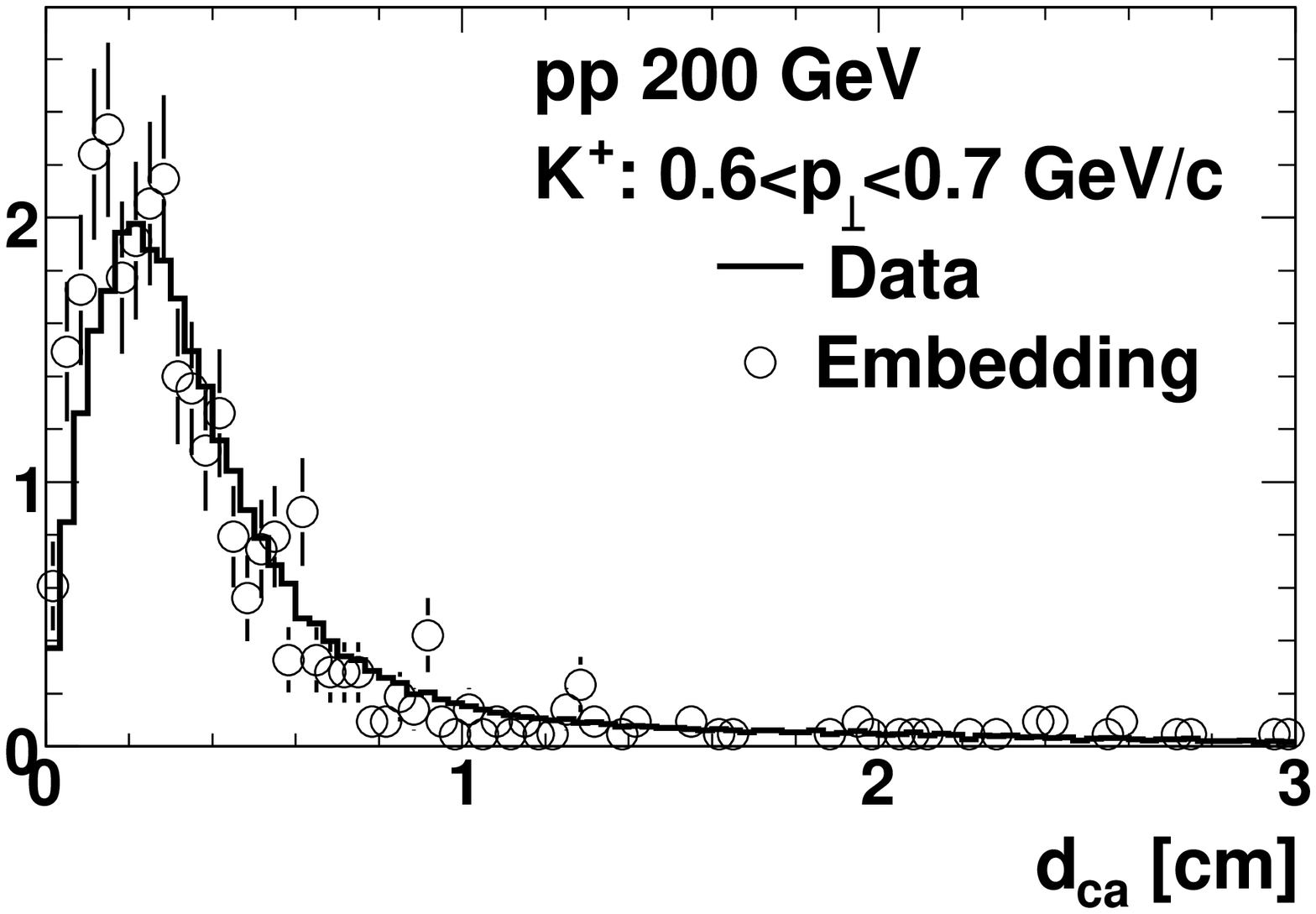}
\includegraphics[width=0.238\textwidth]{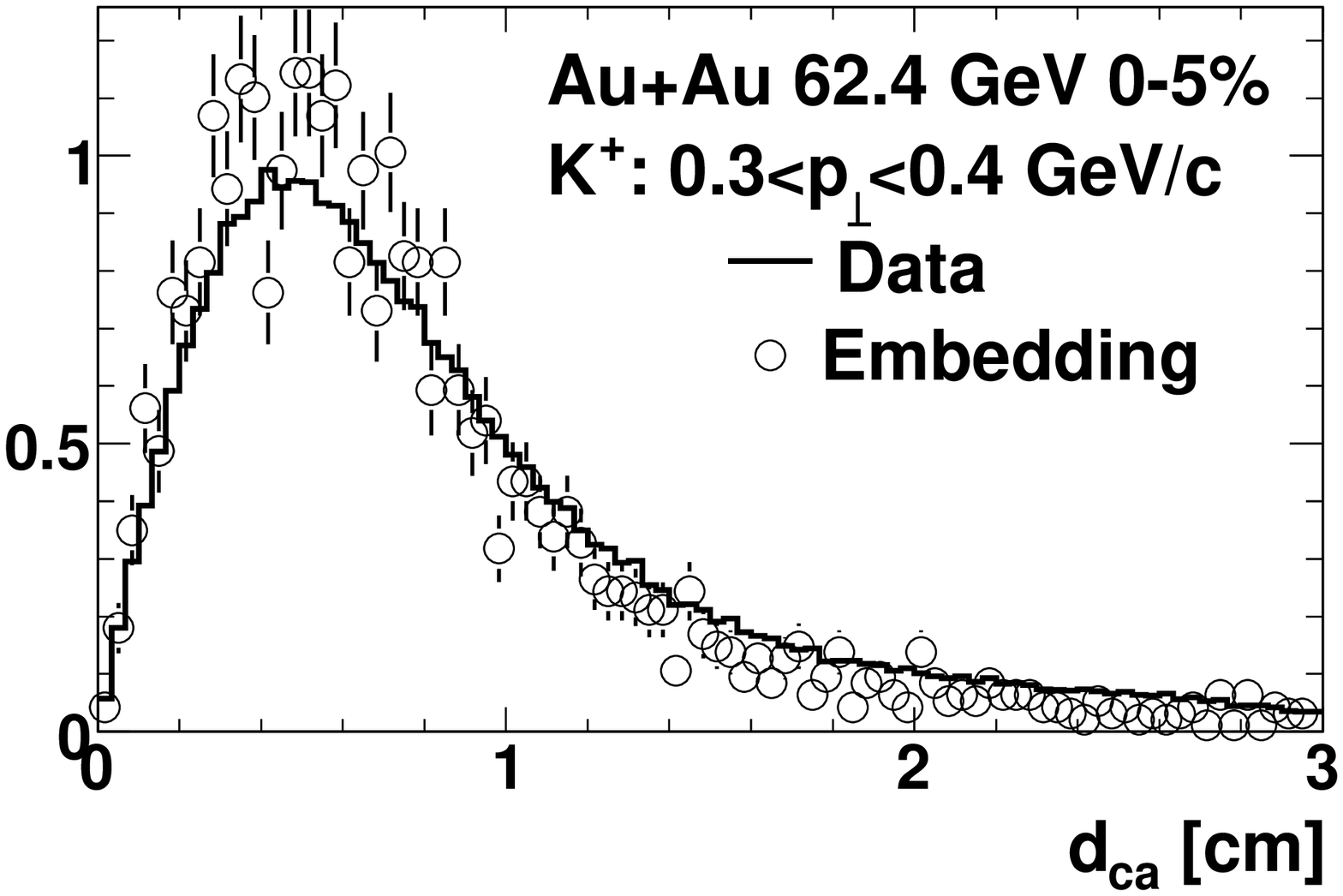}
\includegraphics[width=0.238\textwidth]{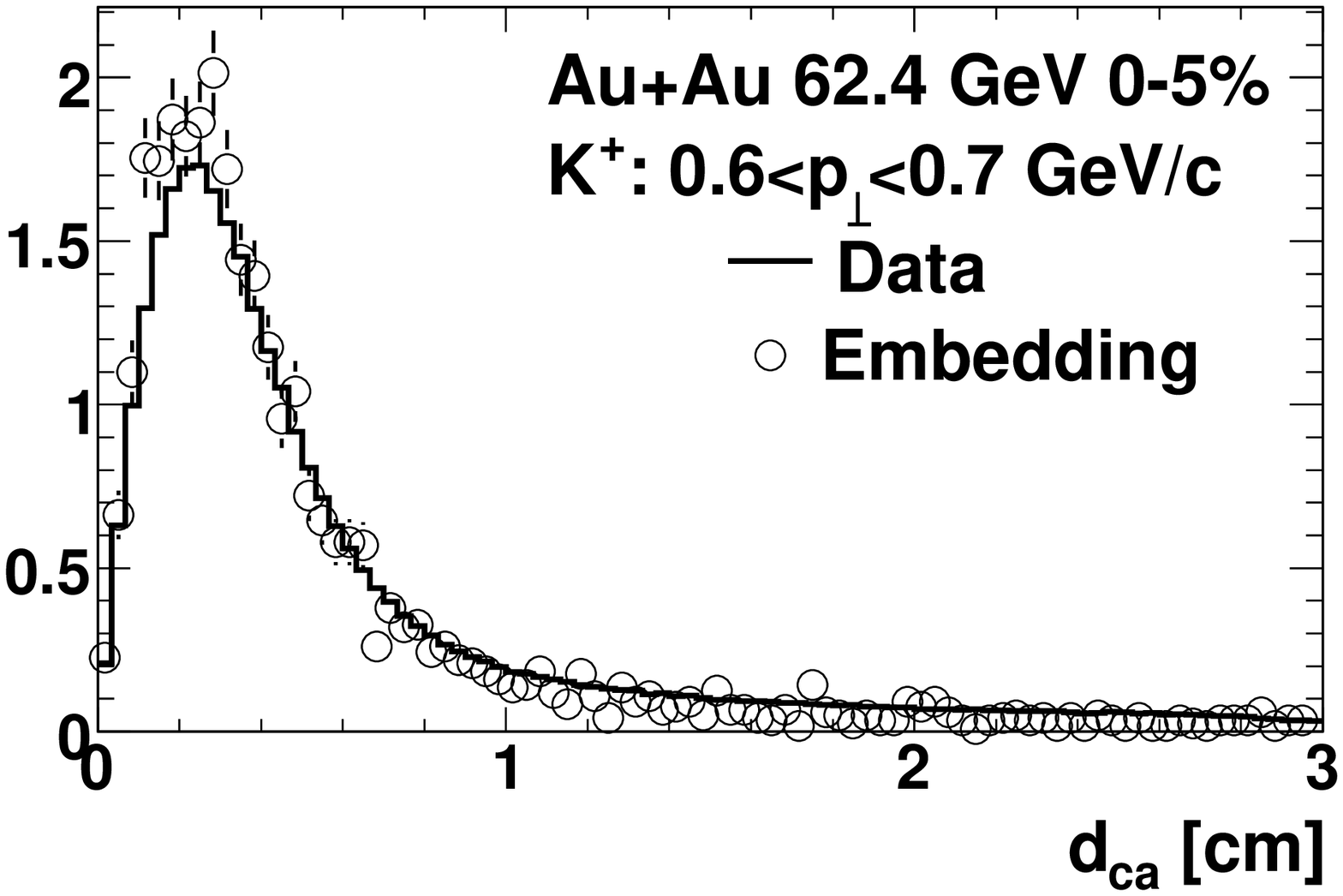}
\caption{Comparison of $\dca$ distributions between $K^+$ candidates from real data and $K^+$ from MC embedding. Two $\pt$ bins are shown for 200~GeV $pp$ collisions (upper panels) and for 62.4~GeV 5\% central Au+Au collisions (lower panels), respectively. Kaon candidates are selected from data by a $\dedx$ cut of $\pm0.5\sigma$ from the Bethe-Bloch expected values. Errors shown are statistical only. The distributions have been normalized to unit area to only compare the shapes.}
\label{fig:dcakp_embeddatacomp}
\end{center}
\end{figure}

\begin{figure}[hbt]
\begin{center}	
\includegraphics[width=0.238\textwidth]{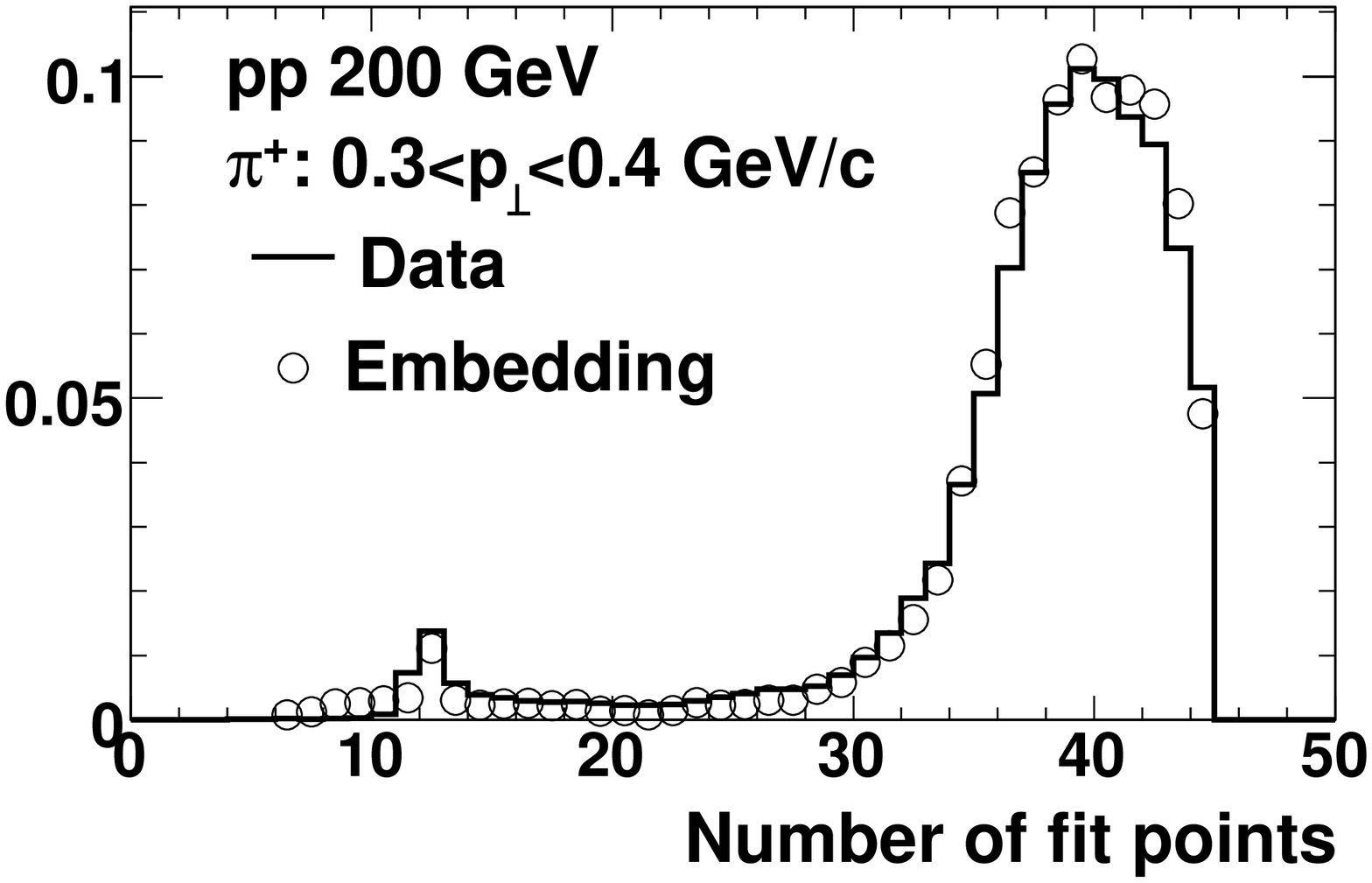}
\includegraphics[width=0.238\textwidth]{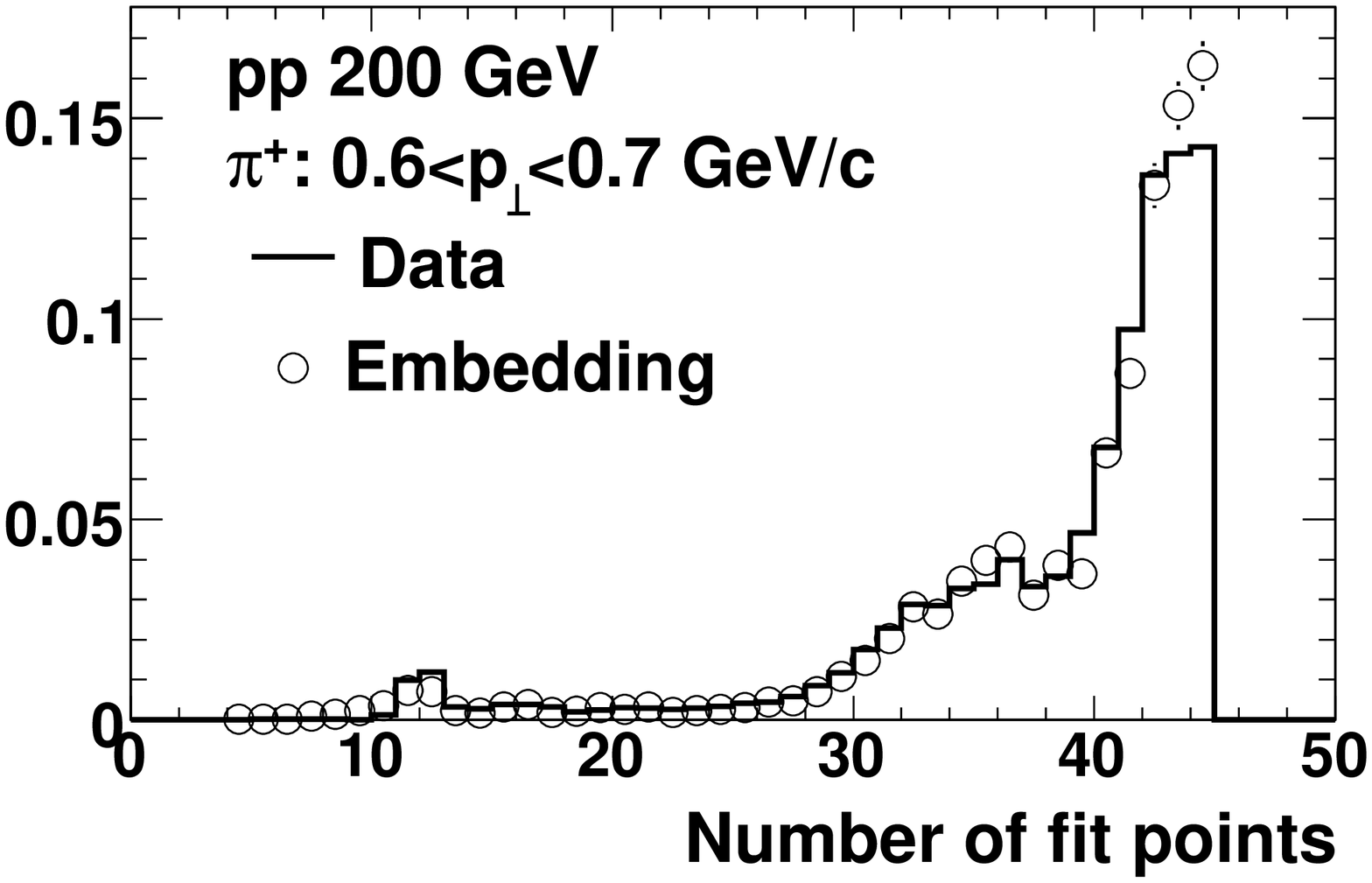}
\includegraphics[width=0.238\textwidth]{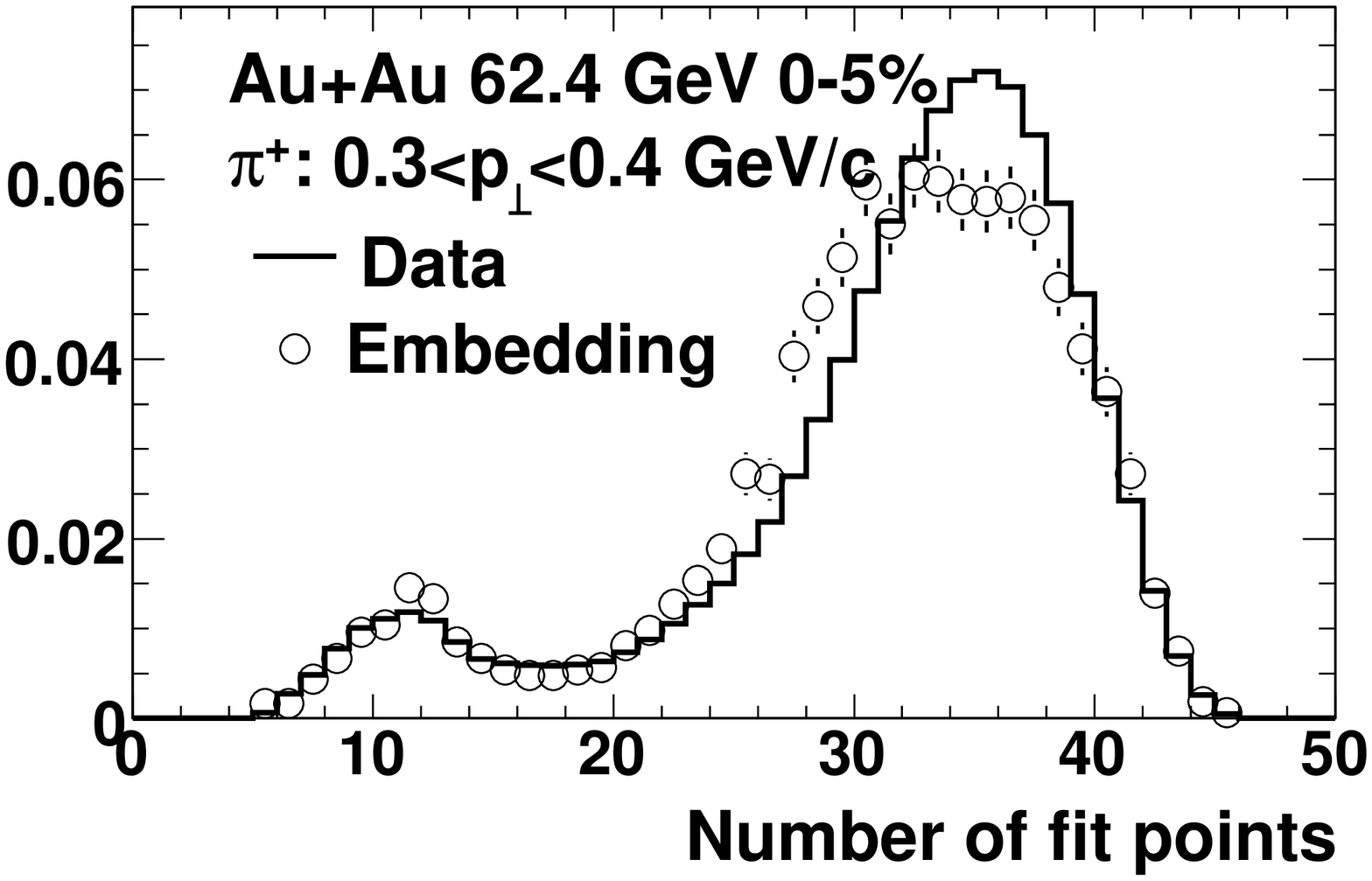}
\includegraphics[width=0.238\textwidth]{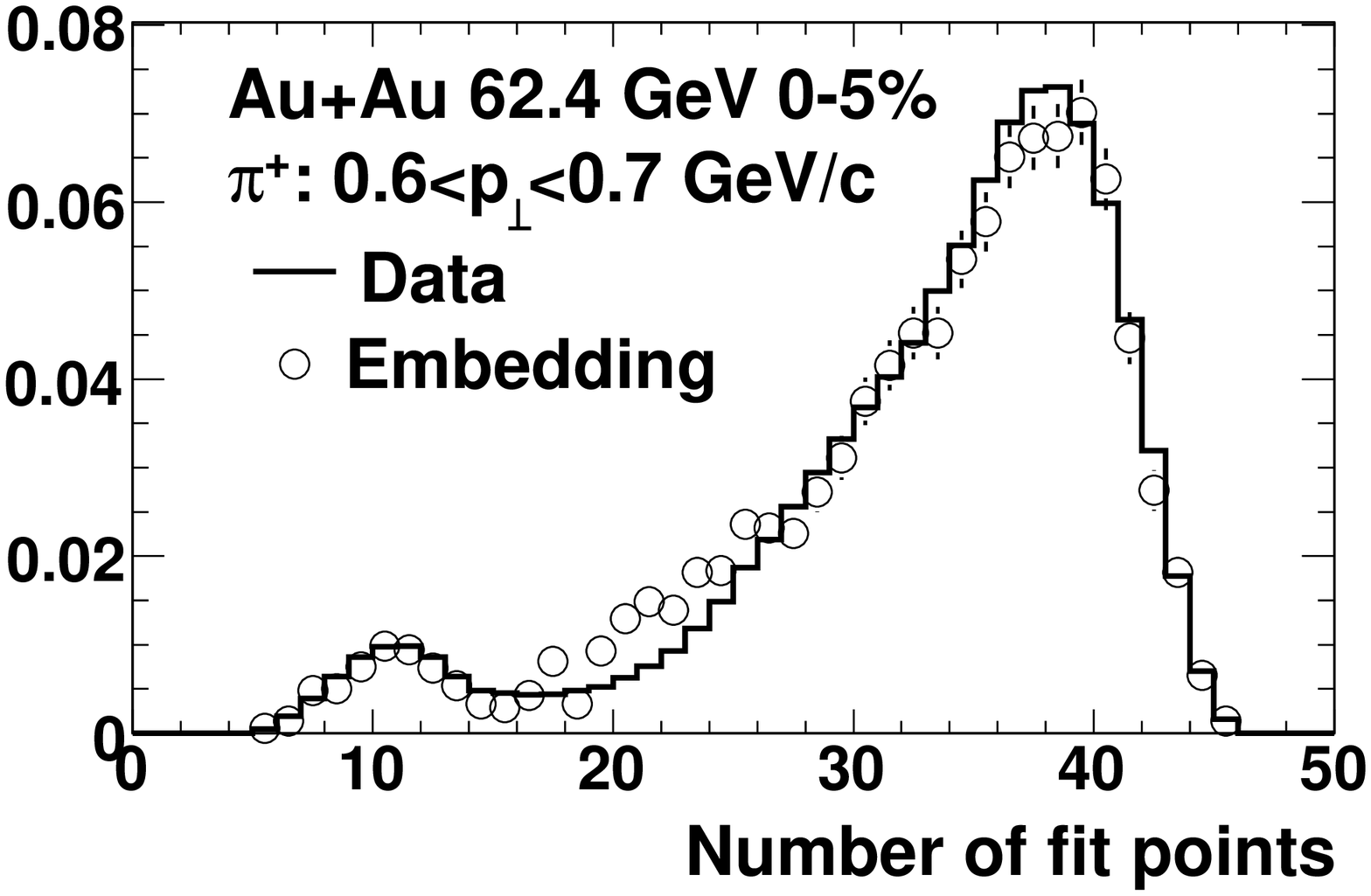}
\caption{Comparison of distributions of the number of fit points between $\pi^-$ candidates from real data and $\pi^-$ from MC embedding. Two $\pt$ bins are shown for 200~GeV $pp$ collisions (upper panels) and for 62.4~GeV 5\% central Au+Au collisions (lower panels), respectively. Pion candidates are selected by a $\dedx$ cut of $|z_{\pi}|<0.3$. Errors shown are statistical only. The distributions have been normalized to unit area to only compare the shapes.}
\label{fig:nfitpim_embeddatacomp}
\end{center}
\end{figure}

The most critical quality assurance is to make sure that the MC simulation reproduces the characteristics of the real data. This is carried out by comparing various distributions from real data and from embedding MC.
\begin{itemize}
\item Figure~\ref{fig:trackresid} shows the longitudinal and transverse hit residuals for matched MC tracks from embedding and for real data tracks. The hit residuals are compared as a function of the dip angle (the angle between the particle momentum and the $z$-direction), the crossing angle (the angle between the particle momentum and the TPC pad-row direction~\cite{TPC1,TPC2}) and the hit $z$ position. Good agreement is found as seen from Fig.~\ref{fig:trackresid}. The observed differences are small relative to the typical TPC occupancy and do not affect the obtained reconstruction efficiency.
\item Figure~\ref{fig:dcakp_embeddatacomp} shows the $\dca$ distributions of kaons reconstructed from matched MC kaon tracks and kaon candidates from real data. Kaon candidates are selected from real data by applying a tight $\dedx$ cut of $\pm 0.5\sigma$ around the kaon Bethe-Bloch curve. Kaons are used because they contain minimal weak decay contributions and other background so that their $\dca$ distributions give a good assessment of the quality of the embedding data. Good agreement is found between embedding MC and real data.
\item Figure~\ref{fig:nfitpim_embeddatacomp} shows a comparison of the number of hits distributions between reconstructed pions from MC embedding and pion candidates from real data. Pion candidates are selected by applying a $\dedx$ cut of $|z_{\pi}|<0.3$. Good agreement is found. The small differences found at large number of hits do not affect the calculated reconstruction efficiency because the cut on the number of hits is 25 which is significantly below the peak of the hit distribution.
\end{itemize}

\subsection{Energy Loss Correction}

\begin{figure*}[hbtp]
\centering
\includegraphics[width=0.95\textwidth]{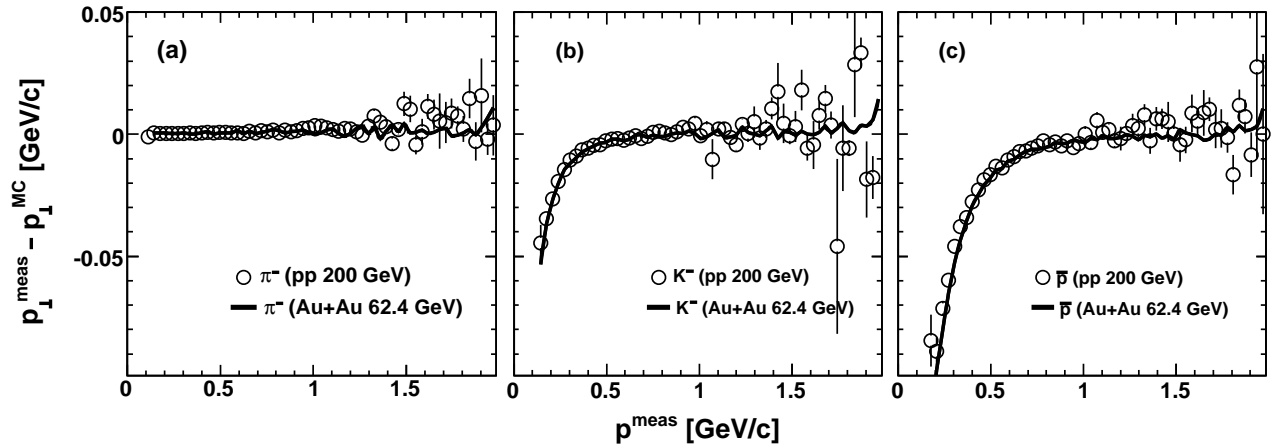}
\caption{Energy loss effect for $\pi^{\pm}$ (a), $K^{\pm}$ (b), and $p$ and $\pbar$ (c) at mid-rapidity ($|y|<0.1$) as a function of particle momentum magnitude in 200~GeV $pp$ and 62.4~GeV central 0-5\% Au+Au collisions. Only negative particles are shown; energy loss for particles and antiparticles are the same. Errors shown are statistical only. The pion energy loss is already corrected by the track reconstruction algorithm.}
\label{fig:energyloss}
\end{figure*}

Low momentum particles lose significant amounts of energy while traversing the detector material~\cite{PDG}. The track reconstruction algorithm takes into account the Coulomb scattering and the energy loss, while assuming pion mass for each particle. A correction for the energy loss in the momenta of the heavier particles ($K^{\pm}$, $p$ and $\pbar$) is needed. The correction is obtained from embedding MC. Figure~\ref{fig:energyloss} shows the difference between the measured transverse momentum and the MC input transverse momentum, $\pt^{\rm meas}-\pt^{\rm MC}$, versus the measured momentum magnitude, $p^{\rm meas}$, for particles within $|y|<0.1$. The profile can be parameterized to provide the correction function for the measured momentum:
\begin{equation} 
\pt-\pt^{\rm meas}=\delta_0 + \delta\left(1+\frac{m^2}{(p^{\rm meas})^2}\right)^{\alpha} .
\end{equation}
Here $m$ is the mass of the particle and $\delta_0$, $\delta$, and $\alpha$ are the fit parameters. The fit values are 
$\delta_0 = 0.006 (0.013)$~GeV/$c$, $\delta = -0.0038 (-0.0081)$~GeV/$c$, and $\alpha = 1.10 (1.03)$ for $K^{\pm}$ ($p$ and $\pbar$), respectively.

The energy loss correction shows little centrality dependence as expected. It only depends on the detector geometry of a given run. Although the SSD (Silicon Strip Detector) detector~\cite{SSDNIM} was installed in STAR after the 200~GeV $pp$ and Au+Au runs, there is no observable change in the magnitude of the correction for the subsequent 200~GeV d+Au and 62.4~GeV Au+Au runs. We have also investigated energy loss in different rapidity windows to assess possible systematic effects. No evidence for rapidity dependence of the energy loss is found; the energy loss correction is observed to be the same for symmetric rapidity cuts within $|y| < 0.5$.

The energy loss correction is applied off-line to all tracks using the correction formula for the given particle type of interest (i.e. the particle type being analyzed with the $z_i$ distribution). For all the results presented in this paper, the corrected $\pt$ is used.

\subsection{Vertex Inefficiency and Fake Vertex}

Several labels are used in this section to refer to tracks used for different purposes: global tracks, {\em good} global tracks, primary tracks, {\em good} primary tracks, and primary tracks used in the particle spectra analysis. They are listed in Table~\ref{tab:tracks} with the corresponding definitions and cuts.

In high multiplicity Au+Au collisions, the primary vertex can be determined accurately. In $pp$ and d+Au collisions where charged particle multiplicity is low, the vertex-finding algorithm (an already improved and better-tuned version~\cite{BBC} than the one used for Au+Au data) occasionally fails to find a primary vertex. In addition, at high luminosity, the vertex finder can fail due to the confusion from pile-up events, and in some cases, provides a wrong reconstructed vertex.

\begin{table}
\caption{Various track definitions and the corresponding cuts.}
\label{tab:tracks}
\begin{ruledtabular}
\begin{tabular}{ccc}
Track definition	& $\dca$ cut	& Min. number of hits \\ \hline
global			& --		& 10	\\
good global		& --		& 15	\\
primary			& $\dca < 3$ cm	& 10	\\
good primary		& $\dca < 3$ cm	& 15	\\
used in analysis	& $\dca < 3$ cm	& 25	\\ 
\end{tabular}
\end{ruledtabular}
\end{table}

\begin{figure*}[htbp]
\centering
\includegraphics[width=0.48\textwidth]{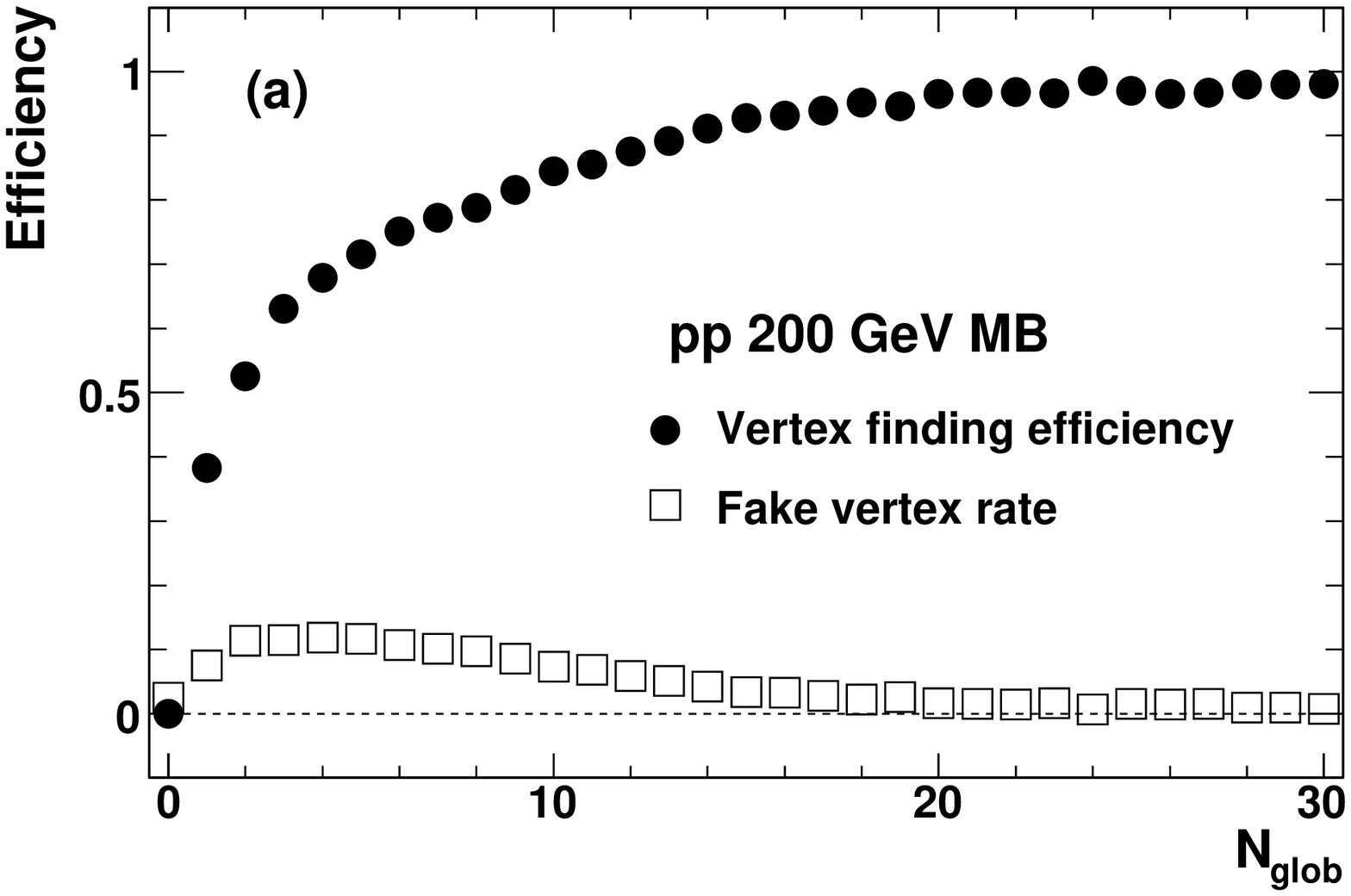}
\includegraphics[width=0.48\textwidth]{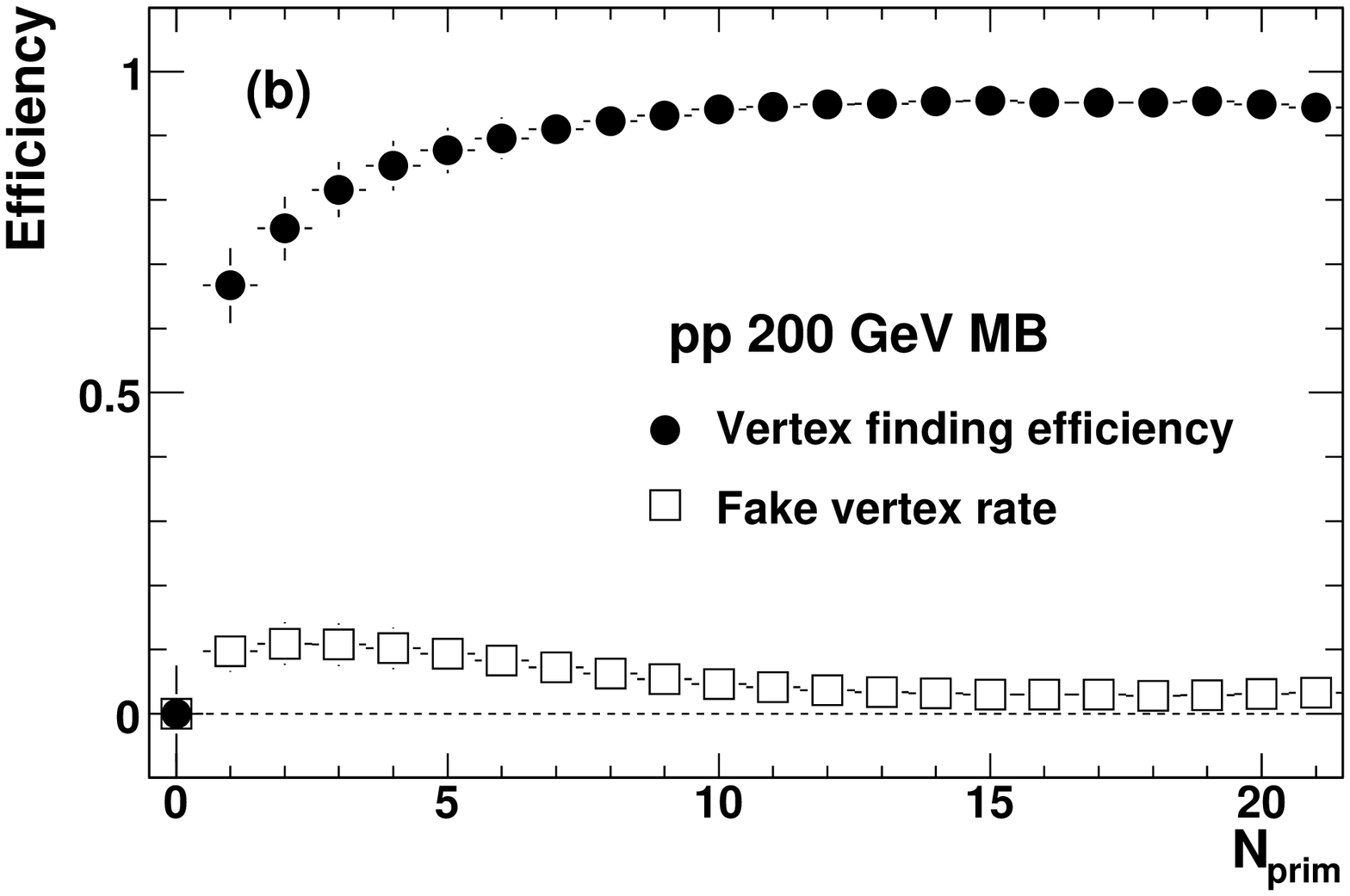}
\caption{Vertex-finding efficiency ($\epsilon_{\rm vtx}$) and fake vertex rate ($\delta_{\rm fake}$) as a function of the number of good global tracks (a) and the number of good primary tracks (b). Errors shown or smaller than the point size are statistical only.}
\label{fig:ppvtxeff}
\end{figure*}		

In order to study the pile-up effect, two simulated sets of $pp$ events are mixed at the raw data level, and the combined set is reconstructed by the full reconstruction chain. The first set is considered the real event and the other set is used as the pile-up background event. The pile-up level is varied from 0 to 100\%, where 100\% means each real event has a pile-up event in it. (The real $pp$ data from Run II has a much smaller rate than this.) The reconstructed numbers of {\em good} global tracks and {\em good} primary tracks of the mixed event are examined as a function of the pile-up level. The number of good primary tracks, $N_{\rm prim}$, is found to be independent of the degree of pile-up level. The number of good global tracks, $N_{\rm glob}$, increases with increasing pile-up level. Therefore, $N_{\rm prim}$ is chosen to characterize the vertex-finding efficiency. This is also desirable because only the number of primary tracks represents the true event of interest.

The vertex-finding efficiencies in $pp$ and d+Au collisions are studied by HIJING MC events~\cite{Hijing} embedded into abort-gap events (events triggered and reconstructed at empty bunch crossings). The abort-gap events represent the background in the real collision environment. The embedded event is subsequently reconstructed by the full reconstruction chain. In every MC event there is a well defined primary vertex. With the embedded event reconstructed and the MC information in hand, the vertex-finding efficiency can be obtained. The overall vertex-finding efficiency, $\epsilon_{\rm vtx}(N_{\rm glob})$, is determined as the ratio of the number of reconstructed events with the correct vertex position (within 2~cm of the input MC event vertex) to the number of input MC events. The obtained $\epsilon_{\rm vtx}(N_{\rm glob})$ is shown in Fig.~\ref{fig:ppvtxeff}(a). The vertex-finding efficiency here is expressed in terms of $N_{\rm glob}$ because the number of primary tracks cannot be readily obtained -- those MC events that fail the vertex-finding program do not have primary tracks defined. 

A reconstructed vertex that is farther than 2~cm (3-dimensional distance) away from that of the input MC event is considered as a fake vertex. The fake vertex rate, $\delta_{\rm fake}(N_{\rm glob})$, is obtained by the ratio of the number of fake vertex events to the number of input MC events. The obtained $\delta_{\rm fake}(N_{\rm glob})$ is also shown in Fig.~\ref{fig:ppvtxeff}(a). 

The extracted vertex-finding efficiency and fake vertex rate are expressed as a function of $N_{\rm glob}$. However, as mentioned earlier, the number of good primary tracks should be used as the variable because it is not affected by pile-up. In order to use $N_{\rm prim}$ as the variable, a map of $N_{\rm prim}$ versus $N_{\rm glob}$ is used: for each $N_{\rm prim}$ bin the vertex-finding efficiency and the fake vertex rate are obtained by convoluting $\epsilon_{\rm vtx}(N_{\rm glob})$ and $\delta_{\rm fake}(N_{\rm glob})$, respectively. The obtained $\epsilon_{\rm vtx}(N_{\rm prim})$ and $\delta_{\rm fake}(N_{\rm prim})$ are shown in Fig.~\ref{fig:ppvtxeff}(b).

The vertex-finding efficiency and the fake vertex rate are corrected by weighting the particles in each event by the factor
\begin{equation}
[\epsilon_{\rm vtx}(N_{\rm prim}) + \delta_{\rm fake}(N_{\rm prim})]^{-1} .
\end{equation}
Each event is weighted by the same factor when counting events for normalization. The overall correction factor is found to be nearly one for the two central bins of d+Au collisions, so the correction is only applied to the peripheral bin of d+Au collisions and minimum bias $pp$ collisions.

\begin{figure}[htbp]
\centering
\includegraphics[width=0.48\textwidth]{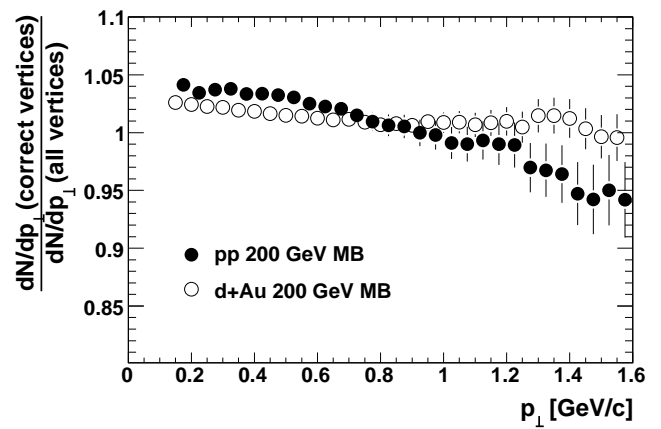}
\caption{The $\pt$ dependent correction to particle spectra due to fake vertex events, $\epsilon_{\rm fake}(\pt)$, in 200~GeV minimum bias $pp$ and d+Au collisions. Errors shown are statistical only.}
\label{fig:FakeVertexCorrection}
\end{figure}

From the MC study, the particle spectra from fake vertex events are extracted and compared to those from good events (with a correctly reconstructed vertex). It is found that particles from the fake vertex events have somewhat harder $\pt$ spectra than those in good events, presumably due to the wrongly assigned primary vertex position in final track fitting and the fact that higher $\pt$ particles are assigned larger weight in the vertex-fitting algorithm~\cite{BBC}. Figure~\ref{fig:FakeVertexCorrection} shows the ratio of the charged hadron $\pt$ spectrum in good vertex events to that in all events with a reconstructed vertex (i.e. sum of good and fake vertex events) for minimum bias $pp$ and d+Au collisions. The spectra are normalized per event before the ratio is taken. This ratio is parameterized, and the parameterization, $\epsilon_{\rm fake}(\pt)$, is multiplied with all $\pt$ spectra to correct for the $\pt$-dependent effect of the fake vertex events. The correction is found to be rather insensitive to the particle type, so a single correction function is applied to all particle species. Again the correction is found to be negligible in the two central bins of d+Au collisions, so it is only applied to the peripheral bin of d+Au collisions and minimum bias $pp$ collisions.

\subsection{Tracking Efficiency\label{sec:efficiency}}

\begin{figure*}[thbp]
\centering
\includegraphics[width=0.48\textwidth]{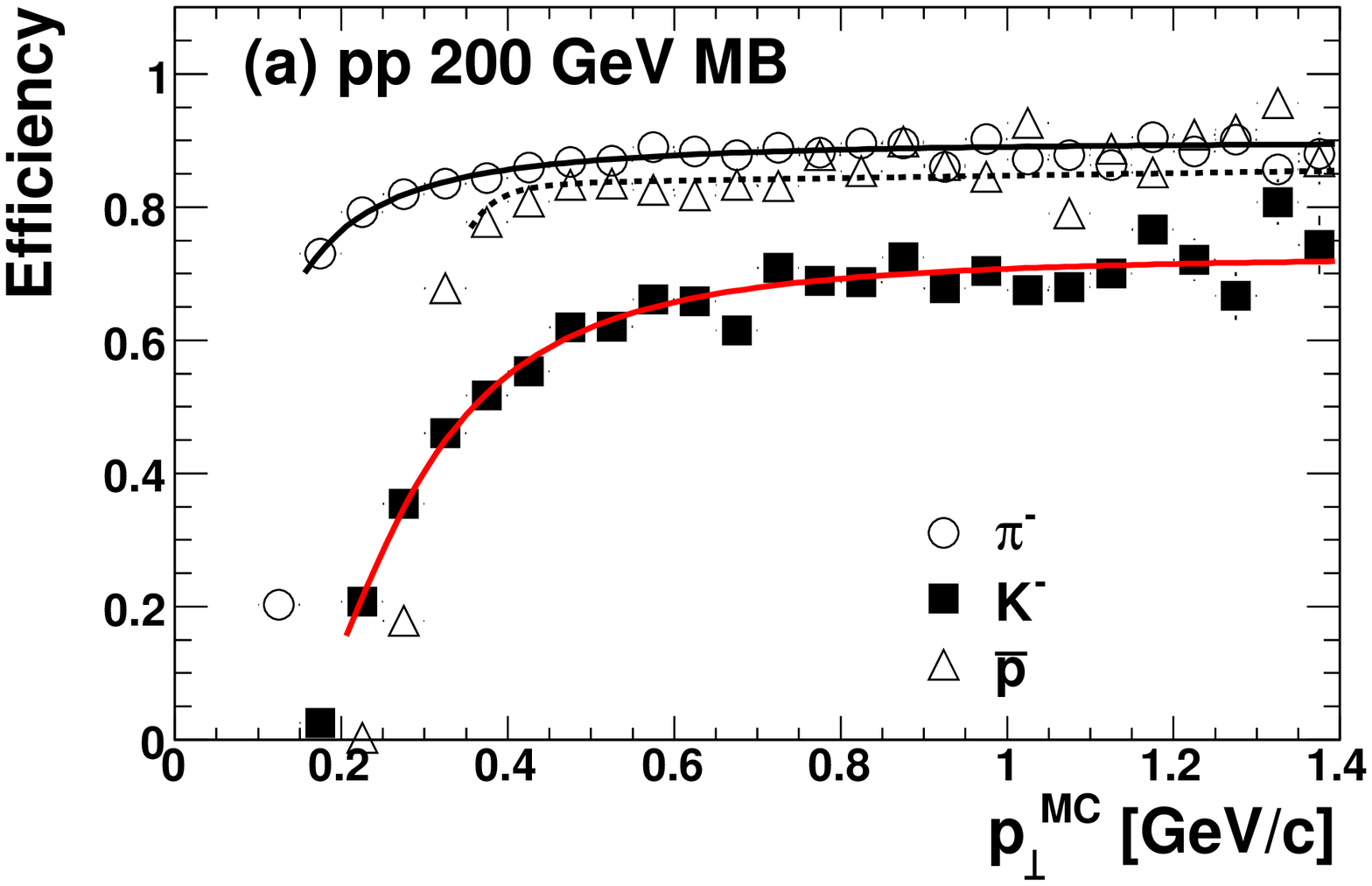}
\includegraphics[width=0.48\textwidth]{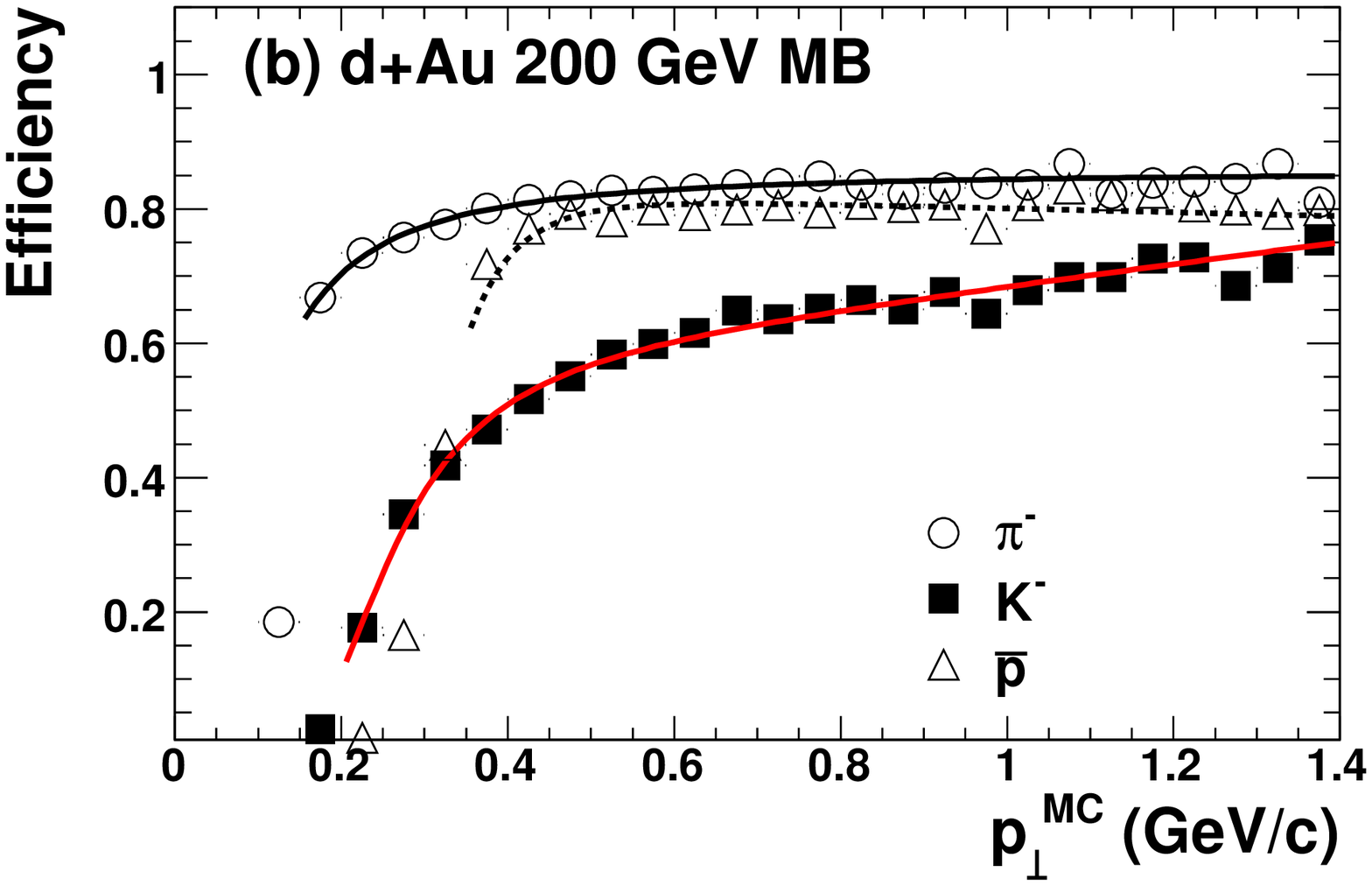}
\caption{Efficiency (product of tracking efficiency and detector acceptance) of $\pi^-$, $K^-$, and $\pbar$ in $pp$ (a) and d+Au collisions (b) at 200~GeV as a function of input MC $\pt$. Errors shown are binomial errors. The curves are parameterizations to the efficiency data and are used for corrections in the analysis.}
\label{fig:treffppdAu}
\vspace{0.2in}
\centering
\includegraphics[width=0.48\textwidth]{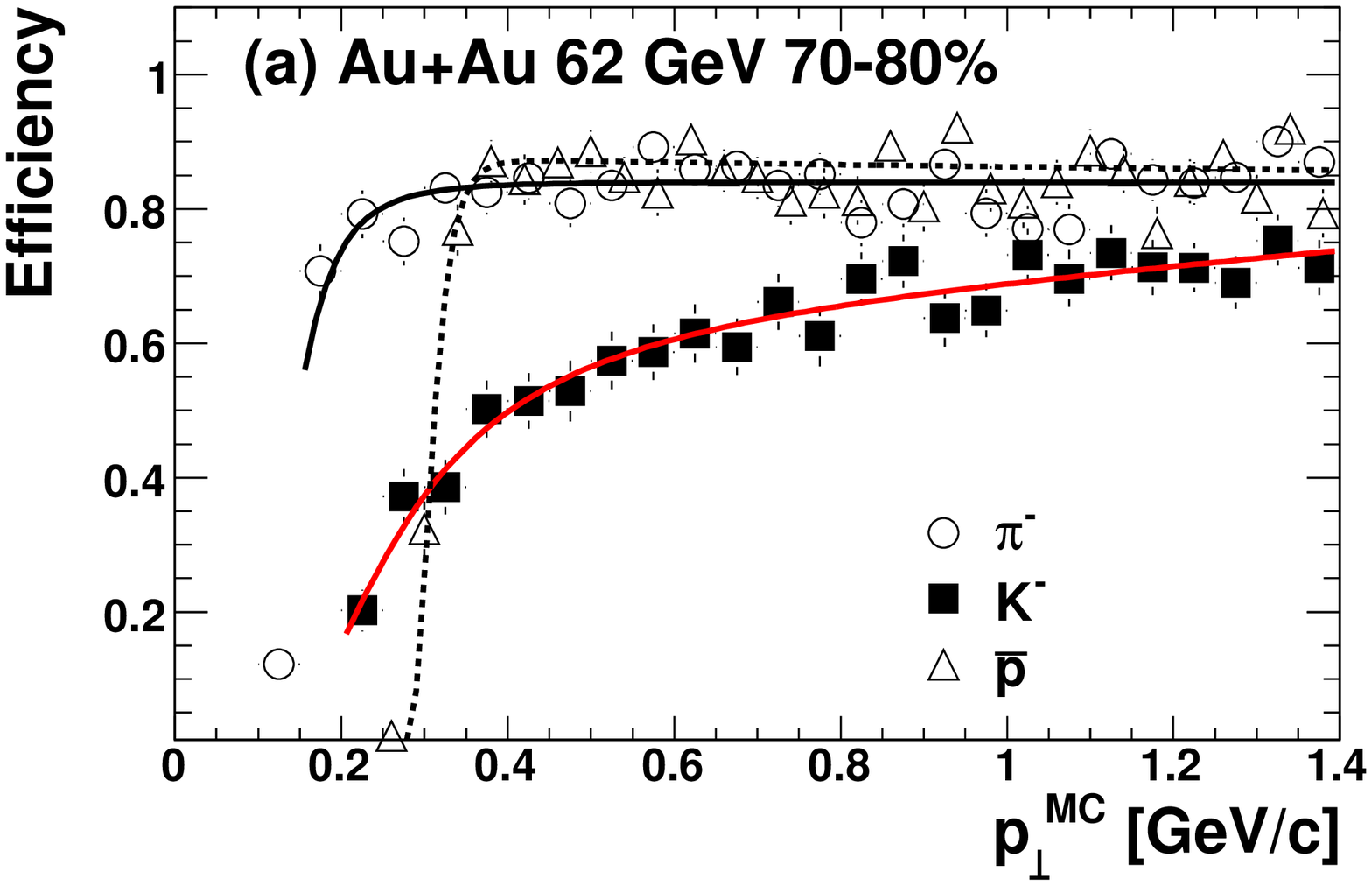}
\includegraphics[width=0.48\textwidth]{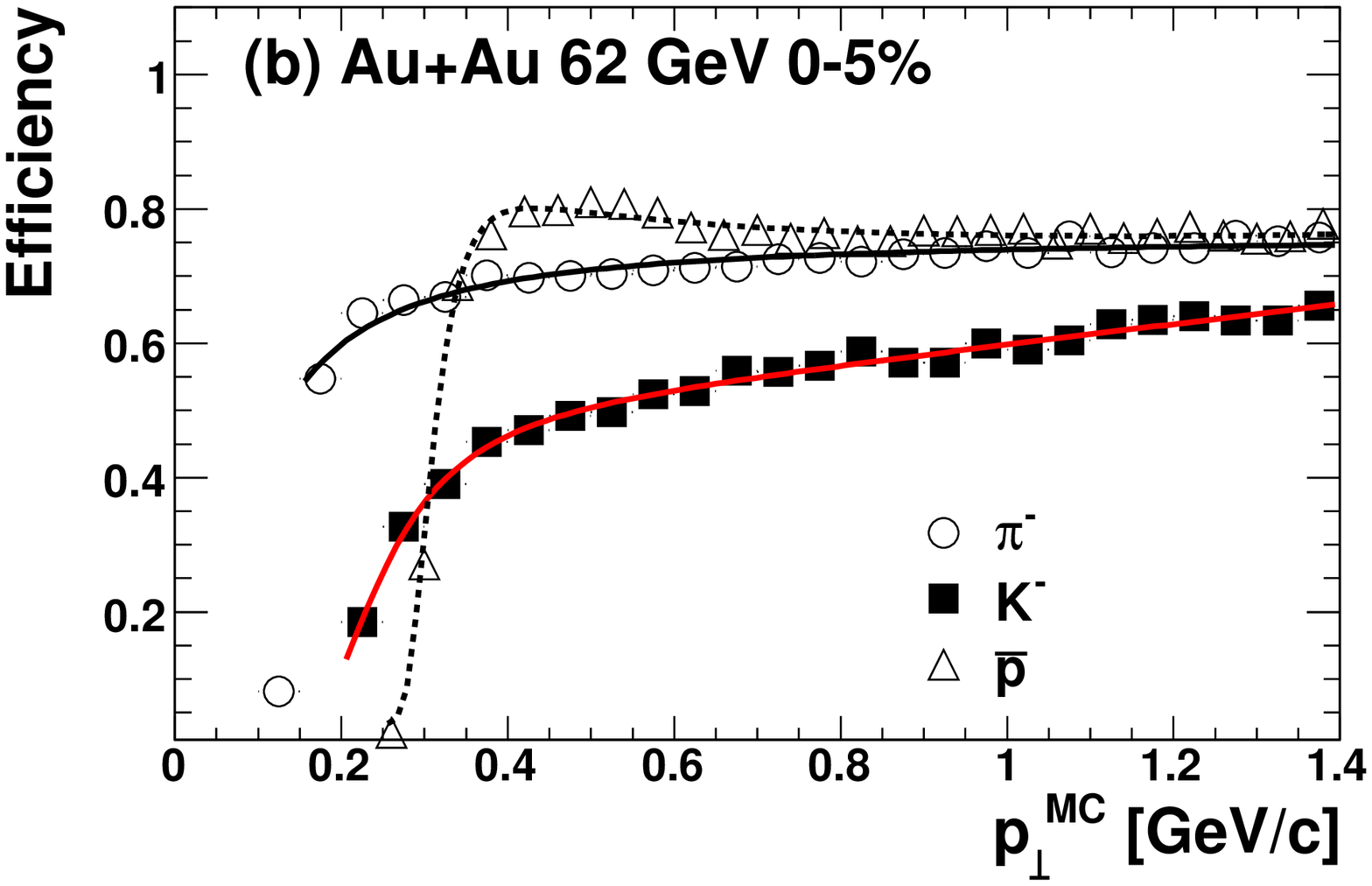}
\caption{Efficiency (product of tracking efficiency and detector acceptance) of $\pi^-$, $K^-$, and $\pbar$ in 70-80\% peripheral Au+Au (a) and 0-5\% central Au+Au collisions (b) at 62.4~GeV as a function of input MC $\pt$. Errors shown are binomial errors. The curves are parameterizations to the efficiency data and are used for corrections in the analysis.}
\label{fig:treffAuAu}
\end{figure*}

The raw spectra are corrected for detector acceptance and tracking efficiency which are obtained from the MC embedding method. Figures~\ref{fig:treffppdAu} and~\ref{fig:treffAuAu} show the obtained efficiency, the product of tracking efficiency and detector acceptance, for minimum bias $pp$ and d+Au collisions and for peripheral and central Au+Au collisions, respectively. All efficiencies are expressed as a function of the input MC $\pt$. The $\pt$ dependences are the same in $pp$ and d+Au collisions and similar in Au+Au collisions. The pion efficiency is independent of $\pt$ for $\pt>0.2$~GeV/$c$, but falls steeply at lower $\pt$ because particles below $\pt=0.15$~GeV/$c$ cannot traverse the entire TPC due to their large track curvature inside the solenoidal magnetic field. The efficiency for protons and antiprotons is flat above $\pt\sim 0.35$~GeV/$c$. At lower $\pt$, the efficiency drops steeply because of the large multiple scattering effect due to the large (anti)proton mass. The kaon efficiency shown in Figs.~\ref{fig:treffppdAu} and~\ref{fig:treffAuAu} increases smoothly with $\pt$ and already includes decay loss (which decreases with increasing $\pt$). The significantly smaller kaon efficiency at small momentum than that of pions is caused by the large loss of kaons due to decays.

In $pp$ and d+Au collisions, the difference between the efficiencies for the different multiplicity bins is negligible because the multiplicities are low and the different occupancies have no effect on the track reconstruction performance. In Au+Au collisions, the particle multiplicity (hence the TPC occupancy) is high, resulting in the different reconstruction efficiency magnitudes in peripheral and central Au+Au collisions as seen in Fig.~\ref{fig:treffAuAu}. The change in the efficiency from peripheral to central collisions at 62.4~GeV is smooth and is of the order of 15-20\%. However, this is still a relatively small variation; the 5\% embedded multiplicity used in the embedding MC simulation, which biases the embedded events towards higher multiplicity and TPC occupancy, has negligible effect on the calculated reconstruction efficiency for each centrality bin.

The curves superimposed in Figs.~\ref{fig:treffppdAu} and~\ref{fig:treffAuAu} are parameterizations to the efficiencies. Table~\ref{tab:efficiency} lists the fit parameters for 200~GeV minimum bias d+Au data and five centrality bins of 62.4~GeV Au+Au data. The fit parameters for $\pi^-$, $K^-$, $\pbar$ and $p$ are tabulated. The fit parameters for $\pi^+$ and $\pi^-$ are similar, and also for $K^+$ and $K^-$. These parameterizations are used in the analysis for efficiency corrections.

\begin{table}
\caption{Parameterizations to $\pi^-$, $K^-$, $\pbar$ and $p$ efficiencies for 200~GeV minimum bias d+Au data and five centrality bins of 62.4~GeV Au+Au data.}
\label{tab:efficiency}
\begin{ruledtabular}
\begin{tabular}{rrrrrrr}
& d+Au & \multicolumn{5}{c}{Au+Au} \\ \cline{3-7}
& min bias & 70-80\% & 50-60\% & 30-40\% & 10-20\% & 0-5\% \\ \hline
\multicolumn{7}{c}{$\pi^{-}$: $P_0\exp[-(P_1/\pt)^{P_2}]$} \\ \hline
$P_0$ & $ 0.856$ & $ 0.840$ & $ 0.818$ & $ 0.809$ & $ 0.781$ & $ 0.759$ \\
$P_1$ & $ 0.075$ & $ 0.129$ & $ 0.111$ & $ 0.109$ & $ 0.097$ & $ 0.070$ \\
$P_2$ & $ 1.668$ & $ 4.661$ & $ 3.631$ & $ 3.224$ & $ 2.310$ & $ 1.373$ \\ \hline
\multicolumn{7}{c}{$K^{-}$: $P_0\exp[-(P_1/\pt)^{P_2}]+P_3\pt$} \\ \hline
$P_0$ & $ 0.527$ & $ 0.608$ & $ 0.585$ & $ 0.503$ & $ 0.494$ & $ 0.450$ \\
$P_1$ & $ 0.241$ & $ 0.238$ & $ 0.234$ & $ 0.231$ & $ 0.229$ & $ 0.229$ \\
$P_2$ & $ 3.496$ & $ 2.425$ & $ 3.034$ & $ 3.968$ & $ 3.492$ & $ 3.925$ \\
$P_3$ & $ 0.160$ & $ 0.099$ & $ 0.085$ & $ 0.152$ & $ 0.139$ & $ 0.149$ \\ \hline
\multicolumn{7}{c}{$\bar{p}$: $(P_0\exp[-(P_1/\pt)^{P_2}]+P_3\pt)\exp[(P_4/\pt)^{P_5}]$} \\ \hline
$P_0$ & $ 0.830$ & $ 0.317$ & $ 0.233$ & $ 0.227$ & $ 0.245$ & $ 0.246$ \\
$P_1$ & $ 0.295$ & $ 0.303$ & $ 0.303$ & $ 0.300$ & $ 0.305$ & $ 0.301$ \\
$P_2$ & $ 7.005$ & $19.473$ & $13.480$ & $11.183$ & $15.567$ & $13.054$ \\
$P_3$ & $-0.029$ & $-0.004$ & $ 0.028$ & $ 0.041$ & $ 0.026$ & $ 0.039$ \\
$P_4$ & 0 fixed  & $10.156$ & $ 4.160$ & $ 2.031$ & $ 1.895$ & $ 0.889$ \\
$P_5$ & 0 fixed  & $ 0.006$ & $ 0.104$ & $ 0.153$ & $ 0.107$ & $ 0.160$ \\ \hline
\multicolumn{7}{c}{$p$: $(P_0\exp[-(P_1/\pt)^{P_2}]+P_3\pt)\exp[(P_4/\pt)^{P_5}]$} \\ \hline
$P_0$ & $ 0.921$ & $ 0.189$ & $ 0.201$ & $ 0.193$ & $ 0.187$ & $ 0.221$ \\
$P_1$ & $ 0.291$ & $ 0.306$ & $ 0.310$ & $ 0.303$ & $ 0.308$ & $ 0.307$ \\
$P_2$ & $ 7.819$ & $10.643$ & $16.825$ & $ 9.565$ & $12.826$ & $14.488$ \\
$P_3$ & $-0.057$ & $ 0.023$ & $ 0.026$ & $ 0.041$ & $ 0.042$ & $ 0.033$ \\
$P_4$ & 0 fixed  & $15.212$ & $ 9.331$ & $ 3.585$ & $ 3.380$ & $ 1.944$ \\
$P_5$ & 0 fixed  & $ 0.131$ & $ 0.127$ & $ 0.194$ & $ 0.194$ & $ 0.181$ \\
\end{tabular}
\end{ruledtabular}
\end{table}

\subsection{Proton Background Correction}

The proton sample contains background protons knocked out from the beam pipe and the detector materials by interactions of produced hadrons in these materials~\cite{pionAbsorb}. Most of these protons have large $\dca$ and are not reconstructed as primary particles. However, some of these background protons have small $\dca$ and are therefore included in the primary track sample and a correction is needed.

\begin{figure*}[thbp]
\begin{center}	
\includegraphics[width=0.48\textwidth]{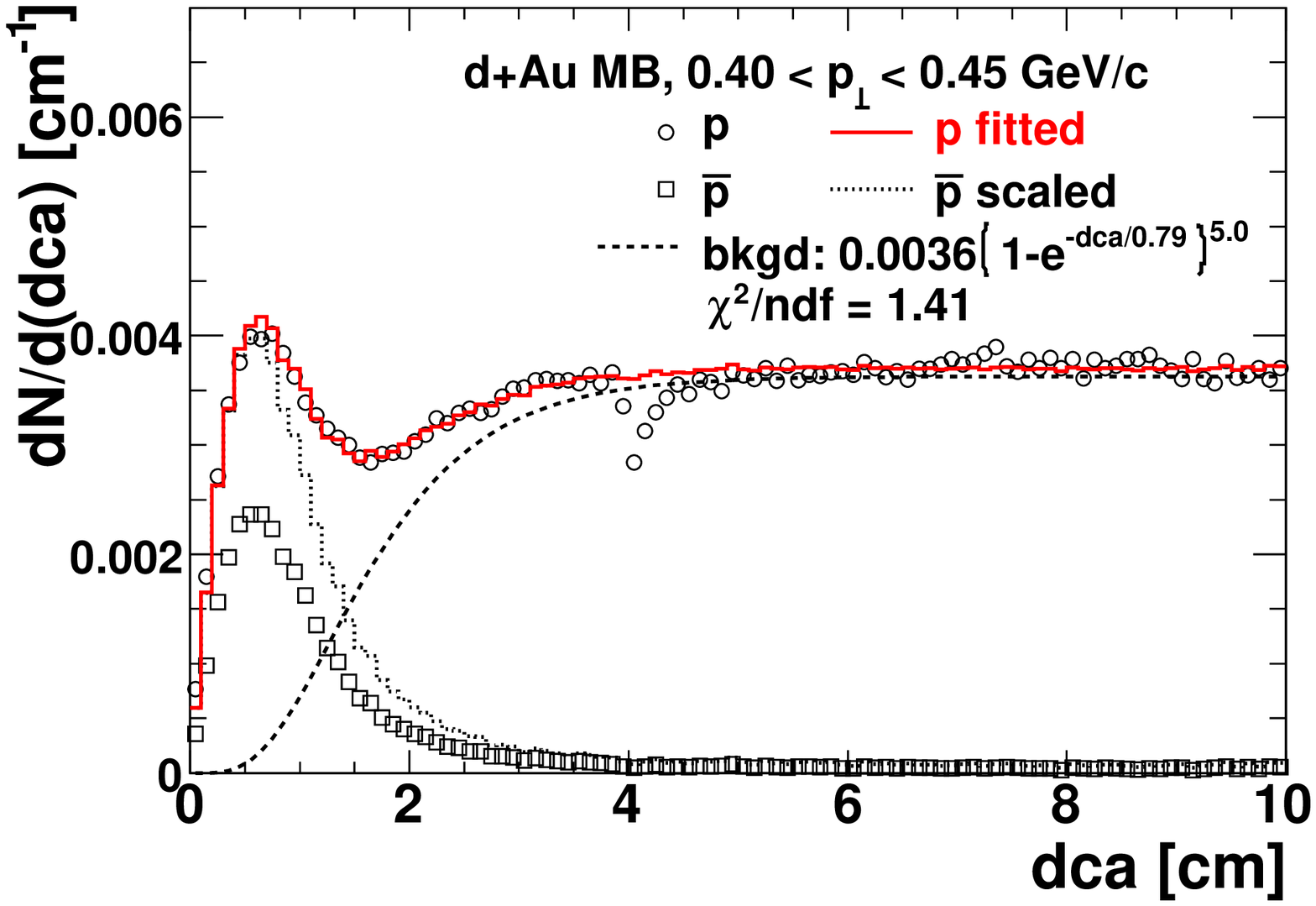}
\includegraphics[width=0.48\textwidth]{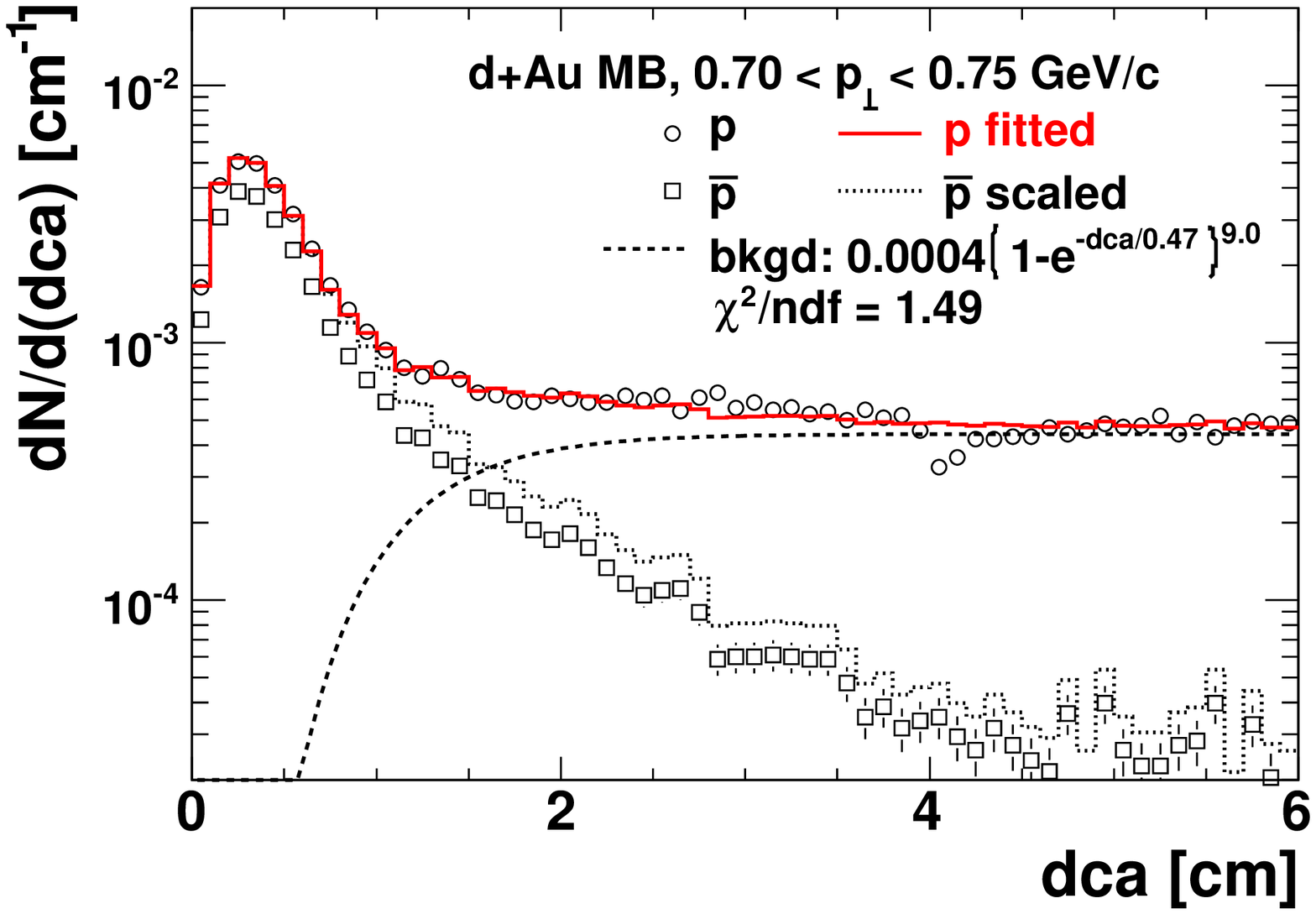}
\includegraphics[width=0.48\textwidth]{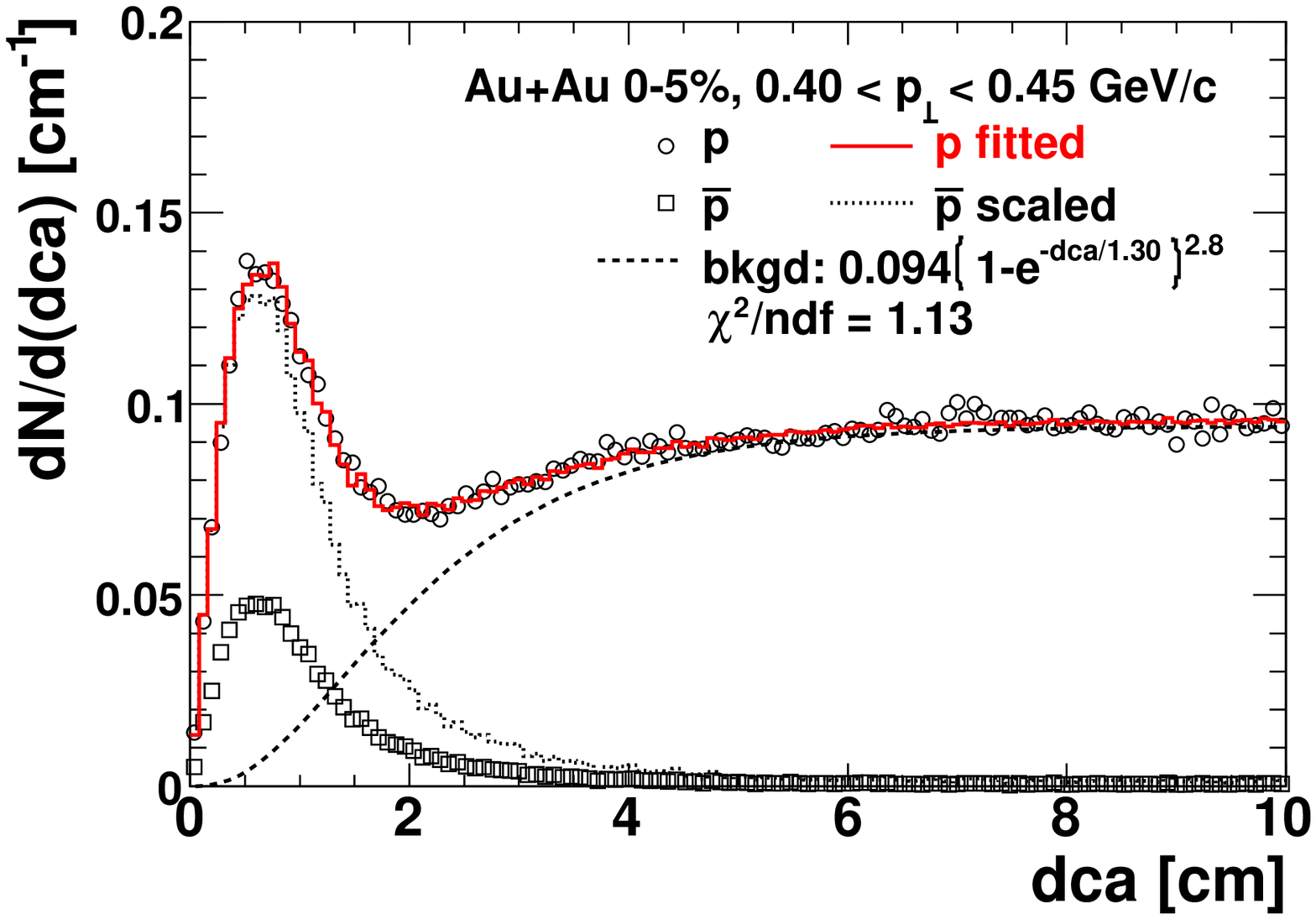}
\includegraphics[width=0.48\textwidth]{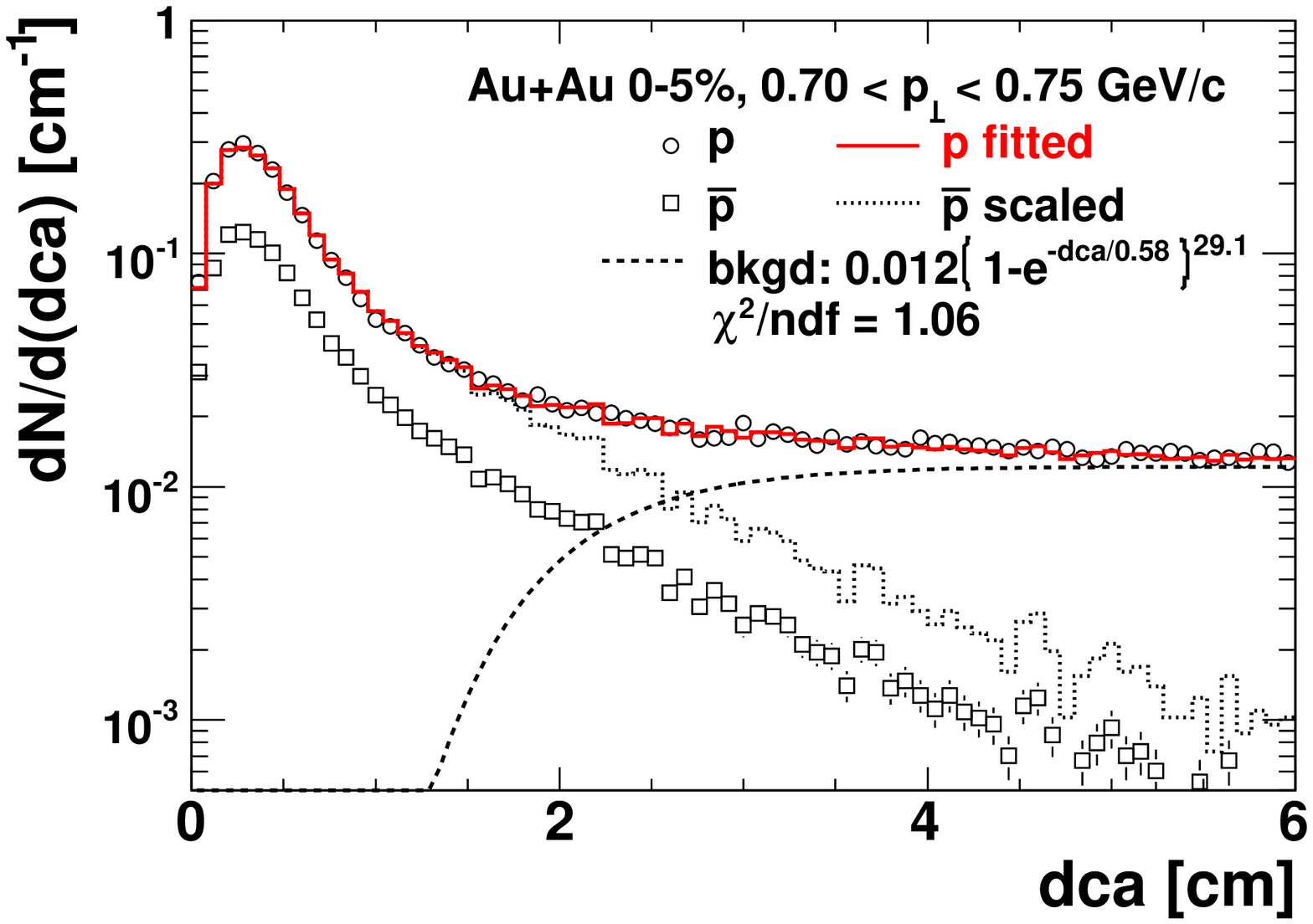}
\caption{The $\dca$ distributions of protons and antiprotons for $0.40<\pt<0.45$~GeV/$c$ and $0.70<\pt<0.75$~GeV/$c$~in 200~GeV minimum bias d+Au (upper panels) and 62.4~GeV 0-5\% central Au+Au collisions (lower panels). Errors shown are statistical only. The dashed curve is the fit proton background; the dotted histogram is the $\pbar$ distribution scaled up by the fit $p/\pbar$ ratio; and the solid histogram is the fit $p$ distribution by Eq.~(\ref{eq:dca_fit}). The range $3.2<\dca<5$~cm is excluded from the fit for the d+Au data. Note the logarithm scale of the right panels.}
\label{fig:dca_protons}
\end{center}
\end{figure*}

Figure~\ref{fig:dca_protons} shows the $\dca$ distributions of protons and antiprotons for two selected $\pt$ bins in 200~GeV d+Au (upper panels) and 62.4~GeV central Au+Au collisions (lower panels), respectively. The protons and antiprotons are selected by a $\dedx$ cut of $|z_p|<0.3$ where $z_p$ is given by Eq.~(\ref{eq:z}). The long, nearly flat $\dca$ tail in the proton distribution comes mainly from knock-out background protons. The effect is large at low $\pt$ and significantly diminished at high $\pt$ (note the logarithm scale for the high-$\pt$ data). Antiprotons do not have knock-out background; the flat $\dca$ tail is absent from their $\dca$ distributions. 

In order to correct for the knock-out background protons, the $\dca$ dependence at $\dca<3$~cm is needed for the knock-out protons. Based on MC simulation studies, we found the following functional form to describe the background protons well~\cite{p130}:
\begin{equation}
p_{\rm bkgd}(\dca) \propto \left[1-\exp(-\dca/d_0)\right]^{\alpha}\; .\label{eq:dca_knockout}
\end{equation}
Assuming that the shape of the background subtracted proton $\dca$ distribution is identical to that for the antiproton $\dca$ distribution, the proton data can be fit by
\begin{equation}
p(\dca) = \pbar(\dca)/r_{\pbar/p} + A\cdot p_{\rm bkgd}(\dca) \; ,\label{eq:dca_fit}
\end{equation}
where the magnitude of the background protons $A$, the parameter $d_0$, the exponent $\alpha$, and the antiproton-to-proton ratio $r_{\pbar/p}$ are free parameters. This assumption is, however, not strictly valid because the weak decay contributions to the proton and antiproton samples are in principle different, and the $\dca$ distribution of the weak decay products differs from that of the primordial protons and antiprotons. However, the measured $\bar{\Lambda}/\Lambda$ ratio is close to the $\pbar/p$ ratio~\cite{STARlambda} and the difference in $\dca$ distributions between protons and antiprotons arising from weak decay contaminations is small. The effect of slightly different proton and antiproton $\dca$ distributions on the extracted background proton fraction is estimated and is within the systematic uncertainty discussed in Section~\ref{sec:systuncer}.

The $\dca$ distributions of protons and antiprotons are fit with Eq.~(\ref{eq:dca_fit}) in each $\pt$ and centrality bin. The $\dca$ distributions up to 10~cm are included in the fit for the Au+Au data. The proton $\dca$ distributions in d+Au collisions, however, have a peculiar dip at $\dca\approx 4$~cm as seen in Fig.~\ref{fig:dca_protons}. We think this dip is related to effects of the beam pipe (whose diameter is 3~inches) and a specific algorithm of the vertex finder in low multiplicity collisions; however, its exact cause is still under investigation. Due to the dip in the d+Au data, we fit the $\dca$ distributions up to 10~cm but exclude the region $3.2<\dca<5$~cm from the fit. The fit results are shown in Fig.~\ref{fig:dca_protons}. The dashed curve is the fit proton background. The dotted curve is the $\pbar$ distribution scaled up by the fit $p/\pbar$ ratio. The solid histogram is the fit of Eq.~(\ref{eq:dca_fit}) to the proton distribution. The fit qualities are good. It is found that the fit power exponent $\alpha$ is larger than one, indicating that the background protons die off faster than the simple $1-\exp(-\dca/d_0)$ form at small $\dca$. The $\alpha$ value is large at high $\pt$; there is practically no background at high $\pt$ at small $\dca$. 

Table~\ref{tab:protonBg} lists the fraction of knock-out background protons out of the total measured proton sample within $\dca<3$~cm as a function of $\pt$ in minimum bias d+Au and three selected centrality bins of Au+Au data. The fraction of knock-out background protons depends on a number of factors, including the amount of detector material, analysis cuts, the total particle multiplicity produced in the collisions and their kinetic energies. Since the ratio of proton multiplicity to total particle multiplicity varies somewhat with centrality, and the particle kinematics change with centrality, the background fraction varies slightly with centrality. 

For $pp$ data~\cite{spec200} and Au+Au data at 130~GeV~\cite{p130} and 200~GeV~\cite{spec200}, the background protons are corrected in a similar way. The fraction of background protons are similar in all collision systems.

\begin{table}
\caption{Fraction of proton background out of total measured proton sample as a function of $\pt$. Minimum bias d+Au collisions at 200~GeV and three centrality bins of Au+Au collisions at 62.4~GeV are listed. The errors are systematic uncertainties.}
\label{tab:protonBg}
\begin{ruledtabular}
\begin{tabular}{c|c|ccc}
$\pt$ & d+Au 200~GeV & \multicolumn{3}{c}{Au+Au 62.4~GeV} \\
(GeV/$c$) & min.~bias & 70-80\% & 30-40\% & 0-5\% \\ \hline
0.425 & $0.49\pm0.07$ & $0.32\pm0.09$ & $0.36\pm0.08$ & $0.36\pm0.08$ \\
0.475 & $0.47\pm0.04$ & $0.29\pm0.06$ & $0.29\pm0.06$ & $0.29\pm0.05$ \\
0.525 & $0.41\pm0.04$ & $0.26\pm0.05$ & $0.22\pm0.04$ & $0.23\pm0.04$ \\
0.575 & $0.36\pm0.04$ & $0.18\pm0.05$ & $0.15\pm0.03$ & $0.16\pm0.03$ \\
0.625 & $0.28\pm0.04$ & $0.12\pm0.05$ & $0.12\pm0.03$ & $0.11\pm0.02$ \\
0.675 & $0.23\pm0.04$ & $0.09\pm0.05$ & $0.08\pm0.02$ & $0.07\pm0.02$ \\
0.725 & $0.17\pm0.04$ & $0.06\pm0.05$ & $0.05\pm0.02$ & $0.04\pm0.01$ \\
0.775 & $0.12\pm0.05$ & $0.05\pm0.05$ & $0.04\pm0.02$ & $0.03\pm0.01$ \\
0.825 & $0.10\pm0.05$ & $0.03\pm0.03$ & $0.02\pm0.01$ & $0.02\pm0.01$ \\
0.875 & $0.06\pm0.04$ & $0.03\pm0.03$ & $0.02\pm0.01$ & $0.02\pm0.01$ \\
0.925 & $0.06\pm0.04$ & $0.02\pm0.02$ & $0.01\pm0.01$ & $0.01\pm0.01$ \\
0.975 & $0.04\pm0.04$ & $0.02\pm0.02$ & $0.01\pm0.01$ & $0.01\pm0.01$ \\
\end{tabular}
\end{ruledtabular}
\end{table}

\subsection{Pion Background Correction}

\begin{figure}[htbp]
\centering
\includegraphics[width=0.48\textwidth]{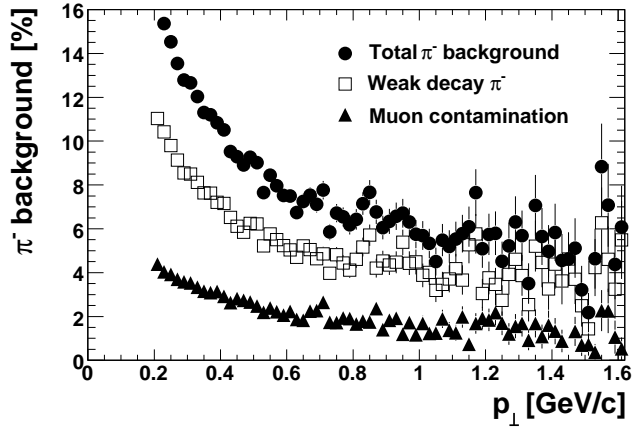}
\caption{Pion background fraction from weak decays ($\Lambda$, $K^{0}_{S}$) and $\mu^{\pm}$ contamination as a function of $\pt$ in minimum bias d+Au collisions at 200~GeV. Errors shown are statistical only.}
\label{fig:PiBG_dAu_mb}
\end{figure}

The pion spectra are corrected for feed-downs from weak decays, muon contamination, and background pions produced in the detector materials. The corrections are obtained from MC simulations of HIJING events, with the STAR geometry and a realistic description of the detector response. The simulated events are reconstructed in the same way as for real data. The weak-decay daughter pions are mainly from $K_S^0$ and $\Lambda$ and are identified by the parent particle information accessible from the simulation. The pion decay muons can be mis-identified as primordial pions because of the similar masses of muon and pion. This contamination is obtained from MC by identifying the decay, which is also accessible from the simulation. The obtained weak-decay pion background and muon contamination are shown in Fig.~\ref{fig:PiBG_dAu_mb} as a function of $\pt$. The total background rate, which is dominated by these two sources, is also shown.

The pion background fraction is independent of event multiplicity in 200~GeV $pp$ and d+Au collisions; therefore, a single correction is applied. In 62.4~GeV Au+Au collisions the multiplicity dependence of the pion background is weak (within 1.5\% over the entire centrality range); a single, averaged correction is applied to all centralities, similar to the approach in Ref.~\cite{spec200}.

\section{Systematic Uncertainties\label{sec:systuncer}}

\subsection{On Transverse Momentum Spectra\label{sec:syst_spectra}}

The point-to-point systematic uncertainties on the spectra are estimated by varying event and track selection and analysis cuts and by assessing sample purity from the $\dedx$ measurement. In addition, the Gaussian fit ranges are varied to estimate the systematic uncertainty on the extracted raw spectra. The estimated uncertainties are less than 4$\%$ for $\pi^{\pm}$, $p$ and $\pbar$. Those for $K^{\pm}$ are less than 12$\%$ for $\pt$ bins with significant overlap in $\dedx$ with $e^{\pm}$ or $\pi^{\pm}$, and less than 4$\%$ for other bins. These point-to-point systematic errors are similar for $pp$, d+Au, and Au+Au collisions. The point-to-point systematic errors are combined with statistical errors in quadrature in the plotted spectra in Figs.~\ref{fig:dAu_spectra},~\ref{fig:AuAu62_spectra},~\ref{fig:AuAu130_spectra} and~\ref{fig:all_spectra}. The combined errors are treated as random errors and are included in the fitting of the spectra.

For proton spectra, an additional systematic error is estimated due to background subtraction. The estimated uncertainty at $\pt=0.45$-0.50~GeV/$c$ is about 8\% and drops rapidly with $\pt$~\cite{pbarp130,p130} (see Table~\ref{tab:protonBg}). The $\pt$ dependence of background contribution varies somewhat with centrality, presumably due to the combined effect of the rapid change in the proton $\pt$ spectral shape with centrality and little change in the pion's. The proton background uncertainties for $pp$ and d+Au collisions are similar. The systematic uncertainties on the pion spectra due to background correction are negligible.

A correlated overall systematic uncertainty of 5$\%$ is estimated for all spectra and is dominated by uncertainties in the MC determination of reconstruction efficiencies. This systematic uncertainty is estimated by varying parameters in the MC simulation.

\subsection{On $dN/dy$\label{sec:dNdySystErr}}

The particle yield measured at mid-rapidity ($|y|<0.1$) for each identified particle spectrum is calculated from the measured $\pt$ range and extrapolated to the unmeasured regions with various parameterizations. For kaons and protons the extrapolation is done by the hydrodynamics-motivated blast-wave model fit (described in section~\ref{sec:Model}). Our default blast-wave fit does not include resonance decays (the effect of which is studied in detail in Appendix~\ref{app:Resonance}). The fit is done to all six spectra of $\pi^{\pm}$, $K^{\pm}$, $p$ and $\pbar$ simultaneously. However, because the low $\pt$ region of the pion spectra are affected by resonance decays, the $\pt<0.5$~GeV/$c$ part of the pion spectra is excluded from the blast-wave fit. Instead, the Bose-Einstein distribution,
\begin{equation}
\frac{dN}{m_\perp dm_\perp} \propto 1\left/\left[\exp(m_\perp/T_{BE})-1\right]\right.\label{eq:BE}
\end{equation}
is found to describe the pion spectra well and is employed to extrapolate the pion spectra, with $T_{BE}$ a fit parameter. The point-to-point systematic errors on the spectra are included in the fits. 

\begin{table}
\caption{Fraction of measured and extrapolated yield for negatively charged particles for selected collision systems and centralities. For extrapolation, Bose-Einstein fit is used for pions and blast-wave fit is used for kaons and (anti)protons.}
\label{tab:extrapolation}
\begin{ruledtabular}
\begin{tabular}{c|ccc}
	& \hspace{0.3in}measured\hspace{0.3in} & \multicolumn{2}{c}{extrapolated $dN/dy$} \\
system	& $dN/dy$ & low $\pt$ & high $\pt$ \\ \hline
	& \multicolumn{3}{c}{$\pi^-$, measured range $\pt$=0.2-0.7~GeV/$c$} \\
d+Au min.~bias	& 59\% & 30\% & 11\% \\
Au+Au 70-80\%	& 58\% & 32\% & 10\% \\
Au+Au 30-40\%	& 58\% & 28\% & 14\% \\
Au+Au 0-5\%	& 58\% & 28\% & 14\% \\ \hline
	& \multicolumn{3}{c}{$K^-$, measured range $\pt$=0.2-0.75~GeV/$c$} \\
d+Au min.~bias	& 60\% & 12\% & 28\% \\    
	& \multicolumn{3}{c}{$K$, measured range $\pt$=0.25-0.75~GeV/$c$} \\
Au+Au 70-80\%	& 58\% & 21\% & 21\% \\
Au+Au 30-40\%	& 56\% & 15\% & 29\% \\
Au+Au 0-5\%	& 54\% & 13\% & 33\% \\ \hline
	& \multicolumn{3}{c}{$\pbar$, measured range $\pt$=0.4-1.10~GeV/$c$} \\
d+Au min.~bias	& 53\% & 21\% & 26\% \\
	& \multicolumn{3}{c}{$\pbar$, measured range $\pt$=0.35-1.15~GeV/$c$} \\
Au+Au 70-80\%	& 65\% & 21\% & 14\% \\
Au+Au 30-40\%	& 64\% & 12\% & 24\% \\
Au+Au 0-5\%	& 60\% & 9\% & 31\% \\ 
\end{tabular}
\end{ruledtabular}
\end{table}

Table~\ref{tab:extrapolation} shows the fractional yields of $dN/dy$ extrapolated to the unmeasured $\pt$ regions. The systematic uncertainties on the extrapolated yields are estimated by comparing the extrapolation to those using other fit functions. Those fit functions are:
\begin{equation}
\begin{array}{lrcl}
\pt{\rm -exponential:}& \frac{dN}{\pt d\pt} &\propto& \exp(-\pt/T_{\pt})\ , \\
\pt{\rm -Gaussian:}& \frac{dN}{\pt d\pt} &\propto& \exp(-\pt^2/T^2_{\pt})\ ,\\
\pt^3{\rm -exponential:}& \frac{dN}{\pt d\pt} &\propto& \exp(-\pt^3/T^3_{\pt})\ ,\\
\mt{\rm -exponential:}& \frac{dN}{m_\perp dm_\perp} &\propto& \exp(-m_\perp/T_{\mt})\ ,\\
{\rm Boltzmann:}& \frac{dN}{m_\perp dm_\perp} &\propto& m_{T} \exp(-m_\perp/T_B)\ .
\end{array}
\end{equation}
where $T_{\pt}$, $T_{\mt}$, and $T_B$ are fit parameters. The fit functions used for pion $dN/dy$ systematic uncertainty assessment are the blast-wave function and the $\pt$-exponential. Those used for kaons are the $\mt$-exponential and the Boltzmann function. Those used for proton and antiproton are the $\pt$-Gaussian and $\pt^3$-exponential; also used for Au+Au 20-80\% centrality bins and for d+Au collisions are the Boltzmann function, the $\mt$-exponential, and the $\pt$-exponential.

The systematic uncertainties on the extrapolated total particle yields are dominated by the uncertainties in the extrapolation, which are estimated to be of the order of 15\% of the extrapolated part of the integrated yields for pions and kaons, and 15-40\% for antiprotons and protons depending on centrality. The 5\% overall MC uncertainty is added in quadrature. For protons, the $\pt$-dependent systematic uncertainty on background subtraction leads to an overall systematic uncertainty in the yields. This systematic uncertainty is estimated and included in quadrature in the total systematic uncertainties on $dN/dy$.

\begin{figure}[hbt]
\centering
\includegraphics[width=0.48\textwidth]{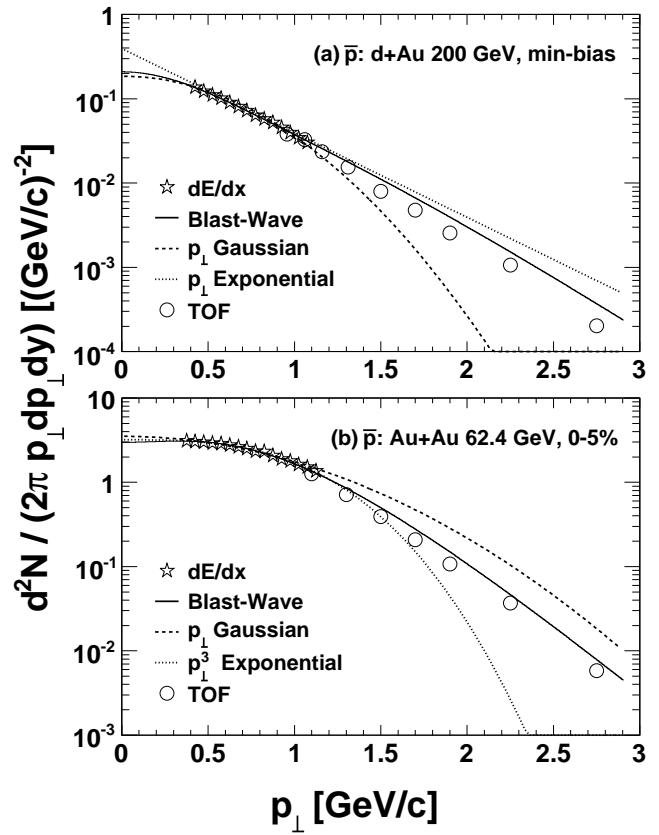}
\caption{Mid-rapidity identified antiproton spectra in 200~GeV minimum bias d+Au (a) and 62.4~GeV central Au+Au collisions (b) measured by $\dedx$ together with those by TOF~\cite{TOF_ppdAu,TOF_AuAu}. The $\dedx$ data are from $|y|<0.1$ and the TOF data are from $|y|<0.5$. The curves are various fits to the $\dedx$ data for extrapolation. The quadratic sum of statistical errors and point-to-point systematic errors are plotted, but are smaller than the point size.}
\label{fig:TOF}
\end{figure}

Identified particle spectra in $pp$ and d+Au collisions at 200~GeV and Au+Au collisions at 62.4~GeV are measured at relatively high $\pt$ by the TOF detector in STAR~\cite{TOF_ppdAu,TOF_AuAu}. In the overlap region in $\pt$, the TOF measurements and the TPC measurements reported here are consistent within systematic uncertainties. The TOF measurement is a good systematic check on our extrapolation. As an example, Fig.~\ref{fig:TOF} shows the measured antiproton spectra by $\dedx$ in d+Au and central Au+Au collisions and their various parameterizations, together with the TOF measurements. The TOF measurements are well within the range of the extrapolations. Blast-wave fits are also performed including the large $\pt$ ranges from TOF~\cite{TOF_ppdAu,TOF_AuAu} and the spectra obtained by the extended particle identification method (r$\dedx$)~\cite{rdEdx}. The blast-wave fit parameters thus obtained are consistent with our results within the systematic uncertainties. To keep consistency and fair comparisons of the various datasets, only the TPC measurements are studied here, since TOF was only installed as a prototype test for a full TOF system and was absent in many collision systems reported here.

The total charged particle density $\dNchdy$, calculated from the sum of the individual $dN/dy$ yields of $\pi^{\pm}$, $K^{\pm}$, $p$ and $\pbar$, is used as one of the centrality measures in this paper. The systematic uncertainties on $\dNchdy$ are calculated assuming that the extrapolation uncertainties are completely correlated between particles and antiparticles and completely uncorrelated between different particle species. In addition, the efficiency uncertainty is common for all particle species, and the proton background uncertainty is uncorrelated with the rest.

\subsection{On Particle Ratios and $\meanpt$}

Systematic uncertainties on particle yield ratios come from those on the extrapolated total yields, estimated as above. The efficiency uncertainties are canceled in the ratios. The extrapolation uncertainties are canceled to a large degree in the antiparticle-to-particle ratios; a common systematic uncertainty of 2\%, 3\%, and 5\% is assigned to $\pi^-/\pi^+$, $K^-/K^+$, and $\pbar/p$, respectively~\cite{spec200}. The extrapolation uncertainties are treated as uncorrelated in the unlike particle ratios ($K^-/\pi^-, \pbar/\pi^-, p/\pi^-$, etc.). The uncertainty due to proton background substraction is added in quadrature for the ratios involved with the proton yield. 

The average transverse momentum, $\meanpt$, is extracted from the measured spectra and the extrapolations (blast-wave model fits for kaons and protons and Bose-Einstein function for pions as described above). Systematic uncertainties on $\meanpt$ are also estimated by using the various functional forms mentioned before for extrapolation of the spectra. For protons an additional systematic uncertainty on $\meanpt$ due to the $\pt$-dependent proton background subtraction is estimated and included in quadrature in the total systematic uncertainties. 

\subsection{On Chemical Freeze-Out Parameters}

Chemical freeze-out parameters (chemical freeze-out temperature $\Tch$, baryon and strangeness chemical potentials $\mu_B$ and $\mu_S$, and strangeness suppression factor $\gamma_S$) are extracted from the measured particle ratios obtained from the six particle spectra within the framework of a statistical model. The systematic uncertainties on the particle ratios are included in the statistical model fit and are treated as independent. These uncertainties propagate to the systematic uncertainties on the chemical freeze-out parameters.

Our measured protons are inclusive of all protons from primordial $\Sigma^+$ and $\Lambda$ (and $\Sigma^0$-decay $\Lambda$) decays, and likewise for the antiparticles. To assess the systematic uncertainties on the fit chemical freeze-out parameters, we vary the detection probability of weak-decay (anti)protons from 100\% to 50\%. The chemical freeze-out temperature thus obtained is larger by about 8~MeV and is included in the systematic uncertainty estimate. The effects on baryon and strangeness chemical potentials are negligible. Due to the decay kinematics, $\Lambda$'s from $\Xi$ and $\Omega$ decays mostly follow the parent direction~\cite{STARlambda}, and the decay protons from most of those decay $\Lambda$'s are reconstructed as primordial protons in the STAR TPC; likewise for the antiparticles. In our statistical model fit, we assume 50\% of the (anti)protons from multi-step decays are included in our measured primary (anti)proton samples. To assess the systematic uncertainty due to the multi-step decay products, we include either all the multi-step decay (anti)protons or none of them in the statistical model fit. We found that this systematic uncertainty is small.

The other source of systematic uncertainty is due to the relatively limited set of particle ratios used in this analysis. While the $\Tch$, $\mu_{B}$ and $\mu_{S}$ should be well constrained because of the high statistics data for pions, kaons, and (anti)protons, the ad-hoc strangeness suppression factor $\gamma_{S}$ is not well constrained because the single-strangeness $K^{\pm}$ are the only strangeness species used in this work. STAR has measured a variety of strange and multi-strange particles, including $K^{*\pm}$, $K_S^0$, $\Lambda$ and $\overline{\Lambda}$, $\Lambda_{1520}$, $\Xi$ and $\overline{\Xi}$, and $\Omega$ and $\overline{\Omega}$ at 130~GeV~\cite{STARlambda,STAR130kstar,STAR130ratio,STAR130msb} and 200~GeV~\cite{STAR200kstar,STAR200strange,STAR200resonance}. The chemical freeze-out parameters have also been studied by particle ratios including these particles~\cite{STAR130msb}. It is found that the extracted chemical freeze-out temperature and baryon and strangeness chemical potentials are similar to those obtained from this work using the limited set of particle ratios. However, the $\gamma_S$ parameter differs: in central Au+Au collisions, $\gamma_S\sim0.9$ from this work and $\sim1.0$ from the fit including the extended list of strange and multi-strange particles~\cite{STAR130msb}. This difference gives a reasonable estimate of the systematic uncertainty on $\gamma_S$.

\subsection{On Kinetic Freeze-Out Parameters}

The kinetic freeze-out parameters are extracted from the simultaneous blast-wave parameterization of the measured particle spectra. The kinetic freeze-out temperature $\Tkin$, the average transverse radial flow velocity $\langle\beta\rangle$, and the flow velocity profile exponent $n$ are treated as free parameters. The point-to-point systematic errors on the spectra are included in the blast-wave fit. The $\pt$-dependent systematic uncertainty due to proton background correction is taken into account in evaluating the systematic uncertainties of the blast-wave parameters. 

The measured pions contain large contributions from resonance decays; the contributions vary with the pion $\pt$. Our default blast-wave fit does not include resonance decays. In order to reduce the systematic uncertainty due to resonance decays, the low $\pt$ part ($\pt<0.5$~GeV/$c$) of the pion spectra is excluded from the blast-wave fit. The remaining systematic uncertainty is estimated by varying the $\pt$ range of the pion spectra included in the blast-wave fit. The resonance decay effect on the blast-wave fit is also thoroughly studied in Appendix~\ref{app:Resonance}. Comparisons between the blast-wave parameters obtained including or excluding resonance decays also give a good estimate of the systematic uncertainties.

Due to the large mass of (anti)protons and kaons, the (anti)proton and kaon spectra constrain the transverse flow velocity well. Thus the systematic uncertainties on the kinetic freeze-out parameters are also assessed by excluding the $K^\pm$ spectra, the $p$ spectrum, or the $\pbar$ spectrum from the blast-wave fit. 

While the spectra are mainly determined by $\Tkin$ and $\langle\beta\rangle$, the shape of the flow velocity profile also has some effect on the spectra, due to the non-linearity in the dependence of the spectral shape on the flow velocity. However, the effect is fairly weak, as indicated by the large fitting errors on the velocity profile exponent $n$ for some of the spectra. Nevertheless, to assess the systematic uncertainty from this effect, we fit the spectra by fixing $n$ to unity. The fit qualities are significantly degraded for some of the spectra. However, we use the changes in the fit parameters as our conservative estimates of the systematic uncertainties due to the flow velocity profile used.

We note that the blast-wave model assumes a simple picture of local particle sources of a common temperature in a transverse radial velocity field to describe the flattening of particle transverse spectra. The extracted kinetic freeze-out parameters are within the framework of this picture. However, it is possible that other effects may also contribute to the spectra flattening: semi-hard scatterings may even be the main contributer in $pp$ collisions~\cite{trainor_pp}; the possible effect of statistical global energy and momentum conservation on particle spectra is recently studied in Ref.~\cite{lisa}. Such effects are not included in our systematic uncertainties on the extracted values of the kinetic freeze-out parameters.

\section{Results and Discussions\label{sec:Results}}

\begin{figure*}[thbp]
\centering
\includegraphics[width=0.95\textwidth]{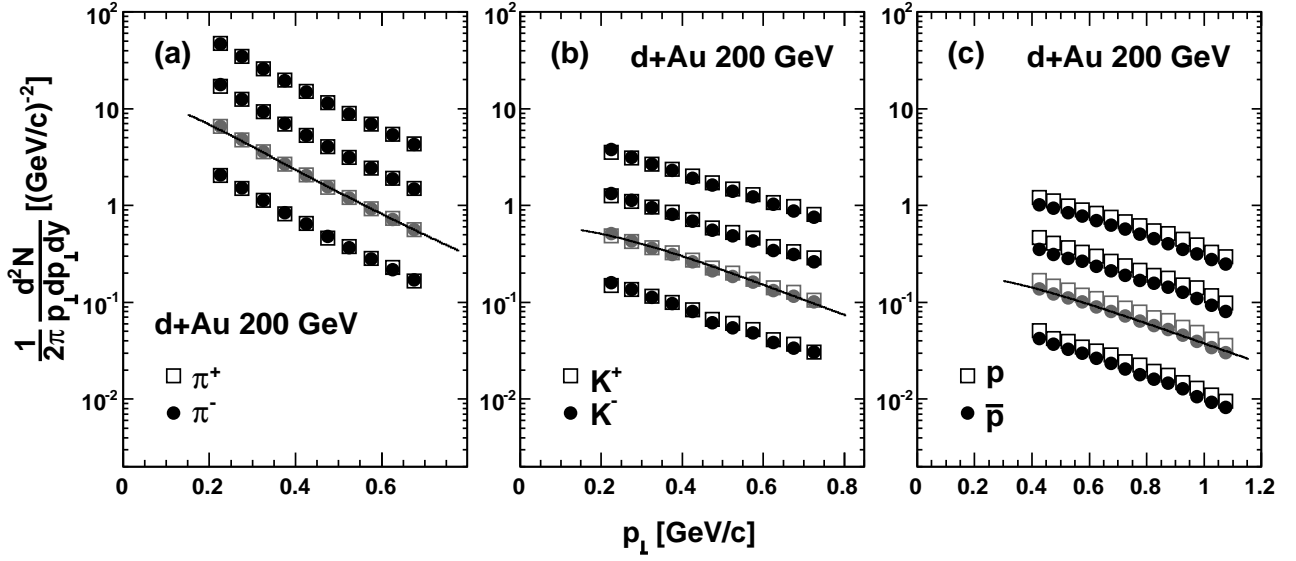}
\caption{Mid-rapidity ($|y|<$ 0.1) identified particle spectra in d+Au collisions at 200~GeV. The $p$ and $\pbar$ spectra are inclusive, including weak decay products. Spectra are plotted for three centrality bins and for minimum bias events. Spectra from top to bottom are for 0-20\% scaled by 4, 20-40\% scaled by 2, minimum bias not scaled, and 40-100\% scaled by 1/2. Errors plotted are statistical and point-to-point systematic errors added in quadrature, but are smaller than the point size. The curves are the blast-wave model fits to the minimum bias data; the normalizations of the curves are fixed by the corresponding negative particle spectra.}
\label{fig:dAu_spectra}
\end{figure*}

\begin{figure*}[thbp]
\centering
\includegraphics[width=0.95\textwidth]{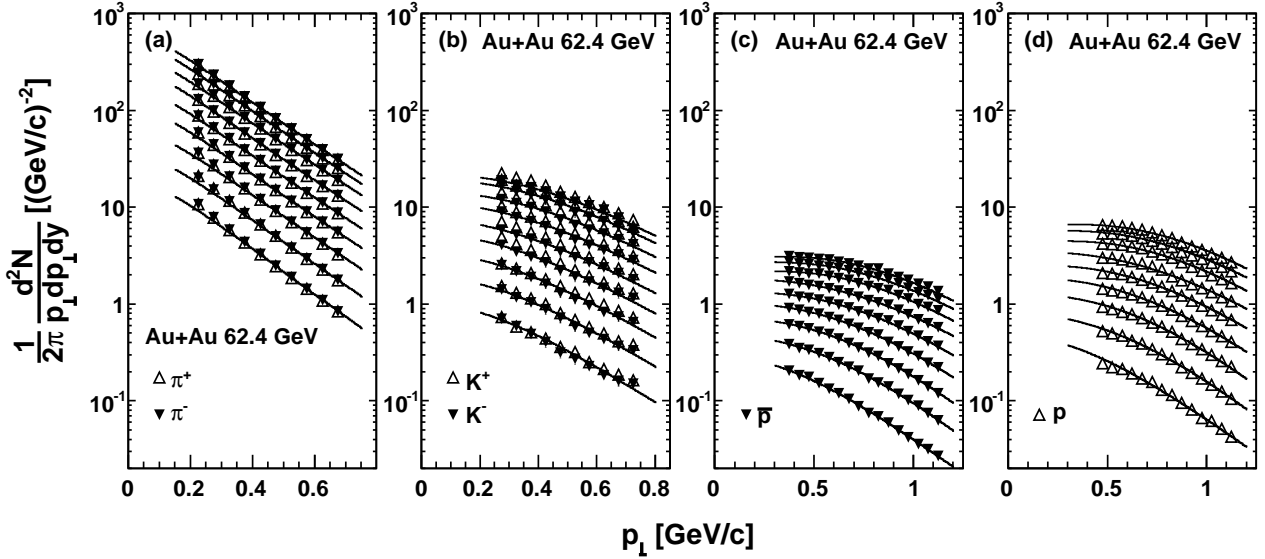}
\vspace{-0.2in}
\caption{Mid-rapidity ($|y|<$ 0.1) identified particle spectra in Au+Au collisions at 62.4~GeV. The $p$ and $\pbar$ spectra are inclusive, including weak decay products. Spectra are plotted for nine centrality bins, from top to bottom, 0-5\%, 5-10\%, 10-20\%, 20-30\%, 30-40\%, 40-50\%, 50-60\%, 60-70\%, and 70-80\%. Errors plotted are statistical and point-to-point systematic errors added in quadrature, but are smaller than the point size. The curves are the blast-wave model fits to the spectra; the normalizations of the curves in (a,b) are fixed by the corresponding negative particle spectra.}
\label{fig:AuAu62_spectra}
\end{figure*}

\begin{figure}[hbtp]
\centering
\includegraphics[width=0.48\textwidth]{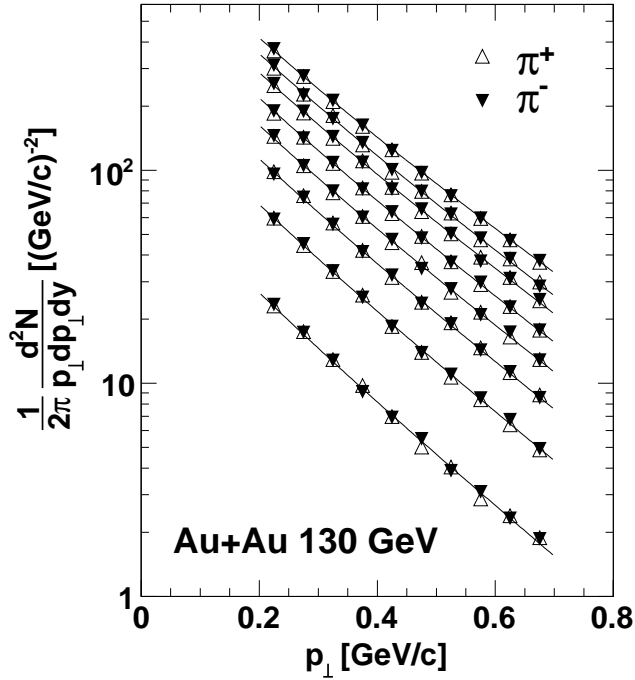}
\caption{Mid-rapidity ($|y|<$ 0.1) identified pion spectra in Au+Au collisions at 130~GeV. Spectra are plotted for eight centrality bins, from top to bottom, 0-6\%, 6-11\%, 11-18\%, 18-26\%, 26-34\%, 34-45\%, 45-58\%, and 58-85\%. Errors plotted are statistical and point-to-point systematic errors added in quadrature, but they are smaller than the data point size. The curves are the Bose-Einstein fits to the spectra; the normalizations of the curves are fixed by the corresponding negative particle spectra.}
\label{fig:AuAu130_spectra}
\end{figure}

\begin{figure*}[thbp]
\centering
\includegraphics[width=0.95\textwidth]{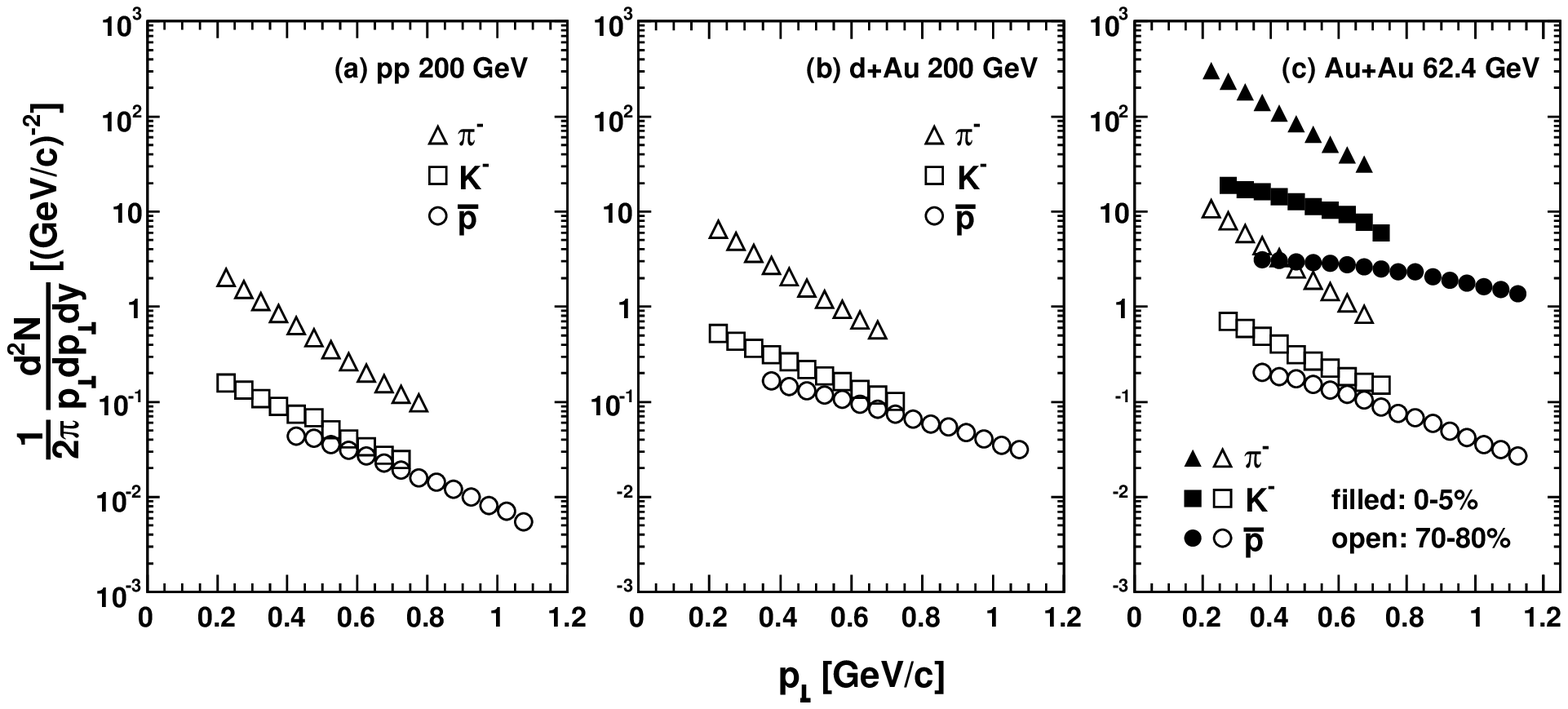}
\caption{Comparisons of $\pi^-$, $K^-$, and $\pbar$ transverse momentum spectra for (a) minimum bias $pp$ collisions at 200~GeV, (b) minimum bias d+Au collisions at 200~GeV, and (c) Au+Au collisions at 62.4~GeV. Two centralities are shown: 0-5\% central collisions in black filled symbols and 70-80\% peripheral collisions in open symbols. Errors are statistical and point-to-point systematic errors added in quadrature.}
\label{fig:all_spectra}
\end{figure*}

In this section, results on identified $\pi^{\pm}$, $K^{\pm}$, $p$ and $\pbar$ in d+Au collisions at 200~GeV and Au+Au collisions at 62.4~GeV~\cite{Molnar} are presented and discussed. The results are measured at mid-rapidity in the range $\left|y\right|<$ 0.1. The charged pion spectra in Au+Au collisions at 130~GeV are also presented. The results are discussed together with previously published identified $\pi^{\pm}$, $K^{\pm}$, $p$ and $\pbar$ results in $pp$ and Au+Au collisions at 200~GeV~\cite{spec200}, and charged kaons~\cite{kaon130} and (anti)protons results~\cite{p130} at 130~GeV.

The identified particle spectra are presented first, followed by the average transverse momenta $\meanpt$, the integrated particle multiplicity densities $dN/dy$ and ratios, and baryon and strangeness production rates. The $\meanpt$ and $dN/dy$ are extracted from the measured spectra and the extrapolations from the blast-wave model fits for kaons and protons and the Bose-Einstein function for pions. In order to have the same procedure to obtain $dN/dy$ and $\meanpt$, the identified particle spectra from 130~GeV Au+Au collisions are fit by the blast-wave model parameterization in this work. The extracted $\meanpt$ and $dN/dy$ are listed in Table~\ref{tab:meanpt} and Table~\ref{tab:dndy}, respectively. The quoted errors are the quadratic sum of statistical and systematic uncertainties and are dominated by the latter. Since the systematic uncertainties on particle ratios cannot be readily obtained from the individual particle $dN/dy$ yields, Table~\ref{tab:ratios} lists particle ratios together with the total uncertainties. 

\subsection{Transverse Momentum Spectra}

Figure~\ref{fig:dAu_spectra} shows the centrality dependent and the minimum bias $\pi^{\pm}$, $K^{\pm}$, $p$ and $\bar{p}$ spectra in d+Au collisions at 200~GeV. The minimum bias spectra are obtained from the cross-section weighted sum of the corresponding spectra in each centrality bin. The minimum bias d+Au spectra are in good agreement with the previously published results~\cite{TOF_ppdAu}. Spectra from different centralities are similar. 

Figure~\ref{fig:AuAu62_spectra} shows the centrality dependence of the $\pi^{\pm}$, $K^{\pm}$, $p$ and $\bar{p}$ spectra measured in Au+Au collisions at 62.4~GeV. Pion spectral shapes are similar in all centrality bins. Kaon and (anti)proton spectra show a significant flattening with increasing centrality with the effect being stronger for proton.

Figure~\ref{fig:AuAu130_spectra} shows the centrality dependent pion spectra measured in Au+Au collisions at 130~GeV. All spectra are parallel indicating no significant centrality dependence of the shape. The kaon spectra at 130~GeV are published in Ref.~\cite{kaon130}, and the proton and antiproton spectra are published in Ref.~\cite{pbar130}.

Spectra results from $pp$ and Au+Au collisions at 200~GeV are published in Ref.~\cite{spec200}. Spectra shapes from 62.4~GeV, 130~GeV, and 200~GeV Au+Au collisions are all similar. Hardening of the spectra is more pronounced with increasing centrality and increasing particle mass at all three energies.

Figure~\ref{fig:all_spectra} compares pion, kaon, and antiproton spectra in $pp$, d+Au, and Au+Au collisions. The $pp$, d+Au, and peripheral Au+Au spectra are similar in shape. The central Au+Au spectra of kaons and (anti)protons are significantly flatter.

\subsection{Average Transverse Momenta}

\begin{figure*}[thbp]
\centering
\includegraphics[width=0.48\textwidth]{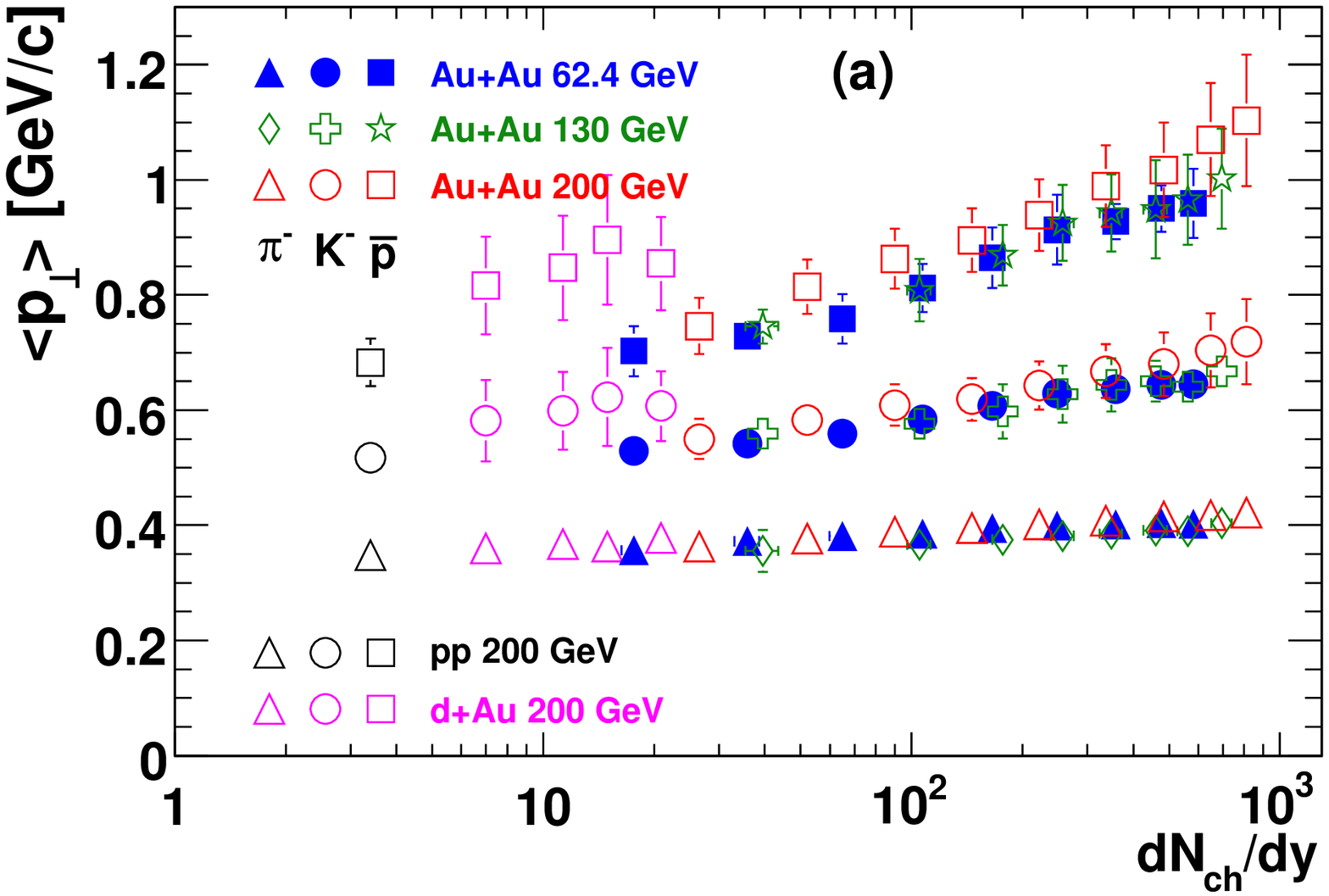}
\includegraphics[width=0.48\textwidth]{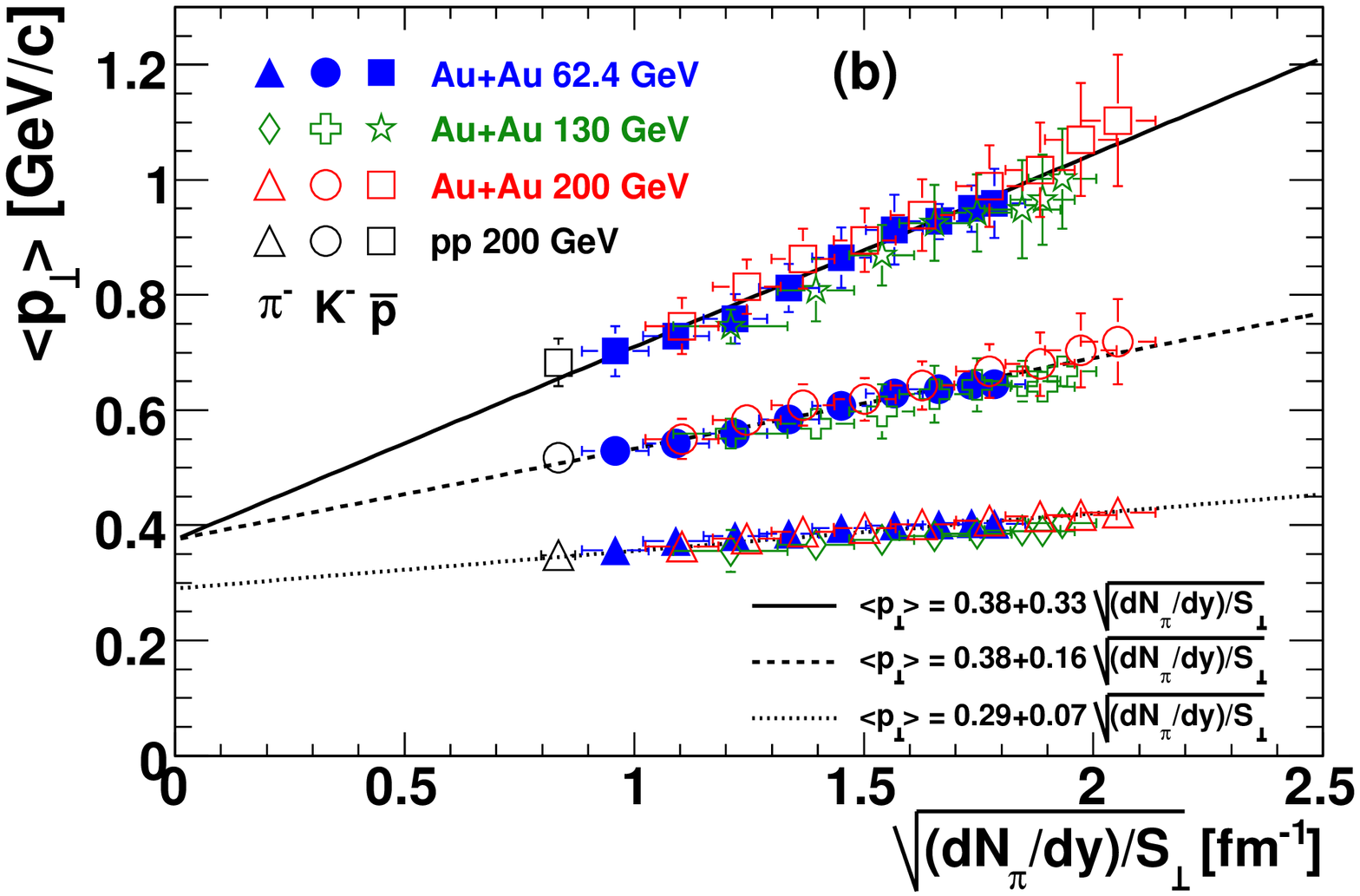}
\caption{(color online) Average transverse momenta as a function of $\dNchdy$ (a) and $\sqrt{\dNdyS}$ (b) for Au+Au collisions at 62.4~GeV, 130~GeV, and 200~GeV. The minimum bias $pp$ data are also shown. The d+Au data are shown in panel (a). Errors shown are systematic errors and statistical errors added in quadrature.}
\label{fig:meanpt}
\end{figure*}

The spectra shape can be quantified by the average transverse momentum, $\meanpt$. In Fig.~\ref{fig:meanpt}, the evolution of $\meanpt$ is shown as a function of the charged particle multiplicity. The pion $\meanpt$ increases slightly with centrality in Au+Au collisions. For kaons, protons and antiprotons the $\meanpt$ increases significantly with centrality. No obvious centrality dependence is observed for d+Au collisions.

One interesting observation is that the $\meanpt$ in central d+Au collisions is larger than that in peripheral Au+Au collisions. This can be due to jets, $\kt$ broadening, and multiple scattering~\cite{randomwalk}. These effects can be stronger in d+Au collisions than in peripheral Au+Au collisions, because nucleons in the deuteron suffer multiple collisions traversing the incoming Au nucleus in central d+Au collisions while peripheral Au+Au collisions are close to simple superposition of multiple $pp$ collisions. In fact, the $\meanpt$ in peripheral Au+Au collisions is similar to that in $pp$.

On the other hand, the $\meanpt$ in central d+Au collisions is smaller than that in central Au+Au collisions. Central d+Au collisions likely have larger effects from initial state multiple scattering and $\kt$ broadening. Although jet contribution is larger in central Au+Au than in d+Au, it is likely softened due to jet energy loss in central Au+Au collisions. Consequently jet contribution to the flattening of the low $\pt$ spectra in Au+Au collisions may not be much larger than that in d+Au collisions. The larger $\meanpt$ in central Au+Au collisions cannot be only due to the effects already present in d+Au collisions, as random-walk models argue~\cite{randomwalk}, but also be due to other effects including transverse radial flow, due to thermodynamic pressure, and remaining contributions from (semi-)hard scatterings. Transverse radial flow suggested by these data will be discussed in more detail in Sec.~\ref{sec:Model}.

For Au+Au collisions, $\meanpt$ increases significantly with increasing centrality. The trends are similar at 62.4~GeV, 130~GeV, and 200~GeV, and $\meanpt$ qualitatively agree with each other at the same $\dNchdy$. This suggests that the kinetic freeze-out properties in Au+Au collisions are rather energy independent for the measured collision energies. 

In the Color Glass Condensate (gluon saturation) picture, small $x$ gluons overlap and recombine, reducing the total number of gluons and increasing their transverse energy~\cite{dima01,dima02}. These gluons hadronize into mostly soft hadrons. Thus, a lower particle multiplicity and larger $\meanpt$ is predicted. In the gluon saturation picture, the only relevant scale is $\dNdyS$, and the $\meanpt$ is predicted to scale with $\sqrt{\dNdyS}$~\cite{dima01,dima02}. In Fig.~\ref{fig:meanpt}(b), the $\meanpt$ is shown as a function of $\sqrt{\dNdyS}$ for minimum bias $pp$ and for Au+Au collisions of the various centralities. A linear dependence of the $\meanpt$ on $\sqrt{\dNdyS}$ is observed for all three particle species, as shown by the lines in Fig.~\ref{fig:meanpt}. It is interesting to note that the slope, characterizing the rate of increase in the $\meanpt$, is a factor 2 larger for $\pbar$ than for kaons which is in turn a factor 2 larger than for pions. The intercepts of the linear fits for $\pbar$ and kaons are the same, but are larger than that for pions.

\begin{table*}{}
\caption{Extrapolated average transverse momenta, $\meanpt$ in GeV/$c$, of identified particles for various collision systems and centralities. Quoted errors are the quadratic sum of statistical and systematic uncertainties, and are dominated by the latter.}
\label{tab:meanpt}
\begin{ruledtabular}
\begin{tabular}{cr|c|c|c|c|c|c} 
System 	& Centrality & $\pi^{-}$ & $\pi^{+}$ & $K^{-}$ & $K^{+}$ & $\overline{p}$ & $p$ \\ \hline 
$pp$ 200~GeV	& min.~bias & $0.348\pm0.018$ & $0.348\pm0.018$ & $0.517\pm0.030$ & $0.517\pm0.030$ & $0.683\pm0.041$ & $0.686\pm0.041$ \\
\hline
	& min.~bias & $0.367\pm0.027$ & $0.369\pm0.027$ & $0.599\pm0.068$ & $0.599\pm0.068$ & $0.847\pm0.090$ & $0.847\pm0.093$ \\
d+Au	& 40-100\% & $0.359\pm0.024$ & $0.364\pm0.025$ & $0.582\pm0.071$ & $0.582\pm0.071$ & $0.816\pm0.085$ & $0.817\pm0.087$ \\
200~GeV	& 20-40\% & $0.363\pm0.031$ & $0.370\pm0.031$ & $0.623\pm0.085$ & $0.623\pm0.085$ & $0.896\pm0.112$ & $0.895\pm0.116$ \\
	& 0-20\% & $0.378\pm0.028$ & $0.378\pm0.028$ & $0.607\pm0.061$ & $0.608\pm0.061$ & $0.855\pm0.081$ & $0.855\pm0.085$ \\
\hline
	& 70-80\% & $0.363\pm0.018$ & $0.367\pm0.018$ & $0.550\pm0.035$ & $0.553\pm0.035$ & $0.746\pm0.049$ & $0.749\pm0.049$ \\
	& 60-70\% & $0.377\pm0.019$ & $0.377\pm0.019$ & $0.583\pm0.033$ & $0.583\pm0.033$ & $0.814\pm0.047$ & $0.817\pm0.047$ \\
	& 50-60\% & $0.389\pm0.020$ & $0.389\pm0.020$ & $0.609\pm0.036$ & $0.608\pm0.036$ & $0.863\pm0.052$ & $0.864\pm0.052$ \\
Au+Au	& 40-50\% & $0.395\pm0.020$ & $0.395\pm0.020$ & $0.619\pm0.037$ & $0.619\pm0.037$ & $0.895\pm0.055$ & $0.897\pm0.055$ \\
	& 30-40\% & $0.402\pm0.021$ & $0.404\pm0.021$ & $0.643\pm0.042$ & $0.643\pm0.042$ & $0.939\pm0.062$ & $0.939\pm0.062$ \\
200~GeV	& 20-30\% & $0.408\pm0.021$ & $0.411\pm0.021$ & $0.668\pm0.047$ & $0.668\pm0.047$ & $0.989\pm0.071$ & $0.989\pm0.071$ \\
	& 10-20\% & $0.416\pm0.021$ & $0.421\pm0.021$ & $0.680\pm0.055$ & $0.681\pm0.055$ & $1.017\pm0.082$ & $1.017\pm0.082$ \\
	& 5-10\% & $0.418\pm0.021$ & $0.422\pm0.021$ & $0.704\pm0.064$ & $0.703\pm0.064$ & $1.070\pm0.098$ & $1.071\pm0.098$ \\
	& 0-5\% & $0.422\pm0.022$ & $0.427\pm0.022$ & $0.719\pm0.074$ & $0.720\pm0.074$ & $1.103\pm0.114$ & $1.104\pm0.110$ \\
\hline
	& 58-85\% & $0.355\pm0.036$ & $0.351\pm0.035$ & $0.559\pm0.020$ & $0.560\pm0.020$ & $0.745\pm0.030$ & $0.745\pm0.030$ \\
	& 45-58\% & $0.366\pm0.020$ & $0.360\pm0.020$ & $0.576\pm0.030$ & $0.571\pm0.030$ & $0.808\pm0.054$ & $0.808\pm0.054$ \\
	& 34-45\% & $0.375\pm0.014$ & $0.375\pm0.014$ & $0.598\pm0.048$ & $0.604\pm0.048$ & $0.869\pm0.053$ & $0.871\pm0.053$ \\
Au+Au	& 26-34\% & $0.382\pm0.020$ & $0.383\pm0.020$ & $0.628\pm0.049$ & $0.633\pm0.049$ & $0.925\pm0.066$ & $0.926\pm0.066$ \\
130~GeV	& 18-26\% & $0.386\pm0.020$ & $0.388\pm0.020$ & $0.644\pm0.046$ & $0.640\pm0.046$ & $0.942\pm0.067$ & $0.944\pm0.067$ \\
	& 11-18\% & $0.391\pm0.023$ & $0.395\pm0.023$ & $0.650\pm0.036$ & $0.649\pm0.036$ & $0.949\pm0.085$ & $0.949\pm0.085$ \\
	& 6-11\% & $0.390\pm0.011$ & $0.393\pm0.011$ & $0.640\pm0.034$ & $0.642\pm0.034$ & $0.965\pm0.078$ & $0.966\pm0.078$ \\
	& 0-6\% & $0.404\pm0.013$ & $0.404\pm0.013$ & $0.667\pm0.030$ & $0.666\pm0.030$ & $1.002\pm0.087$ & $1.003\pm0.087$ \\
\hline
	& 70-80\% & $0.357\pm0.021$ & $0.356\pm0.021$ & $0.529\pm0.023$ & $0.531\pm0.023$ & $0.702\pm0.044$ & $0.706\pm0.045$ \\
	& 60-70\% & $0.372\pm0.019$ & $0.364\pm0.019$ & $0.542\pm0.015$ & $0.542\pm0.015$ & $0.728\pm0.028$ & $0.729\pm0.031$ \\
	& 50-60\% & $0.381\pm0.018$ & $0.379\pm0.018$ & $0.560\pm0.022$ & $0.560\pm0.022$ & $0.759\pm0.042$ & $0.761\pm0.046$ \\
Au+Au	& 40-50\% & $0.385\pm0.017$ & $0.385\pm0.017$ & $0.584\pm0.020$ & $0.583\pm0.020$ & $0.812\pm0.042$ & $0.814\pm0.049$ \\
	& 30-40\% & $0.395\pm0.015$ & $0.394\pm0.015$ & $0.607\pm0.021$ & $0.607\pm0.021$ & $0.864\pm0.053$ & $0.864\pm0.061$ \\
62.4~GeV	& 20-30\% & $0.400\pm0.012$ & $0.403\pm0.013$ & $0.629\pm0.023$ & $0.629\pm0.023$ & $0.913\pm0.060$ & $0.910\pm0.070$ \\
	& 10-20\% & $0.402\pm0.014$ & $0.402\pm0.014$ & $0.636\pm0.029$ & $0.636\pm0.029$ & $0.928\pm0.031$ & $0.925\pm0.050$ \\
	& 5-10\% & $0.404\pm0.010$ & $0.407\pm0.011$ & $0.644\pm0.027$ & $0.643\pm0.027$ & $0.950\pm0.040$ & $0.948\pm0.059$ \\
	& 0-5\% & $0.403\pm0.011$ & $0.406\pm0.011$ & $0.645\pm0.029$ & $0.646\pm0.029$ & $0.959\pm0.060$ & $0.956\pm0.075$ \\
\end{tabular} 
\end{ruledtabular}
\end{table*}

\begin{table*}{}
\caption{Integrated multiplicity rapidity density, $dN/dy$, of identified particles and net-protons for various collision systems and centralities. Quoted errors are the quadratic sum of statistical and systematic uncertainties, and are dominated by the latter.}
\label{tab:dndy}
\begin{ruledtabular}
\begin{tabular}{cr|c|c|c|c|c|c|c}
System & Centrality & $\pi^{-}$ & $\pi^{+}$ & $K^{-}$ & $K^{+}$ & $\overline{p}$ & $p$ & $p - \pbar$ \\ \hline 
$pp$ 200~GeV 	& min.~bias	& $1.42\pm0.11$ & $1.44\pm0.11$ & $0.145\pm0.013$ & $0.150\pm0.013$ & $0.113\pm0.010$ & $0.138\pm0.012$ & $0.025\pm0.004$	\\
\hline
	& min.~bias & $4.63\pm0.31$ & $4.62\pm0.31$ & $0.582\pm0.052$ & $0.595\pm0.054$ & $0.412\pm0.053$ & $0.500\pm0.069$ & $0.088\pm0.029$ \\
d+Au	& 40-100\% & $2.89\pm0.20$ & $2.87\pm0.21$ & $0.348\pm0.032$ & $0.356\pm0.033$ & $0.236\pm0.030$ & $0.281\pm0.039$ & $0.045\pm0.018$ \\
200~GeV	& 20-40\% & $6.06\pm0.41$ & $6.01\pm0.41$ & $0.783\pm0.085$ & $0.803\pm0.087$ & $0.569\pm0.082$ & $0.72\pm0.11$ & $0.154\pm0.050$ \\
	& 0-20\% & $8.42\pm0.57$ & $8.49\pm0.58$ & $1.09\pm0.09$ & $1.11\pm0.10$ & $0.793\pm0.087$ & $0.95\pm0.11$ & $0.159\pm0.049$ \\
\hline
	&70-80\%	& $10.9\pm0.8$ & $10.8\pm0.8$ & $1.38\pm0.13$ & $1.41\pm0.13$ & $0.915\pm0.081$ & $1.09\pm0.10$ & $0.170\pm0.030$ \\
	&60-70\%	& $21.1\pm1.6$ & $21.1\pm1.6$ & $2.89\pm0.26$ & $2.98\pm0.27$ & $1.84\pm0.16$ & $2.20\pm0.20$ & $0.361\pm0.061$ \\
	&50-60\%	& $36.3\pm2.8$ & $36.2\pm2.7$ & $5.19\pm0.47$ & $5.40\pm0.49$ & $3.16\pm0.29$ & $3.88\pm0.35$ & $0.72\pm0.11$ \\
Au+Au	&40-50\%	& $58.9\pm4.5$ & $58.7\pm4.5$ & $8.37\pm0.78$ & $8.69\pm0.81$ & $4.93\pm0.46$ & $6.17\pm0.57$ & $1.24\pm0.18$ \\
	&30-40\%	& $89.6\pm6.8$ & $89.2\pm6.8$ & $13.2\pm1.3$ & $13.6\pm1.3$ & $7.46\pm0.72$ & $9.30\pm0.89$ & $1.85\pm0.30$ \\
200~GeV	&20-30\%	& $136\pm10$ & $135\pm10$ & $19.7\pm2.0$ & $20.5\pm2.0$ & $11.2\pm1.1$ & $14.4\pm1.4$ & $3.22\pm0.51$ \\
	&10-20\%	& $196\pm15$ & $194\pm15$ & $28.7\pm3.1$ & $30.0\pm3.2$ & $15.7\pm1.7$ & $20.1\pm2.2$ & $4.42\pm0.77$ \\
	& 5-10\%	& $261\pm20$ & $257\pm20$ & $39.8\pm4.6$ & $40.8\pm4.7$ & $21.4\pm2.5$ & $28.2\pm3.3$ & $6.8\pm1.3$ \\
	& 0- 5\%	& $327\pm25$ & $322\pm25$ & $49.5\pm6.2$ & $51.3\pm6.5$ & $26.7\pm3.4$ & $34.7\pm4.4$ & $8.0\pm1.8$ \\
\hline
	& 58-85\% & $16.0\pm2.1$ & $16.0\pm1.9$ & $2.23\pm0.14$ & $2.31\pm0.15$ & $1.31\pm0.09$ & $1.65\pm0.11$ & $0.347\pm0.040$ \\
	& 45-58\% & $42.4\pm3.5$ & $42.2\pm3.5$ & $5.81\pm0.41$ & $6.83\pm0.48$ & $3.33\pm0.30$ & $4.38\pm0.39$ & $1.05\pm0.14$ \\
	& 34-45\% & $70.9\pm4.9$ & $71.8\pm5.0$ & $10.1\pm0.9$ & $11.2\pm1.0$ & $5.51\pm0.45$ & $7.35\pm0.60$ & $1.85\pm0.20$ \\
Au+Au	& 26-34\% & $104\pm8$ & $103\pm8$ & $15.0\pm1.3$ & $16.4\pm1.4$ & $8.02\pm0.81$ & $10.9\pm1.1$ & $2.91\pm0.35$ \\
130~GeV	& 18-26\% & $140\pm11$ & $140\pm11$ & $20.5\pm1.8$ & $22.3\pm1.9$ & $10.5\pm1.0$ & $14.4\pm1.3$ & $3.94\pm0.41$ \\
	& 11-18\% & $187\pm16$ & $186\pm16$ & $26.6\pm1.9$ & $29.0\pm2.1$ & $12.8\pm1.6$ & $17.9\pm2.2$ & $5.09\pm0.70$ \\
	& 6-11\% & $228\pm16$ & $228\pm16$ & $33.1\pm2.4$ & $35.6\pm2.6$ & $15.7\pm1.6$ & $21.9\pm2.3$ & $6.25\pm0.75$ \\
	& 0-6\% & $280\pm20$ & $278\pm20$ & $42.7\pm2.8$ & $46.3\pm3.0$ & $20.0\pm2.2$ & $28.2\pm3.1$ & $8.24\pm0.93$ \\
\hline
	& 70-80\% & $7.43\pm0.62$ & $7.34\pm0.62$ & $0.813\pm0.055$ & $0.868\pm0.058$ & $0.464\pm0.047$ & $0.745\pm0.086$ & $0.280\pm0.050$ \\
	& 60-70\% & $14.7\pm1.3$ & $14.8\pm1.3$ & $1.74\pm0.12$ & $1.95\pm0.13$ & $0.960\pm0.059$ & $1.60\pm0.12$ & $0.639\pm0.078$ \\
	& 50-60\% & $26.8\pm2.4$ & $26.5\pm2.3$ & $3.31\pm0.23$ & $3.64\pm0.25$ & $1.68\pm0.12$ & $2.98\pm0.22$ & $1.30\pm0.11$ \\
Au+Au	& 40-50\% & $43.7\pm3.5$ & $43.2\pm3.5$ & $5.68\pm0.39$ & $6.62\pm0.46$ & $2.77\pm0.19$ & $5.07\pm0.36$ & $2.30\pm0.19$ \\
	& 30-40\% & $67.4\pm5.2$ & $66.5\pm5.1$ & $8.89\pm0.62$ & $10.4\pm0.7$ & $4.27\pm0.35$ & $8.08\pm0.67$ & $3.81\pm0.33$ \\
62.4~GeV	& 20-30\% & $101\pm7$ & $98.9\pm6.9$ & $14.0\pm1.0$ & $15.9\pm1.1$ & $6.39\pm0.55$ & $12.2\pm1.1$ & $5.86\pm0.52$ \\
	& 10-20\% & $146\pm11$ & $144\pm11$ & $19.8\pm1.4$ & $23.0\pm1.6$ & $8.77\pm0.78$ & $17.8\pm1.6$ & $9.07\pm0.85$ \\
	& 5-10\% & $192\pm13$ & $191\pm13$ & $27.2\pm1.9$ & $31.2\pm2.2$ & $11.4\pm1.1$ & $23.8\pm2.4$ & $12.4\pm1.3$ \\
	& 0-5\% & $237\pm17$ & $233\pm17$ & $32.4\pm2.3$ & $37.6\pm2.7$ & $13.6\pm1.7$ & $29.0\pm3.8$ & $15.4\pm2.1$ \\
\end{tabular} 
\end{ruledtabular}
\end{table*}

\begin{table*}{}
\caption{Particle $dN/dy$ ratios for various collision systems and centralities. Quoted errors are the quadratic sum of statistical and systematic uncertainties, and are dominated by the latter (except some of the antiparticle-to-particle ratios).}
\label{tab:ratios}
\begin{ruledtabular}
\begin{tabular}{cr|c|c|c|c|c|c|c} 
System & Centrality & $\pi^{-}/\pi^{+}$ & $K^{-}/K^{+}$ & $\pbar/p$ & $K^{-}/\pi^{-}$ & $\pbar/\pi^{-}$ & $K^{+}/\pi^{+}$ & $p/\pi^{+}$ \\ \hline 
$pp$ 200~GeV	& min.~bias & $0.988\pm0.043$ & $0.967\pm0.040$ & $0.819\pm0.047$ & $0.102\pm0.008$ & $0.080\pm0.006$ & $0.104\pm0.008$ & $0.096\pm0.008$ \\
\hline
	& min.~bias & $1.003\pm0.031$ & $0.979\pm0.036$ & $0.824\pm0.061$ & $0.126\pm0.011$ & $0.089\pm0.011$ & $0.129\pm0.011$ & $0.108\pm0.015$ \\
d+Au	& 40-100\% & $1.008\pm0.042$ & $0.977\pm0.037$ & $0.841\pm0.067$ & $0.120\pm0.011$ & $0.082\pm0.010$ & $0.124\pm0.012$ & $0.098\pm0.014$ \\
200~GeV	& 20-40\% & $1.007\pm0.035$ & $0.976\pm0.041$ & $0.787\pm0.064$ & $0.129\pm0.014$ & $0.094\pm0.013$ & $0.134\pm0.014$ & $0.120\pm0.018$ \\
	& 0-20\% & $0.993\pm0.035$ & $0.982\pm0.036$ & $0.833\pm0.058$ & $0.130\pm0.011$ & $0.094\pm0.010$ & $0.131\pm0.011$ & $0.112\pm0.013$ \\
\hline
	& 70-80\% & $1.003\pm0.044$ & $0.981\pm0.049$ & $0.843\pm0.048$ & $0.127\pm0.010$ & $0.084\pm0.007$ & $0.130\pm0.011$ & $0.100\pm0.008$ \\
	& 60-70\% & $1.003\pm0.043$ & $0.971\pm0.040$ & $0.836\pm0.047$ & $0.137\pm0.011$ & $0.087\pm0.007$ & $0.141\pm0.011$ & $0.104\pm0.008$ \\
	& 50-60\% & $1.002\pm0.044$ & $0.961\pm0.040$ & $0.815\pm0.047$ & $0.143\pm0.011$ & $0.087\pm0.007$ & $0.149\pm0.012$ & $0.107\pm0.009$ \\
Au+Au	& 40-50\% & $1.003\pm0.044$ & $0.963\pm0.039$ & $0.799\pm0.046$ & $0.142\pm0.012$ & $0.084\pm0.007$ & $0.148\pm0.012$ & $0.105\pm0.009$ \\
	& 30-40\% & $1.005\pm0.045$ & $0.969\pm0.040$ & $0.801\pm0.047$ & $0.147\pm0.013$ & $0.083\pm0.007$ & $0.152\pm0.013$ & $0.104\pm0.009$ \\
200~GeV	& 20-30\% & $1.008\pm0.046$ & $0.961\pm0.039$ & $0.777\pm0.047$ & $0.145\pm0.013$ & $0.082\pm0.008$ & $0.152\pm0.014$ & $0.107\pm0.010$ \\
	& 10-20\% & $1.012\pm0.049$ & $0.959\pm0.041$ & $0.780\pm0.048$ & $0.147\pm0.015$ & $0.080\pm0.008$ & $0.155\pm0.016$ & $0.104\pm0.011$ \\
	& 5-10\% & $1.014\pm0.050$ & $0.975\pm0.046$ & $0.759\pm0.051$ & $0.153\pm0.017$ & $0.082\pm0.009$ & $0.159\pm0.017$ & $0.110\pm0.012$ \\
	& 0-5\% & $1.015\pm0.051$ & $0.965\pm0.048$ & $0.769\pm0.055$ & $0.151\pm0.018$ & $0.082\pm0.010$ & $0.159\pm0.019$ & $0.108\pm0.013$ \\
\hline
	& 58-85\% & $0.996\pm0.066$ & $0.963\pm0.050$ & $0.790\pm0.043$ & $0.140\pm0.018$ & $0.082\pm0.010$ & $0.144\pm0.016$ & $0.103\pm0.012$ \\
	& 45-58\% & $1.004\pm0.040$ & $0.850\pm0.047$ & $0.760\pm0.043$ & $0.137\pm0.011$ & $0.078\pm0.008$ & $0.162\pm0.013$ & $0.104\pm0.010$ \\
	& 34-45\% & $0.988\pm0.037$ & $0.900\pm0.044$ & $0.749\pm0.040$ & $0.142\pm0.013$ & $0.078\pm0.006$ & $0.156\pm0.014$ & $0.102\pm0.008$ \\
Au+Au	& 26-34\% & $1.003\pm0.039$ & $0.912\pm0.045$ & $0.734\pm0.039$ & $0.145\pm0.014$ & $0.077\pm0.008$ & $0.159\pm0.015$ & $0.106\pm0.011$ \\
130~GeV	& 18-26\% & $1.002\pm0.037$ & $0.920\pm0.045$ & $0.727\pm0.038$ & $0.146\pm0.014$ & $0.075\pm0.007$ & $0.159\pm0.015$ & $0.103\pm0.010$ \\
	& 11-18\% & $1.003\pm0.037$ & $0.915\pm0.046$ & $0.716\pm0.039$ & $0.142\pm0.012$ & $0.069\pm0.009$ & $0.156\pm0.013$ & $0.096\pm0.013$ \\
	& 6-11\% & $1.003\pm0.043$ & $0.929\pm0.045$ & $0.715\pm0.039$ & $0.145\pm0.010$ & $0.069\pm0.007$ & $0.156\pm0.011$ & $0.096\pm0.010$ \\
	& 0-6\% & $1.008\pm0.029$ & $0.923\pm0.037$ & $0.708\pm0.036$ & $0.153\pm0.010$ & $0.071\pm0.008$ & $0.167\pm0.011$ & $0.101\pm0.011$ \\
\hline
	& 70-80\% & $1.012\pm0.031$ & $0.936\pm0.036$ & $0.623\pm0.047$ & $0.109\pm0.009$ & $0.063\pm0.007$ & $0.118\pm0.010$ & $0.101\pm0.013$ \\
	& 60-70\% & $0.990\pm0.031$ & $0.894\pm0.037$ & $0.600\pm0.039$ & $0.119\pm0.010$ & $0.065\pm0.005$ & $0.132\pm0.011$ & $0.108\pm0.010$ \\
	& 50-60\% & $1.011\pm0.032$ & $0.907\pm0.038$ & $0.563\pm0.031$ & $0.123\pm0.011$ & $0.063\pm0.006$ & $0.137\pm0.012$ & $0.113\pm0.010$ \\
Au+Au	& 40-50\% & $1.012\pm0.032$ & $0.858\pm0.036$ & $0.546\pm0.030$ & $0.130\pm0.010$ & $0.063\pm0.005$ & $0.153\pm0.012$ & $0.117\pm0.009$ \\
	& 30-40\% & $1.014\pm0.033$ & $0.854\pm0.036$ & $0.529\pm0.028$ & $0.132\pm0.010$ & $0.063\pm0.006$ & $0.156\pm0.012$ & $0.121\pm0.011$ \\
62.4~GeV	& 20-30\% & $1.023\pm0.034$ & $0.883\pm0.036$ & $0.522\pm0.027$ & $0.138\pm0.010$ & $0.063\pm0.005$ & $0.160\pm0.011$ & $0.124\pm0.011$ \\
	& 10-20\% & $1.013\pm0.033$ & $0.862\pm0.037$ & $0.492\pm0.026$ & $0.136\pm0.010$ & $0.060\pm0.006$ & $0.160\pm0.012$ & $0.124\pm0.012$ \\
	& 5-10\% & $1.007\pm0.033$ & $0.870\pm0.036$ & $0.481\pm0.026$ & $0.141\pm0.010$ & $0.060\pm0.006$ & $0.164\pm0.011$ & $0.125\pm0.012$ \\
	& 0-5\% & $1.018\pm0.033$ & $0.860\pm0.035$ & $0.469\pm0.026$ & $0.137\pm0.010$ & $0.057\pm0.007$ & $0.162\pm0.012$ & $0.125\pm0.016$ \\
\end{tabular} 
\end{ruledtabular}
\end{table*}

\subsection{Total Particle Production\label{sec:total_dndy}}

The total particle multiplicity reflects the total entropy generated in the collision system. There has been renewed interest in total particle production as its centrality dependence could distinguish between different models of particle production~\cite{Wang00}. Models based on the assumption of final-state gluon saturation advocate a decrease of the charged particle multiplicities per participant nucleon with increasing centrality. For example, the EKRT model~\cite{EKRT00} parameterizes the multiplicity rapidity density as 
\begin{equation}
\left.\frac{d\Nch}{d\eta}\right|_{b=0}=C\ \frac{2}{3}\ 1.16\ \left(\frac{\Npart}{2}\right)^{0.92}(\sqrt{s})^{0.40}\ .\label{eq:ekrt}
\end{equation}

Models based on initial state gluon saturation (e.g. the color glass condensate model~\cite{McLerran1,McLerran2}) or pQCD inspired models (e.g. the HIJING~\cite{Hijing,Hijing91} or the soft/hard scattering model used in Ref.~\cite{dima01}) predict an increase of the rapidity density per participant nucleon with centrality. In both the HIJING and the soft/hard model, particle production arises from two major contributions: (a) a soft component scaling with the number of participants $\Npart$, and (b) a hard component from mini-jet production, which is directly proportional to the number of binary collisions $\Ncoll$ and the average inclusive jet cross-section. Reference~\cite{dima01} expresses these two components as 
\begin{equation}
d\Nch/d\eta=(1-x_{\rm hard})\ \npp\ \frac{\Npart}{2}+x_{\rm hard}\npp \Ncoll \label{eq:KN}
\end{equation}
where $x_{\rm hard}$ is the fraction of hard collisions. The basic assumption here is that the average particle multiplicity produced per hard process in heavy-ion collisions is identical to that in $pp$ collisions. In Eq.~(\ref{eq:KN}), $\npp$ is the charged particle pseudo-rapidity density in NSD $pp$ interactions. We have measured $\npp$ in $pp$ collisions only at 200~GeV. In order to apply the two-component model to data at other energies, we use a parameterization from $p\bar{p}$ measurements~\cite{UA5_dNdeta1,UA5_dNdeta2,UA5_dNdeta3,pbarp90} by
\begin{equation}
\npp=(2.5\pm1.0)-(0.25\pm0.19)\ln(s)+(0.023\pm0.008)\ln^2(s) \label{eq:npp}
\end{equation}
where $s$ is the squared center-of-mass energy in GeV$^2$. The parameterized value of $\npp=2.43$ at 200~GeV differs from our measurement in $pp$ collisions because of the numerical difference between our measured NSD cross-section of $30.0\pm3.5$~mb~\cite{highpt200} and the measurement in Ref.~\cite{NSD_UA1} of $35\pm1$~mb.

In the following, we call Eq.~(\ref{eq:ekrt}) and~(\ref{eq:KN}) the EKRT and K-N parameterizations, respectively, and use them to study the discrimination power of our data against the two opposing models of particle production. Unfortunately, neither $\Npart$ nor $\Ncoll$ can be directly measured in the experiment. They can only be derived by calculating the nuclear overlap integral with the help of the Glauber model. However, two different implementations of the Glauber calculation, the optical and the MC Glauber calculations, lead to different values of $\Npart$ and $\Ncoll$ with rather large uncertainties for peripheral collisions (for details see Appendix~\ref{app:Glauber}).

Figure \ref{fig:dNchdEtaNpart} shows the pseudo-rapidity multiplicity density per participant pair, $\frac{\dNdeta}{\Npart/2}$, versus the number of participants $\Npart$ for Au+Au collisions at 62.4 and 200~GeV, where we have used $\Npart$ and $\Ncoll$ from the optical Glauber calculation in the left panel and the MC Glauber calculation in the right panel. The $\dNdeta$ data are from Table~\ref{tab:collision_prop}. In both panels, the vertical error bars represent the quadratic sum of the systematic uncertainties on $\dNdeta$ and $\Npart$. The latter dominates the uncertainties for peripheral collisions. 

\begin{figure*}[thbp]
\centering
\includegraphics[width=0.95\textwidth]{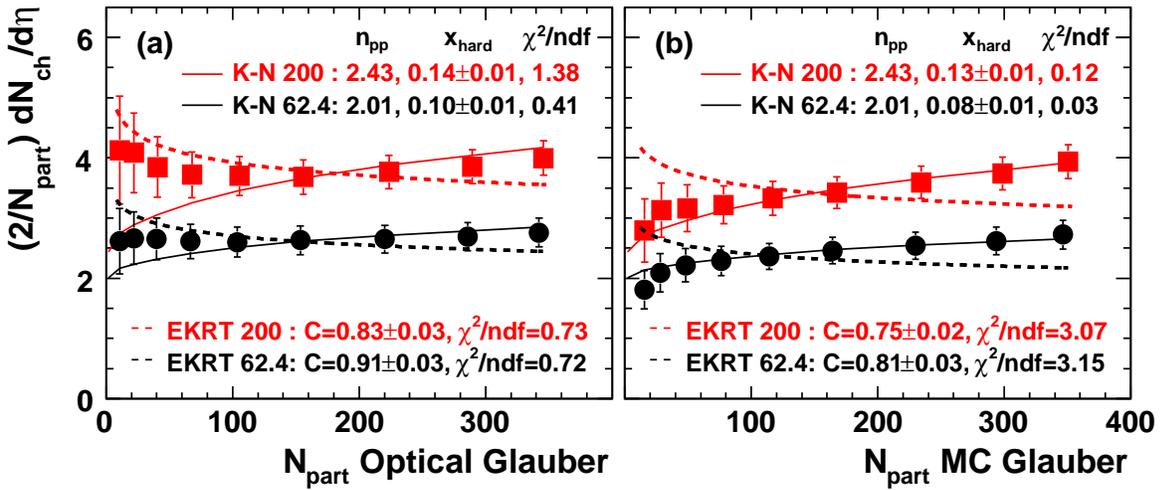}
\caption{(color online) The pseudo-rapidity multiplicity density per participant nucleon pair $\frac{\dNdeta}{\Npart/2}$ versus the number of participants $\Npart$, with $\Npart$ calculated from the optical Glauber model (a) and from the MC Glauber model (b). Data are presented for Au+Au collisions at 62.4~GeV (black dots) and 200~GeV (red squares). The vertical errors are total uncertainties including uncertainties on $\Npart$. The uncertainties on $\Npart$ (horizontal error bars) are smaller than the data point size. The solid curves are the K-N fit by Eq.~(\ref{eq:KN}) where $x_{\rm hard}$ is a fit parameter and $\npp$ is fixed from Eq.~(\ref{eq:npp}). The dashed curves are the EKRT fit by Eq.~(\ref{eq:ekrt}) where $C$ is a fit parameter.}
\label{fig:dNchdEtaNpart}
\end{figure*}

As seen in Fig.~\ref{fig:dNchdEtaNpart}(a) (using the optical Glauber calculation), we observe no significant change in charged hadron production as a function of centrality within the large uncertainties (mainly from the optical Glauber calculations). Superimposed for comparison are the EKRT and K-N parameterizations in the dashed and solid curves, respectively. The EKRT parameterization is obtained from the best fit to the data by Eq.(\ref{eq:ekrt}), treating $C$ as the single fit parameter. The K-N parameterization is obtained from the best fit to the data by Eq.(\ref{eq:KN}), treating $\npp$ as fixed from Eq.(\ref{eq:npp}) and $x_{\rm hard}$ as the single fit parameter. Neither our data nor the EKRT parameterization seems to approach the parameterized $\npp$ by Eq.(\ref{eq:npp}) in the limit of $\Npart=2$. The K-N parameterization recovers $\npp$ for $\Npart=2$ by construction of the model. Both models do a modest job in describing the data.

When using the MC Glauber model to evaluate $\Npart$ and $\Ncoll$ as done in Fig.~\ref{fig:dNchdEtaNpart}(b), our data clearly exhibit a centrality dependence rising from the most peripheral to the most central collisions, by about $(50\pm20)$\% and $(40\pm20)$\% for 62.4~GeV and 200~GeV, respectively. The data are fit by Eq.~(\ref{eq:ekrt}) treating $C$ as the single fit parameter. The obtained EKRT parameterizations, shown in the dashed curves, clearly fail to describe our data due to the opposite centrality dependence. The fit $\chi^2$/{\sc ndf} is printed on the plot and is fairly large, especially considering that the systematic uncertainties are included in the fit as random errors. On the other hand, shown in the solid curves are the K-N parameterizations obtained from fitting Eq.~(\ref{eq:KN}) to the data fixing $\npp$ by Eq.(\ref{eq:npp}) and treating $x_{\rm hard}$ as the single fit parameter. As can be seen, the K-N parameterization fits the data better. We obtain the fit fraction of hard collisions to be $x_{\rm hard}=(7.8\pm1.3)\%$ and $(12.8\pm1.3)\%$ for Au+Au collisions at 62.4~GeV and 200~GeV, respectively. We may evaluate the fraction of produced particles originating from hard collisions, within the framework of the K-N two-component model, as
\begin{equation}
F_{\rm hard}=\frac{x_{\rm hard}\ \npp\ \Ncoll}{d\Nch/d\eta}\ ,
\end{equation}
yielding $F_{\rm hard}=(30\pm5)\%$ and $(46\pm5)\%$ for the top 5\% central Au+Au collisions at 62.4~GeV and 200~GeV, respectively.

In our K-N two-component model study, we have used the charged particle multiplicity from NSD $pp$ interactions in Eq.~\ref{eq:npp} and the Glauber model results calculated with the total $pp$ cross-section. This is because singly diffractive nucleon-nucleon interactions also contribute to the total charged particle multiplicities in Au+Au collisions. If we use instead the Glauber data of $\sigmapp=36$~mb from Table~\ref{tab:collision_prop} for 200~GeV, we obtain $x_{\rm hard}=(15\pm2)\%$.

It should be noted that the K-N two-component model assumes the same average particle multiplicity per hard process in $pp$ and Au+Au collisions. This assumption is likely invalid because jet-medium interactions induce a larger average multiplicity per hard process in Au+Au collisions with a softer energy distribution~\cite{jetspec}. The size of this effect is dependent on centrality. This relative increase in particle multiplicity from hard processes would result in an overestimate of the fraction of hard component, especially for $pp$ collisions. A two-component model study based on the multiplicity dependence of transverse rapidity spectra from $pp$ collisions, assuming most of the charged particles are pions, has revealed a significantly smaller fraction of hard component~\cite{trainor_pp}. It remains an open question how realistic the simple K-N two-component model is for heavy-ion collisions. An improved two-component model would be to use the total transverse energy instead of the total particle multiplicity as the total transverse energy likely remains the same with jet modification processes. However, such a model would need as input the total transverse energy in inelastic $pp$ collisions which is not well measured.

It is worth noting that the normalized pseudo-rapidity density $\frac{\dNdeta}{\Npart/2}$ in the EKRT parameterization has only the overall scale $C$ as a free parameter. The centrality dependence is fixed by $\Npart^{0.92}$. In the K-N parameterization, on the other hand, the overall scale is fixed by $\npp$, while the centrality dependence changes with the free parameter $x_{\rm hard}$. However, the $\npp$ value is obtained from parameterization to elementary collision data, and thus is designed to describe the overall scale of the heavy-ion data. As shown in Fig.~\ref{fig:dNchdEtaNpart}, due to the uncertainties from the Glauber calculations, we cannot explicitly rule out either of the models. However, recent developments in analyzing the small systems (Cu+Cu) indicate that MC Glauber model is preferred, albeit with its own caveats as mentioned before. This in turn favors the two-component model and initial state gluon saturation~\cite{GlauberMiller} over the EKRT model.

\subsection{Bjorken Energy Density Estimate\label{bjorken}}

The central rapidity region is approximately boost invariant~\cite{spec200}. Under boost invariance, the energy density of the central rapidity region in the collision zone at formation time $\tau$ can be estimated by the Bjorken energy density~\cite{Bj}:
\begin{equation}
\ebj=\frac{d\Et}{dy}\frac{1}{\Sovlp\tau}\;\;\;,\label{eq:Bj}
\end{equation}
where $\Et$ is the total transverse energy and $\Sovlp$ is the transverse overlap area of the colliding nuclei. Since we do not measure transverse energy, but only charged particle transverse momenta, we use the approximation
\begin{equation}
\frac{d\langle\Et\rangle}{dy}\approx\frac{3}{2}\left(\langle\mt\rangle\frac{dN}{dy}\right)_{\pi^\pm}+2\left(\langle\mt\rangle\frac{dN}{dy}\right)_{K^\pm,p,\bar{p}}\ .
\end{equation}
Here, we calculate $\langle\mt\rangle=\sqrt{\langle\pt\rangle^2+m^2}$ from the $\pi^\pm$, $K^\pm$, $p$, and $\bar{p}$ average transverse momenta presented in this work and in Refs.~\cite{spec200,kaon130,pbar130,p130}. The factors $3/2$ and 2 compensate for the neutral particles. Isospin effects are estimated to be less than 2\% and are neglected. Propagation of systematic uncertainties is done in the same way as for the total $\dNchdy$ discussed in Section~\ref{sec:dNdySystErr}, i.e. the extrapolation uncertainties are correlated between particle and antiparticle and uncorrelated between different particle species, and the overall reconstruction efficiency is correlated for all particle species. The uncertainties on the $\meanpt$ are not included because they come from extrapolation of the spectra, similar to those on the $\dNchdy$, and are already applied to the $\dNchdy$.

\begin{figure}[thtp]
\centering
\includegraphics[width=0.48\textwidth]{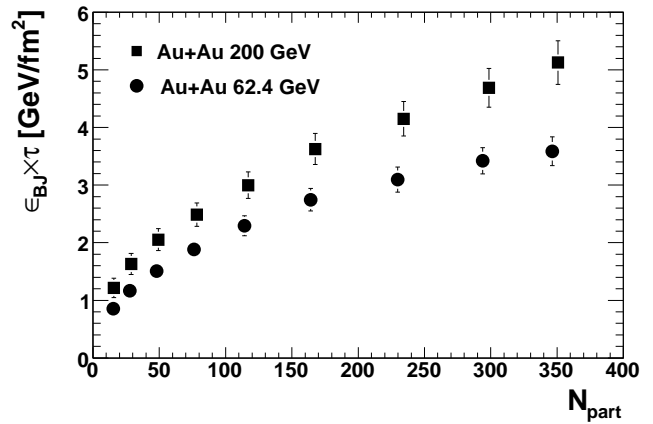}
\caption{Estimate of the product of the Bjorken energy density and the formation time ($\ebj\cdot\tau$) as a function of centrality $\Npart$. Errors shown are the quadratic sum of statistical and systematic uncertainties.}
\label{fig:Bj}
\end{figure}

Figure~\ref{fig:Bj} shows the product of the Bjorken energy density and the formation time as a function of $\Npart$. For the top 5\% central collisions, $\ebj\cdot\tau=3.7\pm0.3$~GeV/fm$^2$ at collision energy 62.4~GeV, $4.4\pm0.3$~GeV/fm$^2$ at 130~GeV (not shown), and $5.2\pm0.4$~GeV/fm$^2$ at 200~GeV. Our 130~GeV value is in good agreement with the value $\ebj\cdot\tau=4.6$~GeV/fm$^2$ quoted in Ref.~\cite{Adcox:2002uc} for the most central 2\% inelastic collisions. These estimated Bjorken energy densities are at least several~GeV/fm$^3$ with a formation time $\tau<1$~fm/$c$. They well exceed the phase transition energy density of 1~GeV/fm$^3$ predicted by Lattice QCD~\cite{Karsch02}.

At the top SPS energy, the formation time is traditionally taken as $\tau=1$~fm/$c$ resulting in $\ebj=3.2$~GeV/fm$^3$ for central Pb+Pb collisions~\cite{NA49Bj}. At RHIC, the choice of $\tau$ is still a matter of debate. While Ref.~\cite{Bass00} uses $\tau=0.6$~fm/$c$ for their hydrodynamic model ($\snn=200$~GeV), Ref.~ \cite{Wang02} uses $\tau=0.2$~fm/$c$, evaluated from the energy loss of high-$\pt$ $\pi^0$ in $\snn=130$~GeV Au+Au collisions. Because of these uncertainties in $\tau$, the Bjorken energy density estimate should be taken with caution, in addition to the assumptions of Bjorken longitudinal boost invariance and formation of a thermalized central region at an initial time $\tau$. It should be noted that due to final state interactions, measured (final) total transverse energies are expected to be less than initial ones~\cite{Eskola01}. 

\subsection{Antiparticle-to-Particle Ratios}

\begin{figure}[thbp]
\centering
\includegraphics[width=0.48\textwidth]{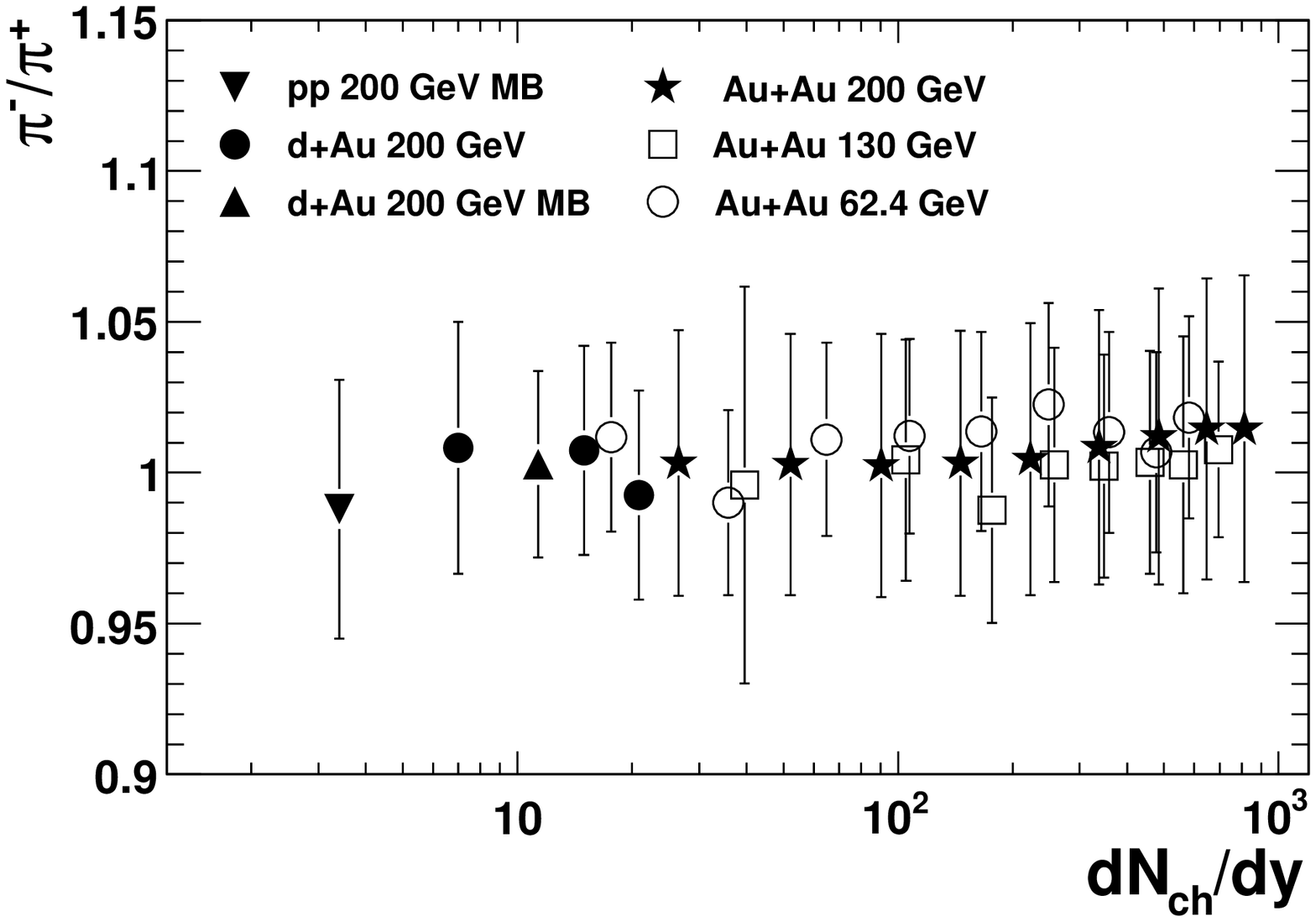}
\includegraphics[width=0.48\textwidth]{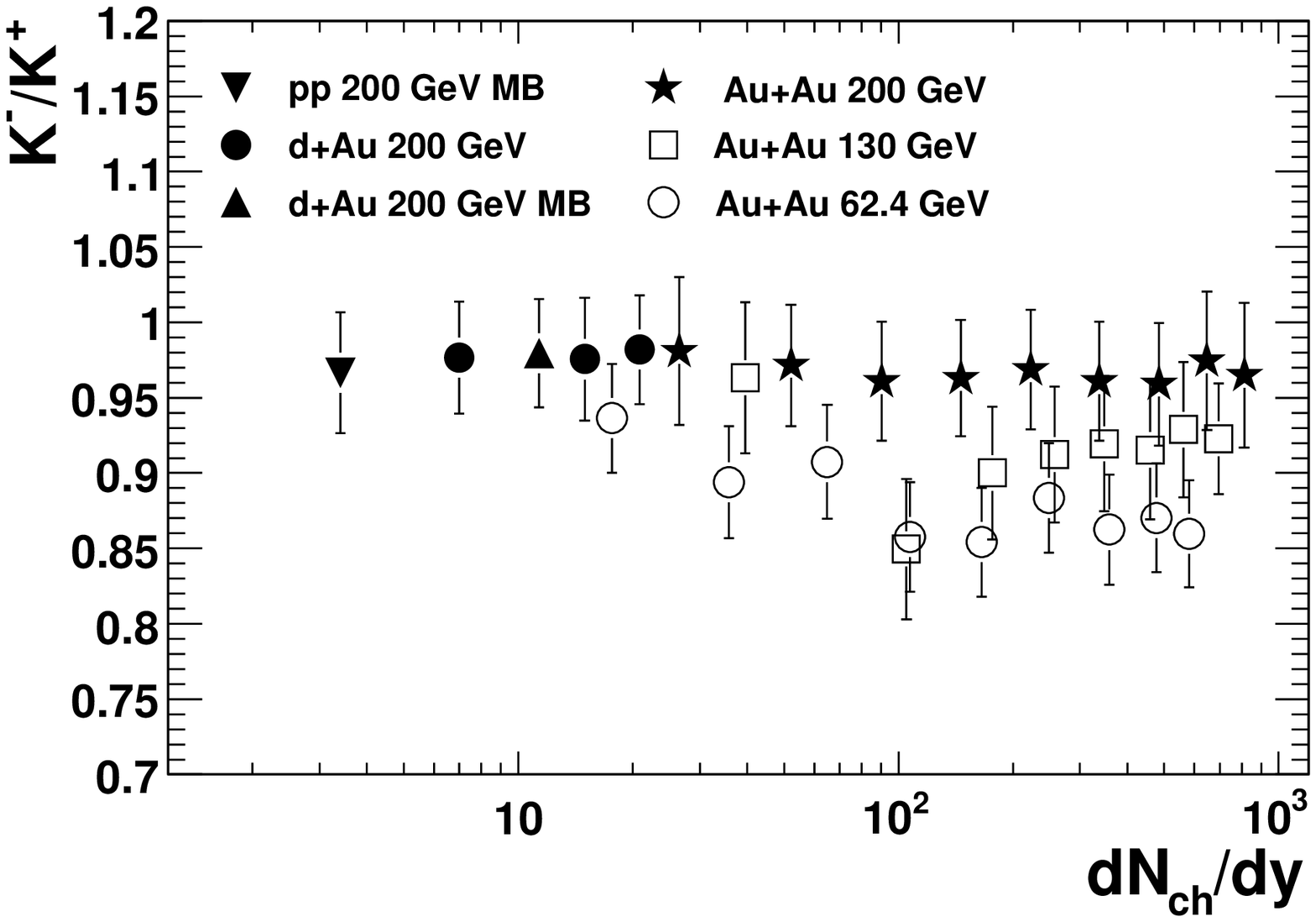}
\includegraphics[width=0.48\textwidth]{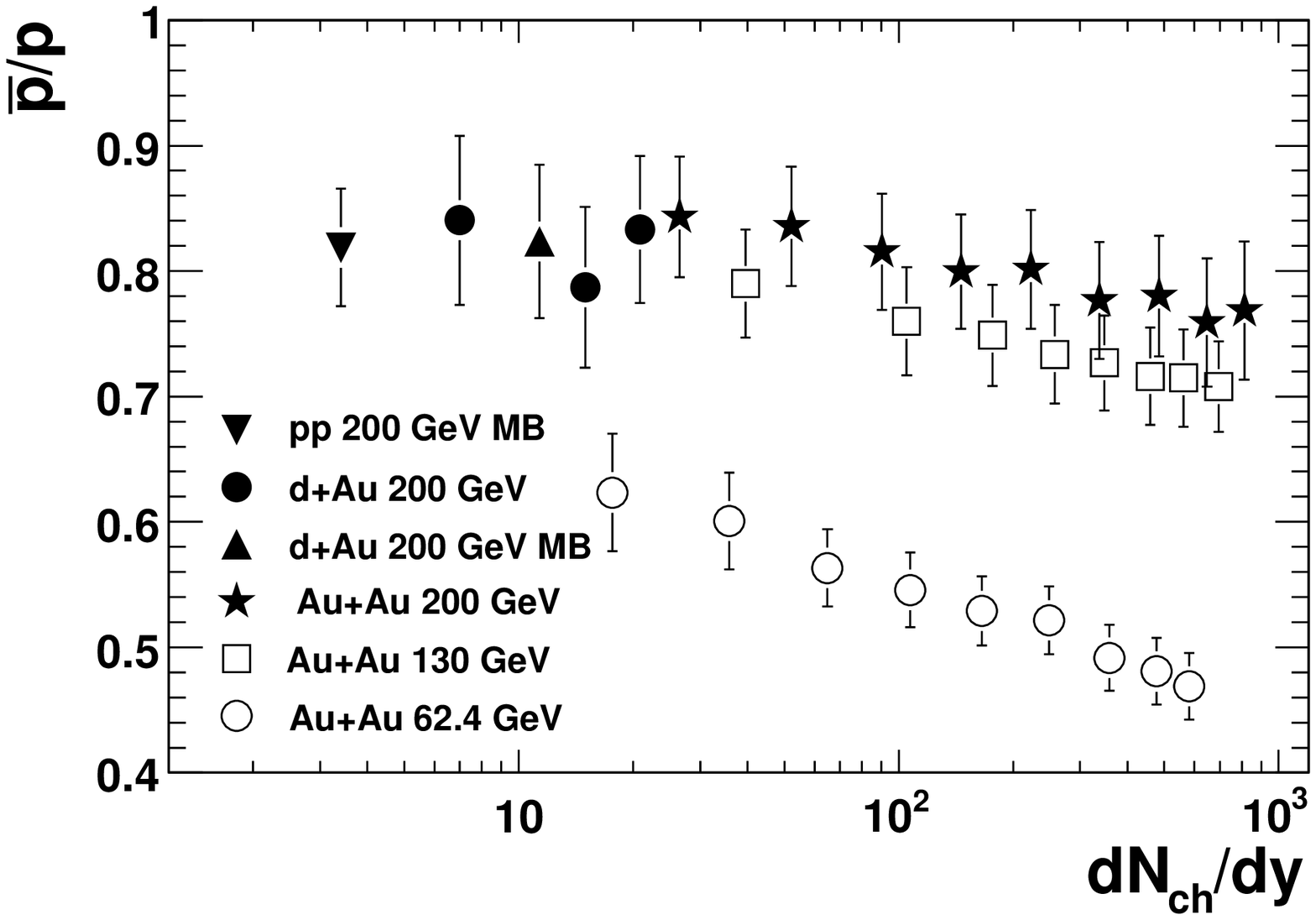}
\caption{Antiparticle-to-particle ratios as a function of $\dNchdy$ for $pp$ and d+Au collisions at 200~GeV and Au+Au collisions at 62.4~GeV, 130~GeV, and 200~GeV. Errors shown are the quadratic sum of statistical and systematic uncertainties.}
\label{fig:papratios}
\end{figure}

Relative particle production can be studied by particle ratios of the integrated $dN/dy$ yields. Figure~\ref{fig:papratios} shows the antiparticle-to-particle ratios ($\pi^{-}/\pi^{+}$, $K^-/K^+$, and $\pbar/p$) as a function of the charged particle multiplicity in $pp$, d+Au at 200~GeV and Au+Au collisions at 62.4~GeV, 130~GeV, and 200~GeV. The 200~GeV and some of the 130~GeV data have been presented before~\cite{spec200,pbarp130,STAR130ratio}. The $\pi^{-}/\pi^{+}$ ratio is approximately 1 for all measured collision systems and collision energies. The ratios are independent of multiplicity and centrality. Similar behavior has been observed at lower collision energies as well.

The $K^-/K^+$ ratios are close to 1 in $pp$, d+Au and Au+Au collisions at 200~GeV. The ratio decreases slightly from 200~GeV to 62.4~GeV Au+Au data. This may be due to the increasing net baryon density in the collision zone which leads to differences in associated production of kaons. There appears to be a decreasing trend with centrality in the 62.4~GeV data, presumably due to a significant increase in the net baryon density.

The $\pbar/p$ ratio appears to be independent of multiplicity in $pp$ and d+Au collisions at 200~GeV. The ratio in peripheral Au+Au at 200~GeV is similar to that in $pp$ and d+Au collisions at the same energy. A slight decrease is observed with increasing centrality in Au+Au collisions at 200~GeV and 130~GeV. The ratio is significantly lower at 62.4~GeV and shows a considerable drop with increasing centrality. The drop of the $\pbar/p$ ratio with increasing centrality is consistent with larger baryon stopping in central collisions.

\begin{figure}[htbp]
\centering
\includegraphics[width=0.48\textwidth]{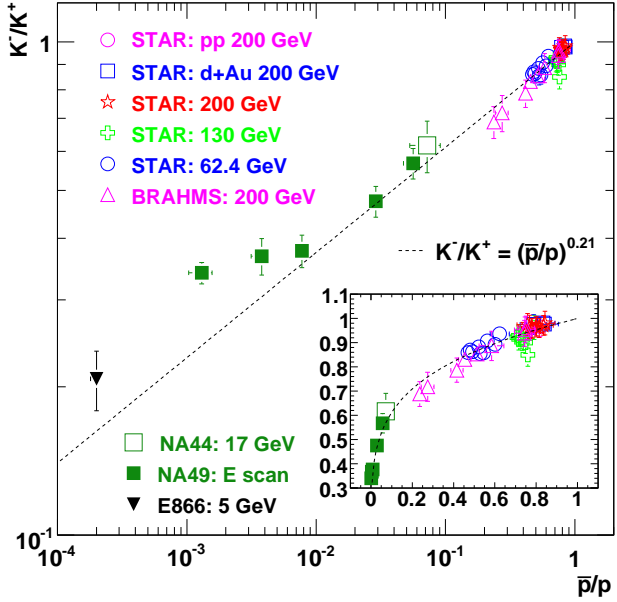}
\caption{(color online) Ratio of charged kaons versus that of antiprotons to protons at various energies. Errors shown are the quadratic sum of statistical and systematic uncertainties. The line is a power-law fit to the data except the AGS data point (the inverse solid triangle) and the two lowest energy SPS data points (solid squares) which have significant baryon absorption effect. The insert is a linear plot.}
\label{fig:kktopbarp}
\end{figure}

Figure~\ref{fig:kktopbarp} shows the $K^-/K^+$ ratio versus the $\pbar/p$ ratio, together with results from other energies~\cite{pbarpBRAHMS,pbar802,E866kaon,NA49horn,NA49spec,NA44spec}. Both ratios are affected by the net baryon content; they show a strong correlation as seen in Fig.~\ref{fig:kktopbarp}. This can be simply understood in the chemical equilibrium model where particle ratios are governed by only a few parameters. This aspect will be discussed in section~\ref{sec:Model}. It is worth to note that at low energies, the absorption of antiprotons in the baryon-rich environment plays a vital role.

\subsection{Baryon Production and Transport}

The antiproton is the lightest antibaryon. Most high mass antibaryons decay into antiprotons. The $\pbar/\pi^{-}$ ratio, therefore, characterizes well antibaryon production relative to total particle multiplicity. As mentioned earlier, the inclusive $\pbar$ yield reported here is the sum of the primordial $\pbar$ yield and the weak-decay contributions. Because all decay (anti)protons are measured in the data sample, the weak-decay contribution can be estimated as $0.64\cdot(\overline{\Lambda}+\overline{\Sigma}^{0}+\overline{\Xi}+\overline{\Omega}^{+})+0.52\cdot\overline{\Sigma}^{-}$. With the assumption of isospin symmetry with $\overline{n} \approx \pbar$ and $\overline{\Sigma}^{0} \approx \overline{\Sigma}^{+} \approx \overline{\Sigma}^{-}$, one may estimate the total antibaryon rapidity density to be approximately twice the measured antiproton rapidity density~\cite{pbar130}, and the total net-baryon density to be approximately twice the total net-proton density. The assumption of isospin symmetry is fairly good for Au+Au collisions, and should be good for $pp$ collisions at high energy because of the efficient charge exchange reactions to convert between protons and neutrons~\cite{Videbaek}.

\begin{figure}[htbp]
\centering
\includegraphics[width=0.48\textwidth]{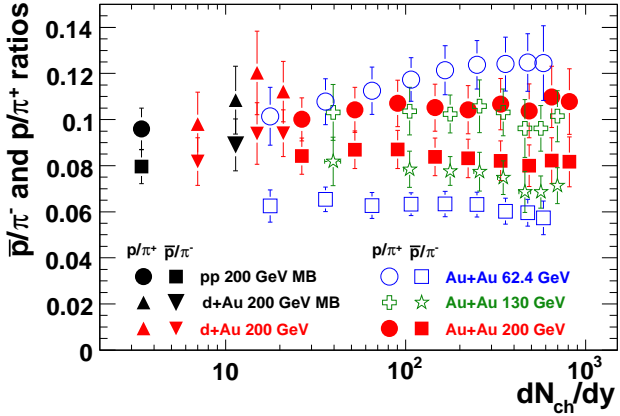}
\caption{(color online) The $p/\pi^+$ and $\pbar/\pi^-$ ratios as a function of the charged particle multiplicity in $pp$, d+Au and Au+Au collisions. Errors shown are the quadratic sum of statistical and systematic uncertainties.}
\label{fig:ppiratios}
\end{figure}

Figure~\ref{fig:ppiratios} shows the $\pbar/\pi^{-}$ ratio as a function of event multiplicity in $pp$, d+Au and Au+Au collisions. The ratio at 200~GeV is found to be independent of centrality and is the same for $pp$, d+Au, and Au+Au collisions within the experimental uncertainties. The values of the $\pbar/\pi^{-}$ ratio at 62.4~GeV are lower than those at 200~GeV at all centralities, indicating the significant effect of collision energy on the production of heavy particles even at these high energies. Although the net-baryon density increases with centrality, especially at 62.4~GeV with narrower rapidity gap between the beams, the $\pbar/\pi^{-}$ ratio does not seem to be affected much by the net-baryon density, suggesting that antibaryon absorption is not a significant effect at these energies. At the lower AGS and SPS energies the $\pbar/\pi^{-}$ has a much stronger decreasing trend with increasing centrality~\cite{WangAGS}; baryon stopping and the effect of net-baryon density are much stronger at low energies.

It has been argued that production of antibaryons, due to their large masses, is sensitive to energy density. An increased antibaryon production relative to total entropy with increasing centrality at the same collision energy could indicate formation of high energy density, or QGP in central collisions. On the hadronic level, at high pion density, multiple-pion fusion into baryon-antibaryon pairs could contribute significantly to the antibaryon yield~\cite{rapp}. Such an increase with centrality is not observed in data, but could be canceled by the effect of positive net-baryon density, resulting in antibaryon absorption. On the other hand, antibaryon production does increase with the collision energy. However, this cannot be taken as evidence of QGP formation as antibaryon production is very sensitive to the available energy for production due to their large mass. Indeed, antibaryon production in elementary collisions is found to be a sensitive function of the collision energy.

Figure~\ref{fig:ppiratios} also shows the $p/\pi^+$ ratio as a function of the charged particle multiplicity. The $p/\pi^+$ ratio is found to be constant over centrality at 130~GeV and 200~GeV, and shows an increasing trend with centrality at 62.4~GeV. The $p/\pi^+$ ratio is found to be the same in $pp$, d+Au, and Au+Au collisions at 200~GeV within our experimental uncertainties. Unlike antibaryons, baryons come from two sources: pair production together with antibaryons and transport from the initial colliding nuclei at beam rapidities. The latter can be obtained from the difference between baryon and antibaryon yields. Figure~\ref{fig:ppiratios} indicates a finite net-baryon number is present at mid-rapidity in all collisions. A finite baryon number has been transported over $\sim$ 3-5.4 units of rapidity in these collisions. How baryons are transported over many units of rapidity has been a long-standing theoretical issue~\cite{Btrans1,Btrans2,Btrans3}. Baryon transport occurs very early in the collision and affects the subsequent evolution of the collision system. Further understanding of baryon transport can shed more light on the evolution of heavy-ion collisions.

\begin{figure}[htbp]
\centering
\includegraphics[width=0.48\textwidth]{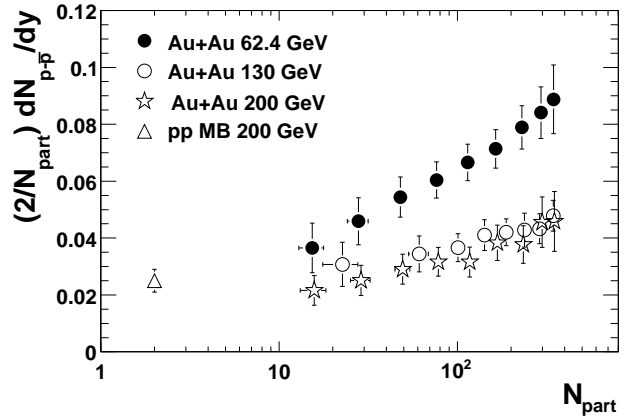}
\caption{The ratio of mid-rapidity net-protons to half of the number of participants versus the number of participants in $pp$ collisions at 200~GeV and in Au+Au collisions at 62.4~GeV, 130~GeV, and 200~GeV. Errors shown are the quadratic sum of statistical and systematic uncertainties.}
\label{fig:netproton_vs_Npart}
\end{figure}

Figure~\ref{fig:netproton_vs_Npart} shows the ratio of the number of net-protons ($p-\pbar$) to half the number of participant nucleons, i.e. the approximate probability of each incoming nucleon to be transported to mid-rapidity, as a function of $\Npart$. The probability is non-zero even in $pp$ collisions at 200~GeV. Compared to $pp$, the probability is larger in central heavy-ion collisions at the same energy by a factor $\sim2$. The probability of baryon transport to mid-rapidity is larger in the lower 62.4~GeV collisions, due to the smaller beam rapidity. 

Our data demonstrate that baryon-antibaryon pair production and baryon stopping are two independent processes: The baryon-antibaryon pair production rate does not depend on the collision centrality and increases with the collision energy, whereas the baryon stopping increases with the collision centrality and decreases with the collision energy. The net-baryon density due to baryon stopping may have an effect on the final observed yield of antibaryons because of absorption. However, this effect does not seem to be significant at our measured energies.

\begin{figure}[htbp]
\includegraphics[width=0.48\textwidth]{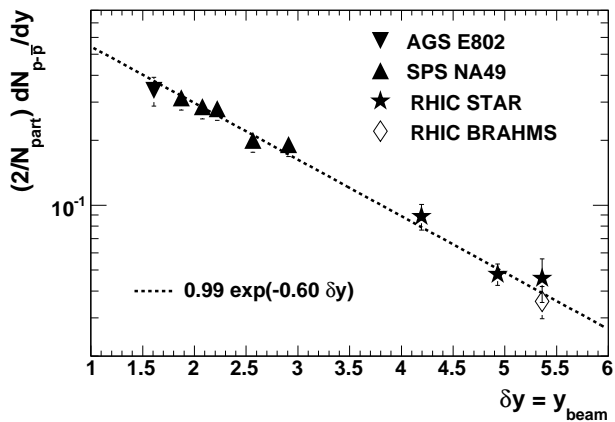}
\caption{The ratio of mid-rapidity inclusive net-protons to half of the number of participants in central heavy-ion collisions as a function of the rapidity shift. The AGS data is taken from Refs.~\cite{pbar802,E802netp2}, SPS data from Refs.~\cite{NA49netp,NA49netp2,NA49netp3}, and BRAHMS data from Ref.~\cite{BRAHMSnetp}. The published SPS data have already been corrected for weak-decays, the size of which is of the order 20-25\%~\cite{NA49netp2}, so we have added 25\% to the published net-proton yields to obtain the inclusive ones. Errors shown are total statistical and systematic uncertainties. The dashed line is an exponential fit to the data.}
\label{fig:netproton_vs_dy}
\end{figure}

Proton and antiproton production has been measured in heavy-ion collisions at lower energies. Figure~\ref{fig:netproton_vs_dy} shows the ratio of mid-rapidity inclusive net-proton density to half of the number of participants in central Au+Au collisions as a function of the beam rapidity (i.e. the rapidity shift suffered by those net-protons). The measured NA49 data have been corrected for weak-decays, which is dominated by weak-decay protons, the size of which is of the order 20-25\%~\cite{NA49netp2}. In order to obtain the inclusive net-proton yield, we multiplied the measured NA49 data by a factor 1.25. All other data are inclusive measurements already including weak-decay products. The ratio (or the approximate probability of each nucleon to be transported to mid-rapidity) drops rapidly with increasing rapidity shift. The dashed line is an exponential fit to the data, yielding $\frac{dN_{p-\pbar}/dy}{\Npart/2}=0.99\exp(-0.60\delta y)$.

One may view the net-proton density versus rapidity shift, obtained from central collisions at different energies, as a ``measure" of the rapidity distribution of net-protons in central Au+Au collisions at the top RHIC energy. Since the net-protons shown in Fig.~\ref{fig:netproton_vs_dy} contain equal contributions from the two colliding nuclei, the net-proton rapidity distribution in Au+Au collisions at the top RHIC energy is the data points in Fig.~\ref{fig:netproton_vs_dy} multiplied by a factor varying between 1/2 and 1. At small $\delta y\sim0$ the net-proton density should be close to 1/2 of those shown in Fig.~\ref{fig:netproton_vs_dy}, and at large $\delta y$ (i.e. nearly mid-rapidity) the factor should be close to 1. Assuming an exponential variation in this factor between 1/2 and 1, i.e. a net-proton rapidity distribution of $\frac{dN_{p-\pbar}/dy}{\Npart/2}=2^{\delta y/5.36-1}\times0.99\exp(-0.60\delta y)=0.50\exp(-0.47\delta y)$ in 200~GeV Au+Au collisions (where 5.36 is the beam rapidity for 100~GeV beams), we estimate a rapidity shift of $\langle\delta y\rangle=1/0.47\approx 2.1$. It is interesting to note that the integral of the above rapidity distribution between 0 and 5.36 comes out to be rather close to unity as required by proper normalization. Clearly the exponential form we used is a simplification. BRAHMS has measured the rapidity distribution of net-protons in the range $0<y<3$ in central Au+Au collisions at 200~GeV, and used a more sophisticated functional form to estimate the average rapidity shift to be approximately $2.06\pm0.16$~\cite{BRAHMSnetp}. 

\subsection{Strangeness Production}

Strangeness has a special place in heavy-ion physics. Enhanced production of strangeness has long been predicted as a prominent signature of QGP formation. In a hadron gas strangeness has to be produced via strange hadron pairs which require a large energy, while in QGP it can be produced via a strange quark-antiquark pair, which is energetically favored~\cite{Rafelski,Rafelski2,Rafelski3}. Elementary $pp$ collisions, where QGP formation is unlikely, are important as a reference: an enhanced strangeness production in heavy-ion collisions relative to $pp$ could signal QGP formation. However, other processes can also enhance strangeness production as shown by many studies~\cite{Sorge,Sorge2}. Although not a sufficient signature for QGP formation, strangeness enhancement is a necessary condition which QGP formation requires. 

\begin{figure}[htbp]
\centering
\includegraphics[width=0.48\textwidth]{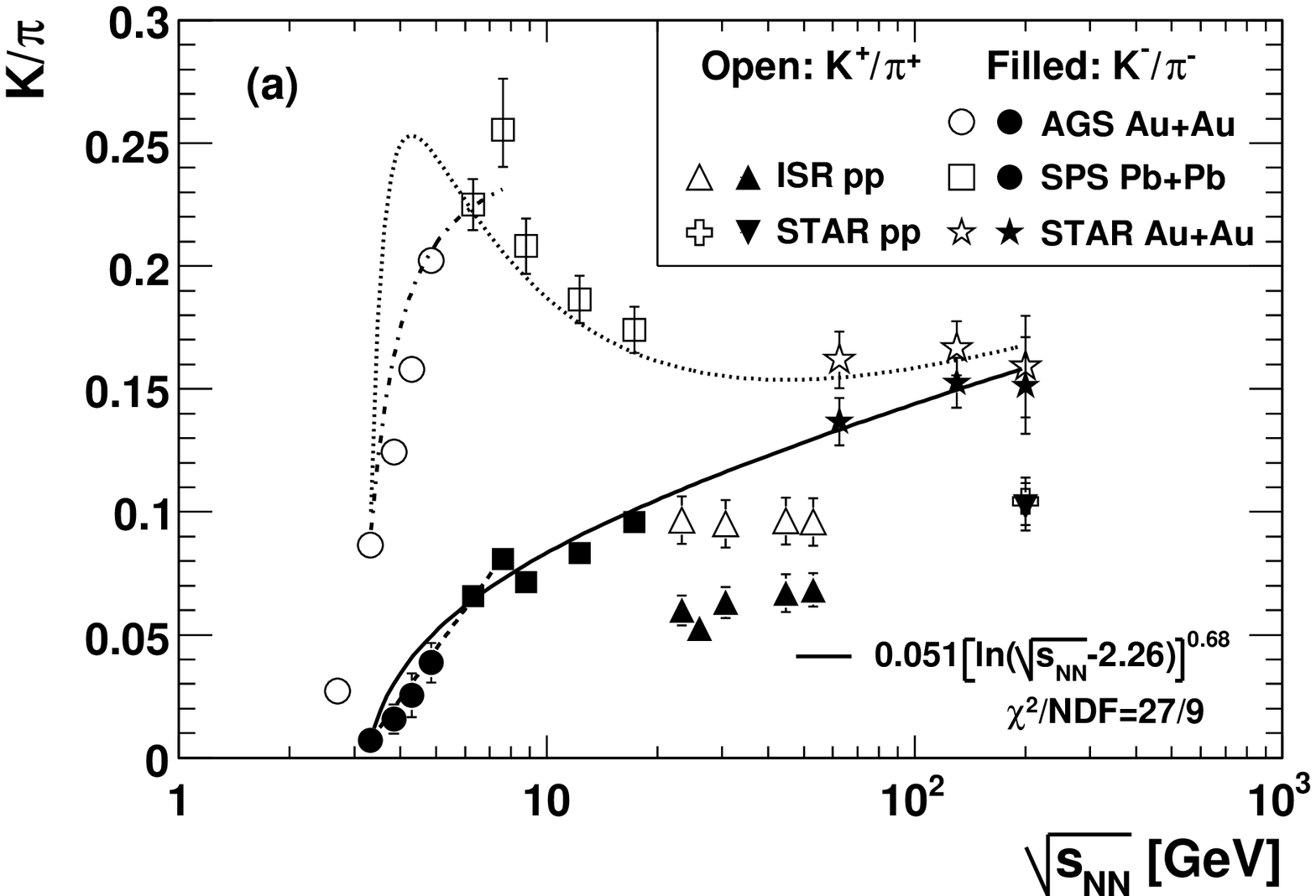}
\includegraphics[width=0.48\textwidth]{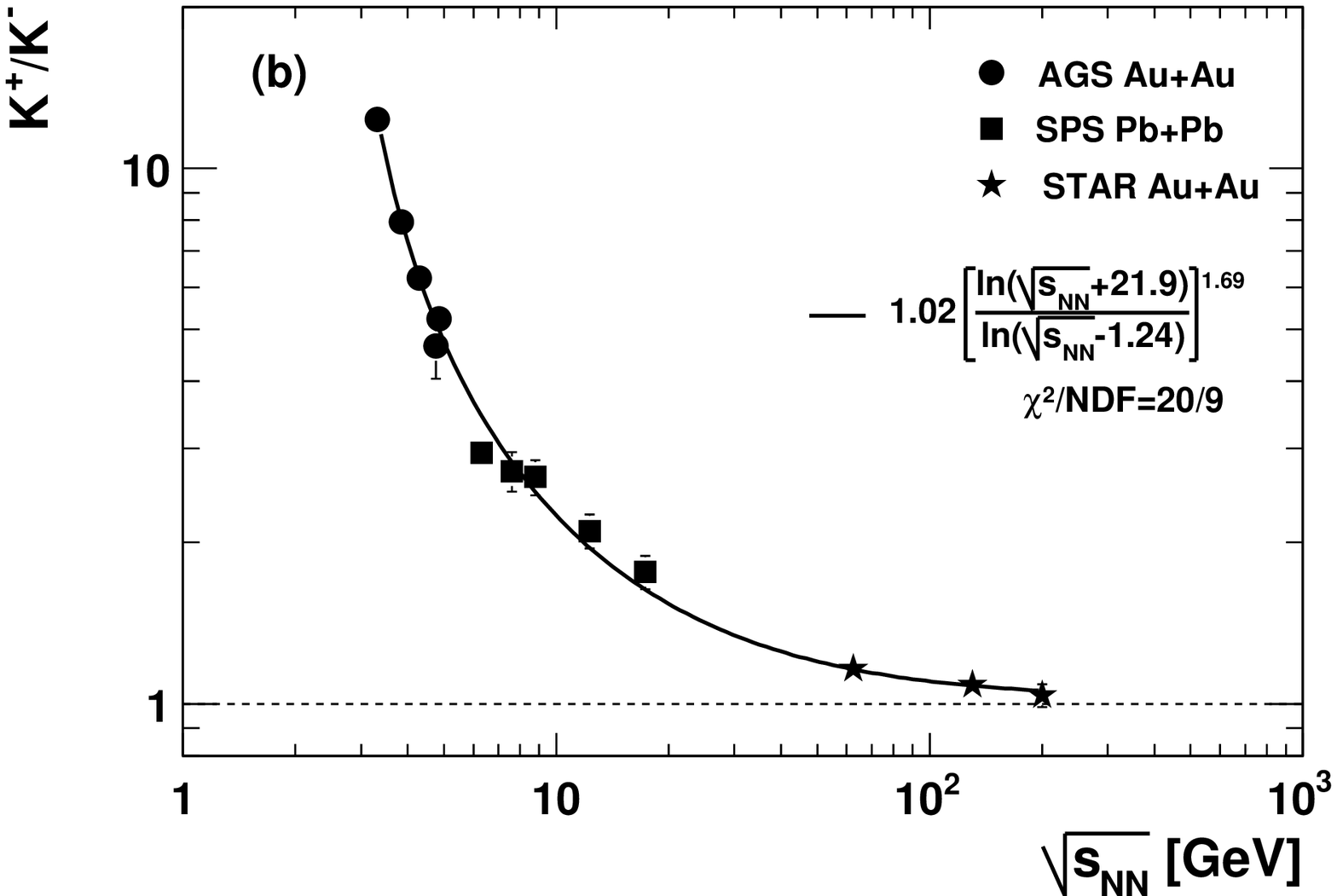}
\caption{(a) The $K^+/\pi^+$ and $K^-/\pi^-$ ratios as a function of the collision energy in $pp$~\cite{pp1,pp2,pbarp1,pbarp2} and central heavy-ion collisions. (b) The $K^+/K^-$ ratio as a function of the collision energy in central heavy-ion collisions. The heavy-ion data not covered in this work are taken from Refs.~\cite{E802kaon,E866kaon,E866E917kpi,E866E917kaon,NA35Bartke,NA35Alber,NA49Sikler,NA44Bearden,NA44Boggild,NA49horn,NA49lowE,NA49system,spec200}. The error bars on the heavy-ion data are the quadratic sum of statistical and systematic uncertainties, and are statistical only on the elementary collision data. The curves going through the heavy-ion $K^-/\pi^-$ and $K^+/K^-$ data are phenomenological fits. The curves going through the heavy-ion $K^+/\pi^+$ data are the product of the fit curves. See text for details.}
\label{fig:kpi_roots}
\end{figure}

Strangeness production and the $K/\pi$ ratios have been intensively studied in heavy-ion collisions at the AGS~\cite{E802kaon,E866kaon,E866E917kpi,E866E917kaon} and the SPS~\cite{NA35Bartke,NA35Alber,NA49Sikler,NA44Bearden,NA44Boggild,NA49horn,NA49lowE,NA49system}, and in elementary interactions of $pp$~\cite{pp1,pp2} and $\pbar p$~\cite{pbarp1,pbarp2}, prior to RHIC~\cite{kaon130,PHENIXspec,spec200}. Figure~\ref{fig:kpi_roots}(a) compiles the $K/\pi$ ratios in $pp$ collisions and central heavy-ion collisions as a function of the collision energy $\snn$. The 200~GeV $pp$ and Au+Au data are from Ref.~\cite{spec200}, and the Au+Au data at 62.4~GeV and 130~GeV are from this work. The $K/\pi$ ratio was already studied in Ref.~\cite{kaon130}, but there the pion yield was not measured but estimated from negatively charged hadrons, kaons, and antiprotons. In this work the measured pion yield is used to obtain the $K/\pi$ ratio. The other data in Fig.~\ref{fig:kpi_roots} are taken from Refs.~\cite{pp1,pp2,pbarp1,pbarp2} for $pp$ collisions and Refs.~\cite{E802kaon,E866kaon,E866E917kpi,E866E917kaon,NA35Bartke,NA35Alber,NA49Sikler,NA44Bearden,NA44Boggild,NA49horn,NA49lowE,NA49system,spec200} for central heavy-ion collisions, as also compiled in Ref.~\cite{kaon130}.

One obvious feature in Fig.~\ref{fig:kpi_roots}(a) is that the $K^-/\pi^-$ ratio in heavy-ion collisions steadily increases with $\snn$, while $K^+/\pi^+$ sharply increases at low energies. The addition of the $K^+/\pi^+$ measurements at RHIC energies clearly demonstrates that $K^+/\pi^+$ drops at high energies. A maximum $K^+/\pi^+$ value is reached at about $\snn\approx 10$~GeV. This behavior of $K^+/\pi^+$ can be partially attributed to the net-baryon density which changes significantly with $\snn$, as noted previously~\cite{netBaryonDensity1,netBaryonDensity4,netBaryonDensity5}. It is instructive to consider the two possible kaon production mechanisms: pair production of $K$ and $\overline{K}$ which is sensitive to $\snn$, and associated production of $K$ ($\overline{K}$) with a hyperon (antihyperon) which is sensitive to the baryon (antibaryon) density\footnote{These mechanisms also apply at the quark level.}. The excess of $K$ over $\overline{K}$ is due to the finite net-baryon density. To visualize the relative contributions from these two mechanisms, Fig.~\ref{fig:kpi_roots}(b) shows the ratio of $K^+/K^-$ as function of $\snn$ in central heavy-ion collisions. The ratio sharply drops with energy, demonstrating the transition from associated production of $K^+$ dominant at low energies to the dominance of equal production of $K^+$ and $K^-$ via either pair production of $K^+K^-$ or associated production of $K^+$ ($K^-$) with hyperon (antihyperon) at high energies. The $K^+/K^-$ dependence on $\snn$ is relatively smooth, and can be fit reasonably well by the functional form shown in the figure. On the other hand, the rate of symmetric production of $K^+$ and $K^-$ increases with $\snn$ as seen in the $K^-/\pi^-$ ratio in Fig.~\ref{fig:kpi_roots}(a). We fit the $K^-/\pi^-$ ratio by the functional form shown in the figure as the solid curve. The curve describes the data points well except at low $\snn$, where the $K^-/\pi^-$ ratio can be better described by a linear increase in log($\snn$) as shown by the dashed line. The product of the curve in Fig.~\ref{fig:kpi_roots}(b) and the solid curve (dashed line) in Fig.~\ref{fig:kpi_roots}(a) yields the dotted (dash-dotted) curve in Fig.~\ref{fig:kpi_roots}(a). It suggests that the smooth dropping of $K^+/K^-$ with $\snn$ in Fig.~\ref{fig:kpi_roots}(b) and the seemingly smooth increase of $K^-/\pi^-$ with $\snn$ can generate a maximum in $K^+/\pi^+$ at $\snn\sim10$~GeV. In fact, model studies~\cite{netBaryonDensity4,netBaryonDensity5} have indeed shown a maximum in the $K^+/\pi^+$ excitation function. However, the maximum peak from model studies is broad and smooth, not as sharp as Fig.~\ref{fig:kpi_roots}(a) shows.

NA49 has first observed the sharp maximum peak structure in the $K^+/\pi^+$ ratio~\cite{NA49horn}, and referred to it as the ``horn". They attribute the horn to a phase-transition between hadrons and the QGP, because ordinary physics (involving production rate and baryon density) does not seem to explain the data. The smooth dependence of the $K^+/K^-$ ratio on $\snn$ indicates that the horn is not $K^+/\pi^+$ specific, but is also present in the $K^-/\pi^-$ ratio as can be seen in Fig.~\ref{fig:kpi_roots}(a). In order to shed light on the horn, more precise measurements are needed for which the RHIC energy scan program should help.

\begin{figure}[htbp]
\centering
\includegraphics[width=0.48\textwidth]{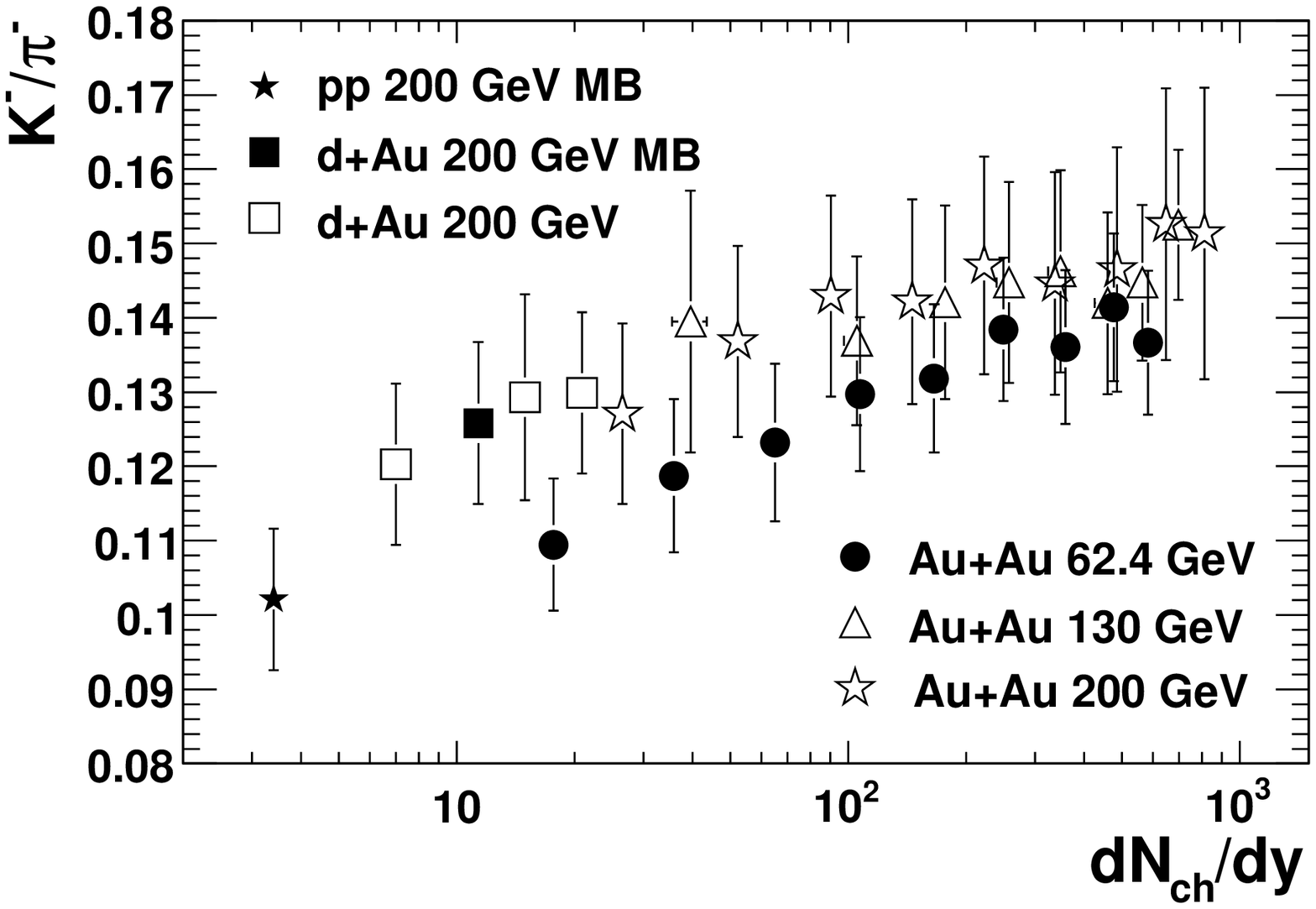}
\caption{The $K^-/\pi^-$ ratio as a function of the charged particle rapidity density in $pp$, d+Au and Au+Au collisions at RHIC. Errors shown are the quadratic sum of statistical and systematic uncertainties.}
\label{fig:kmpiratio}
\includegraphics[width=0.48\textwidth]{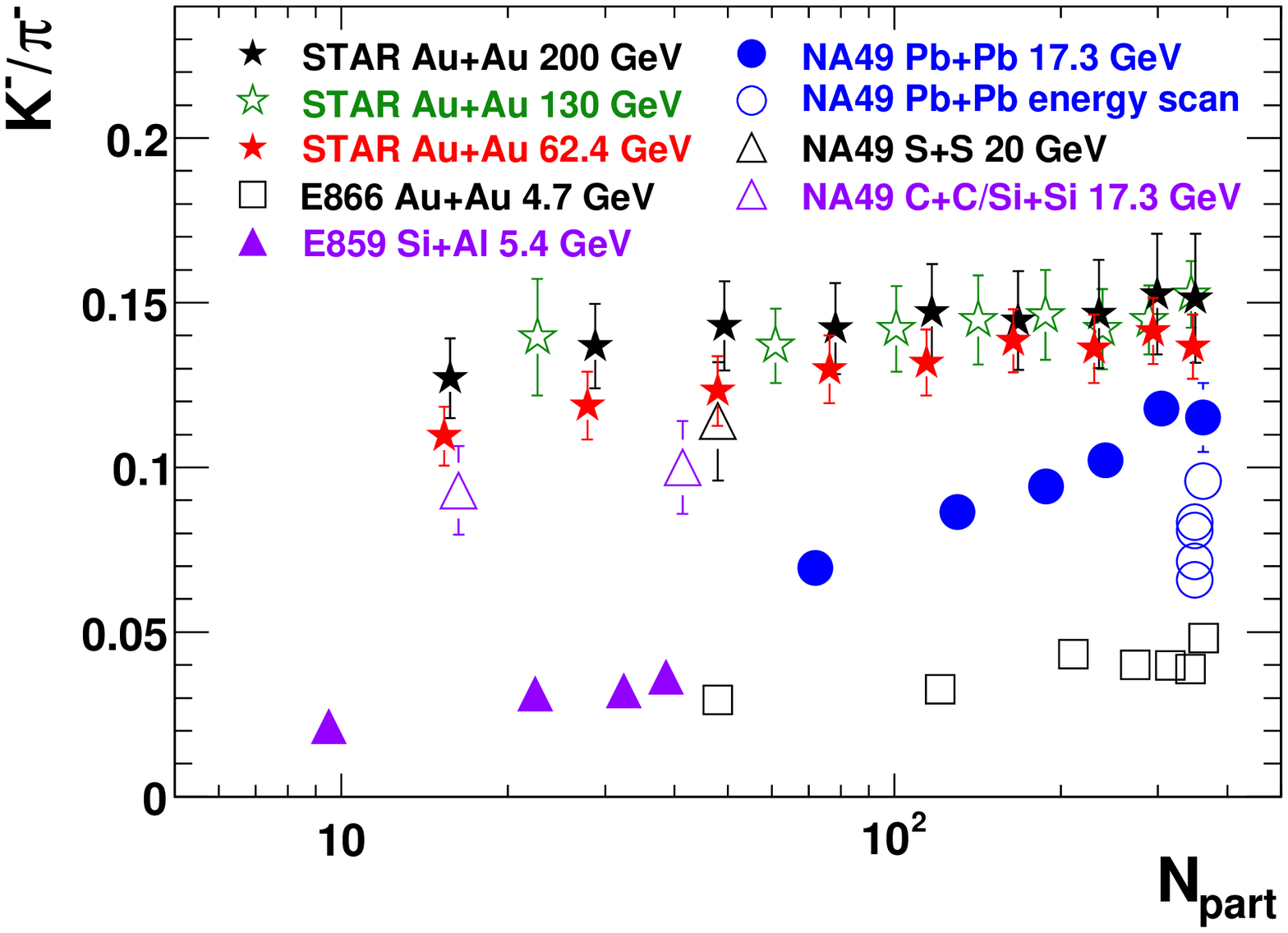}
\caption{(color online) The $K^-/\pi^-$ ratio as a function of the number of participants $\Npart$ in heavy-ion collisions at the AGS~\cite{E866kaon,E866E917kpi}, the SPS~\cite{NA35Alber,NA49Sikler,NA44Bearden,NA44Boggild,NA49horn,NA49lowE,NA49system}, and RHIC. Errors shown are the quadratic sum of statistical and systematic uncertainties for the RHIC data, and only statistical for the AGS and SPS data.}
\label{fig:kmpi_npart}
\end{figure}

Figure~\ref{fig:kpi_roots}(a) indicates that the enhancement in $K^-/\pi^-$ from elementary $pp$ to central heavy-ion collisions is about 50\% and is similar at the SPS and RHIC, while that in $K^+/\pi^+$ is larger at lower energies due to the large net-baryon density in heavy-ion collisions. The increase in $K/\pi$ ratios from $pp$ to central heavy-ion collisions has been argued as due to canonical suppression of strangeness production in small-volume $pp$ collisions~\cite{canonicalSup1,canonicalSup2,canonicalSup3,canonicalSup4}. Although the increase in the $K/\pi$ ratios from $pp$ to central heavy-ion collisions cannot be readily taken as evidence for QGP formation, it is interesting to study how and where the increase happens as a function of centrality. Figure~\ref{fig:kmpiratio} shows the $K^-/\pi^-$ ratio as a function of the charged hadron multiplicity in $pp$, d+Au, and Au+Au collisions at RHIC energies. The $K^+/\pi^+$ ratio shows similar dependence on centrality. The $K^-/\pi^-$ ratio appears to increase approximately linearly with log$(\dNchdy)$.

Experiments at the AGS and SPS have also studied the centrality dependence of kaon production in heavy-ion collisions. Figure~\ref{fig:kmpi_npart} shows those results as a function of the number of participants, $\Npart$, together with our results at RHIC. The $K^-/\pi^-$ ratio increases with $\Npart$ within the same collision system\footnote{Systematic uncertainties on the $K/\pi$ ratio are largely correlated. The 130~GeV data may miss a very peripheral but crucial data point.}. The increase happens rather quickly at RHIC, restricted to very peripheral collisions; little variation with centrality is found from medium-central to central collisions. At lower energies, the $K^-/\pi^-$ ratio increases steadily with $\Npart$. However at the same value of $\Npart$, the ratio differs in different systems at similar energies as shown in Refs.~\cite{E866kaon,NA49Sikler}, indicating that $\Npart$ is not an appropriate variable to describe $K^-/\pi^-$. This has been noted and emphasized before~\cite{E866kaon,NA49Sikler}. 

\begin{figure}[htbp]
\centering
\includegraphics[width=0.48\textwidth]{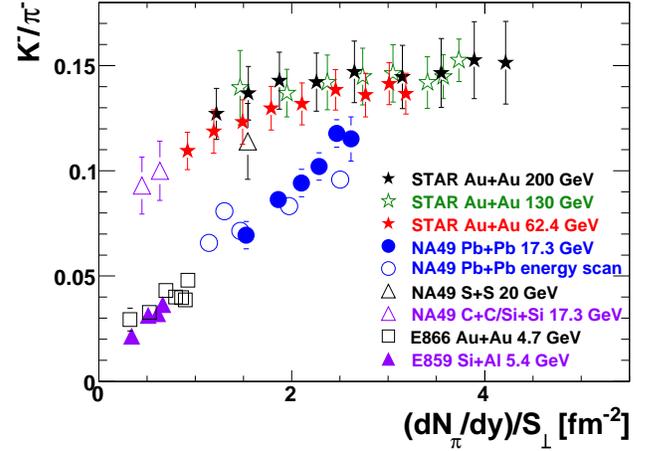}
\caption{(color online) The $K^-/\pi^-$ ratio as a function of $\dNdyS$ in heavy-ion collisions at the AGS~\cite{E866kaon,E866E917kpi}, the SPS~\cite{NA35Alber,NA49Sikler,NA44Bearden,NA44Boggild,NA49horn,NA49lowE,NA49system}, and RHIC. Errors shown are the quadratic sum of statistical and systematic uncertainties for the RHIC data, and statistical only for the AGS and SPS data.}
\label{fig:kmpiratio_dNdyS}
\end{figure}

Neither charged hadron multiplicity nor the number of participants can satisfactorily describe the systematics of the $K^-/\pi^-$ ratio. It is desirable to search for a quantity that better describes the systematics. We first note that strangeness production may be enhanced due to the fast and energetically favorable process of gluon-gluon fusion into strange quark-antiquark pairs, and therefore may be sensitive to the initial gluon density. Indeed, it has been argued that particle production at RHIC (and perhaps at SPS) is dominated by the gluon saturation region~\cite{saturation1,saturation3}. At high energies, the only relevant quantity in the gluon saturation picture is $\dNdyS$, which is approximately proportional to the number of nucleon-nucleon collisions per participant as mentioned earlier. Motivated by these considerations, Fig.~\ref{fig:kmpiratio_dNdyS} shows the $K^-/\pi^-$ ratio as a function of $\dNdyS$. It is interesting to note that the $K^-/\pi^-$ ratio linearly increases with $\dNdyS$ in the AGS and SPS energy regime. The RHIC data show a different behavior: the $K^-/\pi^-$ ratio increases from $pp$ to peripheral Au+Au collisions, but quickly saturates in medium central to central collisions.

In the gluon saturation picture, it is possible that the initial gluon density is saturated at RHIC energies~\cite{saturation3}. As the saturation scale becomes large, the difference between kaon and pion masses becomes less important, resulting in a roughly constant $K^-/\pi^-$. Gluon saturation may already be relevant in central Pb+Pb collisions at the top SPS energy~\cite{saturation3}. Gluon saturation should be irrelevant at AGS energies, as gluons can be distinguished longitudinally and quark contribution to particle production is significant. However, the fact that Si+Al and Au+Au data are on top of each other in Fig.~\ref{fig:kmpiratio_dNdyS} indicates that $\dNdyS$ may be the relevant quantity for $K^-/\pi^-$ at the AGS, although the interpretation may be different from that at high energies.

\section{Freeze-Out Properties\label{sec:Model}}

In this section, particle ratios are used in the context of a thermal equilibrium model~\cite{PBM95,PBM96,PBM99,Kaneta} to extract chemical freeze-out properties. The extracted blast-wave model fit parameters are investigated to learn about the kinetic freeze-out properties. The systematics of the chemical and kinetic freeze-out properties extracted from data within the model frameworks are studied, and implications of these results in terms of the system created in heavy-ion collisions are discussed.

\subsection{Chemical Freeze-out Properties}

In the chemical equilibrium model, particle abundance in a thermal system of volume $V$ is governed by only a few parameters,
\begin{equation}
N_i/V=\frac{g_i}{(2\pi)^3}\gamma_{S}^{S_i}\int\frac{1}{\exp\left(\frac{E_i-\mu_BB_i-\mu_SS_i}{\Tch}\right)\pm1}d^3p\,,\label{eq:chemical}
\end{equation}
where $N_i$ is the abundance of particle species $i$, $g_i$ is the spin degeneracy, $B_i$ and $S_i$ are the baryon number and strangeness number, respectively, $E_i$ is the particle energy, and the integral is over the whole momentum space. The model parameters are the chemical freeze-out temperature (the temperature of the system), $\Tch$, the baryon and strangeness chemical potentials, $\mu_{B}$ and $\mu_{S}$, respectively, and the ad-hoc strangeness suppression factor, $\gamma_{S}$.

The measured particle abundance ratios are fit by the chemical equilibrium model. The ratios included in the fit are: $\pi^{-}/\pi^{+}$, $K^-/K^+$, $\pbar/p$, $K^-/\pi^-$, $\pbar/\pi^{-}$. The fit is performed for each collision system and each multiplicity or centrality class. The extracted chemical freeze-out parameters are summarized in Table~\ref{tab:model_pars}. The 200~GeV $pp$ and Au+Au results are from Ref.~\cite{spec200}.

\begin{figure}[htbp]
\includegraphics[width=0.48\textwidth]{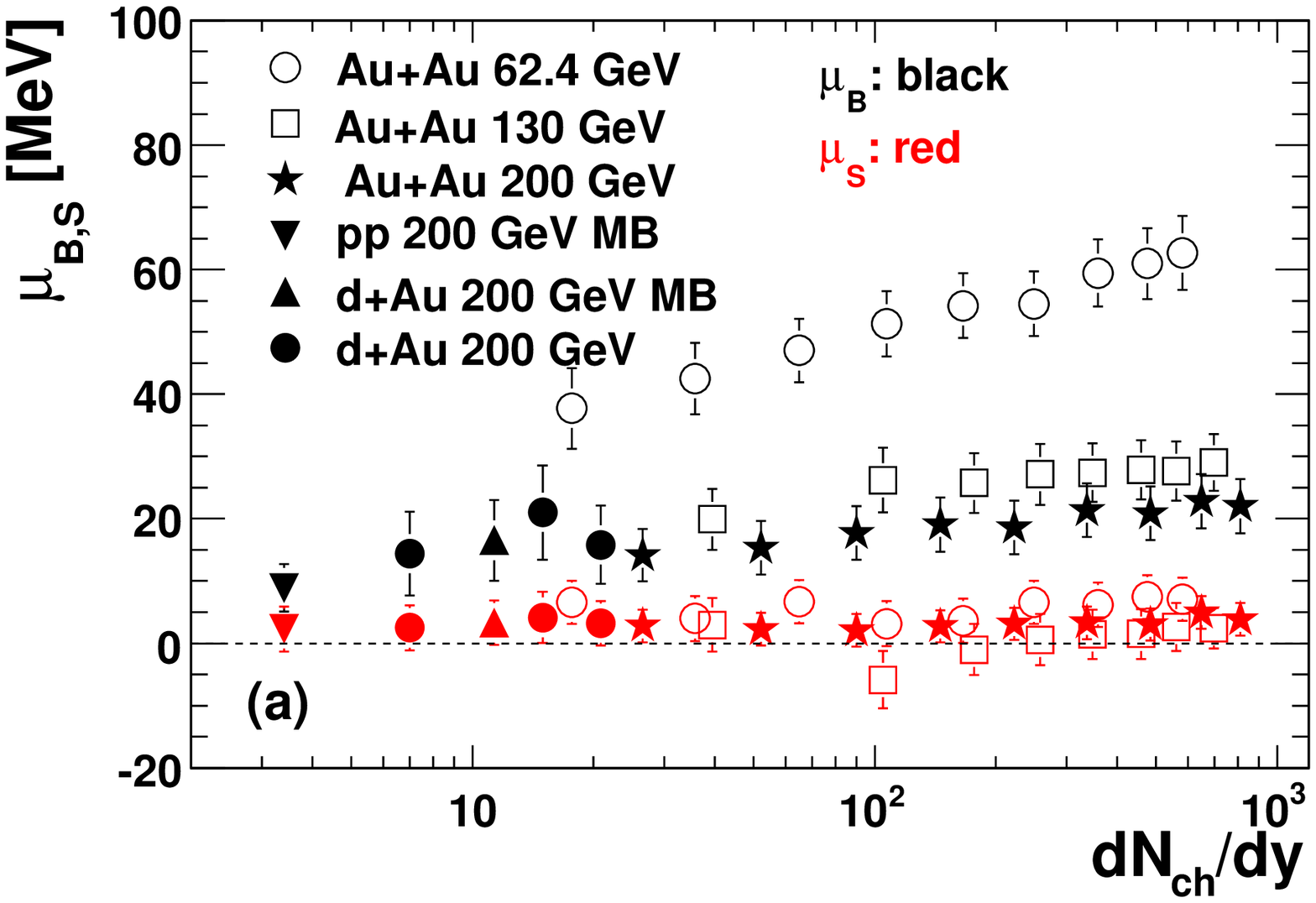}
\includegraphics[width=0.48\textwidth]{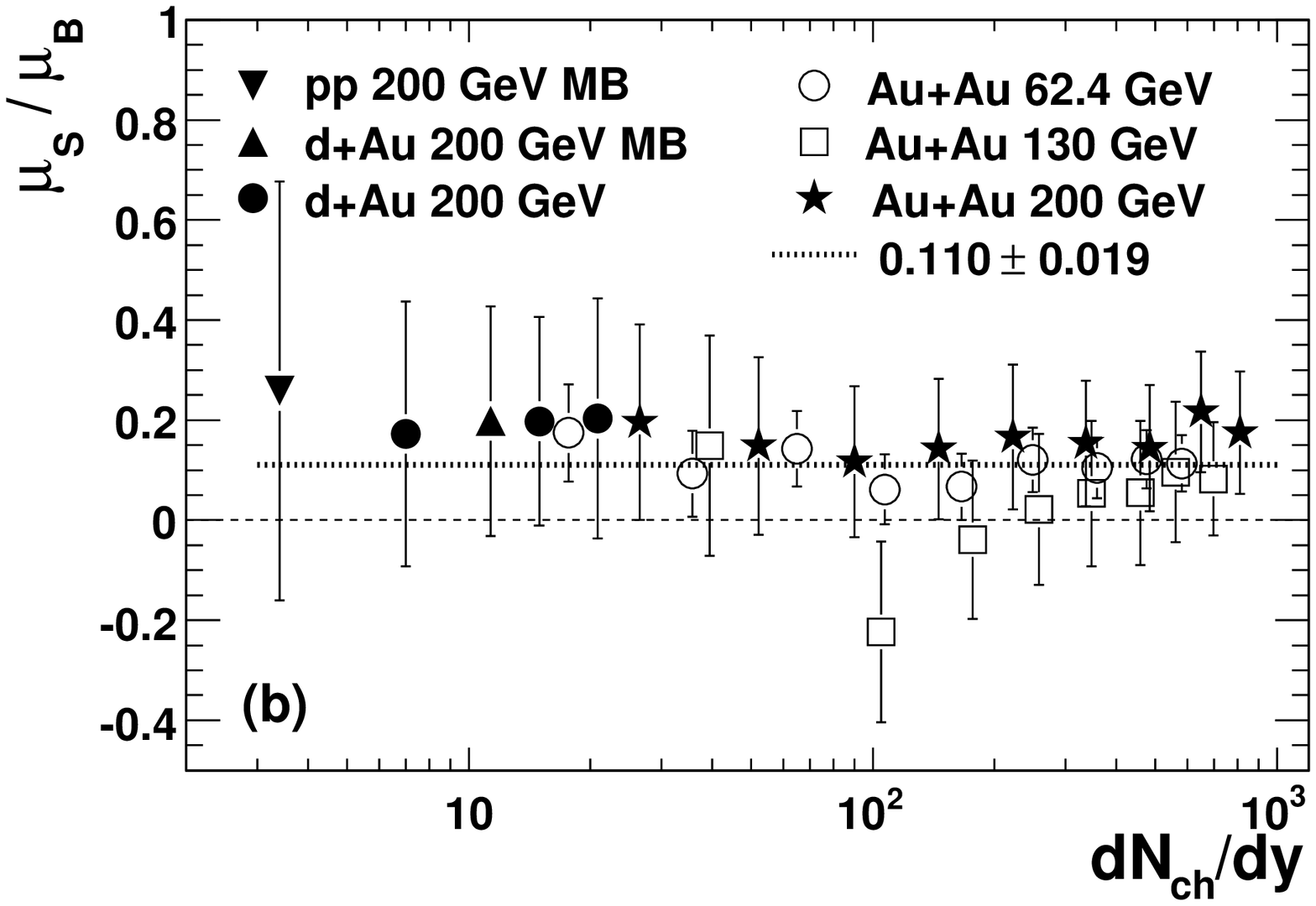}
\caption{(color online) (a) Baryon ($\mu_B$) and strangeness ($\mu_S$) chemical potentials extracted from chemical equilibrium model fits to $pp$ and d+Au data at 200~GeV, and Au+Au data at 62.4~GeV, 130~GeV, and 200~GeV. (b) Ratio $\mu_S/\mu_B$ of the extracted chemical potentials. Errors shown are the total statistical and systematic errors. The 200~GeV $pp$ and Au+Au fit results are taken from Ref.~\cite{spec200}.}
\label{fig:muBmuS}
\end{figure}

Figure~\ref{fig:muBmuS}(a) shows the extracted baryon and strangeness chemical potentials as a function of the charged particle multiplicity in $pp$ and d+Au at 200~GeV, and Au+Au collisions at 62.4~GeV, 130~GeV, and 200~GeV. The baryon chemical potential increases with centrality in heavy-ion collisions, especially at 62.4~GeV. This is already indicated by the $\pbar/p$ ratio in Fig.~\ref{fig:papratios}. The strangeness chemical potential is small and close to zero. It is mainly reflected in the $K/\pi$ and $K^-/K^+$ ratios. As already shown in Fig.~\ref{fig:kktopbarp}, the $K^-/K^+$ ratio is correlated with the $\pbar/p$ ratio by a universal curve. In the chemical equilibrium picture without considering resonance decays, these ratios are simply equal to $K^-/K^+=\exp[(-2\mu_B/3+2\mu_S)/\Tch]$ and $\pbar/p=\exp(-2\mu_B/\Tch)$, respectively. Weak decays and resonance decays complicate the situation, but the effects of decays are small for the $K^-/K^+$ and $\pbar/p$ ratios. A power-law fit to all data points in Fig.~\ref{fig:kktopbarp} (except the AGS data point and the two lowest SPS data points) yields $K^-/K^+ \propto (\pbar/p)^{0.21}$. This gives $\mu_S/\mu_B\approx 0.12$ in the chemical equilibrium picture. We show in Fig.~\ref{fig:muBmuS}(b) the ratio of the extracted $\mu_S$ to $\mu_B$. A fit to a constant indeed shows $\mu_S/\mu_B=0.110\pm0.019$. Analyses of chemical freeze-out parameters in heavy-ion collisions at other energies indicate a similar relationship~\cite{Murray}. The strong correlation between $\mu_S$ and $\mu_B$ should not come as a surprise, as the (anti)hyperons couple these two parameters naturally. However, the same relationship holding for different energies is not expected a priori. 

\begin{figure}[htbp]
\includegraphics[width=0.48\textwidth]{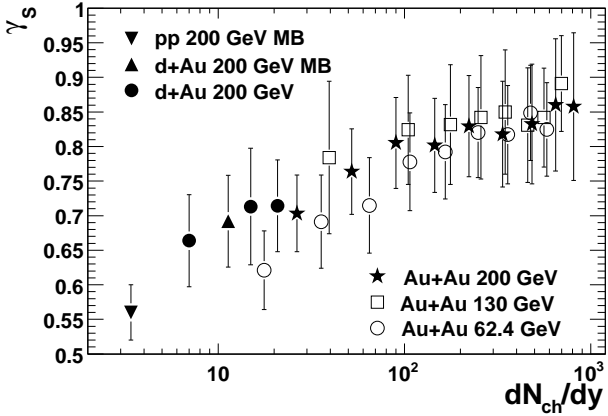}
\caption{Strangeness suppression factor extracted from chemical equilibrium model fit to $pp$ and d+Au data at 200~GeV, and Au+Au data at 62.4~GeV, 130~GeV, and 200~GeV. Errors shown are the total statistical and systematic uncertainties. The 200~GeV $pp$ and Au+Au fit results are taken from Ref.~\cite{spec200}.}
\label{fig:gammaS}
\end{figure}

Figure~\ref{fig:gammaS} shows the extracted strangeness suppression factor $\gamma_S$ as a function of the charged particle multiplicity. The $\gamma_S$ in $pp$, d+Au, and peripheral Au+Au collisions is significantly smaller than unity, suggesting that strangeness production is strongly suppressed in these collisions. The $\gamma_S$ factor increases with centrality, reaching a value in central Au+Au collisions that is not much smaller than unity. This suggests that strangeness production in central collisions is no longer strongly suppressed; strangeness is nearly chemically equilibrated with the light flavors.

The extracted chemical freeze-out temperature is shown in Fig.~\ref{fig:Tfo}. A striking feature is that the chemical freeze-out temperature is independent of collision system or centrality. In each system investigated the extracted chemical freeze-out temperature is $\Tch\approx$ 156 MeV which is close to the Lattice QCD calculation of the cross-over temperature between the deconfined phase and the hadronic phase for three flavors ($154\pm8$~MeV)~\cite{Karsch04}. On the other hand, the initial conditions in Au+Au collisions of different centralities (and at different energies) are very different. In other words, systems starting off with different initial conditions always evolve toward a `universal' condition at chemical freeze-out, independent of the initial conditions~\cite{spec200}. The proximity of the fit $\Tch$ and the predicted phase-transition temperature strongly suggests that chemical freeze-out happens at the phase-transition boundary, or hadronization. Indeed, hadronization should be universal.

The success of the chemical equilibrium model in describing the data should not be readily taken as a proof of chemical equilibrium of each individual collision~\cite{Becattini03}. In $pp$ (and other elementary) collisions the compositions of most particles are described well by the chemical equilibrium model but with the ad-hoc strangeness suppression factor significantly smaller than unity. This has been argued as due, in part, to canonical suppression from conservation of strangeness in small volumes~\cite{canonicalSup1,canonicalSup2,Becattini03}. Canonical suppression appears to explain elementary $e^+e^-$ data, while additional suppression seems needed to account for strangeness production in $pp$ collisions. The apparent success of the chemical equilibrium model in describing elementary collisions, despite the strangeness suppression factor, in all likelihood suggests that particle production in these collisions is a statistical process, and the chemical temperature is a parameter governing the statistical production processes~\cite{Becattini03}.

On the other hand, the stringent constrains of conservation laws are largely lifted in heavy-ion collisions as they only need to be satisfied globally over a large volume. As a result particle ensembles can be treated in a grand canonical framework. The chemical equilibrium model can describe the abundances of all stable hadrons. The ad-hoc strangeness suppression factor extracted from central heavy-ion collisions is close to unity, implying that strangeness is as equally equilibrated as light quarks. Moreover, many experimental results indicate that the medium created at RHIC is strongly interacting~\cite{whitepaper}, which will naturally lead to thermalization. Thus the success of the chemical equilibrium model may indeed suggest that the individual Au+Au collisions are largely thermalized. 

\begin{table*}{}
\caption{Chemical and kinetic freeze-out properties in $pp$ and d+Au collisions at 200~GeV, and Au+Au collisions at 62.4~GeV, 130~GeV, and 200~GeV. Quoted errors are the total statistical and systematic uncertainties. The 200~GeV $pp$ and Au+Au data are taken from Ref.~\cite{spec200}.}
\label{tab:model_pars}
\begin{ruledtabular}
\begin{tabular}{cc|ccccc|cccc} 
System & Centrality & $\Tch$ (MeV) & $\mu_B$ (MeV) & $\mu_S$ (MeV) & $\gamma_S$ & $\chi^{2}$/{\sc ndf} & $\Tkin$ (MeV) & $\langle\beta\rangle$ & $n$ & $\chi^{2}$/{\sc ndf} \\ \hline 
$pp$ 200~GeV & min.~bias & $157.5\pm3.6$ & $8.9\pm3.8$ & $2.3\pm3.6$ & $0.56\pm0.04$ & 0.81 & $127\pm13$ & $0.244\pm0.081$ & $4.3\pm1.7$ & 1.18 \\ 
\hline
	& min.~bias & $164^{+11}_{-8}$ & $16.5\pm6.5$ & $3.3\pm3.6$ & $0.69\pm0.07$ & $0.013$ & $112\pm26$ & $0.407\pm0.033$ & $1.9\pm0.9$ & $ 0.89$ \\
d+Au	& 40-100\% & $159^{+10}_{-7}$ & $14.4\pm6.7$ & $2.5\pm3.6$ & $0.66\pm0.07$ & $0.051$ & $112\pm24$ & $0.377\pm0.031$ & $2.2\pm1.3$ & $ 1.55$ \\
200~GeV	& 20-40\% & $168^{+14}_{-10}$ & $21.0\pm7.6$ & $4.2\pm4.1$ & $0.71\pm0.08$ & $0.068$ & $107\pm33$ & $0.428\pm0.067$ & $1.9\pm0.9$ & $ 1.32$ \\
	& 0-20\% & $167^{+12}_{-7}$ & $15.8\pm6.3$ & $3.2\pm3.6$ & $0.71\pm0.07$ & $0.069$ & $116\pm21$ & $0.420\pm0.030$ & $1.6\pm0.7$ & $ 0.72$ \\
\hline
& 70-80\% & $157.9\pm3.9$ & $14.1\pm4.2$ & $2.8\pm2.6$ & $0.70\pm0.06$ & 0.51 & $129\pm14$ & $0.358\pm0.084$ & $1.50\pm0.28$ & 0.70 \\ 
& 60-70\% & $158.7\pm4.1$ & $15.3\pm4.2$ & $2.3\pm2.6$ & $0.76\pm0.06$ & 0.51 & $118\pm13$ & $0.405\pm0.071$ & $1.57\pm0.11$ & 0.42 \\ 
& 50-60\% & $158.8\pm4.1$ & $17.7\pm4.2$ & $2.1\pm2.6$ & $0.81\pm0.07$ & 0.35 & $115\pm12$ & $0.456\pm0.071$ & $1.16\pm0.08$ & 0.45 \\ 
Au-Au & 40-50\% & $155.8\pm4.0$ & $18.9\pm4.2$ & $2.7\pm2.6$ & $0.80\pm0.07$ & 0.17 & $108\pm12$ & $0.499\pm0.071$ & $0.98\pm0.06$ & 0.48 \\ 
& 30-40\% & $156.5\pm4.2$ & $18.6\pm4.2$ & $3.1\pm2.6$ & $0.83\pm0.07$ & 0.04 & $109\pm11$ & $0.514\pm0.061$ & $0.90\pm0.05$ & 0.45 \\ 
200~GeV & 20-30\% & $156.7\pm4.8$ & $21.3\pm4.2$ & $3.2\pm2.6$ & $0.82\pm0.08$ & 0.03 & $102\pm11$ & $0.539\pm0.061$ & $0.90\pm0.04$ & 0.37 \\ 
& 10-20\% & $155.1\pm4.8$ & $21.0\pm4.2$ & $3.0\pm2.6$ & $0.83\pm0.09$ & 0.02 & $99\pm12$ & $0.560\pm0.061$ & $0.80\pm0.03$ & 0.36 \\ 
& 5-10\% & $156.5\pm5.3$ & $22.8\pm4.5$ & $4.9\pm2.6$ & $0.86\pm0.10$ & 0.02 & $91\pm12$ & $0.577\pm0.051$ & $0.86\pm0.02$ & 0.51 \\ 
& 0-5\% & $159.3\pm5.8$ & $21.9\pm4.5$ & $3.9\pm2.6$ & $0.86\pm0.11$ & 0.03 & $89\pm12$ & $0.592\pm0.051$ & $0.82\pm0.02$ & 0.25 \\ 
\hline 
	& 58-85\% & $159^{+11}_{-7}$ & $19.9\pm4.9$ & $3.0\pm4.3$ & $0.78\pm0.11$ & $0.004$ & $136\pm32$ & $0.400\pm0.027$ & $0.0\pm10.1$ & $ 0.96$ \\
	& 45-58\% & $158^{+10}_{-6}$ & $26.2\pm5.2$ & $-5.9\pm4.6$ & $0.82\pm0.08$ & $0.015$ & $113\pm16$ & $0.465\pm0.010$ & $0.6\pm0.5$ & $ 0.78$ \\
	& 34-45\% & $158^{+9}_{-5}$ & $25.8\pm4.8$ & $-1.0\pm4.1$ & $0.83\pm0.09$ & $0.153$ & $103\pm11$ & $0.502\pm0.013$ & $0.8\pm0.3$ & $ 1.01$ \\
Au+Au	& 26-34\% & $158^{+10}_{-6}$ & $27.1\pm4.9$ & $0.6\pm4.1$ & $0.84\pm0.09$ & $0.006$ & $103\pm16$ & $0.526\pm0.017$ & $0.8\pm0.3$ & $ 0.81$ \\
130~GeV	& 18-26\% & $156^{+10}_{-6}$ & $27.4\pm4.7$ & $1.5\pm4.0$ & $0.85\pm0.09$ & $0.005$ & $103\pm15$ & $0.531\pm0.017$ & $0.8\pm0.2$ & $ 0.81$ \\
	& 11-18\% & $153^{+9}_{-6}$ & $27.9\pm4.8$ & $1.5\pm4.0$ & $0.83\pm0.08$ & $0.012$ & $106\pm20$ & $0.538\pm0.023$ & $0.7\pm0.4$ & $ 0.52$ \\
	& 6-11\% & $153^{+9}_{-5}$ & $27.7\pm4.7$ & $2.7\pm3.8$ & $0.84\pm0.07$ & $0.005$ & $ 93\pm12$ & $0.558\pm0.019$ & $0.7\pm0.3$ & $ 0.70$ \\
	& 0-6\% & $154^{+10}_{-6}$ & $29.0\pm4.6$ & $2.4\pm3.3$ & $0.89\pm0.07$ & $0.136$ & $ 96\pm8$ & $0.567\pm0.020$ & $0.7\pm0.3$ & $ 0.67$ \\
\hline
	& 70-80\% & $154^{+8}_{-6}$ & $37.7\pm6.5$ & $6.6\pm3.5$ & $0.62\pm0.06$ & $0.244$ & $130\pm15$ & $0.306\pm0.065$ & $2.1\pm1.8$ & $ 0.93$ \\
	& 60-70\% & $156^{+8}_{-5}$ & $42.5\pm5.8$ & $4.0\pm3.6$ & $0.69\pm0.07$ & $0.178$ & $130\pm15$ & $0.389\pm0.019$ & $0.4\pm0.9$ & $ 0.55$ \\
	& 50-60\% & $155^{+8}_{-5}$ & $47.0\pm5.1$ & $6.7\pm3.5$ & $0.71\pm0.07$ & $0.197$ & $129\pm16$ & $0.426\pm0.021$ & $0.0\pm9.8$ & $ 0.59$ \\
Au+Au	& 40-50\% & $156^{+9}_{-5}$ & $51.3\pm5.2$ & $3.2\pm3.6$ & $0.78\pm0.07$ & $0.237$ & $120\pm13$ & $0.459\pm0.009$ & $0.6\pm0.5$ & $ 0.54$ \\
	& 30-40\% & $157^{+9}_{-5}$ & $54.2\pm5.2$ & $3.6\pm3.6$ & $0.79\pm0.07$ & $0.275$ & $113\pm12$ & $0.494\pm0.008$ & $0.6\pm0.4$ & $ 0.45$ \\
62.4~GeV	& 20-30\% & $157^{+9}_{-5}$ & $54.5\pm5.2$ & $6.6\pm3.4$ & $0.82\pm0.07$ & $0.715$ & $105\pm10$ & $0.517\pm0.020$ & $0.8\pm0.3$ & $ 0.52$ \\
	& 10-20\% & $156^{+9}_{-5}$ & $59.4\pm5.4$ & $6.2\pm3.5$ & $0.82\pm0.07$ & $0.261$ & $105\pm14$ & $0.535\pm0.017$ & $0.6\pm0.4$ & $ 0.52$ \\
	& 5-10\% & $155^{+9}_{-5}$ & $61.0\pm5.7$ & $7.4\pm3.5$ & $0.85\pm0.07$ & $0.066$ & $100\pm12$ & $0.546\pm0.019$ & $0.7\pm0.3$ & $ 0.77$ \\
	& 0-5\% & $154^{+10}_{-7}$ & $62.7\pm6.0$ & $7.1\pm3.5$ & $0.82\pm0.07$ & $0.480$ & $ 99\pm10$ & $0.554\pm0.018$ & $0.6\pm0.4$ & $ 0.75$ \\
\end{tabular} 
\end{ruledtabular}
\end{table*} 

\subsection{Kinetic Freeze-out Properties\label{sec:kineticFO}}

The measured $\pt$ spectral shape flattens significantly with increasing particle mass in central Au+Au collisions. This suggests the presence of a collective transverse radial flow field, although other physics mechanisms such as (semi-)hard scatterings also contribute. As shown in Figs.~\ref{fig:dAu_spectra} and \ref{fig:AuAu62_spectra}, the spectra are well described by the hydrodynamics-motivated blast-wave model~\cite{bw,hydro1a,hydro1b,hydro2a,hydro2b,hydro3,LisaBW}. The blast-wave model makes the simple assumption that particles are locally thermalized at a kinetic freeze-out temperature and are moving with a common collective transverse radial flow velocity field. The common flow velocity field results in a larger transverse momentum of heavier particles, leading to the change in the observed spectral shape with increasing particle mass. 

Assuming a hard-sphere uniform density particle source with a kinetic freeze-out temperature $\Tkin$ and a transverse radial flow velocity $\beta$, the particle transverse momentum spectral shape is given by~\cite{bw}
\begin{equation}
\frac{dN}{\pt d\pt}\propto\int_0^Rrdr\,\mt I_0\left(\frac{\pt\sinh\rho}{\Tkin}\right)K_1\left(\frac{\mt\cosh\rho}{\Tkin}\right)\,,
\end{equation}
where $\rho=\tanh^{-1}\beta$, and $I_0$ and $K_1$ are the modified Bessel functions. We use a flow velocity profile of the form 
\begin{equation}
\beta=\beta_{S}\left(r/R\right)^n\,,\label{eq:flowProfile}
\end{equation}
where $\beta_{S}$ is the surface velocity and $r/R$ is the relative radial position in the thermal source. The choice of the value of $R$ bears no effect in the model.

Six particle spectra ($\pi^{\pm}$, $K^{\pm}$, $p$ and $\pbar$) of a given centrality bin are fit simultaneously with the blast-wave model. The free parameters are: the kinetic freeze-out temperature, $\Tkin$, the average transverse flow velocity, $\langle\beta\rangle=\frac{2}{2+n}\beta_S$, and the exponent of the assumed flow velocity profile, $n$. The low momentum part of the pion spectra ($\pt<0.5$~GeV/$c$) are excluded from the fit, due to significant contributions from resonance decays. 

The blast-wave fit results for Au+Au collisions are listed in Table~\ref{tab:model_pars}. The $\chi^2$/{\sc ndf} is smaller than unity because the point-to-point systematic errors, which are included in the fit and dominate over statistical ones, are estimated on the conservative side and might not be completely random. If the $\chi^2$/{\sc ndf} is scaled such that the minimum is unity, then somewhat smaller statistical errors on the fit parameters are obtained. 

\begin{figure}[htbp]
\includegraphics[width=0.48\textwidth]{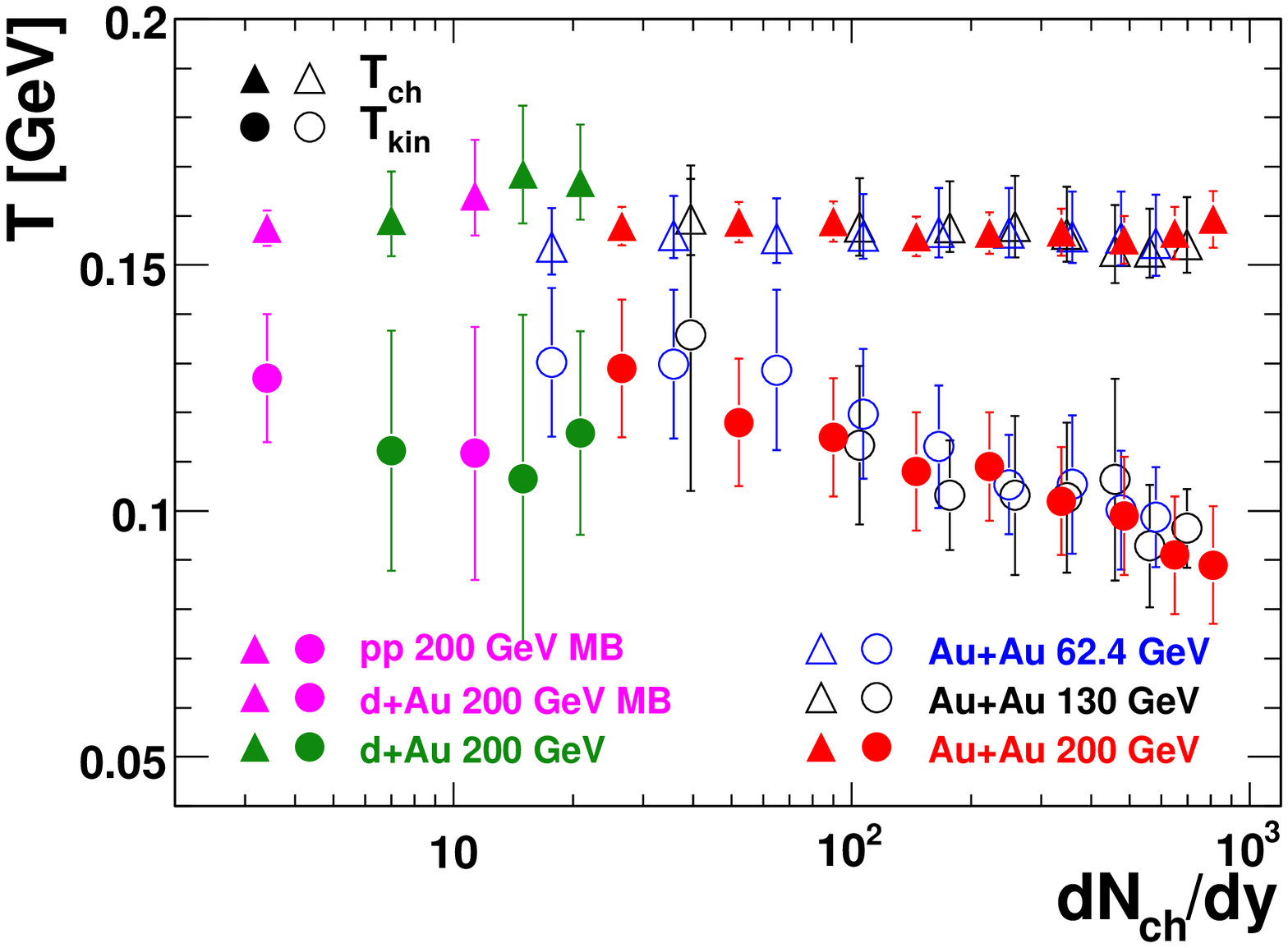}
\caption{(color online) Chemical and kinetic freeze-out temperatures as a function of the charged hadron multiplicity. Errors shown are the total statistical and systematic uncertainties. The 200~GeV $pp$ and Au+Au data are taken from Ref.~\cite{spec200}.}
\label{fig:Tfo}
\centering
\includegraphics[width=0.48\textwidth]{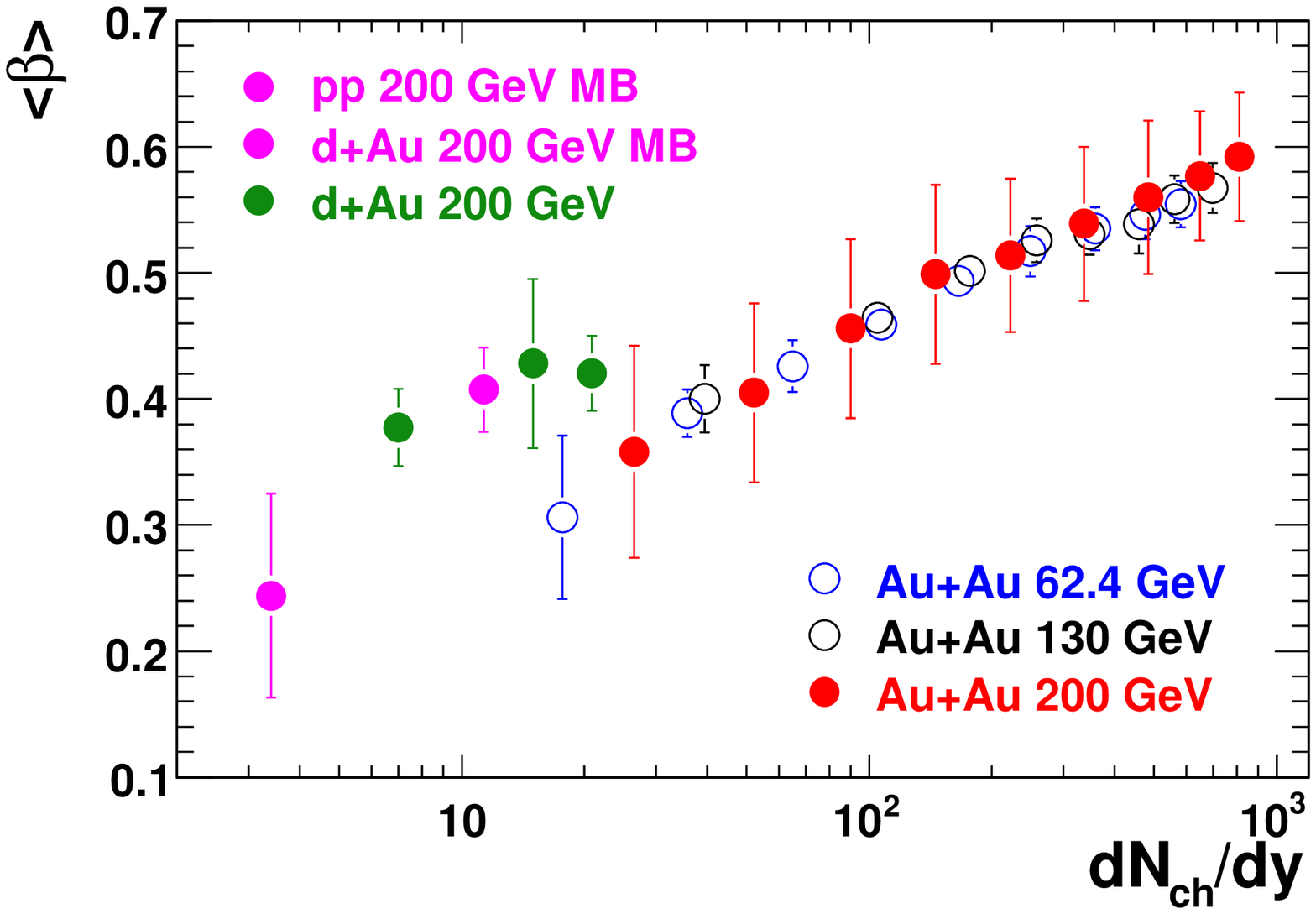}
\caption{(color online) Average transverse radial flow velocity extracted from blast-wave model fit to $pp$ and d+Au at 200~GeV, and to Au+Au collisions at 62.4~GeV, 130~GeV, and 200~GeV as a function of the charged hadron multiplicity. Errors shown are the total statistical and systematic uncertainties. The 200~GeV $pp$ and Au+Au data are taken from Ref.~\cite{spec200}.}
\label{fig:auaubeta}
\end{figure}

Figure~\ref{fig:Tfo} shows the extracted kinetic freeze-out temperature as a function of the event multiplicity for $pp$ and d+Au collisions at 200~GeV and for Au+Au collisions at 62.4~GeV, 130~GeV, and 200~GeV, together with the chemical freeze-out temperature. As opposed to $\Tch$, the kinetic freeze-out temperature $\Tkin$ shows a notable decreasing trend with centrality in Au+Au collisions. The $\Tkin$ values from $pp$ and d+Au collisions are similar to those in peripheral Au+Au, although the systematic uncertainties are large.

Figure~\ref{fig:auaubeta} shows the extracted average transverse radial flow velocity $\langle\beta\rangle$ as a function of the event multiplicity. The $\langle\beta\rangle$ increases dramatically with increasing centrality in Au+Au collisions. The effect of the $\langle\beta\rangle$ increase on the transverse spectra is significantly stronger than the counter effect of the $\Tkin$ drop. The combination of the $\pi$, $K$, $p$ and $\pbar$ spectra favor an increase of $\langle\beta\rangle$ with centrality rather than a similar increase in $\Tkin$. 

In order to have the same base for comparison, the $pp$ and d+Au data are also fit by the blast-wave model. The fit results are listed in Table~\ref{tab:model_pars} and shown as a function of the event multiplicity in Figs.~\ref{fig:Tfo} and~\ref{fig:auaubeta} together with the Au+Au results. The model is found to give a fairly good description of the measured $\pi^{\pm}$, $K^{\pm}$, $p$ and $\pbar$ spectra. Surprisingly, the fit average flow velocities from $pp$ and d+Au collisions are not small, and certainly not zero as one would naively expect. This should not be taken as a proof that there is collective flow in $pp$ and d+Au collisions, because hard scatterings and jet production, generating relatively more high-$\pt$ hadrons, can mimic collective flow and give rise to the extracted finite $\langle\beta\rangle$~\cite{trainor_pp}. In d+Au collisions, there is an additional effect of initial state scattering, which broadens the transverse momentum of the colliding constituents and hence the produced hadrons in the final state. Meanwhile, statistical global energy and momentum conservation can deplete large momentum particles shown in recent studies~\cite{lisa}, and the effect can be large in low multiplicity collisions. In the same framework, large initial energy fluctuation available for mid-rapidity particle production tends to harden the transverse spectrum~\cite{STAR200kstar,Wilk}. The interplay, as well as the relevance of statistical global energy and momentum conservation in high energy collisions, needs further quantitative studies.

In Au+Au collisions the contribution from hard (and semi-hard) scatterings is larger than in $pp$ collisions because hard scatterings scale with the number of binary nucleon-nucleon collisions while soft processes scale with the number of participant nucleons. From the two-component model study in Section~\ref{sec:total_dndy}, the hard-scattering contribution in $pp$ collisions at 200~GeV is 13\%, while in the top 5\% central Au+Au collisions it is 46\%, a factor of 3.5 times that in $pp$. From the blast-wave model with a linear flow velocity profile, the increase in average $\meanpt$ or $\langle\mt\rangle$ due to radial flow velocity $\langle\beta\rangle$ is approximately proportional to $\langle\beta\rangle^3$. Assuming the apparent finite flow velocity extracted from $pp$ data, $\langle\beta\rangle_{pp}=0.24\pm0.08$, is solely due to the energy excess of produced particles from hard processes over soft processes, and assuming the particle production from hard processes is identical in $pp$ and central Au+Au collisions, then the hard processes in central Au+Au collisions would generate an apparent flow velocity of $3.5^{1/3}\langle\beta\rangle_{pp}=0.36$. However, the extracted flow velocity from the blast-wave model for central Au+Au collisions is significantly larger, $\langle\beta\rangle_{\rm AA}=0.59\pm0.05$. One may take the additional excess in central Au+Au collisions as the effect of collective transverse radial flow, and estimate the collective flow velocity in central Au+Au collisions by $\langle\beta\rangle_{\rm flow}\sim\sqrt[3]{\langle\beta\rangle^3_{\rm AA}-3.5\langle\beta\rangle^3_{pp}}=0.54\pm0.08$. As discussed in section~\ref{sec:total_dndy}, the Kharzeev-Nardi two-component model likely overestimates the fraction of the hard component in $pp$ collisions. However, using the hard-component fraction obtained from Ref.~\cite{trainor_pp}, with the same assumptions as stated above, the estimate of the collective flow velocity in central Au+Au collisions is not significantly altered. We note, however, that the preceding estimate is simplistic. The full understanding of the effects on transverse spectra from radial flow, (semi-)hard scatterings, interactions between (semi-)hard scatterings and the medium~\cite{jetspec,estruct130,estruct200}, and the interplay between these effects will need rigorous study which is outside the scope of this paper. It should be understood that the extracted values of the radial flow velocity in this paper is under the framework of the Blast-wave model.

Despite the different physical processes, the extracted $\Tkin$ and $\langle\beta\rangle$ evolve smoothly from $pp$ to central heavy-ion collisions. In $pp$ and peripheral Au+Au collisions, the kinetic freeze-out temperature is close to the chemical freeze-out temperature. As the multiplicity increases the $\Tkin$ decreases and the $\langle\beta\rangle$ increases. This trend continues through d+Au and Au+Au collisions. 

The extracted kinetic freeze-out temperature and the radial flow velocity are similar for Au+Au collisions at the three measured energies. As shown by Figs.~\ref{fig:Tfo} and~\ref{fig:auaubeta}, the magnitudes of the freeze-out parameters extracted from Au+Au collisions seem to be correlated only with the charged particle multiplicity, $\dNchdy$. This may suggest that the expansion rates, both before and after chemical freeze-out, are determined by the total event multiplicity, or the initial energy density as expressed through the energy density estimate in Eq.~(\ref{eq:Bj}). In other words, a higher initial energy density results in a larger expansion rate and longer expansion time, yielding a larger flow velocity and lower kinetic freeze-out temperature. 

The blast-wave fit so far treated all particles as primordial, ignoring resonance decays which are contained in the measured inclusive spectra. To assess the effect of resonance decays on the extracted kinetic freeze-out parameters, we extended the blast-wave model to include resonance decays as described in detail in Appendix~\ref{app:Resonance}. We found that the thus extracted kinetic freeze-out parameters agree with those obtained without including resonances within systematic uncertainties. This is because the resonance decay contributions are relatively $\pt$-independent within the $\pt$ ranges of our measurements. In addition, our study including short-lived resonances lends support to the picture of regeneration of short-lived resonances~\cite{Torrieri,Bleicher,STAR130kstar,STAR200kstar} during a relatively long time span from chemical to kinetic freeze-out.

\subsection{Excitation Functions}

The thermal model has been very successful in describing heavy-ion collisions and elementary particle collisions over a wide range of collision energies. Heavy-ion data from many energies have also been successfully fit by the blast-wave model. We compile results from some of these previous investigations~\cite{PBM95,PBM96,PBM99,SPS40chem,SISchem,SISchem2,Becattini01,Becattini03,Kaneta,Cleymans05}, together with RHIC data to study the excitation functions of the extracted chemical and kinetic freeze-out parameters. We note that the thermal model studies in Refs.~\cite{PBM95,PBM96,PBM99} do not include $\gamma_S$ as a free parameter; strangeness is treated as equilibrated with light flavors, i.e. $\gamma_S=1$.

\begin{figure}[htbp]
\includegraphics[width=0.48\textwidth]{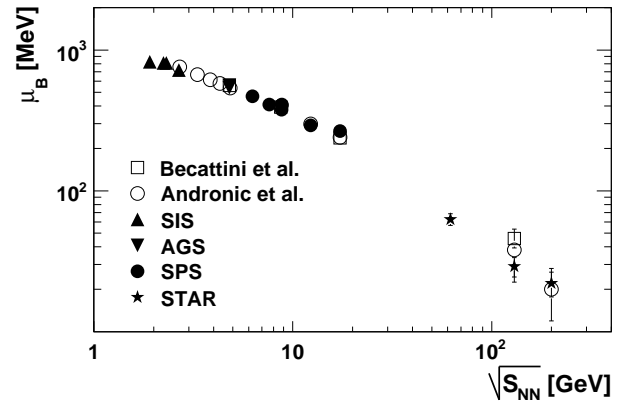}
\caption{Baryon chemical potential extracted for central heavy-ion collisions as a function of the collision energy. STAR 62.4~GeV and 130~GeV data are from this work; the 200~GeV data are from Ref.~\cite{spec200}. Other data are from SIS~\cite{SISchem,SISchem2}, AGS~\cite{PBM95,PBM99,Becattini01,Becattini03}, SPS~\cite{PBM96,PBM99,SPS40chem,Becattini01,Becattini03} and compilation by Refs.~\cite{Andronic05,Cleymans05}. Errors shown are the total statistical and systematic errors.}
\label{fig:muB_vs_roots}
\end{figure}

Figure~\ref{fig:muB_vs_roots} shows the baryon chemical potential extracted from chemical equilibrium model fits to central heavy-ion (Au+Au/Pb+Pb) data at various energies. The extracted $\mu_B$ falls monotonically from low to high energies. There are fewer net-baryons at mid-rapidity at higher energy because fewer baryons can transport over the larger rapidity gap.

\begin{figure}[hbtp]
\includegraphics[width=0.48\textwidth]{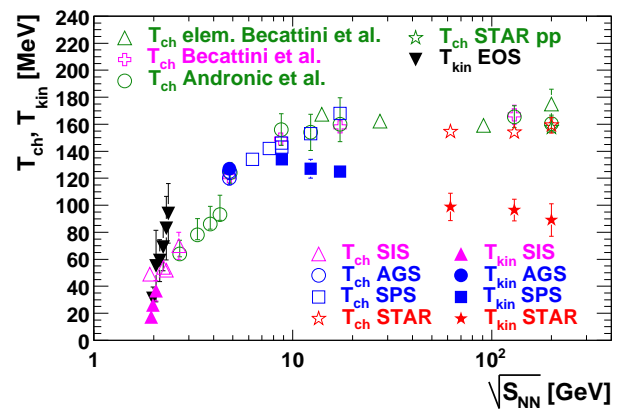}
\caption{(color online) The extracted chemical (open symbols) and kinetic (filled symbols) freeze-out temperatures for central heavy-ion collisions as a function of the collision energy. The STAR 62.4~GeV and 130~GeV data are from this work; the STAR 200~GeV data are from Ref.~\cite{spec200}. The other kinetic freeze-out results are from FOPI~\cite{FOPIbw2}, EOS~\cite{EOSbw}, E866~\cite{E802bw}, and NA49~\cite{NA49bw}. The other chemical freeze-out data are from SIS~\cite{SISchem,SISchem2}, AGS~\cite{PBM95,PBM99,Becattini01,Becattini03}, SPS~\cite{PBM96,PBM99,SPS40chem,Becattini01,Becattini03} and compilation by Refs.~\cite{Andronic05,Cleymans05}. Errors shown are the total statistical and systematic errors.}
\label{fig:T_vs_roots}
\end{figure}

\begin{figure}[htbp]
\includegraphics[width=0.48\textwidth]{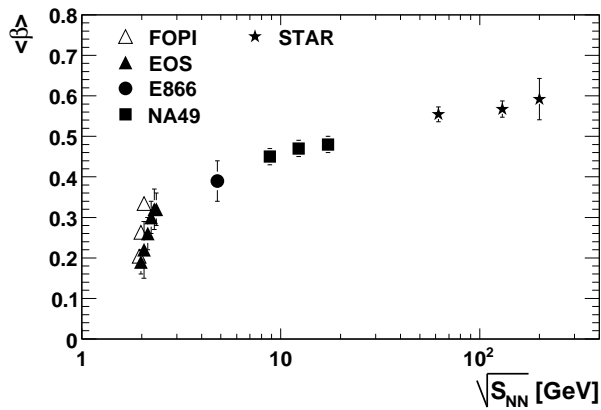}
\caption{Average transverse radial flow velocity extracted from the blast-wave model for central heavy-ion collisions as a function of the collision energy. The STAR 62.4~GeV and 130~GeV data are from this work, and the STAR 200~GeV $pp$ and Au+Au data from Ref.~\cite{spec200}. The other data are from FOPI~\cite{FOPIbw2}, EOS~\cite{EOSbw}, E866~\cite{E802bw}, and NA49~\cite{NA49bw}. Errors shown are the total statistical and systematic errors.}
\label{fig:beta_vs_roots}
\end{figure}

\begin{figure}[htbp]
\includegraphics[width=0.48\textwidth]{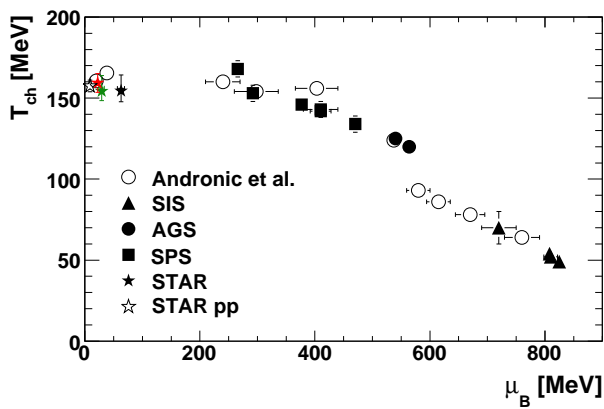}
\caption{(color online) Phase diagram plot of chemical freeze-out temperature versus baryon chemical potential extracted from chemical equilibrium models. Low energy data are taken from Refs.~\cite{SISchem,SISchem2,PBM95,PBM96,PBM99,SPS40chem,Becattini01,Becattini03} and compilations in Refs.~\cite{Andronic05,Cleymans05}. Errors shown are the total statistical and systematic errors.}
\label{fig:phase_diagram}
\end{figure}

Figure~\ref{fig:T_vs_roots} shows the evolution of the extracted chemical (open symbols) and kinetic (filled symbols) freeze-out temperature as a function of the collision energy in central heavy-ion collisions. The extracted $\Tch$ rapidly rises at SIS and AGS energy range and saturates at SPS and RHIC energies. In other words, central heavy-ion collisions at high energies can be characterized by a unique, energy independent chemical freeze-out temperature. The value of $\Tch$ is close to the phase transition temperature predicted by Lattice QCD. This suggests the collision system at high energies decouples chemically at the phase boundary.

On the other hand, the extracted kinetic freeze-out temperature rises at SIS and AGS energies, and decreases at higher energies, especially at RHIC energies. At low energies, the extracted $\Tkin$ is similar to $\Tch$. This suggests that kinetic freeze-out happens relatively quickly after or concurrently with chemical freeze-out. The two measured temperatures begin to separate at a collision energy around $\snn=10$~GeV, above which $\Tkin$ decreases with increasing energy, while $\Tch$ remains relatively constant. This suggests a prolonging of the period between chemical and kinetic freeze-outs, during which the particles scatter elastically, building up additional collective motion in the system while it undergoes further expansion and cooling.

Figure~\ref{fig:beta_vs_roots} shows the evolution of the extracted average flow velocity as a function of the collision energy. The extracted $\langle\beta\rangle$ steeply increases from SIS to AGS energies, and continues to increase at a lower rate at higher energies. Collective flow is an integral of all collective flow contributions over the entire evolution of the collision system. Part of it comes from the early stage of the collisions before chemical freeze-out, built up by the high pressure in the core of the collision zone. After chemical freeze-out, particles continue to interact elastically in central collisions, building up further transverse radial flow. This late-stage transverse expansion cools down the system and results in a lower kinetic freeze-out temperature in central collisions as discussed above. One should note that the extracted average flow velocity can be generated by different underlying physics at very low (SIS, AGS) and high (SPS, RHIC) incident energies. 

It is valuable to study collective radial flow at chemical freeze-out, as it comes from the early stage of the collision and hence is more sensitive to the initial condition than the final measured radial flow. The radial flow at chemical freeze-out may be assessed by analyzing $\pt$ spectra of particles with small hadronic interaction cross sections; some rare particles like $\phi$, $\Xi$, and $\Omega$ must develop most of their flow early (perhaps pre-hadronization) because their interaction cross sections are much lower than for the common $\pi$, $K$, $p$ and $\pbar$. It is found that the extracted radial flow for these rare particles is substantial in central heavy-ion collisions at RHIC, perhaps suggesting strong partonic flow in these collisions~\cite{STAR130msb,rareFlow}.

Figure~\ref{fig:phase_diagram} shows the chemical freeze-out temperature versus baryon chemical potential extracted from chemical equilibrium model fits to central Au+Au data. Low energy data points (SIS, AGS, SPS) are from the chemical equilibrium model fits~\cite{PBM99,SPS40chem,SISchem,SISchem2,Andronic05,Cleymans05} and references therein. At RHIC energies the chemical freeze-out points appear to be in the vicinity of the hadron-QGP phase-transition (hadronization) predicted by lattice gauge theory~\cite{LGT,Stachel}.

\section{Summary\label{sec:Summary}}

Charged particles of $\pi^{\pm}$, $K^{\pm}$, $p$ and $\pbar$ are identified by the specific ionization energy loss ($\dedx$) method in STAR at low transverse momenta and mid-rapidity ($|y|<0.1$) in $pp$ and d+Au collisions at $\snn=200$~GeV and in Au+Au collisions at 62.4~GeV, 130~GeV, and 200~GeV. Transverse momentum spectra of the identified particles are reported. Spectra of heavy particles are flatter than those of light particles in all collision systems. This effect becomes more prominent in more central Au+Au collisions. In $pp$ and d+Au collisions processes such as semi-hard scattering and $\kt$ broadening should play an important role. In central Au+Au collisions the flattening of the spectra is likely dominated by collective transverse radial flow, developed due to the large pressure buildup in the early stage of heavy-ion collisions. 

The transverse momentum spectra are extrapolated to the unmeasured regions by the hydrodynamics-motivated blast-wave model parameterization for kaons, protons and antiprotons and by the Bose-Einstein function for pions. The total integrated particle yields are reported. The Bjorken energy density estimated from the total transverse energy is at least several~GeV/fm$^3$ at a formation time of less than 1~fm/$c$. The extrapolated $\meanpt$ increases with particle mass in each collision system, and increases with centrality for each particle species. The $\meanpt$ systematics are similar for the three measured energies at RHIC, and appear to be strongly correlated with the total particle multiplicity density or the ratio of the multiplicity density over the transverse overlap area of the colliding nuclei.

Ratios of the integrated particle yields are presented and discussed. While rather independent of centrality for 130~GeV and 200~GeV, the $\pbar/p$ ratio drops significantly with centrality in 62.4~GeV Au+Au collisions. This indicates a more significant net-baryon content at mid-rapidity in Au+Au collisions at 62.4~GeV. On the other hand, antibaryon production relative to the total particle multiplicity, while lower at the lower energy, is independent of centrality for all three collision energies at RHIC, despite the increasing net-baryon density at the low 62.4~GeV energy.

Strangeness production relative to the total particle multiplicity is similar at the different RHIC energies. The effect of collision energy on the production rate is significantly smaller on strangeness production than on antibaryon production. Relative strangeness production increases quickly with centrality in peripheral Au+Au collisions, and remains the same above medium-central collisions at RHIC. The increase in relative strangeness production in central Au+Au collisions from $pp$ is approximately 50\%. 

The particle yield ratios are fit in the framework of the thermal equilibrium model. The extracted chemical freeze-out temperature is the same in $pp$, d+Au, and Au+Au collisions at all measured energies at RHIC, and shows little centrality dependence in Au+Au collisions. The extracted value of chemical freeze-out temperature is close to the Lattice QCD predicted phase transition temperature between hadronic matter and the Quark-Gluon Plasma, suggesting that chemical freeze-out happens in the vicinity of the phase boundary shortly after hadronization. The extracted strangeness suppression factor is substantially below unity in $pp$, d+Au, and peripheral Au+Au collisions; strangeness production is significantly suppressed in these collisions. The strangeness suppression factor in medium-central to central Au+Au collisions is not much below unity; the strangeness and light flavor are nearly equilibrated, which may suggest a fundamental change from peripheral to central collisions.

The extracted kinetic freeze-out temperature from the blast-wave fit to the transverse momentum spectra, on the other hand, decreases from $pp$ and d+Au to central Au+Au collisions. At the same time, the extracted collective flow velocity increases significantly with increasing centrality. While the apparent finite flow velocity fit in $pp$ and d+Au collisions may be due to semi-hard scatterings and jets, the extracted large flow velocity in central Au+Au collisions is likely dominated by collective transverse radial flow. The significant difference between the extracted chemical and kinetic freeze-out temperature suggests the presence of an elastic rescattering phase between the two freeze-outs. The variations of the extracted freeze-out properties are smooth from $pp$ and d+Au to Au+Au collisions and over the measured energies for the Au+Au collision system; the trends seem to be tied to the event multiplicity. Resonance decays are found to have little effect on the extracted kinetic freeze-out parameters due to the fact that the resonance decay products have similar kinematics as the primordial particles in our measured transverse momentum ranges. The study including different contributions from short-lived resonances lends support to the regeneration picture of those resonances with a long time span from chemical to kinetic freeze-out.

The identified particle spectra at RHIC energies and the equilibrium model studies presented here suggest that the collision systems chemically decouple at a universal temperature, independent of the vastly different initial conditions at different centralities. The apparent different collective flow strengths in the final state of non-peripheral heavy-ion collisions likely are dominated by transverse radial flow and stem out of the different amount of pressure build-up at the initial stage. Part of the collective flow in central collisions appears to be built up after chemical freeze-out, during which the collision zone undergoes further expansion and cooling through particle elastic scatterings, resulting in a lower kinetic decoupling temperature in more central collisions.

\appendix
\section{The Glauber Model\label{app:Glauber}}

To describe heavy-ion collisions, geometric quantities are often used, such as the number of participant nucleons ($\Npart$), the number of nucleon-nucleon binary collisions ($\Ncoll$), and the transverse overlap area of the colliding nuclei ($\Sovlp$). Unfortunately these quantities cannot be measured directly from experiments~\footnote{An exception is that in fixed target experiments the number of participants can be experimentally measured by zero degree calorimeters.}. Their values can only be derived by mapping the measured data, such as the $dN/d\Nch$ distribution, to the corresponding distribution obtained from phenomenological calculations, thus relating $\Npart$, $\Ncoll$, and $\Sovlp$ to the measured $dN/d\Nch$ distribution. These types of calculations are generally called Glauber model calculations and come in two implementation schemes, the {\it optical} and the {\it Monte-Carlo} Glauber calculations.

\begin{figure}[htbp]
\includegraphics[width=0.48\textwidth]{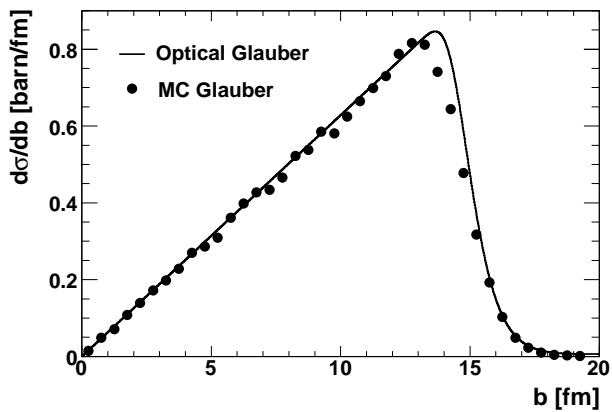}
\caption{Differential cross-sections obtained from the optical and MC Glauber calculations for Au+Au collisions at 200~GeV. Statistical errors are smaller than the point size.}
\label{fig:dsigmadb}
\end{figure}

The optical model is based on an analytic consideration of continuously overlapping nuclei~\cite{Wong,Baltz,Antinori00,dima01}. The MC approach is based on a computer simulation of billiard ball-like colliding nucleons~\cite{CorrNch1,PHENIX130,BRAHMS130,Back01,Back03,GlauberMiller}. Figure~\ref{fig:dsigmadb} shows the differential cross-sections versus $b$ of minimum bias Au+Au collisions at 200~GeV calculated by the optical and MC Glauber models. As seen from the figure, the differential cross-sections agree between the two calculations except at large impact parameters, or in very peripheral collisions. The disagreement in very peripheral collisions is understood because the optical approach loses its validity in these collisions. Due to this disagreement, the integrated total cross-sections differ between the optical and MC calculations, by about 5\%.

To relate Glauber calculations to experimental measurements, one first obtains the impact parameter range corresponding to the measured centrality bin using the differential cross-section, such as the ones shown in Fig.~\ref{fig:dsigmadb}. The average $\Npart$ and $\Ncoll$ values are then calculated in the Glauber model for the impact parameter range. Table~\ref{tab:optical} lists the $\Npart$ and $\Ncoll$ values obtained from the optical Glauber calculations for our multiplicity classes in 62.4~GeV, 130~GeV, and 200~GeV Au+Au collisions. The MC Glauber results are already listed in Table~\ref{tab:collision_prop} in the main text. As seen from the tables, different implementations of the Glauber model lead to slightly different values for $\Npart$ and $\Ncoll$, as has been noted before in Ref.~\cite{Back01}. The results are different for non-peripheral collisions even though the differential cross-sections match between the two Glauber calculations. This is because the impact parameter ranges corresponding to the same measured centrality bin differ slightly due to the different total cross-sections. The disagreement in the Glauber results is more significant in peripheral collisions due to reasons noted above. Thus, any results reported in terms of Glauber quantities must be carefully interpreted based upon specifics of the underlying calculations. 

\begin{table}[htbp]
\caption{The optical Glauber model results corresponding to the centrality bins used in the 62.4~GeV, 130~GeV, and 200~GeV Au+Au data. The quoted errors are systematic uncertainties.}
\label{tab:optical}
\begin{ruledtabular}
\begin{tabular}{ccccc}
centrality & $b$-range (fm) & $b$ (fm) & $\Npart$ & $\Ncoll$ \\ \hline
\multicolumn{5}{c}{Au+Au 200~GeV ($\sigmapp=41$ mb)}\\ \hline
90-100\%& 14.3-15.7 & $14.8^{+1.1}_{-0.5}$ & $1.43^{+0.73}_{-0.64}$ & $1.02^{+0.57}_{-0.47}$ \\
80-90\% & 13.4-14.3 & $13.8\pm0.4$	  & $4.5^{+1.3}_{-1.1}$  & $3.7^{+1.2}_{-1.0}$ \\
70-80\% & 12.5-13.4 & $13.0\pm0.3$	  & $10.7^{+2.2}_{-2.0}$  & $10.0^{+2.7}_{-2.3}$ \\
60-70\% & 11.6-12.5 & $12.1\pm0.3$	  & $22.0^{+3.3}_{-3.1}$  & $25.1^{+5.3}_{-4.8}$ \\
50-60\% & 10.6-11.6 & $11.1\pm0.3$	  & $40.6\pm4.3$	  & $56.2^{+9.1}_{-8.6}$ \\
40-50\% & 9.48-10.6 & $10.0\pm0.3$	  & $67.8\pm5.0$	  & $113\pm14$ \\
30-40\% & 8.21-9.48 & $8.86\pm0.23$	  & $105.4\pm5.3$	  & $206\pm19$ \\
20-30\% & 6.70-8.21 & $7.48\pm0.19$	  & $155.9\pm5.1$	  & $351\pm26$ \\
10-20\% & 4.74-6.70 & $5.78\pm0.15$	  & $223.6\pm4.2$	  & $571\pm36$ \\
 5-10\% & 3.35-4.74 & $4.08\pm0.11$	  & $289.6^{+2.9}_{-3.1}$ & $807\pm48$ \\
 0-5\% & 0  -3.35 & $2.23\pm0.06$	  & $345.8^{+1.8}_{-2.0}$ & $1027\pm61$ \\
\hline
\multicolumn{5}{c}{Au+Au 130~GeV ($\sigmapp=39$ mb)}\\ \hline
85-100\%& 13.6-15.2 & $14.3^{+1.0}_{-0.5}$ & $2.7^{+1.7}_{-1.3}$  & $2.0^{+1.4}_{-1.0} $ \\
58-85\% & 11.3-13.6 & $12.5\pm0.4$ 	  & $17.0^{+4.6}_{-3.9}$ & $18.6^{+6.6}_{-5.2} $ \\
45-58\% & 9.92-11.3 & $10.6\pm0.4$ 	  & $51.8^{+7.6}_{-7.0}$ & $76^{+16}_{-14} $ \\
34-45\% & 8.62-9.92 & $9.29\pm0.31$ 	  & $89.7^{+8.4}_{-8.0}$ & $161\pm23$ \\
26-34\% & 7.54-8.62 & $8.10\pm0.27$ 	  & $131.0^{+8.3}_{-8.1}$ & $268\pm28$ \\
18-26\% & 6.28-7.54 & $6.93\pm0.23$ 	  & $175.7\pm7.6$ 	  & $398\pm33$ \\
11-18\% & 4.91-6.28 & $5.62\pm0.19$ 	  & $228.2\pm6.3$	  & $564\pm39$ \\
6-11\% & 3.62-4.91 & $4.30\pm0.14$	  & $280.0\pm4.7$	  & $740\pm47$ \\
 0-6\% & 0  -3.62 & $2.42\pm0.08$	  & $339.3\pm2.6$	  & $958\pm59$ \\
\hline
\multicolumn{5}{c}{Au+Au 62.4~GeV ($\sigmapp=36$ mb)}\\ \hline
90-100\%& 14.2-15.6 & $14.7^{+1.0}_{-0.5}$  & $1.47^{+0.73}_{-0.64}$ & $1.02^{+0.56}_{-0.46} $ \\
80-90\% & 13.3-14.2 & $13.7^{+0.4}_{-0.3}$  & $ 4.6^{+1.2}_{-1.1}$ & $3.6^{+1.2}_{-1.0}$ \\
70-80\% & 12.5-13.3 & $12.9\pm0.3$ 	   & $ 10.6^{+2.2}_{-1.9}$ & $9.7^{+2.5}_{-2.2}$ \\
60-70\% & 11.5-12.5 & $12.0\pm0.3$ 	   & $ 21.8^{+3.3}_{-3.1}$ & $23.6^{+4.9}_{-4.4}$ \\
50-60\% & 10.5-11.5 & $11.0\pm0.3$ 	   & $ 40.0^{+4.3}_{-4.1}$ & $52.0^{+8.2}_{-7.7}$ \\
40-50\% & 9.42-10.5 & $9.98\pm0.26$ 	   & $ 66.8\pm4.9$ 	   & $103\pm12$ \\
30-40\% & 8.15-9.42 & $8.80\pm0.22$ 	   & $103.8\pm5.2$ 	   & $186\pm17$ \\
20-30\% & 6.66-8.15 & $7.43\pm0.19$ 	   & $153.5\pm5.0$ 	   & $314\pm24$ \\
10-20\% & 4.71-6.66 & $5.74\pm0.15$ 	   & $220.4^{+4.1}_{-4.3}$ & $506\pm34$ \\
 5-10\% & 3.33-4.71 & $4.06\pm0.10$ 	   & $285.9^{+3.1}_{-3.3}$ & $712\pm46$ \\
 0-5\% & 0  -3.33 & $2.22\pm0.06$ 	   & $342.2\pm2.3$ 	   & $903\pm59$ \\
\end{tabular}
\end{ruledtabular}
\end{table}

In the following we will briefly describe the optical and MC Glauber calculations.

\subsection{The Optical Glauber Model}

For our optical Glauber calculation, we start by assuming a spherically symmetric Woods-Saxon density profile,
\begin{equation}
\rho(r)=\frac{\rho_0}{1+\exp(\frac{r-r_0}{a})},\label{eq:WoodsSaxon}
\end{equation}
with the parameter $a=0.535\pm0.027$~fm as experimentally measured in $e$-Au scattering reported in Refs.~\cite{deJager:74,deJager:87}. From the same publication we extracted the value for $r_0$, but increased it from $6.38$~fm to $6.5\pm0.1$~fm to approximate the effect of the neutron skin. The normalization factor $\rho_0 = 0.161$~fm$^{-3}$ is fixed by $\int_{0}^{\infty}\rho(r)4\pi r^2dr=197$, the total number of nucleons in the Au nucleus. 

We concentrate on symmetric Au+Au collisions. Let the beam-axis be along $\hat{z}$. The nuclear thickness density function is given by
\begin{equation}
T_A(\vec{s})=T_A(s)=\int_{-\infty}^{\infty}\rho(\vec{s},z)dz\ ,\label{eq:TA}
\end{equation}
where $\vec{s}$ is a vector perpendicular to the beam-axis $\hat{z}$, and $s=|\vec{s}|$; $\rho(\vec{s},z)$ is the nuclear density in the volume element $ds^2dz$ at ($\vec{s},z$), and for our spherical nucleus, $\rho(\vec{s},z)=\rho(s,z)=\rho(\sqrt{s^2+z^2})$ as given by Eq.~(\ref{eq:WoodsSaxon}).

For a Au+Au collision with impact parameter $\vec{b}$, the nuclear overlap integral can be calculated as the integral over two density profiles, 
\begin{equation}
T_{AA}(b)=\int T_A(\vec{s})T_A(\vec{s}-\vec{b})ds^2\ .\label{eq:TAA}
\end{equation}
The number of binary nucleon-nucleon collisions is given by
\begin{equation}
\Ncoll(b)=\sigmapp T_{AA}(b)\ .\label{eq:Ncoll}
\end{equation}
Here, we assume that the interaction probability is solely given by the proton-proton cross-section $\sigmapp$, thus neglecting effects like excitation and energy loss. The number of participant nucleons (nucleons suffering at least one collision) is derived from $T_A$ by 
\begin{equation}
\Npart(b)=2\int T_A(\vec{s})\left(1-e^{-\sigmapp T_A(\vec{s}-\vec{b})}\right)ds^2 \ .\label{eq:Npart}
\end{equation}

By definition, there is no fluctuation in the optical Glauber model. For a given $b$, quantities like $T_A$, $T_{AA}$, $\Ncoll$, and $\Npart$ are analytically defined. In order to calculate the cross-section, however, one has to invoke the concept of fluctuation. In this sense, Eq.~(\ref{eq:Ncoll}) gives the average number of binary collisions for Au+Au collisions at impact parameter $b$, and taking Poisson statistics, the probability for no interaction is $e^{-\Ncoll(b)}$. The differential cross-section is thus given by
\begin{equation}
\frac{d\sigmaAA}{db}=2\pi b\left(1-e^{-\sigmapp T_{AA}(b)}\right)\ .\label{eq:dsigmadb}
\end{equation}
The total hadronic cross-section for Au+Au collisions can hence be obtained as
\begin{equation}
\sigmaAA=\int_{0}^{\infty}db\frac{d\sigmaAA}{db}\ .\label{eq:sigmaAA}
\end{equation}
The values of $\sigmapp$ are taken to be $\sigmapp=36\pm2$~mb, $39\pm2$~mb and $41\pm2$~mb for 62.4~GeV, 130~GeV, and 200~GeV, respectively. With these $pp$ cross-sections, the corresponding total cross sections for Au+Au are calculated to be approximately $\sigmaAA=7.18$~barns, 7.24~barns, and 7.27~barns, respectively.

To relate $\Npart$ and $\Ncoll$ to the experimental observable $\Nch$, the mean of the total number of charged tracks in our centrality bin, we use Eqs.~(\ref{eq:dsigmadb}) and~(\ref{eq:sigmaAA}) to obtain the impact parameters corresponding to the fraction of the total geometric cross-section for our centrality bin. For a given impact parameter range $b_1<b<b_2$ for each centrality bin, we then use Eqs.(\ref{eq:dsigmadb}), (\ref{eq:Ncoll}), and (\ref{eq:Npart}) to calculate the $\Ncoll$ and $\Npart$ by 
\begin{eqnarray}
\Ncoll&=&\frac{\int_{b_1}^{b_2}\Ncoll(b)\frac{d\sigmaAA}{db}db}{\int_{b_1}^{b_2}\frac{d\sigmaAA}{db}db}\ ,\label{eq:meanNcoll}\\
\Npart&=&\frac{\int_{b_1}^{b_2}\Npart(b)\frac{d\sigmaAA}{db}db}{\int_{b_1}^{b_2}\frac{d\sigmaAA}{db}db}\ .\label{eq:meanNpart}
\end{eqnarray}

\subsection{The Monte-Carlo Glauber Model}

The MC method simulates a number of independent Au+Au collisions. For each collision, a target and a projectile nucleus are modeled according to the Woods-Saxon nucleon density profile of Eq.~(\ref{eq:WoodsSaxon}). The nucleons are separated by a minimum distance $d_{\rm min}=0.4$~fm which is characteristic of the range of the repulsive nucleon-nucleon force. The target and projectile nuclei are separated by the impact parameter $b$, with $b^2$ chosen randomly from a flat distribution. The nucleons follow straight-line trajectories in collisions. A pair of nucleons along the path is determined to `interact' if they are separated by a transverse distance
\begin{equation}d\le\sqrt{\frac{\sigmapp}{\pi}}\ ,\end{equation}
where $\sigmapp$ is the nucleon-nucleon interaction cross-section. The colliding nuclei are considered to have interacted (resulting in an Au+Au event) if at least one pair of nucleons has interacted. Again, the values of $\sigmapp$ are taken to be $\sigmapp=36\pm2$~mb, $39\pm2$~mb and $41\pm2$~mb for 62.4~GeV, 130~GeV, and 200~GeV, respectively. With these $pp$ cross-sections, the corresponding total cross sections for Au+Au are calculated to be approximately $\sigmaAA=6.84$~barns, 6.89~barns, and 6.93~barns, respectively.

The normalized differential cross-section, $\frac{1}{\sigmaAA}\frac{d\sigmaAA}{db}$, is obtained from the normalized event distribution. The $\frac{1}{\sigmaAA}\frac{d\sigmaAA}{db}$ distribution is divided into bins corresponding to the fractions of the measured total cross-section of the used centrality bins. The number of participants $\Npart$ is defined as the total number of nucleons that undergo at least one interaction. The number of binary collisions $\Ncoll$ is defined as the total number of nucleon-nucleon interactions in the collision. The mean values of $\Npart$ and $\Ncoll$ are determined for each centrality bin in the same way as for the optical Glauber model, by Eq.~(\ref{eq:meanNcoll}) and Eq.~(\ref{eq:meanNpart}).

The transverse overlap area, $\Sovlp$, for $pp$ collisions is taken to be the $pp$ cross-section $\sigmapp$. To calculate the transverse overlap area between the colliding nuclei of Au+Au collisions, the individual $pp$ interaction cross-sections are projected onto the transverse plane. The overlap area in the transverse plane, $\Sovlp$, is then calculated. The overlapping portion of the projected areas from two or more nucleon-nucleon interactions is counted only once. The mean $\Sovlp$, weighted by the differential cross-section, is determined in the same manner as in Eq.~(\ref{eq:meanNcoll}) and Eq.~(\ref{eq:meanNpart}). Table~\ref{tab:collision_prop} lists the obtained $\Sovlp$ along with $\Npart$ and $\Ncoll$. 

\subsection{Uncertainties}

The uncertainties on $\Npart$, $\Ncoll$, and $\Sovlp$ from both the optical and MC Glauber model calculations are evaluated by varying the Woods-Saxon parameters, the values of $\sigmapp$ and $d_{\rm min}$, and by including an uncertainty in the determination of the measured total Au+Au cross-section. 
\begin{itemize}
\item The Woods-Saxon nuclear density profile parameters $a$ and $r_0$ are varied within their respective uncertainties: $a=0.535\pm0.027$~fm and $r=6.50\pm0.12$~fm.
\item The $\sigmapp$ values are 36~mb (62.4~GeV), 39~mb (130~GeV), and 41~mb (200~GeV) as default and are varied within an uncertainty of $\pm 2$~mb.
\item The $d_{\rm min}$ value is 0.4~fm as default and is varied between 0.2~fm and 0.5~fm. This only applies to the MC Glauber calculation.
\item Due to inefficiencies in the online trigger and offline primary vertex reconstruction for peripheral collisions, our measured total (minimum bias) cross-section does not fully account for the total Au+Au hadronic cross-section. The measured fractions of the total cross-section are determined to be $97\pm3$\%~\cite{photon62,TOF_AuAu}, $95\pm5$\%~\cite{hminus130,Manuel}, and $97\pm3$\%~\cite{highpt200,CorrNch3} for minimum bias Au+Au collisions at 62.4~GeV, 130~GeV, and 200~GeV, respectively. These fractions are used in the determination of our Glauber model results, and their uncertainties are included in the quoted uncertainties on the results.
\end{itemize}
The uncertainties from these sources are determined separately and summed in quadrature in the quoted uncertainties on the Glauber results in Table~\ref{tab:collision_prop} and Table~\ref{tab:optical}. In peripheral collisions, the uncertainties are dominated by those in the minimum bias cross-section measurements. In central collisions, the uncertainty in $\Ncoll$ is dominated by the uncertainty in $\sigmapp$, while all sources contribute significantly to the uncertainties in $\Npart$ and $\Sovlp$. The uncertainties on $\Npart$, $\Ncoll$, and $\Sovlp$ are correlated.

\section{Resonance Effect on Blast-Wave Fit\label{app:Resonance}}

\begin{table*}[htb] 
\caption{Extracted kinetic freeze-out parameters and the fit $\chi^2$/{\sc ndf} from the blast-wave model including resonances for minimum bias $pp$ and top 5\% central Au+Au collisions at 200~GeV. Three cases of treating $\rho$ decays are studied. The flow profile $n$ parameter is fixed to 0.82 for the Au+Au fit and is free for the $pp$ fit. All errors are statistical.}
\label{tab:reso_fits}
\begin{ruledtabular}
\begin{tabular}{c|cccc|cccc}
&\multicolumn{4}{c|}{$pp$ minimum bias} & \multicolumn{4}{c}{Au+Au top 5\%} \\
Case & $\Tkin$ $(MeV)$ & $\langle\beta\rangle$ & $n$ &$\chi^2$/{\sc ndf} & $\Tkin$ $(MeV)$ & $\langle\beta\rangle$ & $n$ &$\chi^2$/{\sc ndf} \\ 
\hline
100\% $\rho$ & $117.8\pm3.2$ & $0.29\pm0.01$ & $3.1\pm0.6$ & 1.1 & $77.2^{ +0.8}_{ -0.9}$ & $0.604^{ +0.004}_{ -0.003}$ & 0.82 fixed & 0.60 \\
0\% $\rho$ 	& $		121.9\pm0.9$ & $0.35\pm0.01$ & $1.2\pm0.1$ & 4.4 & $94.6^{ +0.9}_{ -1.0}$ & $0.603^{ +0.004}_{ -0.002}$ & 0.82 fixed & 0.37 \\
50\% $\rho$	&  $122.2\pm1.2$ & $0.35\pm0.01$ & $1.0\pm0.3$ & 2.5 & $87.4^{ +0.9}_{ -1.1}$ & $0.605^{ +0.002}_{ -0.002}$ & 0.82 fixed & 0.45 \\ 
\end{tabular}
\end{ruledtabular}
\end{table*}

The blast-wave fit in Section~\ref{sec:kineticFO} treats all particles as primordial, ignoring resonance decays. However, the measured identified inclusive particle spectra contain contributions from resonance decays. The question arises whether or not resonance decays have a significant effect on the extracted kinetic freeze-out parameters. To answer this question, the blast-wave model fit is extended to include resonance decays. The identified particle $\pt$ spectra measured at mid-rapidity in minimum bias $pp$ and in the most central 5\% Au+Au collisions at 200~GeV~\cite{spec200} are utilized to study the effect of resonance decays on the extracted kinetic freeze-out parameters~\cite{Molnar}.

\subsection{Effect of Resonance Decays}

The study is based on the combination of the chemical equilibrium model~\cite{PBM95,PBM96,PBM99,Kaneta} and the blast-wave model by Wiedemann and Heinz~\cite{HeinzBW}. Several changes have been implemented with respect to the original code~\cite{HeinzBW} to provide the same basis for the calculation as in data~\cite{spec200}. The Wiedemann-Heinz blast-wave model uses the same temperature to determine the relative abundances of particles and resonances and to calculate their kinetic distributions. In this study, two distinct freeze-out temperatures are implemented: the chemical freeze-out temperature and the kinetic freeze-out temperature. The relative abundances of particles and resonances are determined by chemical freeze-out parameters and are fixed in our study. We used the following chemical freeze-out parameters: 
$\Tch$= 159 MeV, $\mu_{B}$= 18 MeV, $\mu_{S}$= 2.3 MeV, and $\gamma$=0.62 for $pp$ collisions at 200~GeV~\cite{spec200,STAR200strange}, and
$\Tch$= 160 MeV, $\mu_{B}$= 24 MeV, $\mu_{S}$= 1.4 MeV, and $\gamma$=0.99 for the top 5\% central Au+Au collisions at 200~GeV~\cite{spec200,SBenhancement}. 
More particles are included than in Ref.~\cite{HeinzBW}: $\rho$, $\omega$, $\eta$, $\eta'$, $K^{*0}$, $K^{*\pm}$, $\phi$ and $\Lambda$, $\Delta$, $\Sigma$, $\Xi$, $\Lambda_{1520}$, $\Sigma_{1385}$, $\Omega$. A box flow profile is chosen, similarly to Ref.~\cite{spec200}: $\beta=\beta_{S}\left(r/R\right)^{n}$, where $n$ is fixed to be 0.82 for Au+Au collisions and set free for $pp$ collisions. A flat rapidity distribution is implemented at mid-rapidity. This is needed because, although the measured spectra are in $|y|<0.1$, resonances outside this region can decay into particles falling within the region. The resonance kinematics are calculated at a given kinetic freeze-out temperature and average flow velocity. The spectra of the decay products are combined with those of the primordial ones. Spin, isospin degeneracies and decay branching ratios are properly taken into account.

\begin{figure*}[hbtp]
\centering
\includegraphics[width=0.32\textwidth]{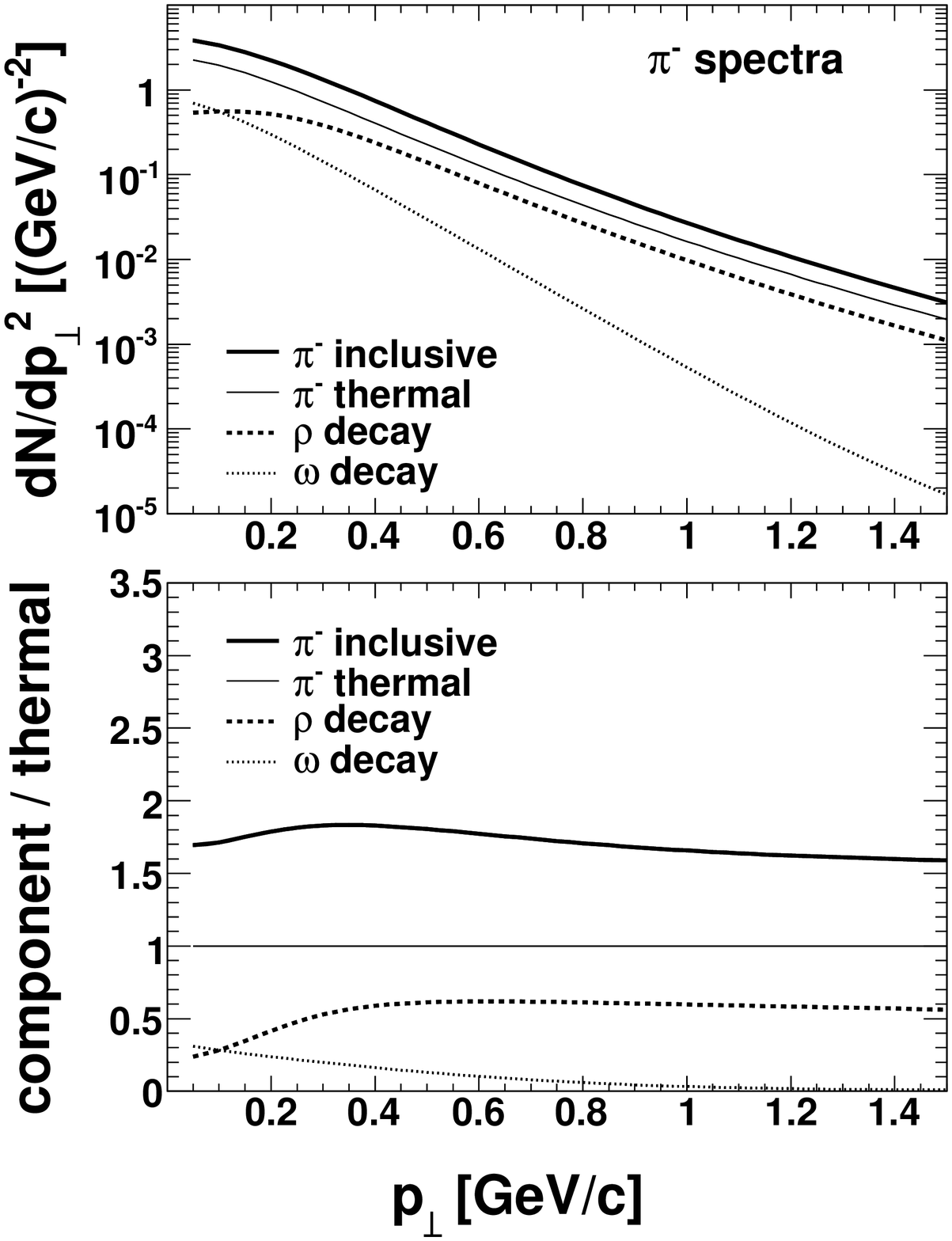}
\includegraphics[width=0.32\textwidth]{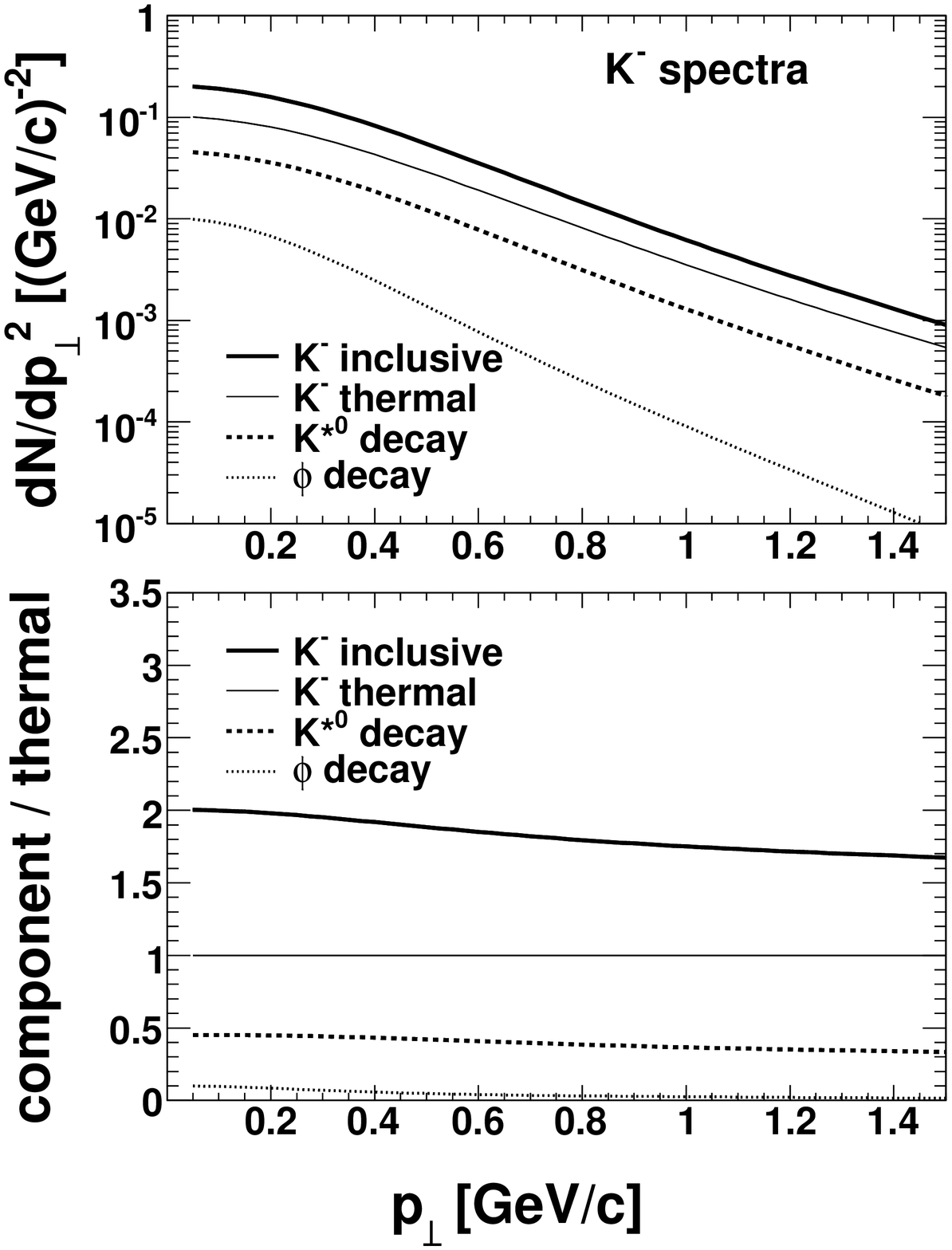}
\includegraphics[width=0.32\textwidth]{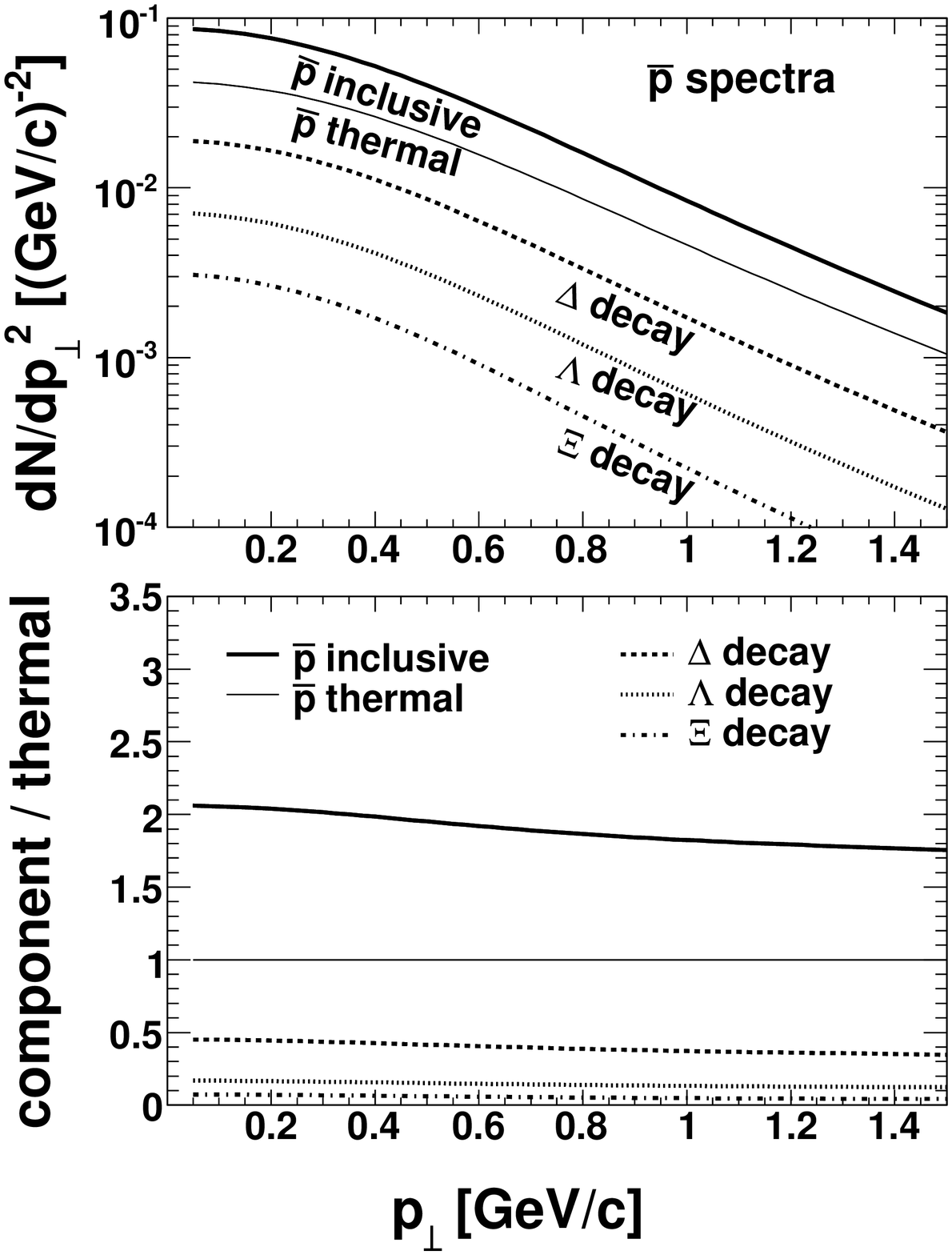}
\caption{Upper panels: Calculated $\pi^{-}$, $K^{-}$, and $\bar{p}$ transverse momentum spectra from the primordial thermal component and major resonance decay contributions for $pp$ collisions at 200~GeV. The kinetic freeze-out parameters fit to data are used for the thermal calculations~\cite{spec200}. Lower panels: Resonance contributions relative to the thermal spectrum. $K^{*-}$ decay (not shown) has the same contribution to $K^-$ (and $K^+$) spectra as $K^{*0}$ decay. $\Sigma$ and $\Sigma_{1385}$ decays (not shown) contribute to the $\pbar$ (and proton) spectra with similar magnitude as $\Lambda$ decays.}
\label{fig:resonance_pp}
\vspace{0.1in}
\centering
\includegraphics[width=0.32\textwidth]{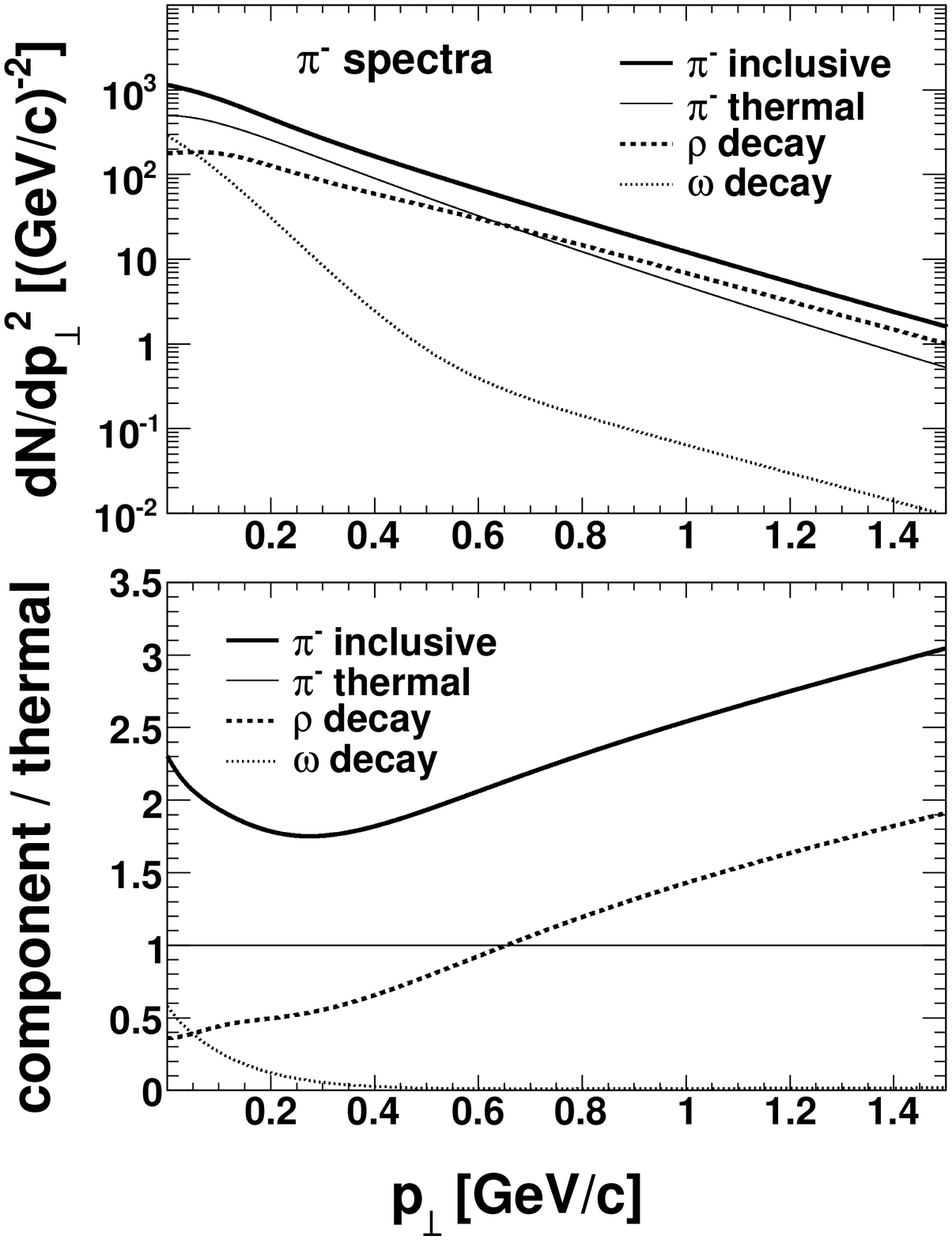}
\includegraphics[width=0.32\textwidth]{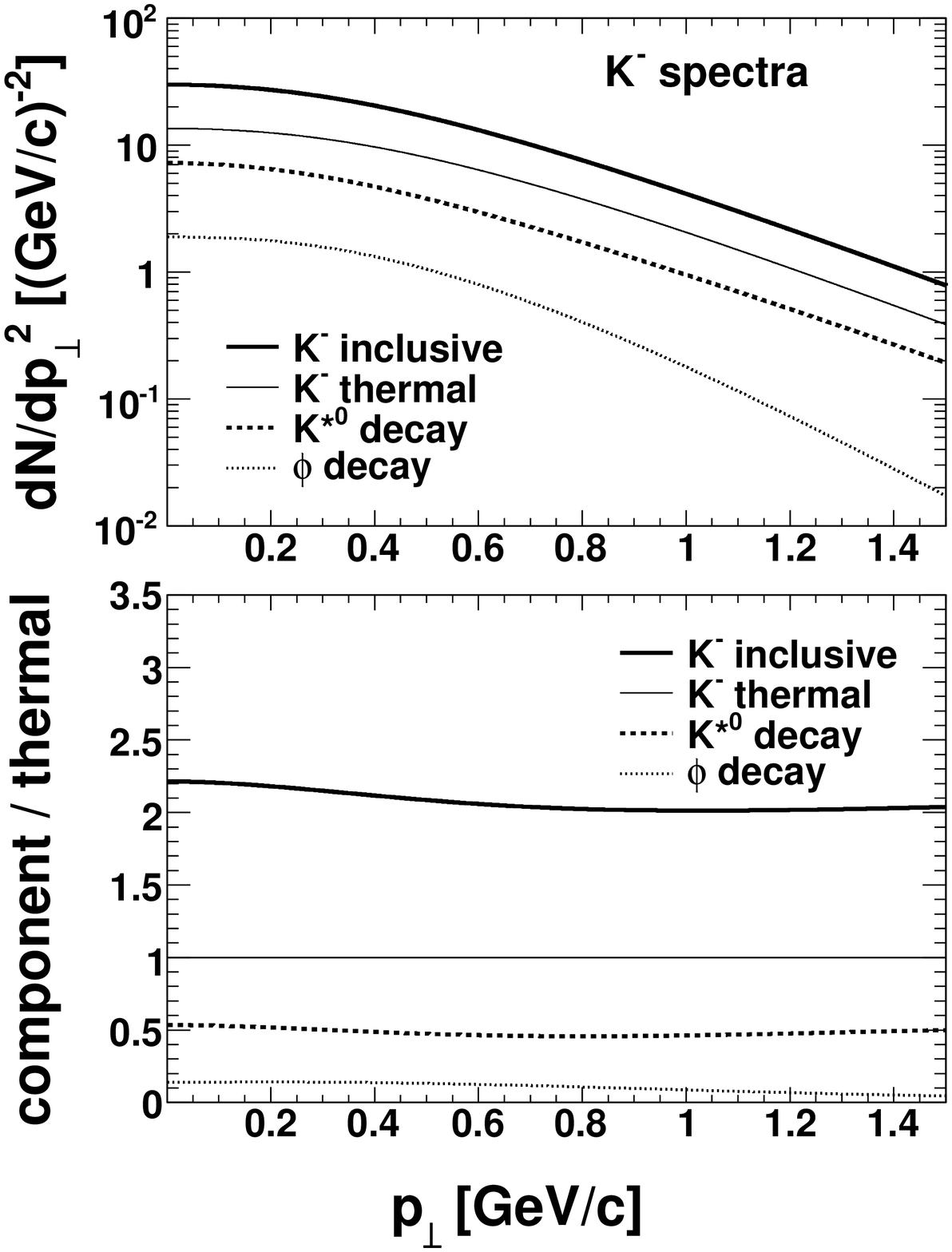}
\includegraphics[width=0.32\textwidth]{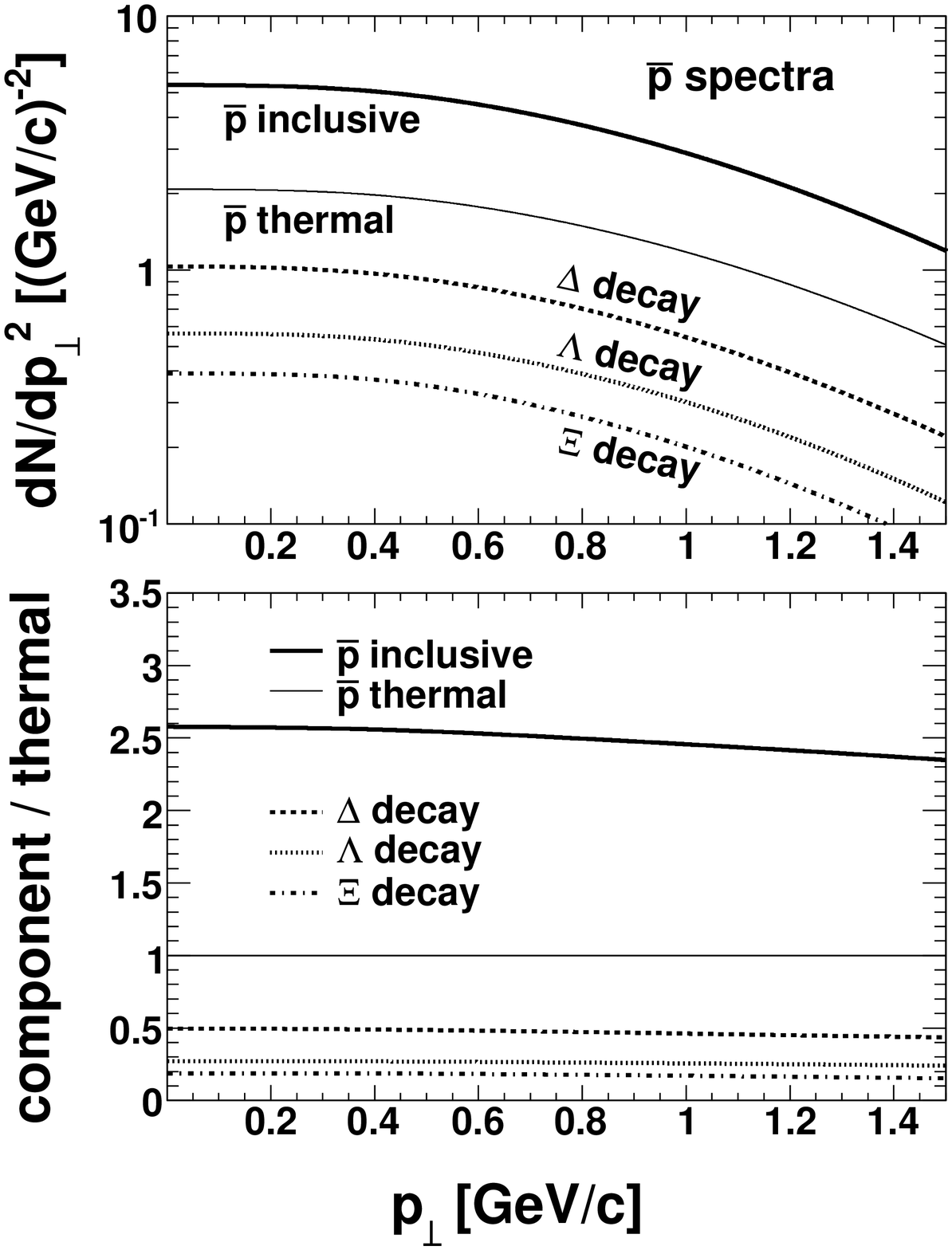}
\caption{Same as Fig.~\ref{fig:resonance_pp} but for the top 5\% central Au+Au collisions at 200~GeV~\cite{spec200}.}
\label{fig:resonance_AuAu}
\end{figure*}

\begin{figure*}[htbp]
\centering
\includegraphics[width=0.245\textwidth]{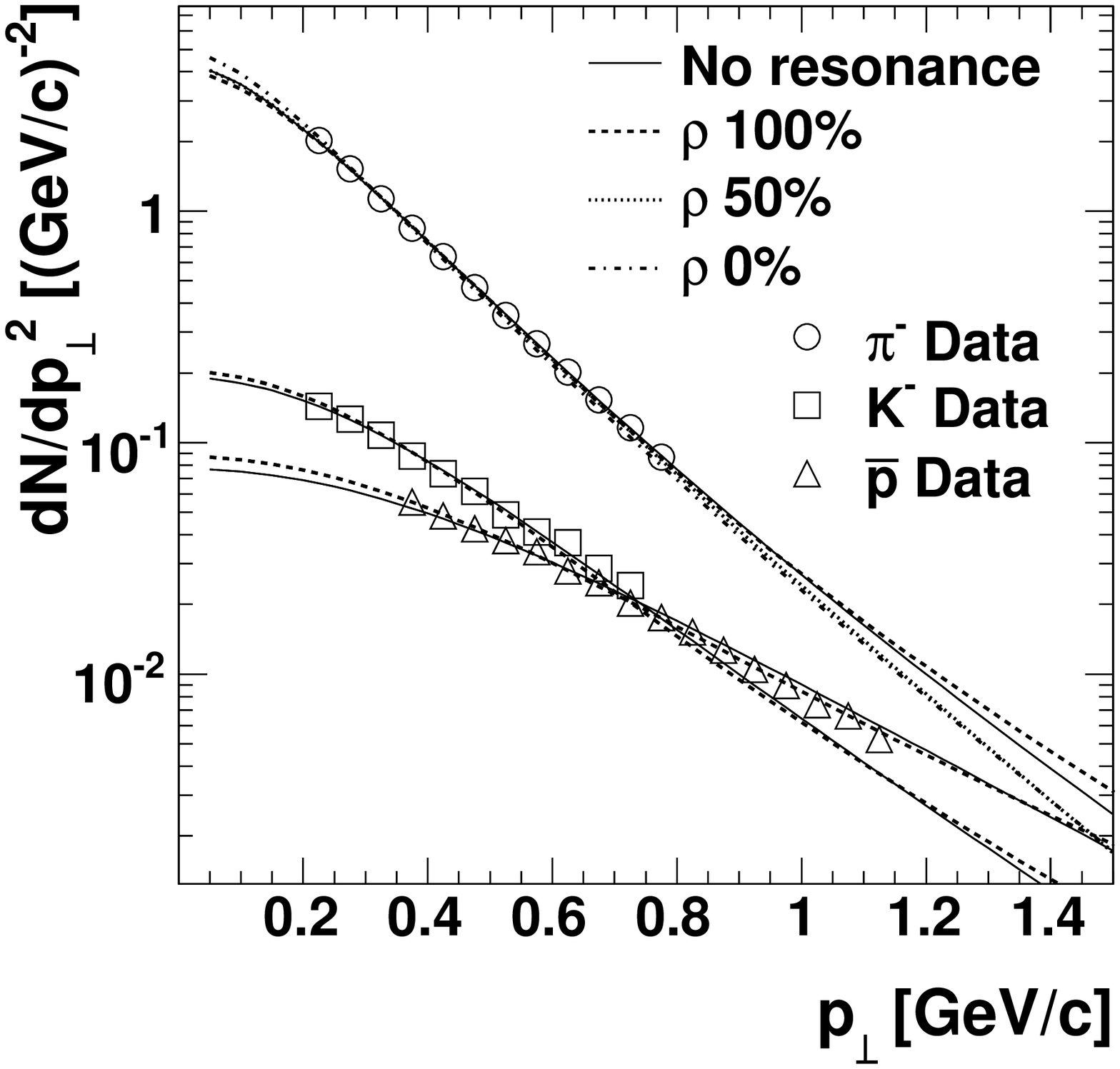}
\includegraphics[width=0.245\textwidth]{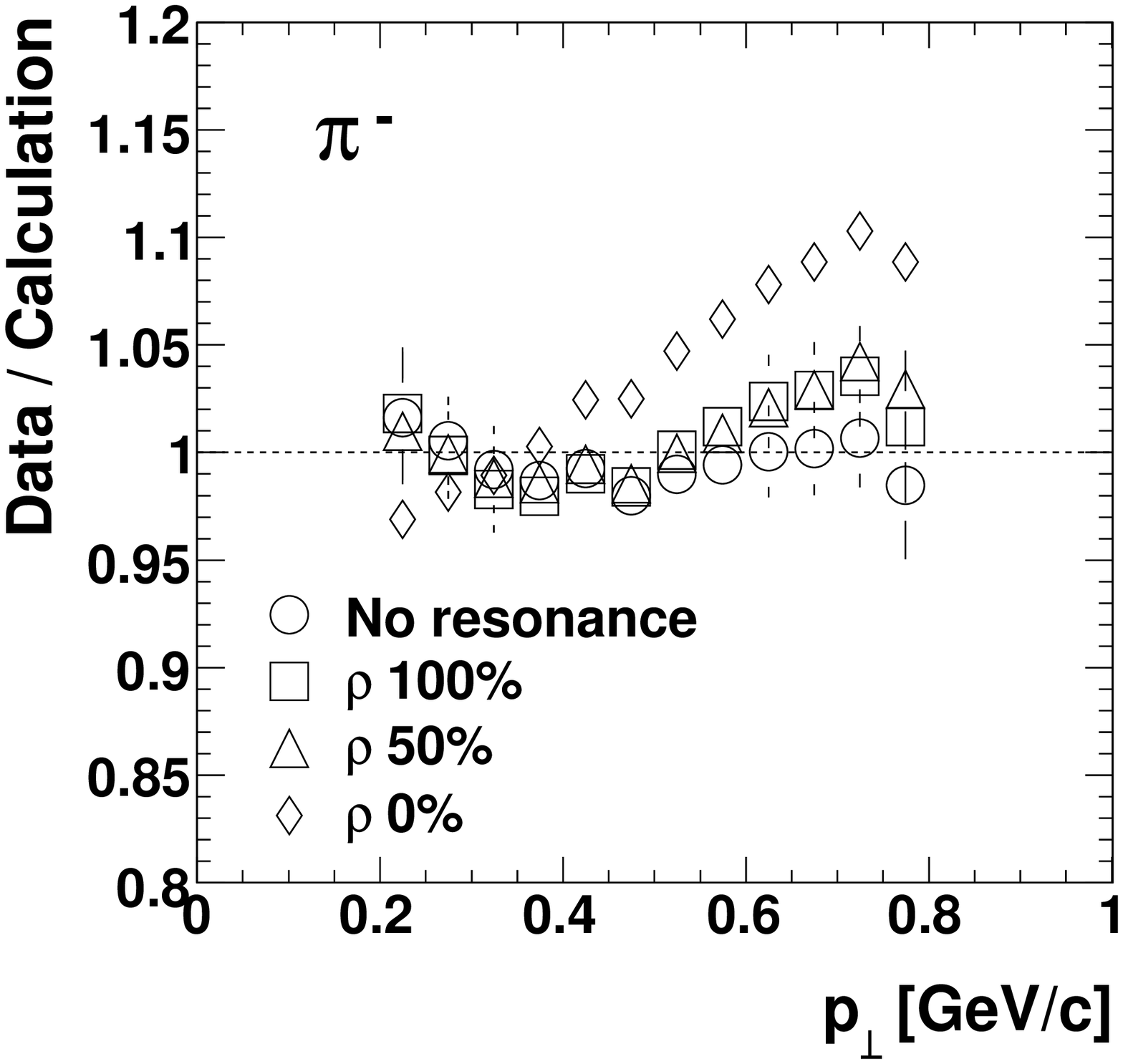}
\includegraphics[width=0.245\textwidth]{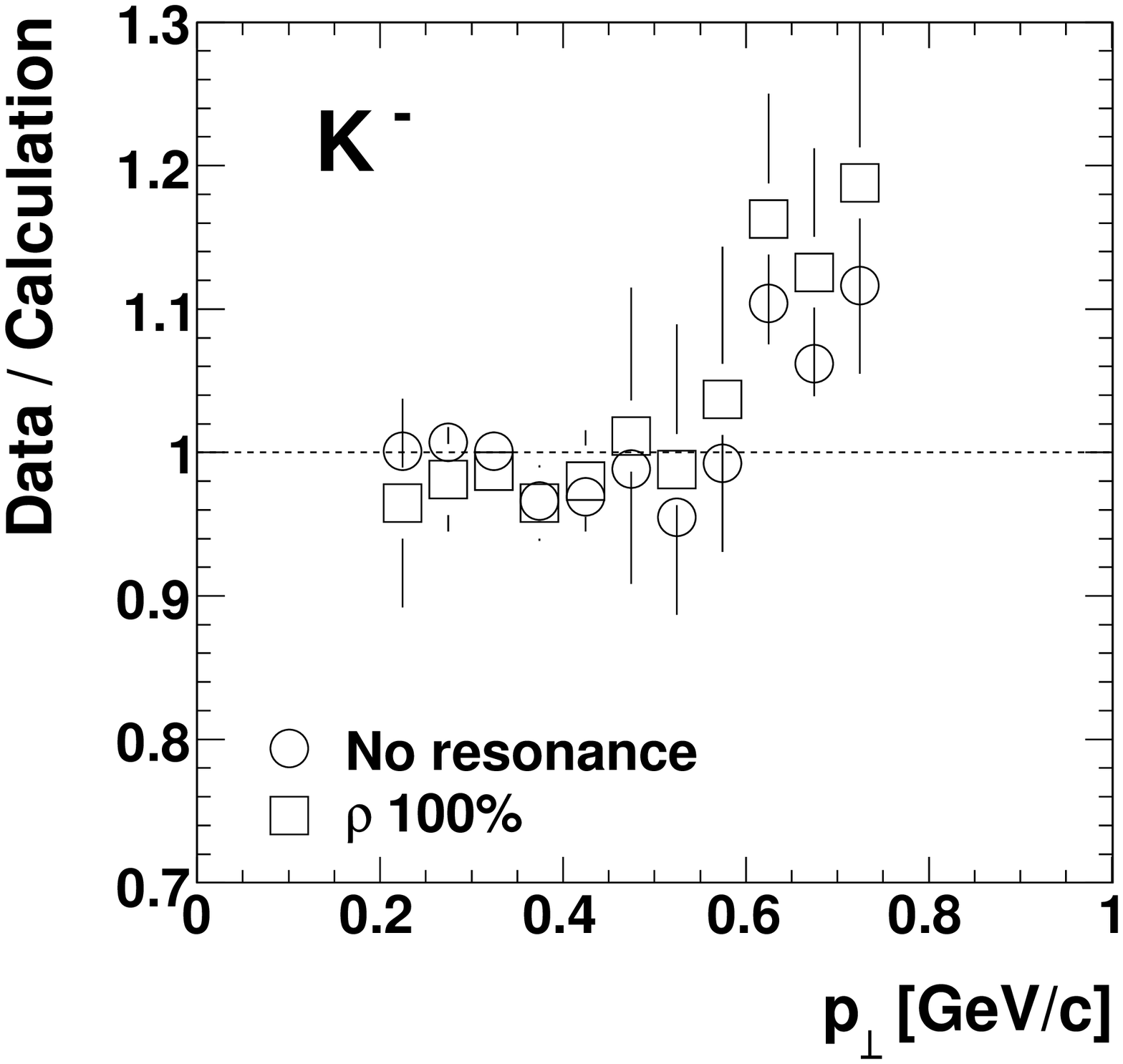}
\includegraphics[width=0.245\textwidth]{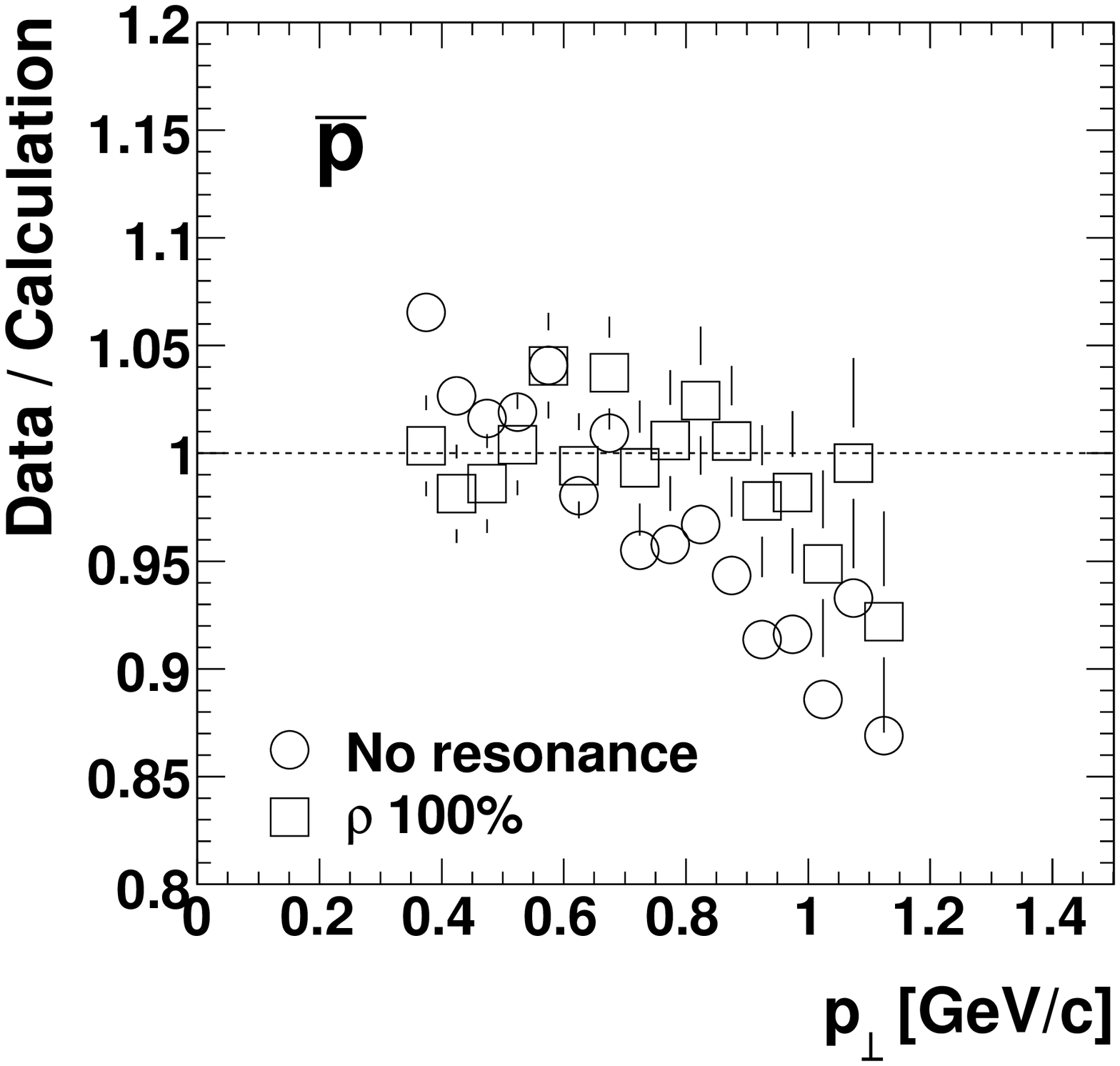}
\caption{Left panel: Fit of the calculated spectra (curves) to the measured ones (data points) in $pp$ collisions at 200~GeV~\cite{spec200}. Four calculated spectra are shown for $\pi^-$ (upper curves): including resonances with three different $\rho$ contributions and excluding resonances. Only two calculated curves are shown for $K^-$ (middle curves) and $\bar{p}$ (lower curves): including resonances with 100\% $\rho$ and excluding resonances. Other panels: Ratios of data spectrum to calculations. Two calculations are shown for $K^-$ and $\pbar$, while four calculations are shown for $\pi^-$. Error bars are the quadratic sum of the statistical and point-to-point systematic errors on the data, and are shown for two sets of the data points for $\pi^-$ and only one set for $K^-$ and $\pbar$.}
\label{fig:rho_pp}
\vspace{0.1in}
\centering
\includegraphics[width=0.245\textwidth]{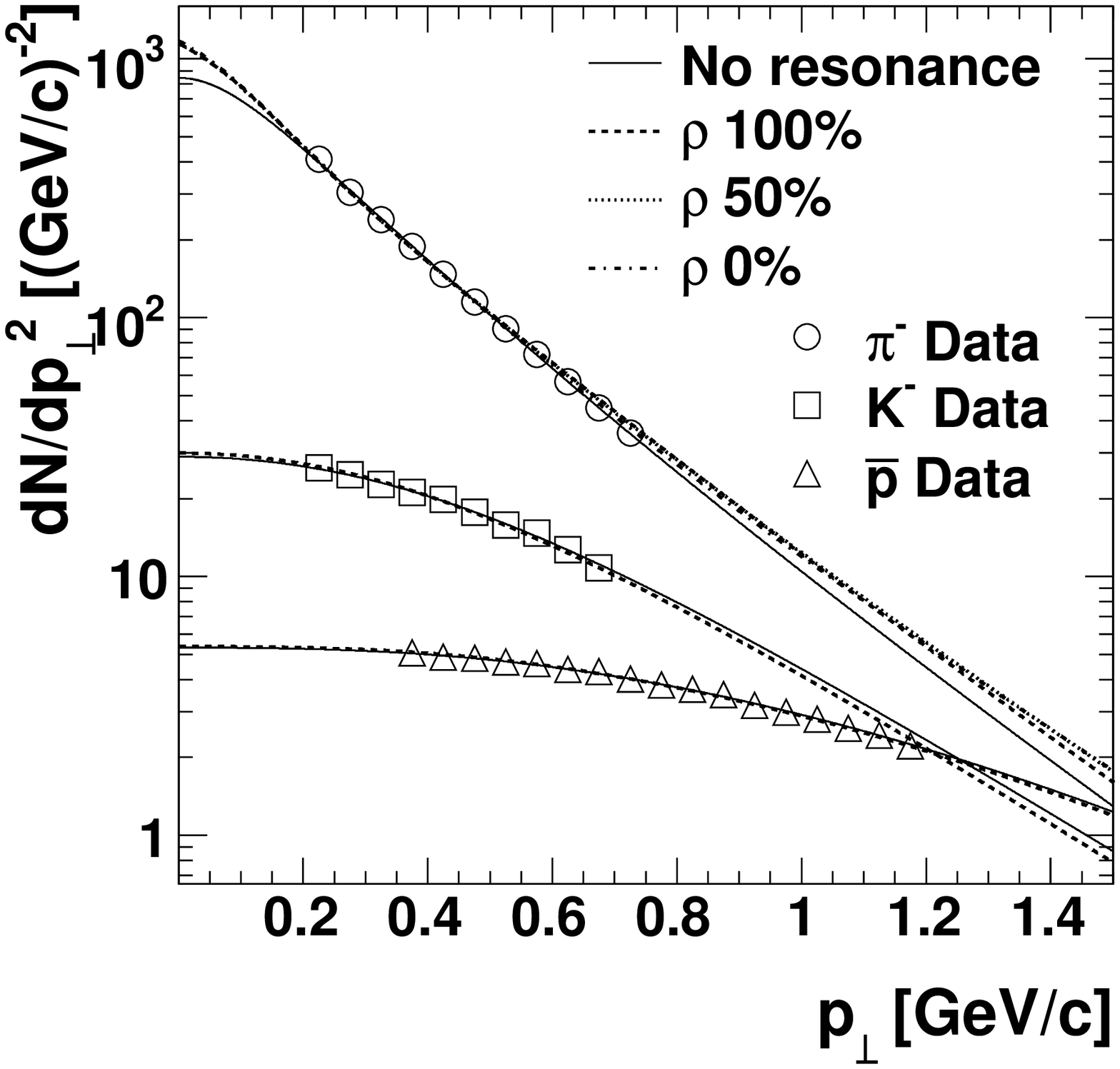}
\includegraphics[width=0.245\textwidth]{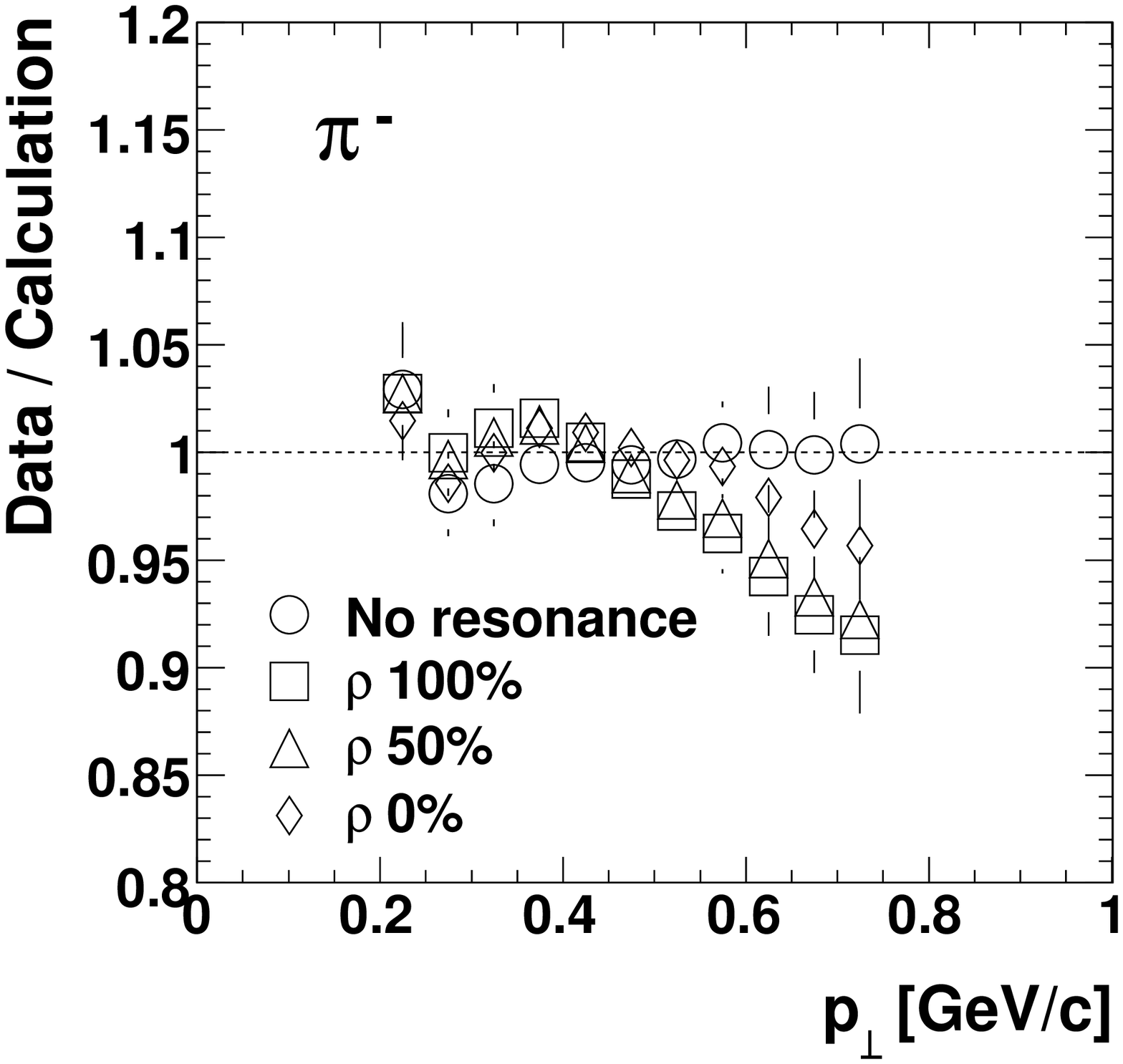}
\includegraphics[width=0.245\textwidth]{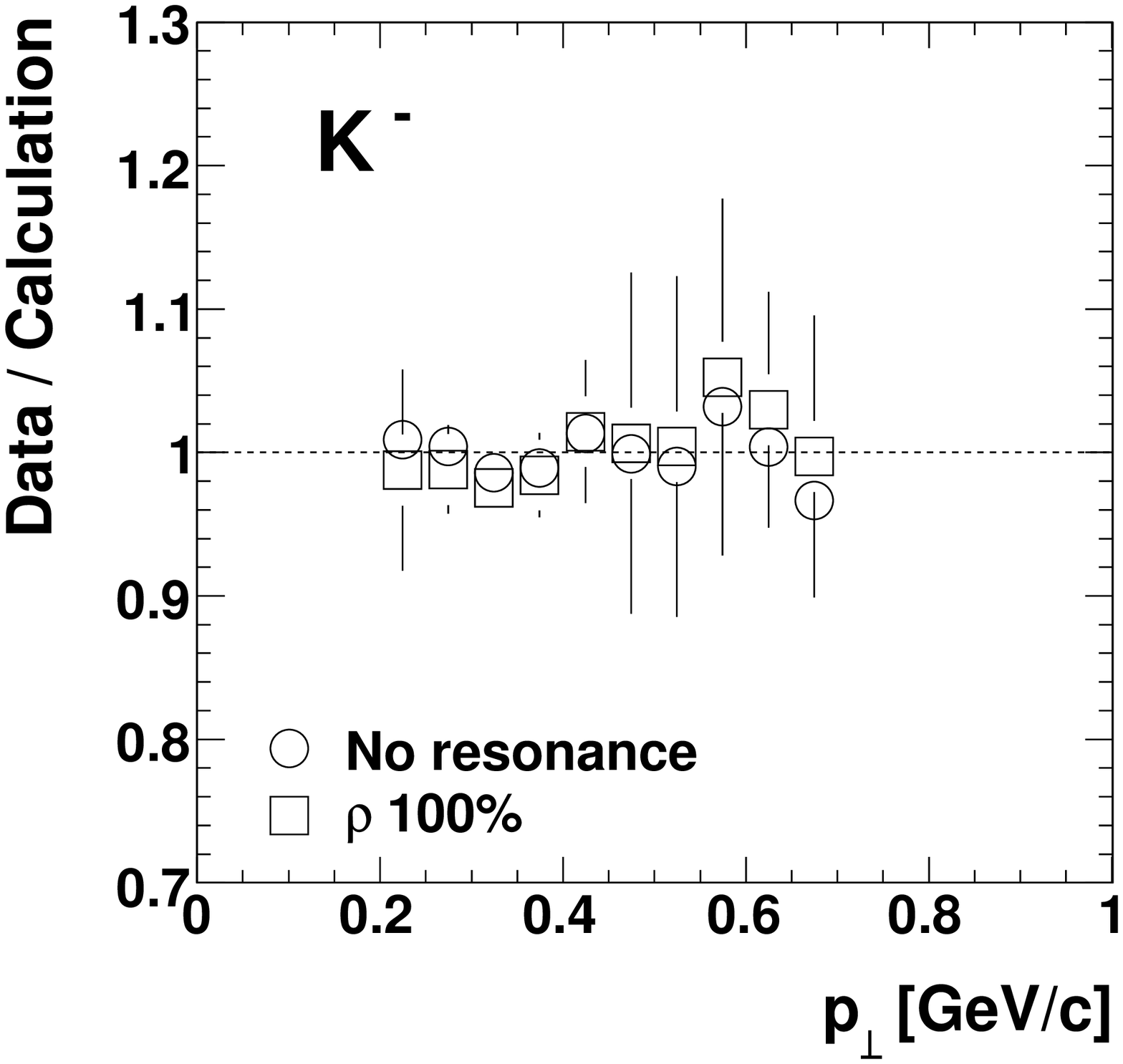}
\includegraphics[width=0.245\textwidth]{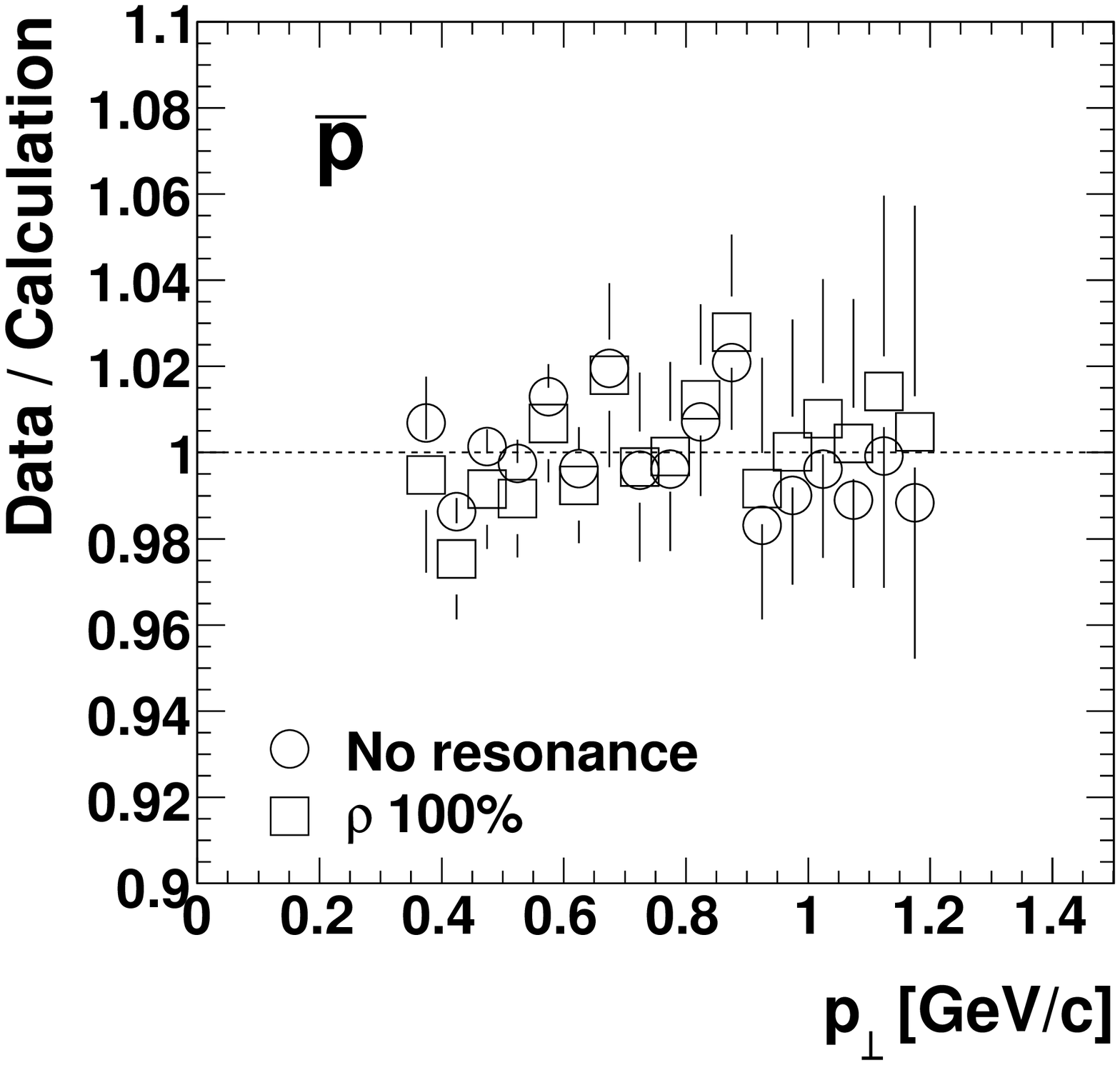}
\caption{Same as Fig.~\ref{fig:rho_pp} but for the top 5\% central Au+Au collisions at 200~GeV~\cite{spec200}.}
\label{fig:rho_AuAu}
\end{figure*}

The calculated particle spectra are fit to the measured identified particle spectra~\cite{spec200}, and the kinetic freeze-out temperature and the transverse flow velocity are extracted for both $pp$ and the 5\% central Au+Au collisions. The extracted parameters are summarized in Table~\ref{tab:reso_fits} (the row labeled by ``100\% $\rho$"). Figures~\ref{fig:resonance_pp} and~\ref{fig:resonance_AuAu} show the calculated, best-fit particle spectra of $\pi^{-}$, $K^{-}$ and $\pbar$ for $pp$ and central Au+Au collisions, respectively. Calculated inclusive pion spectra do not contain weak decay pions, just as in Ref.~\cite{spec200}. Resonance contributions are labeled by the initial resonance particle (e.g. a $\bar{p}$ emerging from the $\bar{\Xi} \rightarrow \bar{\Lambda} \rightarrow \bar{p}$ decays is labeled as ``$\Xi$ decay''). Only major resonance decay contributions are shown, but all contributions including minor ones are included in the calculated inclusive spectra. Those minor contributions include $\eta$, $\eta'$, $\phi$, $\Delta$, $\Sigma$, $\Sigma_{1385}$, and $\Lambda_{1520}$ decays to pions, $\Omega$ and $\Lambda_{1520}$ decays to kaons, and $\Omega$ and $\Lambda_{1520}$ decays to (anti)protons.

The lower panels of Figs.~\ref{fig:resonance_pp} and~\ref{fig:resonance_AuAu} show the resonance contributions to the inclusive spectra relative to the primordial ones for $pp$ and central Au+Au collisions, respectively. The inclusive kaon and antiproton spectra do not show significant changes in the spectral shapes compared to the primordial ones for both $pp$ and central Au+Au collisions in the measured $\pt$ ranges. The shape of the inclusive $\pi$ spectrum in $pp$ is also similar to the primordial one, but is more significantly modified in central Au+Au collisions due to light meson contributions ($\rho$, $\omega$ and $\eta$) at both small and large $\pt$. The largest contribution is from the $\rho$ meson. The shapes of the modifications are different for $pp$ and central Au+Au because of the significant flattening of the spectra in Au+Au but not in $pp$. The $\eta$ and $\omega$ mesons are less significant in $pp$ as compared to those in central Au+Au. However, in our measured $\pt$ range of 0.2-0.7~GeV/$c$, any modification in the shape of the pion spectrum is still not large, as seen from Fig.~\ref{fig:resonance_AuAu}.

For comparison, the default blast-wave fit with no resonances is also shown in Figs.~\ref{fig:resonance_pp} and \ref{fig:resonance_AuAu}. These fits are the spectra for the thermal particles with the corresponding fit parameters. As can be seen from Tables~\ref{tab:model_pars} and~\ref{tab:reso_fits} and the figure, blast-wave fits with and without the resonances give similar quality fits to the data. The extracted kinetic freeze-out temperature and average flow velocity agree within the systematic uncertainties~\cite{spec200}. In other words, resonance decays appear to have no significant effect on the extracted kinetic freeze-out parameters. This is primarily due to the limited $\pt$ ranges of the measured data where resonance decay products have more or less the same spectral shapes as the primordial particles do.

It was claimed in Refs.~\cite{singleT,singleT2,singleT3} that a single freeze-out temperature for chemical and kinetic freeze-out can satisfactorily describe the data. To test this, the spectra are also fit with a single, fixed kinetic freeze-out temperature $\Tkin=\Tch=160$ MeV including resonances. The fit $\langle\beta\rangle$ is $0.520^{ +0.001}_{ -0.002}$ with $\chi^2$/{\sc ndf}=19.6; the quality of our fit is similar to those in Refs.~\cite{singleT,singleT2,singleT3}. Based on this fit quality, a single temperature scenario is ruled out by the data, and we are not able to confirm the conclusion in Refs.~\cite{singleT,singleT2,singleT3}.

\subsection{Regeneration of Short-Lived Resonances}

The blast-wave model assumes that all particles and resonances decouple at the same $\Tkin$ and $\beta$. Short-lived resonances (e.g. $\rho$ and $\Delta$), due to their short lifetimes relative to the system evolution time, decay and are regenerated continuously, hence they might have different flow velocities and temperatures than long-lived resonances (and the bulk itself). Thus, it is reasonable to expect that short-lived resonances do not gain significantly larger flow velocity than the net flow of their decay daughters, as would be naively expected from their large masses.

The $\rho$ meson contributes to the pion spectrum and could alter the inclusive pion spectrum shape significantly. In the default treatment of resonances in the blast-wave parameterization, the $\rho$ acquires $\pt$ as given by the kinetic freeze-out temperature and the common transverse flow velocity (with the corresponding $\rho$ mass), and the decay pions are calculated from decay kinematics. To test the validity of the regeneration picture, two additional cases of different $\rho$ contributions are studied: 
(1) The $\rho$ decay pions have the same $\pt$ spectral shape as the primordial pions. This is equivalent to a zero $\rho$ lifetime; the $\rho$ does not have time to interact with the medium and gain its own flow; the flow it has is from that of the resonant pion pair that flow with the medium. In this case the existence of $\rho$'s does not make differences in the final fit results. This case is referred to as ``0\% $\rho$" below.
(2) Half of the $\rho$ contribution is taken like in (1) and the other half as in the default treatment. This case is referred to as ``50\% $\rho$" below. Since $\rho$ is very efficient at gaining flow compared to the much lighter pions, the default treatment of blast-wave parameterization gives the largest flow to $\rho$, and case (1) gives the smallest flow. 

Table~\ref{tab:reso_fits} shows the fit results for the two cases, together with the default case of resonance treatment (i.e. the ``100\% $\rho$" case), for both $pp$ and central Au+Au collisions. Figures~\ref{fig:rho_pp} and~\ref{fig:rho_AuAu} show the fits of the calculated inclusive pion spectra to the measured ones for $pp$ and central Au+Au, respectively. Fits are performed to the six measured spectra simultaneously, but only negatively charged particles are shown. The fit results from the 0\% and 50\% $\rho$ cases are only shown for the pions. 

As can be seen from the table and the figures, the models with all three cases of $\rho$ contribution describe the spectra data well. The fit $\Tkin$ and $\langle\beta\rangle$ values from all three cases agree; they also agree with those obtained without including resonances within systematic uncertainties (shown in Table~\ref{tab:model_pars}). It is interesting to note, however, that for the three $\rho$ cases, the lowest $\chi^2$/{\sc ndf} is found for the 100\% $\rho$ case in $pp$ collisions and for the 0\% $\rho$ case in central Au+Au collisions. If taken literally, this could imply that $pp$ collisions favor no regeneration, and central Au+Au collisions favor complete regeneration, hence a long time span between chemical freeze-out and kinetic freeze-out, lending support to the similar observation made by the $K^*$ measurement~\cite{STAR130kstar,STAR200kstar}.


\section{Invariant $\pt$ Spectra Data Tables\label{app:spectra}}

The transverse momentum spectra of the invariant yield per event are tabulated in Tables~\ref{tab:spectra_first} - \ref{tab:spectra_last}.

\section*{Acknowledgments}

We thank the RHIC Operations Group and RCF at BNL, and the NERSC Center at LBNL and the resources provided by the Open Science Grid consortium for their support. This work was supported in part by the Offices of NP and HEP within the U.S.~DOE Office of Science, the U.S.~NSF, the Sloan Foundation, the DFG Excellence Cluster EXC153 of Germany, CNRS/IN2P3, RA, RPL, and EMN of France, STFC and EPSRC of the United Kingdom, FAPESP of Brazil, the Russian Ministry of Sci.~and Tech., the NNSFC, CAS, MoST, and MoE of China, IRP and GA of the Czech Republic, FOM of the Netherlands, DAE, DST, and CSIR of the Government of India, Swiss NSF, the Polish State Committee for Scientific Research, and the Korea Sci.~\& Eng.~Foundation.

\begin{table*}[htbp]
\caption{Identified $\pi^{\pm}$, $K^{\pm}$, antiproton and proton invariant transverse momentum spectra at mid-rapidity ($|y|<0.1$) in minimum bias $pp$ collisions at 200~GeV: $d^2N/(2\pi\pt d\pt dy)$ [(GeV/$c$)$^{-2}$] versus $\pt$ [GeV/$c$]. Errors are the quadratic sum of statistical errors and point-to-point systematic errors. For proton, systematic uncertainties due to proton background subtraction are also included in quadrature. See Section~\ref{sec:syst_spectra} for other systematic uncertainties. Data were published in Ref.~\cite{spec200}.}
\label{tab:spectra_first}
\label{tab:spectra_pp200}
\begin{ruledtabular}
\begin{tabular}{lllllll}
$\pt$ & $\pi^-$ & $\pi^+$ & $K^-$ & $K^+$ & $\pbar$ & $p$ \\ \hline
0.225 & \,\,2.02$\pm$0.06 & \,\,2.07$\pm$0.06 & (1.43$\pm$0.11)$\times$10$^{-1}$ & (1.52$\pm$0.11)$\times$10$^{-1}$ & & \\
0.275 & \,\,1.52$\pm$0.03 & \,\,1.54$\pm$0.03 & (1.26$\pm$0.05)$\times$10$^{-1}$ & (1.30$\pm$0.05)$\times$10$^{-1}$ & & \\
0.325 & \,\,1.13$\pm$0.02 & \,\,1.14$\pm$0.02 & (1.08$\pm$0.02)$\times$10$^{-1}$ & (1.08$\pm$0.02)$\times$10$^{-1}$ & & \\
0.375 & (8.44$\pm$0.09)$\times$10$^{-1}$ & (8.57$\pm$0.09)$\times$10$^{-1}$ & (8.77$\pm$0.24)$\times$10$^{-2}$ & (9.16$\pm$0.25)$\times$10$^{-2}$ & (5.54$\pm$0.13)$\times$10$^{-2}$ & \\
0.425 & (6.35$\pm$0.07)$\times$10$^{-1}$ & (6.38$\pm$0.07)$\times$10$^{-1}$ & (7.34$\pm$0.26)$\times$10$^{-2}$ & (7.47$\pm$0.27)$\times$10$^{-2}$ & (4.81$\pm$0.11)$\times$10$^{-2}$ & \\
0.475 & (4.69$\pm$0.05)$\times$10$^{-1}$ & (4.76$\pm$0.05)$\times$10$^{-1}$ & (6.17$\pm$0.63)$\times$10$^{-2}$ & (6.26$\pm$0.64)$\times$10$^{-2}$ & (4.25$\pm$0.10)$\times$10$^{-2}$ & (5.07$\pm$0.25)$\times$10$^{-2}$ \\
0.525 & (3.54$\pm$0.04)$\times$10$^{-1}$ & (3.59$\pm$0.04)$\times$10$^{-1}$ & (4.87$\pm$0.50)$\times$10$^{-2}$ & (5.26$\pm$0.54)$\times$10$^{-2}$ & (3.77$\pm$0.09)$\times$10$^{-2}$ & (4.69$\pm$0.19)$\times$10$^{-2}$ \\
\end{tabular}
\end{ruledtabular}
\end{table*}

\begin{table*}[htbp]
\caption{Identified $\pi^-$ invariant transverse momentum spectra at mid-rapidity ($|y|<0.1$) in d+Au collisions at 200~GeV: $d^2N/(2\pi\pt d\pt dy)$ [(GeV/$c$)$^{-2}$] versus $\pt$ [GeV/$c$]. Errors are the quadratic sum of statistical errors and point-to-point systematic errors. See Section~\ref{sec:syst_spectra} for other systematic uncertainties.}
\label{tab:spectra_dAu_pim}
\begin{ruledtabular}
\begin{tabular}{lllll}
$\pt$ & min.~bias & 40-100\% & 20-40\% & 0-20\% \\ \hline
0.225 & $\,\,6.59\pm0.07$ & $\,\,4.16\pm0.04$ & $\,\,8.84\pm0.09$ & $(1.16\pm0.01)\times10^{1}$ \\
0.275 & $\,\,4.84\pm0.05$ & $\,\,3.07\pm0.03$ & $\,\,6.29\pm0.06$ & $\,\,8.70\pm0.09$ \\
0.325 & $\,\,3.60\pm0.04$ & $\,\,2.28\pm0.02$ & $\,\,4.69\pm0.05$ & $\,\,6.51\pm0.07$ \\
0.375 & $\,\,2.69\pm0.03$ & $\,\,1.69\pm0.02$ & $\,\,3.51\pm0.04$ & $\,\,4.89\pm0.05$ \\
0.425 & $\,\,2.05\pm0.08$ & $\,\,1.28\pm0.05$ & $\,\,2.66\pm0.11$ & $\,\,3.74\pm0.15$ \\
0.475 & $\,\,1.56\pm0.06$ & $(9.66\pm0.39)\times10^{-1}$ & $\,\,2.04\pm0.08$ & $\,\,2.87\pm0.12$ \\
0.525 & $\,\,1.20\pm0.05$ & $(7.31\pm0.29)\times10^{-1}$ & $\,\,1.57\pm0.06$ & $\,\,2.22\pm0.09$ \\
0.575 & $(9.27\pm0.37)\times10^{-1}$ & $(5.64\pm0.23)\times10^{-1}$ & $\,\,1.21\pm0.05$ & $\,\,1.73\pm0.07$ \\
0.625 & $(7.22\pm0.43)\times10^{-1}$ & $(4.38\pm0.26)\times10^{-1}$ & $(9.45\pm0.57)\times10^{-1}$ & $\,\,1.35\pm0.08$ \\
0.675 & $(5.67\pm0.17)\times10^{-1}$ & $(3.40\pm0.10)\times10^{-1}$ & $(7.43\pm0.23)\times10^{-1}$ & $\,\,1.07\pm0.03$ \\
\end{tabular}
\end{ruledtabular}

\caption{Identified $\pi^+$ invariant transverse momentum spectra at mid-rapidity ($|y|<0.1$) in d+Au collisions at 200~GeV: $d^2N/(2\pi\pt d\pt dy)$ [(GeV/$c$)$^{-2}$] versus $\pt$ [GeV/$c$]. Errors are the quadratic sum of statistical errors and point-to-point systematic errors. See Section~\ref{sec:syst_spectra} for other systematic uncertainties.}
\label{tab:spectra_dAu_pip}
\begin{ruledtabular}
\begin{tabular}{lllll}
$\pt$ & min.~bias & 40-100\% & 20-40\% & 0-20\% \\ \hline
0.225 & $\,\,6.54\pm0.07$ & $\,\,4.10\pm0.04$ & $\,\,8.53\pm0.09$ & $(1.19\pm0.01)\times10^{1}$ \\
0.275 & $\,\,4.78\pm0.05$ & $\,\,3.00\pm0.03$ & $\,\,6.23\pm0.06$ & $\,\,8.69\pm0.09$ \\
0.325 & $\,\,3.59\pm0.04$ & $\,\,2.27\pm0.02$ & $\,\,4.65\pm0.05$ & $\,\,6.49\pm0.07$ \\
0.375 & $\,\,2.68\pm0.03$ & $\,\,1.65\pm0.02$ & $\,\,3.48\pm0.04$ & $\,\,4.94\pm0.05$ \\
0.425 & $\,\,2.07\pm0.08$ & $\,\,1.31\pm0.05$ & $\,\,2.66\pm0.11$ & $\,\,3.77\pm0.15$ \\
0.475 & $\,\,1.54\pm0.06$ & $(9.23\pm0.37)\times10^{-1}$ & $\,\,2.04\pm0.08$ & $\,\,2.90\pm0.12$ \\
0.525 & $\,\,1.22\pm0.05$ & $(7.55\pm0.30)\times10^{-1}$ & $\,\,1.58\pm0.06$ & $\,\,2.23\pm0.09$ \\
0.575 & $(9.32\pm0.37)\times10^{-1}$ & $(5.67\pm0.23)\times10^{-1}$ & $\,\,1.22\pm0.05$ & $\,\,1.74\pm0.07$ \\
0.625 & $(7.40\pm0.44)\times10^{-1}$ & $(4.60\pm0.28)\times10^{-1}$ & $(9.54\pm0.57)\times10^{-1}$ & $\,\,1.37\pm0.08$ \\
0.675 & $(5.67\pm0.17)\times10^{-1}$ & $(3.32\pm0.10)\times10^{-1}$ & $(7.49\pm0.23)\times10^{-1}$ & $\,\,1.09\pm0.03$ \\
\end{tabular}
\end{ruledtabular}

\caption{Identified $K^-$ invariant transverse momentum spectra at mid-rapidity ($|y|<0.1$) in d+Au collisions at 200~GeV: $d^2N/(2\pi\pt d\pt dy)$ [(GeV/$c$)$^{-2}$] versus $\pt$ [GeV/$c$]. Errors are the quadratic sum of statistical errors and point-to-point systematic errors. See Section~\ref{sec:syst_spectra} for other systematic uncertainties.}
\label{tab:spectra_dAu_km}
\begin{ruledtabular}
\begin{tabular}{lllll}
$\pt$ & min.~bias & 40-100\% & 20-40\% & 0-20\% \\ \hline
0.225 & $(5.15\pm0.16)\times10^{-1}$ & $(3.21\pm0.11)\times10^{-1}$ & $(6.65\pm0.23)\times10^{-1}$ & $(9.47\pm0.32)\times10^{-1}$ \\
0.275 & $(4.34\pm0.05)\times10^{-1}$ & $(2.73\pm0.04)\times10^{-1}$ & $(5.63\pm0.09)\times10^{-1}$ & $(7.85\pm0.12)\times10^{-1}$ \\
0.325 & $(3.64\pm0.08)\times10^{-1}$ & $(2.25\pm0.05)\times10^{-1}$ & $(4.75\pm0.11)\times10^{-1}$ & $(6.70\pm0.15)\times10^{-1}$ \\
0.375 & $(3.13\pm0.28)\times10^{-1}$ & $(1.96\pm0.18)\times10^{-1}$ & $(4.06\pm0.37)\times10^{-1}$ & $(5.75\pm0.52)\times10^{-1}$ \\
0.425 & $(2.62\pm0.37)\times10^{-1}$ & $(1.61\pm0.23)\times10^{-1}$ & $(3.43\pm0.48)\times10^{-1}$ & $(4.81\pm0.68)\times10^{-1}$ \\
0.475 & $(2.11\pm0.23)\times10^{-1}$ & $(1.24\pm0.14)\times10^{-1}$ & $(2.76\pm0.31)\times10^{-1}$ & $(4.07\pm0.45)\times10^{-1}$ \\
0.525 & $(1.85\pm0.13)\times10^{-1}$ & $(1.09\pm0.08)\times10^{-1}$ & $(2.44\pm0.17)\times10^{-1}$ & $(3.52\pm0.25)\times10^{-1}$ \\
0.575 & $(1.63\pm0.15)\times10^{-1}$ & $(9.63\pm0.87)\times10^{-2}$ & $(2.15\pm0.20)\times10^{-1}$ & $(3.09\pm0.28)\times10^{-1}$ \\
0.625 & $(1.32\pm0.07)\times10^{-1}$ & $(7.67\pm0.40)\times10^{-2}$ & $(1.72\pm0.09)\times10^{-1}$ & $(2.57\pm0.13)\times10^{-1}$ \\
0.675 & $(1.16\pm0.09)\times10^{-1}$ & $(6.76\pm0.55)\times10^{-2}$ & $(1.57\pm0.13)\times10^{-1}$ & $(2.21\pm0.18)\times10^{-1}$ \\
0.725 & $(1.01\pm0.05)\times10^{-1}$ & $(6.16\pm0.36)\times10^{-2}$ & $(1.31\pm0.08)\times10^{-1}$ & $(1.91\pm0.11)\times10^{-1}$ \\
\end{tabular}
\end{ruledtabular}

\caption{Identified $K^+$ invariant transverse momentum spectra at mid-rapidity ($|y|<0.1$) in d+Au collisions at 200~GeV: $d^2N/(2\pi\pt d\pt dy)$ [(GeV/$c$)$^{-2}$] versus $\pt$ [GeV/$c$]. Errors are the quadratic sum of statistical errors and point-to-point systematic errors. See Section~\ref{sec:syst_spectra} for other systematic uncertainties.}
\label{tab:spectra_dAu_kp}
\begin{ruledtabular}
\begin{tabular}{lllll}
$\pt$ & min.~bias & 40-100\% & 20-40\% & 0-20\% \\ \hline
0.225 & $(4.84\pm0.15)\times10^{-1}$ & $(2.99\pm0.10)\times10^{-1}$ & $(6.32\pm0.22)\times10^{-1}$ & $(8.93\pm0.30)\times10^{-1}$ \\
0.275 & $(4.28\pm0.05)\times10^{-1}$ & $(2.71\pm0.04)\times10^{-1}$ & $(5.50\pm0.09)\times10^{-1}$ & $(7.78\pm0.11)\times10^{-1}$ \\
0.325 & $(3.68\pm0.08)\times10^{-1}$ & $(2.31\pm0.05)\times10^{-1}$ & $(4.83\pm0.11)\times10^{-1}$ & $(6.68\pm0.15)\times10^{-1}$ \\
0.375 & $(3.24\pm0.29)\times10^{-1}$ & $(1.99\pm0.18)\times10^{-1}$ & $(4.28\pm0.39)\times10^{-1}$ & $(5.94\pm0.54)\times10^{-1}$ \\
0.425 & $(2.73\pm0.38)\times10^{-1}$ & $(1.68\pm0.24)\times10^{-1}$ & $(3.58\pm0.50)\times10^{-1}$ & $(5.04\pm0.71)\times10^{-1}$ \\
0.475 & $(2.25\pm0.25)\times10^{-1}$ & $(1.34\pm0.15)\times10^{-1}$ & $(2.95\pm0.33)\times10^{-1}$ & $(4.30\pm0.47)\times10^{-1}$ \\
0.525 & $(2.00\pm0.14)\times10^{-1}$ & $(1.22\pm0.09)\times10^{-1}$ & $(2.64\pm0.19)\times10^{-1}$ & $(3.72\pm0.26)\times10^{-1}$ \\
0.575 & $(1.74\pm0.16)\times10^{-1}$ & $(1.05\pm0.09)\times10^{-1}$ & $(2.30\pm0.21)\times10^{-1}$ & $(3.26\pm0.30)\times10^{-1}$ \\
0.625 & $(1.41\pm0.07)\times10^{-1}$ & $(8.23\pm0.42)\times10^{-2}$ & $(1.88\pm0.10)\times10^{-1}$ & $(2.70\pm0.14)\times10^{-1}$ \\
0.675 & $(1.27\pm0.10)\times10^{-1}$ & $(7.36\pm0.60)\times10^{-2}$ & $(1.68\pm0.14)\times10^{-1}$ & $(2.45\pm0.20)\times10^{-1}$ \\
0.725 & $(1.06\pm0.06)\times10^{-1}$ & $(6.11\pm0.35)\times10^{-2}$ & $(1.44\pm0.08)\times10^{-1}$ & $(2.02\pm0.11)\times10^{-1}$ \\
\end{tabular}
\end{ruledtabular}
\end{table*}

\begin{table*}[htbp]
\caption{Identified $\pbar$ invariant transverse momentum spectra at mid-rapidity ($|y|<0.1$) in d+Au collisions at 200~GeV: $d^2N/(2\pi\pt d\pt dy)$ [(GeV/$c$)$^{-2}$] versus $\pt$ [GeV/$c$]. Errors are the quadratic sum of statistical errors and point-to-point systematic errors. See Section~\ref{sec:syst_spectra} for other systematic uncertainties.}
\label{tab:spectra_dAu_pbar}
\begin{ruledtabular}
\begin{tabular}{lllll}
$\pt$ & min.~bias & 40-100\% & 20-40\% & 0-20\% \\ \hline
0.425 & $(1.37\pm0.02)\times10^{-1}$ & $(8.46\pm0.12)\times10^{-2}$ & $(1.76\pm0.03)\times10^{-1}$ & $(2.55\pm0.04)\times10^{-1}$ \\
0.475 & $(1.22\pm0.01)\times10^{-1}$ & $(7.39\pm0.11)\times10^{-2}$ & $(1.55\pm0.02)\times10^{-1}$ & $(2.35\pm0.03)\times10^{-1}$ \\
0.525 & $(1.11\pm0.01)\times10^{-1}$ & $(6.58\pm0.09)\times10^{-2}$ & $(1.43\pm0.02)\times10^{-1}$ & $(2.12\pm0.03)\times10^{-1}$ \\
0.575 & $(1.01\pm0.01)\times10^{-1}$ & $(5.97\pm0.09)\times10^{-2}$ & $(1.33\pm0.02)\times10^{-1}$ & $(1.93\pm0.03)\times10^{-1}$ \\
0.625 & $(9.00\pm0.19)\times10^{-2}$ & $(5.27\pm0.12)\times10^{-2}$ & $(1.18\pm0.03)\times10^{-1}$ & $(1.74\pm0.04)\times10^{-1}$ \\
0.675 & $(8.05\pm0.17)\times10^{-2}$ & $(4.65\pm0.11)\times10^{-2}$ & $(1.06\pm0.02)\times10^{-1}$ & $(1.57\pm0.04)\times10^{-1}$ \\
0.725 & $(7.22\pm0.09)\times10^{-2}$ & $(4.12\pm0.06)\times10^{-2}$ & $(9.47\pm0.15)\times10^{-2}$ & $(1.43\pm0.02)\times10^{-1}$ \\
0.775 & $(6.37\pm0.20)\times10^{-2}$ & $(3.60\pm0.12)\times10^{-2}$ & $(8.48\pm0.27)\times10^{-2}$ & $(1.26\pm0.04)\times10^{-1}$ \\
0.825 & $(5.77\pm0.29)\times10^{-2}$ & $(3.20\pm0.16)\times10^{-2}$ & $(7.88\pm0.41)\times10^{-2}$ & $(1.14\pm0.06)\times10^{-1}$ \\
0.875 & $(5.21\pm0.26)\times10^{-2}$ & $(2.93\pm0.15)\times10^{-2}$ & $(7.18\pm0.37)\times10^{-2}$ & $(1.01\pm0.05)\times10^{-1}$ \\
0.925 & $(4.57\pm0.10)\times10^{-2}$ & $(2.55\pm0.06)\times10^{-2}$ & $(6.37\pm0.15)\times10^{-2}$ & $(8.80\pm0.20)\times10^{-2}$ \\
0.975 & $(3.95\pm0.08)\times10^{-2}$ & $(2.12\pm0.05)\times10^{-2}$ & $(5.47\pm0.13)\times10^{-2}$ & $(7.92\pm0.19)\times10^{-2}$ \\
1.025 & $(3.43\pm0.24)\times10^{-2}$ & $(1.85\pm0.13)\times10^{-2}$ & $(4.69\pm0.34)\times10^{-2}$ & $(6.90\pm0.49)\times10^{-2}$ \\
1.075 & $(3.03\pm0.10)\times10^{-2}$ & $(1.64\pm0.06)\times10^{-2}$ & $(4.03\pm0.14)\times10^{-2}$ & $(6.21\pm0.21)\times10^{-2}$ \\
\end{tabular}
\end{ruledtabular}

\caption{Identified proton invariant transverse momentum spectra at mid-rapidity ($|y|<0.1$) in d+Au collisions at 200~GeV: $d^2N/(2\pi\pt d\pt dy)$ [(GeV/$c$)$^{-2}$] versus $\pt$ [GeV/$c$]. Errors are the quadratic sum of statistical errors and point-to-point systematic errors. See Section~\ref{sec:syst_spectra} for other systematic uncertainties including those due to proton background subtraction.}
\label{tab:spectra_dAu_p}
\begin{ruledtabular}
\begin{tabular}{lllll}
$\pt$ & min.~bias & 40-100\% & 20-40\% & 0-20\% \\ \hline
0.425 & $(1.69\pm0.02)\times10^{-1}$ & $(1.03\pm0.01)\times10^{-1}$ & $(2.35\pm0.03)\times10^{-1}$ & $(3.03\pm0.04)\times10^{-1}$ \\
0.475 & $(1.47\pm0.02)\times10^{-1}$ & $(8.40\pm0.11)\times10^{-2}$ & $(2.04\pm0.03)\times10^{-1}$ & $(2.79\pm0.03)\times10^{-1}$ \\
0.525 & $(1.34\pm0.01)\times10^{-1}$ & $(7.86\pm0.10)\times10^{-2}$ & $(1.84\pm0.02)\times10^{-1}$ & $(2.48\pm0.03)\times10^{-1}$ \\
0.575 & $(1.19\pm0.01)\times10^{-1}$ & $(6.87\pm0.09)\times10^{-2}$ & $(1.64\pm0.02)\times10^{-1}$ & $(2.26\pm0.03)\times10^{-1}$ \\
0.625 & $(1.10\pm0.02)\times10^{-1}$ & $(6.30\pm0.14)\times10^{-2}$ & $(1.51\pm0.03)\times10^{-1}$ & $(2.08\pm0.04)\times10^{-1}$ \\
0.675 & $(9.87\pm0.20)\times10^{-2}$ & $(5.73\pm0.13)\times10^{-2}$ & $(1.33\pm0.03)\times10^{-1}$ & $(1.89\pm0.04)\times10^{-1}$ \\
0.725 & $(8.74\pm0.10)\times10^{-2}$ & $(4.91\pm0.07)\times10^{-2}$ & $(1.19\pm0.02)\times10^{-1}$ & $(1.70\pm0.02)\times10^{-1}$ \\
0.775 & $(7.95\pm0.24)\times10^{-2}$ & $(4.49\pm0.15)\times10^{-2}$ & $(1.08\pm0.03)\times10^{-1}$ & $(1.55\pm0.05)\times10^{-1}$ \\
0.825 & $(6.98\pm0.35)\times10^{-2}$ & $(3.88\pm0.20)\times10^{-2}$ & $(9.45\pm0.48)\times10^{-2}$ & $(1.38\pm0.07)\times10^{-1}$ \\
0.875 & $(6.49\pm0.33)\times10^{-2}$ & $(3.54\pm0.18)\times10^{-2}$ & $(8.93\pm0.46)\times10^{-2}$ & $(1.29\pm0.07)\times10^{-1}$ \\
0.925 & $(5.55\pm0.12)\times10^{-2}$ & $(2.98\pm0.07)\times10^{-2}$ & $(7.57\pm0.17)\times10^{-2}$ & $(1.12\pm0.02)\times10^{-1}$ \\
0.975 & $(4.93\pm0.10)\times10^{-2}$ & $(2.60\pm0.06)\times10^{-2}$ & $(7.08\pm0.16)\times10^{-2}$ & $(9.78\pm0.22)\times10^{-2}$ \\
1.025 & $(4.15\pm0.29)\times10^{-2}$ & $(2.21\pm0.16)\times10^{-2}$ & $(5.84\pm0.42)\times10^{-2}$ & $(8.27\pm0.59)\times10^{-2}$ \\
1.075 & $(3.61\pm0.12)\times10^{-2}$ & $(1.91\pm0.06)\times10^{-2}$ & $(4.90\pm0.22)\times10^{-2}$ & $(7.39\pm0.24)\times10^{-2}$ \\
\end{tabular}
\end{ruledtabular}
\end{table*}

\begin{table*}[htbp]
\caption{Identified $\pi^-$ invariant transverse momentum spectra at mid-rapidity ($|y|<0.1$) in Au+Au collisions at 62.4~GeV: $d^2N/(2\pi\pt d\pt dy)$ [(GeV/$c$)$^{-2}$] versus $\pt$ [GeV/$c$]. Errors are the quadratic sum of statistical errors and point-to-point systematic errors. See Section~\ref{sec:syst_spectra} for other systematic uncertainties.}
\label{tab:spectra_AuAu62_pim}
\begin{ruledtabular}
\begin{tabular}{llllll}
$\pt$ & 70-80\% & 60-70\% & 50-60\% & 40-50\% & \\ \hline
0.225 & $(1.07\pm0.02)\times10^{1}$ & $(2.02\pm0.04)\times10^{1}$ & $(3.63\pm0.07)\times10^{1}$ & $(5.91\pm0.12)\times10^{1}$ & \\
0.275 & $\,\,7.93\pm0.16$ & $(1.53\pm0.03)\times10^{1}$ & $(2.74\pm0.05)\times10^{1}$ & $(4.45\pm0.09)\times10^{1}$ & \\
0.325 & $\,\,5.89\pm0.12$ & $(1.15\pm0.02)\times10^{1}$ & $(2.08\pm0.04)\times10^{1}$ & $(3.36\pm0.07)\times10^{1}$ & \\
0.375 & $\,\,4.39\pm0.05$ & $\,\,8.69\pm0.09$ & $(1.58\pm0.02)\times10^{1}$ & $(2.57\pm0.03)\times10^{1}$ & \\
0.425 & $\,\,3.28\pm0.04$ & $\,\,6.59\pm0.07$ & $(1.20\pm0.01)\times10^{1}$ & $(1.97\pm0.02)\times10^{1}$ & \\
0.475 & $\,\,2.49\pm0.03$ & $\,\,5.02\pm0.05$ & $\,\,9.27\pm0.10$ & $(1.51\pm0.02)\times10^{1}$ & \\
0.525 & $\,\,1.89\pm0.02$ & $\,\,3.85\pm0.04$ & $\,\,7.12\pm0.07$ & $(1.17\pm0.01)\times10^{1}$ & \\
\end{tabular}
\end{ruledtabular}

\caption{Identified $\pi^+$ invariant transverse momentum spectra at mid-rapidity ($|y|<0.1$) in Au+Au collisions at 62.4~GeV: $d^2N/(2\pi\pt d\pt dy)$ [(GeV/$c$)$^{-2}$] versus $\pt$ [GeV/$c$]. Errors are the quadratic sum of statistical errors and point-to-point systematic errors. See Section~\ref{sec:syst_spectra} for other systematic uncertainties.}
\label{tab:spectra_AuAu62_pip}
\begin{ruledtabular}
\begin{tabular}{llllll}
$\pt$ & 70-80\% & 60-70\% & 50-60\% & 40-50\% & \\ \hline
0.225 & $(1.07\pm0.02)\times10^{1}$ & $(2.11\pm0.04)\times10^{1}$ & $(3.66\pm0.07)\times10^{1}$ & $(5.89\pm0.12)\times10^{1}$ & \\
0.275 & $\,\,7.79\pm0.16$ & $(1.56\pm0.03)\times10^{1}$ & $(2.71\pm0.05)\times10^{1}$ & $(4.37\pm0.09)\times10^{1}$ & \\
0.325 & $\,\,5.81\pm0.12$ & $(1.16\pm0.02)\times10^{1}$ & $(2.04\pm0.04)\times10^{1}$ & $(3.31\pm0.07)\times10^{1}$ & \\
0.375 & $\,\,4.32\pm0.05$ & $\,\,8.77\pm0.09$ & $(1.56\pm0.02)\times10^{1}$ & $(2.53\pm0.03)\times10^{1}$ & \\
0.425 & $\,\,3.27\pm0.04$ & $\,\,6.60\pm0.07$ & $(1.19\pm0.01)\times10^{1}$ & $(1.95\pm0.02)\times10^{1}$ & \\
0.475 & $\,\,2.44\pm0.03$ & $\,\,5.04\pm0.05$ & $\,\,9.13\pm0.09$ & $(1.50\pm0.02)\times10^{1}$ & \\
0.525 & $\,\,1.86\pm0.02$ & $\,\,3.81\pm0.04$ & $\,\,7.03\pm0.07$ & $(1.15\pm0.01)\times10^{1}$ & \\
\end{tabular}
\end{ruledtabular}
\end{table*}

\begin{table*}
\caption{Identified $K^-$ invariant transverse momentum spectra at mid-rapidity ($|y|<0.1$) in Au+Au collisions at 62.4~GeV: $d^2N/(2\pi\pt d\pt dy)$ [(GeV/$c$)$^{-2}$] versus $\pt$ [GeV/$c$]. Errors are the quadratic sum of statistical errors and point-to-point systematic errors. See Section~\ref{sec:syst_spectra} for other systematic uncertainties.}
\label{tab:spectra_AuAu62_km}
\begin{ruledtabular}
\begin{tabular}{llllll}
$\pt$ & 70-80\% & 60-70\% & 50-60\% & 40-50\% & \\ \hline
0.275 & $(7.03\pm0.19)\times10^{-1}$ & $\,\,1.41\pm0.03$ & $\,\,2.51\pm0.06$ & $\,\,4.07\pm0.09$ & \\
0.325 & $(5.95\pm0.20)\times10^{-1}$ & $\,\,1.18\pm0.04$ & $\,\,2.18\pm0.07$ & $\,\,3.48\pm0.11$ & \\
0.375 & $(4.89\pm0.26)\times10^{-1}$ & $\,\,1.04\pm0.05$ & $\,\,1.91\pm0.10$ & $\,\,3.08\pm0.16$ & \\
0.425 & $(4.05\pm0.49)\times10^{-1}$ & $(8.79\pm1.06)\times10^{-1}$ & $\,\,1.63\pm0.20$ & $\,\,2.71\pm0.33$ & \\
0.475 & $(3.14\pm0.38)\times10^{-1}$ & $(7.23\pm0.87)\times10^{-1}$ & $\,\,1.31\pm0.16$ & $\,\,2.24\pm0.27$ & \\
0.525 & $(2.70\pm0.33)\times10^{-1}$ & $(5.86\pm0.71)\times10^{-1}$ & $\,\,1.13\pm0.14$ & $\,\,1.92\pm0.23$ & \\
0.575 & $(2.27\pm0.19)\times10^{-1}$ & $(5.04\pm0.41)\times10^{-1}$ & $(9.71\pm0.78)\times10^{-1}$ & $\,\,1.70\pm0.14$ & \\
\end{tabular}
\end{ruledtabular}

\caption{Identified $K^+$ invariant transverse momentum spectra at mid-rapidity ($|y|<0.1$) in Au+Au collisions at 62.4~GeV: $d^2N/(2\pi\pt d\pt dy)$ [(GeV/$c$)$^{-2}$] versus $\pt$ [GeV/$c$]. Errors are the quadratic sum of statistical errors and point-to-point systematic errors. See Section~\ref{sec:syst_spectra} for other systematic uncertainties.}
\label{tab:spectra_AuAu62_kp}
\begin{ruledtabular}
\begin{tabular}{llllll}
$\pt$ & 70-80\% & 60-70\% & 50-60\% & 40-50\% & \\ \hline
0.275 & $(7.29\pm0.19)\times10^{-1}$ & $\,\,1.54\pm0.04$ & $\,\,2.63\pm0.06$ & $\,\,4.90\pm0.10$ & \\
0.325 & $(5.82\pm0.19)\times10^{-1}$ & $\,\,1.31\pm0.04$ & $\,\,2.32\pm0.07$ & $\,\,4.14\pm0.13$ & \\
0.375 & $(5.07\pm0.26)\times10^{-1}$ & $\,\,1.14\pm0.06$ & $\,\,2.09\pm0.11$ & $\,\,3.63\pm0.18$ & \\
0.425 & $(4.40\pm0.53)\times10^{-1}$ & $(9.92\pm1.20)\times10^{-1}$ & $\,\,1.81\pm0.22$ & $\,\,3.13\pm0.38$ & \\
0.475 & $(3.58\pm0.43)\times10^{-1}$ & $(8.04\pm0.97)\times10^{-1}$ & $\,\,1.53\pm0.18$ & $\,\,2.61\pm0.31$ & \\
0.525 & $(3.20\pm0.39)\times10^{-1}$ & $(6.89\pm0.83)\times10^{-1}$ & $\,\,1.31\pm0.16$ & $\,\,2.23\pm0.27$ & \\
0.575 & $(2.52\pm0.21)\times10^{-1}$ & $(5.78\pm0.47)\times10^{-1}$ & $\,\,1.14\pm0.09$ & $\,\,1.93\pm0.16$ & \\
\end{tabular}
\end{ruledtabular}
\end{table*}

\begin{table*}
\caption{Identified $\pbar$ invariant transverse momentum spectra at mid-rapidity ($|y|<0.1$) in Au+Au collisions at 62.4~GeV: $d^2N/(2\pi\pt d\pt dy)$ [(GeV/$c$)$^{-2}$] versus $\pt$ [GeV/$c$]. Errors are the quadratic sum of statistical errors and point-to-point systematic errors. See Section~\ref{sec:syst_spectra} for other systematic uncertainties.}
\label{tab:spectra_AuAu62_pbar}
\begin{ruledtabular}
\begin{tabular}{llllll}
$\pt$ & 70-80\% & 60-70\% & 50-60\% & 40-50\% & \\ \hline
0.375 & $(2.05\pm0.04)\times10^{-1}$ & $(3.87\pm0.06)\times10^{-1}$ & $(6.19\pm0.09)\times10^{-1}$ & $(9.12\pm0.12)\times10^{-1}$ & \\
0.425 & $(1.85\pm0.04)\times10^{-1}$ & $(3.53\pm0.06)\times10^{-1}$ & $(5.93\pm0.08)\times10^{-1}$ & $(8.75\pm0.11)\times10^{-1}$ & \\
0.475 & $(1.76\pm0.03)\times10^{-1}$ & $(3.37\pm0.05)\times10^{-1}$ & $(5.59\pm0.08)\times10^{-1}$ & $(8.31\pm0.11)\times10^{-1}$ & \\
0.525 & $(1.53\pm0.03)\times10^{-1}$ & $(3.11\pm0.05)\times10^{-1}$ & $(5.18\pm0.07)\times10^{-1}$ & $(7.89\pm0.10)\times10^{-1}$ & \\
0.575 & $(1.34\pm0.03)\times10^{-1}$ & $(2.82\pm0.04)\times10^{-1}$ & $(4.87\pm0.07)\times10^{-1}$ & $(7.29\pm0.09)\times10^{-1}$ & \\
0.625 & $(1.20\pm0.03)\times10^{-1}$ & $(2.51\pm0.06)\times10^{-1}$ & $(4.37\pm0.10)\times10^{-1}$ & $(6.85\pm0.15)\times10^{-1}$ & \\
0.675 & $(1.05\pm0.03)\times10^{-1}$ & $(2.24\pm0.05)\times10^{-1}$ & $(3.96\pm0.09)\times10^{-1}$ & $(6.24\pm0.13)\times10^{-1}$ & \\
\end{tabular}
\end{ruledtabular}
\end{table*}

\begin{table*}
\caption{Identified proton invariant transverse momentum spectra at mid-rapidity ($|y|<0.1$) in Au+Au collisions at 62.4~GeV: $d^2N/(2\pi\pt d\pt dy)$ [(GeV/$c$)$^{-2}$] versus $\pt$ [GeV/$c$]. Errors are the quadratic sum of statistical errors and point-to-point systematic errors. See Section~\ref{sec:syst_spectra} for other systematic uncertainties including those due to proton background subtraction.}
\label{tab:spectra_AuAu62_p}
\begin{ruledtabular}
\begin{tabular}{llllll}
$\pt$ & 70-80\% & 60-70\% & 50-60\% & 40-50\% & \\ \hline
0.425 & $(2.66\pm0.04)\times10^{-1}$ & $(5.28\pm0.06)\times10^{-1}$ & $(9.21\pm0.10)\times10^{-1}$ & $\,\,1.43\pm0.02$ & \\
0.475 & $(2.45\pm0.03)\times10^{-1}$ & $(5.25\pm0.06)\times10^{-1}$ & $(9.27\pm0.11)\times10^{-1}$ & $\,\,1.43\pm0.02$ & \\
0.525 & $(2.21\pm0.03)\times10^{-1}$ & $(4.95\pm0.06)\times10^{-1}$ & $(8.87\pm0.10)\times10^{-1}$ & $\,\,1.38\pm0.02$ & \\
0.575 & $(2.12\pm0.03)\times10^{-1}$ & $(4.69\pm0.06)\times10^{-1}$ & $(8.26\pm0.10)\times10^{-1}$ & $\,\,1.32\pm0.01$ & \\
0.625 & $(1.92\pm0.04)\times10^{-1}$ & $(4.17\pm0.09)\times10^{-1}$ & $(7.89\pm0.16)\times10^{-1}$ & $\,\,1.26\pm0.03$ & \\
0.675 & $(1.70\pm0.04)\times10^{-1}$ & $(3.90\pm0.08)\times10^{-1}$ & $(7.13\pm0.15)\times10^{-1}$ & $\,\,1.17\pm0.02$ & \\
0.725 & $(1.56\pm0.04)\times10^{-1}$ & $(3.39\pm0.07)\times10^{-1}$ & $(6.45\pm0.14)\times10^{-1}$ & $\,\,1.07\pm0.02$ & \\
\end{tabular}
\end{ruledtabular}
\end{table*}

\begin{table*}[htbp]
\caption{Identified $\pi^-$ invariant transverse momentum spectra at mid-rapidity ($|y|<0.1$) in Au+Au collisions at 130~GeV: $d^2N/(2\pi\pt d\pt dy)$ [(GeV/$c$)$^{-2}$] versus $\pt$ [GeV/$c$]. Errors are the quadratic sum of statistical errors and point-to-point systematic errors. See Section~\ref{sec:syst_spectra} for other systematic uncertainties.}
\label{tab:spectra_AuAu130_pim}
\begin{ruledtabular}
\begin{tabular}{lllll}
$\pt$ & 58-85\% & 45-58\% & 34-45\% & 26-34\% \\ \hline
0.225 & $(2.34\pm0.08)\times10^{1}$ & $(5.88\pm0.16)\times10^{1}$ & $(9.56\pm0.27)\times10^{1}$ & $(1.45\pm0.04)\times10^{2}$ \\
0.275 & $(1.73\pm0.05)\times10^{1}$ & $(4.52\pm0.12)\times10^{1}$ & $(7.46\pm0.18)\times10^{1}$ & $(1.05\pm0.03)\times10^{2}$ \\
0.325 & $(1.28\pm0.03)\times10^{1}$ & $(3.37\pm0.07)\times10^{1}$ & $(5.56\pm0.10)\times10^{1}$ & $(7.98\pm0.14)\times10^{1}$ \\
0.375 & $\,\,9.11\pm0.21$ & $(2.53\pm0.05)\times10^{1}$ & $(4.13\pm0.08)\times10^{1}$ & $(6.01\pm0.11)\times10^{1}$ \\
0.425 & $\,\,6.92\pm0.17$ & $(1.85\pm0.04)\times10^{1}$ & $(3.22\pm0.06)\times10^{1}$ & $(4.73\pm0.09)\times10^{1}$ \\
0.475 & $\,\,5.49\pm0.14$ & $(1.39\pm0.03)\times10^{1}$ & $(2.37\pm0.05)\times10^{1}$ & $(3.45\pm0.07)\times10^{1}$ \\
0.525 & $\,\,3.88\pm0.11$ & $(1.10\pm0.03)\times10^{1}$ & $(1.90\pm0.04)\times10^{1}$ & $(2.77\pm0.07)\times10^{1}$ \\
\end{tabular}
\end{ruledtabular}

\caption{Identified $\pi^+$ invariant transverse momentum spectra at mid-rapidity ($|y|<0.1$) in Au+Au collisions at 130~GeV: $d^2N/(2\pi\pt d\pt dy)$ [(GeV/$c$)$^{-2}$] versus $\pt$ [GeV/$c$]. Errors are the quadratic sum of statistical errors and point-to-point systematic errors. See Section~\ref{sec:syst_spectra} for other systematic uncertainties.}
\label{tab:spectra_AuAu130_pip}
\begin{ruledtabular}
\begin{tabular}{lllll}
$\pt$ & 58-85\% & 45-58\% & 34-45\% & 26-34\% \\ \hline
0.225 & $(2.31\pm0.07)\times10^{1}$ & $(5.96\pm0.18)\times10^{1}$ & $(9.82\pm0.27)\times10^{1}$ & $(1.44\pm0.04)\times10^{2}$ \\
0.275 & $(1.75\pm0.05)\times10^{1}$ & $(4.43\pm0.12)\times10^{1}$ & $(7.60\pm0.18)\times10^{1}$ & $(1.05\pm0.03)\times10^{2}$ \\
0.325 & $(1.29\pm0.03)\times10^{1}$ & $(3.35\pm0.06)\times10^{1}$ & $(5.66\pm0.10)\times10^{1}$ & $(7.82\pm0.14)\times10^{1}$ \\
0.375 & $\,\,9.76\pm0.22$ & $(2.60\pm0.05)\times10^{1}$ & $(4.21\pm0.08)\times10^{1}$ & $(6.06\pm0.11)\times10^{1}$ \\
0.425 & $\,\,6.97\pm0.17$ & $(1.85\pm0.04)\times10^{1}$ & $(3.14\pm0.06)\times10^{1}$ & $(4.57\pm0.09)\times10^{1}$ \\
0.475 & $\,\,5.01\pm0.14$ & $(1.39\pm0.03)\times10^{1}$ & $(2.40\pm0.05)\times10^{1}$ & $(3.68\pm0.07)\times10^{1}$ \\
0.525 & $\,\,4.04\pm0.12$ & $(1.07\pm0.03)\times10^{1}$ & $(1.92\pm0.05)\times10^{1}$ & $(2.68\pm0.06)\times10^{1}$ \\
\end{tabular}
\end{ruledtabular}
\end{table*}

\begin{table*}[htbp]
\caption{Identified $K^-$ invariant transverse momentum spectra at mid-rapidity ($|y|<0.1$) in Au+Au collisions at 130~GeV: $d^2N/(2\pi\pt d\pt dy)$ [(GeV/$c$)$^{-2}$] versus $\pt$ [GeV/$c$]. Errors are the quadratic sum of statistical errors and point-to-point systematic errors. See Section~\ref{sec:syst_spectra} for other systematic uncertainties. Data were published in Ref.~\cite{kaon130}.}
\label{tab:spectra_AuAu130_km}
\begin{ruledtabular}
\begin{tabular}{lllll}
$\pt$ & 58-85\% & 45-58\% & 34-45\% & 26-34\% \\ \hline
0.175 & $\,\,1.93\pm0.25$ & $\,\,3.78\pm0.50$ & $\,\,7.13\pm0.79$ & $(1.12\pm0.13)\times10^{1}$ \\
0.225 & $\,\,2.03\pm0.18$ & $\,\,4.24\pm0.56$ & $\,\,7.42\pm0.61$ & $(1.08\pm0.09)\times10^{1}$ \\
0.275 & $\,\,1.45\pm0.13$ & $\,\,4.09\pm0.33$ & $\,\,6.52\pm0.50$ & $\,\,9.93\pm0.71$ \\
0.325 & $\,\,1.48\pm0.12$ & $\,\,3.91\pm0.30$ & $\,\,6.09\pm0.42$ & $\,\,7.85\pm0.54$ \\
0.375 & $\,\,1.30\pm0.10$ & $\,\,3.26\pm0.23$ & $\,\,5.46\pm0.36$ & $\,\,7.61\pm0.51$ \\
0.425 & $\,\,1.09\pm0.09$ & $\,\,2.89\pm0.21$ & $\,\,4.58\pm0.32$ & $\,\,6.34\pm0.43$ \\
0.475 & $(8.70\pm0.69)\times10^{-1}$ & $\,\,2.42\pm0.18$ & $\,\,4.49\pm0.32$ & $\,\,6.40\pm0.42$ \\
\end{tabular}
\end{ruledtabular}

\caption{Identified $K^+$ invariant transverse momentum spectra at mid-rapidity ($|y|<0.1$) in Au+Au collisions at 130~GeV: $d^2N/(2\pi\pt d\pt dy)$ [(GeV/$c$)$^{-2}$] versus $\pt$ [GeV/$c$]. Errors are the quadratic sum of statistical errors and point-to-point systematic errors. See Section~\ref{sec:syst_spectra} for other systematic uncertainties. Data were published in Ref.~\cite{kaon130}.}
\label{tab:spectra_AuAu130_kp}
\begin{ruledtabular}
\begin{tabular}{lllll}
$\pt$ & 58-85\% & 45-58\% & 34-45\% & 26-34\% \\ \hline
0.175 & $\,\,1.58\pm0.22$ & $\,\,6.26\pm0.70$ & $\,\,7.02\pm0.79$ & $(1.19\pm0.12)\times10^{1}$ \\
0.225 & $\,\,2.14\pm0.20$ & $\,\,5.56\pm0.46$ & $\,\,8.18\pm0.63$ & $\,\,9.21\pm0.74$ \\
0.275 & $\,\,1.90\pm0.15$ & $\,\,4.91\pm0.37$ & $\,\,7.00\pm0.49$ & $\,\,9.27\pm0.69$ \\
0.325 & $\,\,1.38\pm0.11$ & $\,\,3.82\pm0.28$ & $\,\,6.37\pm0.42$ & $\,\,9.31\pm0.62$ \\
0.375 & $\,\,1.32\pm0.10$ & $\,\,3.69\pm0.26$ & $\,\,5.70\pm0.38$ & $\,\,8.65\pm0.55$ \\
0.425 & $\,\,1.06\pm0.09$ & $\,\,3.23\pm0.22$ & $\,\,5.47\pm0.37$ & $\,\,6.99\pm0.46$ \\
0.475 & $(9.09\pm0.72)\times10^{-1}$ & $\,\,2.88\pm0.22$ & $\,\,4.55\pm0.33$ & $\,\,6.13\pm0.40$ \\
\end{tabular}
\end{ruledtabular}
\end{table*}

\begin{table*}[htbp]
\caption{Identified $\pbar$ invariant transverse momentum spectra at mid-rapidity ($|y|<0.1$) in Au+Au collisions at 130~GeV: $d^2N/(2\pi\pt d\pt dy)$ [(GeV/$c$)$^{-2}$] versus $\pt$ [GeV/$c$]. Errors are the quadratic sum of statistical errors and point-to-point systematic errors. See Section~\ref{sec:syst_spectra} for other systematic uncertainties. Data were published in Ref.~\cite{p130}.}
\label{tab:spectra_AuAu130_pbar}
\begin{ruledtabular}
\begin{tabular}{lllll}
$\pt$ & 58-85\% & 45-58\% & 34-45\% & 26-34\% \\ \hline
0.375 & $(5.05\pm0.29)\times10^{-1}$ & $\,\,1.14\pm0.06$ & $\,\,1.62\pm0.09$ & $\,\,2.01\pm0.11$ \\
0.425 & $(4.84\pm0.20)\times10^{-1}$ & $\,\,1.02\pm0.04$ & $\,\,1.56\pm0.06$ & $\,\,1.99\pm0.07$ \\
0.475 & $(4.37\pm0.13)\times10^{-1}$ & $\,\,1.00\pm0.03$ & $\,\,1.53\pm0.04$ & $\,\,1.98\pm0.05$ \\
0.525 & $(4.02\pm0.12)\times10^{-1}$ & $(9.70\pm0.26)\times10^{-1}$ & $\,\,1.42\pm0.04$ & $\,\,1.93\pm0.05$ \\
0.575 & $(3.61\pm0.11)\times10^{-1}$ & $(8.84\pm0.23)\times10^{-1}$ & $\,\,1.32\pm0.03$ & $\,\,1.77\pm0.04$ \\
0.625 & $(3.37\pm0.10)\times10^{-1}$ & $(8.13\pm0.22)\times10^{-1}$ & $\,\,1.24\pm0.03$ & $\,\,1.72\pm0.04$ \\
0.675 & $(3.24\pm0.09)\times10^{-1}$ & $(7.62\pm0.20)\times10^{-1}$ & $\,\,1.18\pm0.03$ & $\,\,1.64\pm0.04$ \\
\end{tabular}
\end{ruledtabular}

\caption{Identified proton invariant transverse momentum spectra at mid-rapidity ($|y|<0.1$) in Au+Au collisions at 130~GeV: $d^2N/(2\pi\pt d\pt dy)$ [(GeV/$c$)$^{-2}$] versus $\pt$ [GeV/$c$]. Errors are the quadratic sum of statistical errors, point-to-point systematic errors, and systematic uncertainties due to proton background subtraction. See Section~\ref{sec:syst_spectra} for other systematic uncertainties. Data were published in Ref.~\cite{p130}.}
\label{tab:spectra_AuAu130_p}
\begin{ruledtabular}
\begin{tabular}{lllll}
$\pt$ & 58-85\% & 45-58\% & 34-45\% & 26-34\% \\ \hline
0.375 & $(6.40\pm0.66)\times10^{-1}$ & $\,\,1.46\pm0.15$ & $\,\,2.19\pm0.22$ & $\,\,2.78\pm0.28$ \\
0.425 & $(5.99\pm0.47)\times10^{-1}$ & $\,\,1.45\pm0.11$ & $\,\,2.07\pm0.16$ & $\,\,2.74\pm0.21$ \\
0.475 & $(5.54\pm0.32)\times10^{-1}$ & $\,\,1.34\pm0.07$ & $\,\,1.81\pm0.10$ & $\,\,2.59\pm0.14$ \\
0.525 & $(5.12\pm0.24)\times10^{-1}$ & $\,\,1.25\pm0.05$ & $\,\,1.87\pm0.08$ & $\,\,2.54\pm0.10$ \\
0.575 & $(4.50\pm0.21)\times10^{-1}$ & $\,\,1.19\pm0.05$ & $\,\,1.82\pm0.07$ & $\,\,2.44\pm0.10$ \\
0.625 & $(4.61\pm0.16)\times10^{-1}$ & $\,\,1.05\pm0.03$ & $\,\,1.65\pm0.05$ & $\,\,2.27\pm0.07$ \\
0.675 & $(3.72\pm0.13)\times10^{-1}$ & $(9.52\pm0.31)\times10^{-1}$ & $\,\,1.57\pm0.05$ & $\,\,2.24\pm0.06$ \\
\end{tabular}
\end{ruledtabular}
\end{table*}

\begin{table*}[htbp]
\caption{Identified $\pi^-$ invariant transverse momentum spectra at mid-rapidity ($|y|<0.1$) in Au+Au collisions at 200~GeV: $d^2N/(2\pi\pt d\pt dy)$ [(GeV/$c$)$^{-2}$] versus $\pt$ [GeV/$c$]. Errors are the quadratic sum of statistical errors and point-to-point systematic errors. See Section~\ref{sec:syst_spectra} for other systematic uncertainties. Data were published in Ref.~\cite{spec200}.}
\label{tab:spectra_AuAu200_pim}
\begin{ruledtabular}
\begin{tabular}{llllll}
$\pt$ & 70-80\% & 60-70\% & 50-60\% & 40-50\% & \\ \hline
0.225 & $(1.61\pm0.05)\times10^{1}$ & $(3.06\pm0.09)\times10^{1}$ & $(5.09\pm0.15)\times10^{1}$ & $(7.86\pm0.24)\times10^{1}$ & \\
0.275 & $(1.16\pm0.02)\times10^{1}$ & $(2.21\pm0.04)\times10^{1}$ & $(3.69\pm0.07)\times10^{1}$ & $(5.90\pm0.12)\times10^{1}$ & \\
0.325 & $\,\,8.56\pm0.18$ & $(1.63\pm0.03)\times10^{1}$ & $(2.79\pm0.06)\times10^{1}$ & $(4.48\pm0.09)\times10^{1}$ & \\
0.375 & $\,\,6.46\pm0.08$ & $(1.25\pm0.01)\times10^{1}$ & $(2.12\pm0.02)\times10^{1}$ & $(3.45\pm0.04)\times10^{1}$ & \\
0.425 & $\,\,4.82\pm0.06$ & $\,\,9.41\pm0.10$ & $(1.62\pm0.02)\times10^{1}$ & $(2.63\pm0.03)\times10^{1}$ & \\
0.475 & $\,\,3.69\pm0.05$ & $\,\,7.17\pm0.08$ & $(1.25\pm0.01)\times10^{1}$ & $(2.05\pm0.02)\times10^{1}$ & \\
0.525 & $\,\,2.73\pm0.04$ & $\,\,5.56\pm0.06$ & $\,\,9.75\pm0.11$ & $(1.59\pm0.02)\times10^{1}$ & \\
\end{tabular}
\end{ruledtabular}

\caption{Identified $\pi^+$ invariant transverse momentum spectra at mid-rapidity ($|y|<0.1$) in Au+Au collisions at 200~GeV: $d^2N/(2\pi\pt d\pt dy)$ [(GeV/$c$)$^{-2}$] versus $\pt$ [GeV/$c$]. Errors are the quadratic sum of statistical errors and point-to-point systematic errors. See Section~\ref{sec:syst_spectra} for other systematic uncertainties. Data were published in Ref.~\cite{spec200}.}
\label{tab:spectra_AuAu200_pip}
\begin{ruledtabular}
\begin{tabular}{llllll}
$\pt$ & 70-80\% & 60-70\% & 50-60\% & 40-50\% & \\ \hline
0.225 & $(1.62\pm0.05)\times10^{1}$ & $(3.05\pm0.09)\times10^{1}$ & $(5.05\pm0.15)\times10^{1}$ & $(7.85\pm0.24)\times10^{1}$ \\
0.275 & $(1.15\pm0.02)\times10^{1}$ & $(2.20\pm0.04)\times10^{1}$ & $(3.69\pm0.07)\times10^{1}$ & $(5.85\pm0.12)\times10^{1}$ \\
0.325 & $\,\,8.57\pm0.18$ & $(1.64\pm0.03)\times10^{1}$ & $(2.78\pm0.06)\times10^{1}$ & $(4.45\pm0.09)\times10^{1}$ \\
0.375 & $\,\,6.38\pm0.08$ & $(1.23\pm0.01)\times10^{1}$ & $(2.12\pm0.02)\times10^{1}$ & $(3.44\pm0.04)\times10^{1}$ \\
0.425 & $\,\,4.80\pm0.06$ & $\,\,9.45\pm0.10$ & $(1.63\pm0.02)\times10^{1}$ & $(2.63\pm0.03)\times10^{1}$ \\
0.475 & $\,\,3.66\pm0.05$ & $\,\,7.22\pm0.08$ & $(1.25\pm0.01)\times10^{1}$ & $(2.03\pm0.02)\times10^{1}$ \\
0.525 & $\,\,2.81\pm0.04$ & $\,\,5.52\pm0.06$ & $\,\,9.66\pm0.10$ & $(1.59\pm0.02)\times10^{1}$ \\
\end{tabular}
\end{ruledtabular}
\end{table*}

\begin{table*}[htbp]
\caption{Identified $K^-$ invariant transverse momentum spectra at mid-rapidity ($|y|<0.1$) in Au+Au collisions at 200~GeV: $d^2N/(2\pi\pt d\pt dy)$ [(GeV/$c$)$^{-2}$] versus $\pt$ [GeV/$c$]. Errors are the quadratic sum of statistical errors and point-to-point systematic errors. See Section~\ref{sec:syst_spectra} for other systematic uncertainties. Data were published in Ref.~\cite{spec200}.}
\label{tab:spectra_AuAu200_km}
\begin{ruledtabular}
\begin{tabular}{llllll}
$\pt$ & 70-80\% & 60-70\% & 50-60\% & 40-50\% & \\ \hline
0.225 & $\,\,1.35\pm0.11$ & $\,\,2.34\pm0.17$ & $\,\,3.77\pm0.27$ & $\,\,6.47\pm0.47$ & \\
0.275 & $\,\,1.14\pm0.05$ & $\,\,2.18\pm0.08$ & $\,\,3.49\pm0.12$ & $\,\,5.66\pm0.18$ & \\
0.325 & $(9.22\pm0.25)\times10^{-1}$ & $\,\,1.89\pm0.04$ & $\,\,3.05\pm0.06$ & $\,\,4.79\pm0.09$ & \\
0.375 & $(7.44\pm0.24)\times10^{-1}$ & $\,\,1.61\pm0.04$ & $\,\,2.83\pm0.07$ & $\,\,4.31\pm0.11$ & \\
0.425 & $(6.63\pm0.27)\times10^{-1}$ & $\,\,1.37\pm0.05$ & $\,\,2.38\pm0.09$ & $\,\,3.66\pm0.15$ & \\
0.475 & $(5.57\pm0.59)\times10^{-1}$ & $\,\,1.14\pm0.12$ & $\,\,2.11\pm0.23$ & $\,\,3.26\pm0.35$ & \\
0.525 & $(4.78\pm0.50)\times10^{-1}$ & $(9.60\pm1.00)\times10^{-1}$ & $\,\,1.81\pm0.19$ & $\,\,2.80\pm0.30$ & \\
\end{tabular}
\end{ruledtabular}

\caption{Identified $K^+$ invariant transverse momentum spectra at mid-rapidity ($|y|<0.1$) in Au+Au collisions at 200~GeV: $d^2N/(2\pi\pt d\pt dy)$ [(GeV/$c$)$^{-2}$] versus $\pt$ [GeV/$c$]. Errors are the quadratic sum of statistical errors and point-to-point systematic errors. See Section~\ref{sec:syst_spectra} for other systematic uncertainties. Data were published in Ref.~\cite{spec200}.}
\label{tab:spectra_AuAu200_kp}
\begin{ruledtabular}
\begin{tabular}{llllll}
$\pt$ & 70-80\% & 60-70\% & 50-60\% & 40-50\% & \\ \hline
0.225 & $\,\,1.31\pm0.10$ & $\,\,2.43\pm0.18$ & $\,\,3.95\pm0.29$ & $\,\,6.69\pm0.48$ \\
0.275 & $\,\,1.08\pm0.04$ & $\,\,2.22\pm0.08$ & $\,\,3.65\pm0.12$ & $\,\,5.79\pm0.19$ \\
0.325 & $(9.40\pm0.25)\times10^{-1}$ & $\,\,1.90\pm0.04$ & $\,\,3.17\pm0.06$ & $\,\,5.02\pm0.09$ \\
0.375 & $(7.82\pm0.24)\times10^{-1}$ & $\,\,1.66\pm0.05$ & $\,\,2.94\pm0.08$ & $\,\,4.39\pm0.11$ \\
0.425 & $(6.79\pm0.27)\times10^{-1}$ & $\,\,1.43\pm0.05$ & $\,\,2.56\pm0.10$ & $\,\,3.85\pm0.15$ \\
0.475 & $(5.78\pm0.61)\times10^{-1}$ & $\,\,1.19\pm0.13$ & $\,\,2.20\pm0.24$ & $\,\,3.46\pm0.38$ \\
0.525 & $(4.80\pm0.51)\times10^{-1}$ & $\,\,1.04\pm0.11$ & $\,\,1.91\pm0.20$ & $\,\,2.98\pm0.32$ \\
\end{tabular}
\end{ruledtabular}
\end{table*}

\begin{table*}[htbp]
\caption{Identified $\pbar$ invariant transverse momentum spectra at mid-rapidity ($|y|<0.1$) in Au+Au collisions at 200~GeV: $d^2N/(2\pi\pt d\pt dy)$ [(GeV/$c$)$^{-2}$] versus $\pt$ [GeV/$c$]. Errors are the quadratic sum of statistical errors and point-to-point systematic errors. See Section~\ref{sec:syst_spectra} for other systematic uncertainties. Data were published in Ref.~\cite{spec200}.}
\label{tab:spectra_AuAu200_pbar}
\begin{ruledtabular}
\begin{tabular}{llllll}
$\pt$ & 70-80\% & 60-70\% & 50-60\% & 40-50\% & \\ \hline
0.375 & $(3.63\pm0.12)\times10^{-1}$ & $(6.68\pm0.19)\times10^{-1}$ & $(9.92\pm0.26)\times10^{-1}$ & $\,\,1.48\pm0.04$ & \\
0.425 & $(3.42\pm0.10)\times10^{-1}$ & $(6.23\pm0.14)\times10^{-1}$ & $(9.31\pm0.18)\times10^{-1}$ & $\,\,1.37\pm0.02$ & \\
0.476 & $(3.27\pm0.09)\times10^{-1}$ & $(5.72\pm0.12)\times10^{-1}$ & $(8.74\pm0.16)\times10^{-1}$ & $\,\,1.31\pm0.02$ & \\
0.525 & $(2.86\pm0.08)\times10^{-1}$ & $(5.19\pm0.11)\times10^{-1}$ & $(8.32\pm0.15)\times10^{-1}$ & $\,\,1.21\pm0.02$ & \\
0.574 & $(2.61\pm0.07)\times10^{-1}$ & $(4.80\pm0.10)\times10^{-1}$ & $(7.95\pm0.14)\times10^{-1}$ & $\,\,1.16\pm0.02$ & \\
0.624 & $(2.27\pm0.07)\times10^{-1}$ & $(4.27\pm0.09)\times10^{-1}$ & $(7.22\pm0.13)\times10^{-1}$ & $\,\,1.06\pm0.02$ & \\
0.675 & $(2.09\pm0.06)\times10^{-1}$ & $(3.73\pm0.09)\times10^{-1}$ & $(6.43\pm0.13)\times10^{-1}$ & $(9.95\pm0.19)\times10^{-1}$ & \\
\end{tabular}
\end{ruledtabular}
\end{table*}

\begin{table*}[htbp]
\caption{Identified proton invariant transverse momentum spectra at mid-rapidity ($|y|<0.1$) in Au+Au collisions at 200~GeV: $d^2N/(2\pi\pt d\pt dy)$ [(GeV/$c$)$^{-2}$] versus $\pt$ [GeV/$c$]. Errors are the quadratic sum of statistical errors, point-to-point systematic errors, and systematic uncertainties due to proton background subtraction. See Section~\ref{sec:syst_spectra} for other systematic uncertainties. Data were published in Ref.~\cite{spec200}.}
\label{tab:spectra_AuAu200_p}
\label{tab:spectra_last}
\begin{ruledtabular}
\begin{tabular}{llllll}
$\pt$ & 70-80\% & 60-70\% & 50-60\% & 40-50\% & \\ \hline
0.425 & $(3.87\pm0.20)\times10^{-1}$ & $(7.12\pm0.36)\times10^{-1}$ & $\,\,1.17\pm0.06$ & $\,\,1.65\pm0.08$ \\
0.476 & $(3.62\pm0.16)\times10^{-1}$ & $(6.65\pm0.27)\times10^{-1}$ & $\,\,1.11\pm0.04$ & $\,\,1.61\pm0.06$ \\
0.525 & $(3.43\pm0.13)\times10^{-1}$ & $(6.31\pm0.22)\times10^{-1}$ & $\,\,1.01\pm0.03$ & $\,\,1.54\pm0.05$ \\
0.574 & $(3.08\pm0.10)\times10^{-1}$ & $(5.56\pm0.16)\times10^{-1}$ & $(9.46\pm0.26)\times10^{-1}$ & $\,\,1.46\pm0.04$ \\
0.624 & $(2.64\pm0.08)\times10^{-1}$ & $(5.10\pm0.13)\times10^{-1}$ & $(8.50\pm0.20)\times10^{-1}$ & $\,\,1.34\pm0.03$ \\
0.675 & $(2.37\pm0.07)\times10^{-1}$ & $(4.46\pm0.11)\times10^{-1}$ & $(7.85\pm0.18)\times10^{-1}$ & $\,\,1.25\pm0.03$ \\
0.725 & $(2.15\pm0.07)\times10^{-1}$ & $(4.03\pm0.11)\times10^{-1}$ & $(7.11\pm0.18)\times10^{-1}$ & $\,\,1.13\pm0.03$ \\
0.775 & $(1.87\pm0.06)\times10^{-1}$ & $(3.71\pm0.10)\times10^{-1}$ & $(6.34\pm0.16)\times10^{-1}$ & $\,\,1.03\pm0.02$ \\
0.824 & $(1.63\pm0.05)\times10^{-1}$ & $(3.17\pm0.09)\times10^{-1}$ & $(5.79\pm0.14)\times10^{-1}$ & $(9.62\pm0.22)\times10^{-1}$ \\
0.875 & $(1.46\pm0.05)\times10^{-1}$ & $(2.84\pm0.08)\times10^{-1}$ & $(5.28\pm0.13)\times10^{-1}$ & $(8.71\pm0.20)\times10^{-1}$ \\
0.924 & $(1.26\pm0.04)\times10^{-1}$ & $(2.56\pm0.08)\times10^{-1}$ & $(4.93\pm0.13)\times10^{-1}$ & $(7.86\pm0.21)\times10^{-1}$ \\
0.975 & $(1.07\pm0.04)\times10^{-1}$ & $(2.29\pm0.07)\times10^{-1}$ & $(4.28\pm0.12)\times10^{-1}$ & $(7.12\pm0.19)\times10^{-1}$ \\
1.025 & $(9.88\pm0.43)\times10^{-2}$ & $(2.14\pm0.08)\times10^{-1}$ & $(3.89\pm0.13)\times10^{-1}$ & $(6.53\pm0.21)\times10^{-1}$ \\
1.075 & $(8.37\pm0.38)\times10^{-2}$ & $(1.76\pm0.07)\times10^{-1}$ & $(3.38\pm0.12)\times10^{-1}$ & $(5.81\pm0.19)\times10^{-1}$ \\
1.125 & $(7.00\pm0.36)\times10^{-2}$ & $(1.60\pm0.07)\times10^{-1}$ & $(3.14\pm0.12)\times10^{-1}$ & $(5.24\pm0.20)\times10^{-1}$ \\
1.175 & $(6.18\pm0.40)\times10^{-2}$ & $(1.38\pm0.07)\times10^{-1}$ & $(2.82\pm0.14)\times10^{-1}$ & $(4.76\pm0.21)\times10^{-1}$ \\ \hline
$\pt$ & 30-40\% & 20-30\% & 10-20\% & 5-10\% & 0-5\% \\ \hline
0.425 & $\,\,2.33\pm0.11$ & $\,\,3.34\pm0.16$ & $\,\,4.21\pm0.20$ & $\,\,5.70\pm0.28$ & $\,\,6.42\pm0.31$ \\
0.476 & $\,\,2.31\pm0.09$ & $\,\,3.26\pm0.13$ & $\,\,4.28\pm0.17$ & $\,\,5.62\pm0.22$ & $\,\,6.45\pm0.25$ \\
0.525 & $\,\,2.17\pm0.07$ & $\,\,3.13\pm0.10$ & $\,\,4.12\pm0.13$ & $\,\,5.38\pm0.17$ & $\,\,6.18\pm0.20$ \\
0.574 & $\,\,2.07\pm0.05$ & $\,\,2.99\pm0.08$ & $\,\,3.97\pm0.10$ & $\,\,5.09\pm0.13$ & $\,\,5.96\pm0.16$ \\
0.624 & $\,\,1.91\pm0.04$ & $\,\,2.79\pm0.06$ & $\,\,3.77\pm0.08$ & $\,\,4.89\pm0.11$ & $\,\,5.75\pm0.13$ \\
0.675 & $\,\,1.78\pm0.04$ & $\,\,2.63\pm0.06$ & $\,\,3.58\pm0.08$ & $\,\,4.56\pm0.11$ & $\,\,5.44\pm0.13$ \\
0.725 & $\,\,1.67\pm0.04$ & $\,\,2.42\pm0.06$ & $\,\,3.32\pm0.08$ & $\,\,4.30\pm0.10$ & $\,\,5.07\pm0.12$ \\
0.775 & $\,\,1.54\pm0.04$ & $\,\,2.23\pm0.05$ & $\,\,3.13\pm0.07$ & $\,\,4.09\pm0.09$ & $\,\,4.89\pm0.11$ \\
0.824 & $\,\,1.40\pm0.03$ & $\,\,2.11\pm0.05$ & $\,\,2.93\pm0.06$ & $\,\,3.83\pm0.08$ & $\,\,4.67\pm0.10$ \\
0.875 & $\,\,1.30\pm0.03$ & $\,\,1.94\pm0.04$ & $\,\,2.78\pm0.06$ & $\,\,3.66\pm0.08$ & $\,\,4.44\pm0.09$ \\
0.924 & $\,\,1.18\pm0.03$ & $\,\,1.82\pm0.05$ & $\,\,2.59\pm0.07$ & $\,\,3.42\pm0.10$ & $\,\,4.16\pm0.13$ \\
0.975 & $\,\,1.09\pm0.03$ & $\,\,1.66\pm0.05$ & $\,\,2.37\pm0.07$ & $\,\,3.17\pm0.09$ & $\,\,3.95\pm0.12$ \\
1.025 & $(9.97\pm0.32)\times10^{-1}$ & $\,\,1.53\pm0.05$ & $\,\,2.21\pm0.07$ & $\,\,2.95\pm0.09$ & $\,\,3.68\pm0.12$ \\
1.075 & $(8.82\pm0.28)\times10^{-1}$ & $\,\,1.38\pm0.04$ & $\,\,2.06\pm0.06$ & $\,\,2.71\pm0.09$ & $\,\,3.36\pm0.11$ \\
1.125 & $(8.21\pm0.31)\times10^{-1}$ & $\,\,1.27\pm0.05$ & $\,\,1.85\pm0.07$ & $\,\,2.57\pm0.10$ & $\,\,3.24\pm0.14$ \\
1.175 & $(7.36\pm0.32)\times10^{-1}$ & $\,\,1.16\pm0.05$ & $\,\,1.70\pm0.07$ & $\,\,2.34\pm0.11$ & $\,\,2.89\pm0.14$ \\
\end{tabular}
\end{ruledtabular}
\end{table*}

\bibliography{Bibliography}

\end{document}

%% file: authors_20090210.tex
\affiliation{Argonne National Laboratory, Argonne, Illinois 60439, USA}
\affiliation{University of Birmingham, Birmingham, United Kingdom}
\affiliation{Brookhaven National Laboratory, Upton, New York 11973, USA}
\affiliation{University of California, Berkeley, California 94720, USA}
\affiliation{University of California, Davis, California 95616, USA}
\affiliation{University of California, Los Angeles, California 90095, USA}
\affiliation{Universidade Estadual de Campinas, Sao Paulo, Brazil}
\affiliation{Carnegie Mellon University, Pittsburgh, Pennsylvania 15213, USA}
\affiliation{University of Illinois at Chicago, Chicago, Illinois 60607, USA}
\affiliation{Creighton University, Omaha, Nebraska 68178, USA}
\affiliation{Nuclear Physics Institute AS CR, 250 68 \v{R}e\v{z}/Prague, Czech Republic}
\affiliation{Laboratory for High Energy (JINR), Dubna, Russia}
\affiliation{Particle Physics Laboratory (JINR), Dubna, Russia}
\affiliation{Institute of Physics, Bhubaneswar 751005, India}
\affiliation{Indian Institute of Technology, Mumbai, India}
\affiliation{Indiana University, Bloomington, Indiana 47408, USA}
\affiliation{Institut de Recherches Subatomiques, Strasbourg, France}
\affiliation{University of Jammu, Jammu 180001, India}
\affiliation{Kent State University, Kent, Ohio 44242, USA}
\affiliation{University of Kentucky, Lexington, Kentucky, 40506-0055, USA}
\affiliation{Institute of Modern Physics, Lanzhou, China}
\affiliation{Lawrence Berkeley National Laboratory, Berkeley, California 94720, USA}
\affiliation{Massachusetts Institute of Technology, Cambridge, MA 02139-4307, USA}
\affiliation{Max-Planck-Institut f\"ur Physik, Munich, Germany}
\affiliation{Michigan State University, East Lansing, Michigan 48824, USA}
\affiliation{Moscow Engineering Physics Institute, Moscow Russia}
\affiliation{City College of New York, New York City, New York 10031, USA}
\affiliation{NIKHEF and Utrecht University, Amsterdam, The Netherlands}
\affiliation{Ohio State University, Columbus, Ohio 43210, USA}
\affiliation{Panjab University, Chandigarh 160014, India}
\affiliation{Pennsylvania State University, University Park, Pennsylvania 16802, USA}
\affiliation{Institute of High Energy Physics, Protvino, Russia}
\affiliation{Purdue University, West Lafayette, Indiana 47907, USA}
\affiliation{Pusan National University, Pusan, Republic of Korea}
\affiliation{University of Rajasthan, Jaipur 302004, India}
\affiliation{Rice University, Houston, Texas 77251, USA}
\affiliation{Universidade de Sao Paulo, Sao Paulo, Brazil}
\affiliation{University of Science \& Technology of China, Hefei 230026, China}
\affiliation{Shanghai Institute of Applied Physics, Shanghai 201800, China}
\affiliation{SUBATECH, Nantes, France}
\affiliation{Texas A\&M University, College Station, Texas 77843, USA}
\affiliation{University of Texas, Austin, Texas 78712, USA}
\affiliation{Tsinghua University, Beijing 100084, China}
\affiliation{United States Naval Academy, Annapolis, MD 21402, USA}
\affiliation{Valparaiso University, Valparaiso, Indiana 46383, USA}
\affiliation{Variable Energy Cyclotron Centre, Kolkata 700064, India}
\affiliation{Warsaw University of Technology, Warsaw, Poland}
\affiliation{University of Washington, Seattle, Washington 98195, USA}
\affiliation{Wayne State University, Detroit, Michigan 48201, USA}
\affiliation{Institute of Particle Physics, CCNU (HZNU), Wuhan 430079, China}
\affiliation{Yale University, New Haven, Connecticut 06520, USA}
\affiliation{University of Zagreb, Zagreb, HR-10002, Croatia}

\author{B.~I.~Abelev}\affiliation{University of Illinois at Chicago, Chicago, Illinois 60607, USA}
\author{M.~M.~Aggarwal}\affiliation{Panjab University, Chandigarh 160014, India}
\author{Z.~Ahammed}\affiliation{Variable Energy Cyclotron Centre, Kolkata 700064, India}
\author{B.~D.~Anderson}\affiliation{Kent State University, Kent, Ohio 44242, USA}
\author{D.~Arkhipkin}\affiliation{Particle Physics Laboratory (JINR), Dubna, Russia}
\author{G.~S.~Averichev}\affiliation{Laboratory for High Energy (JINR), Dubna, Russia}
\author{Y.~Bai}\affiliation{NIKHEF and Utrecht University, Amsterdam, The Netherlands}
\author{J.~Balewski}\affiliation{Massachusetts Institute of Technology, Cambridge, MA 02139-4307, USA}
\author{O.~Barannikova}\affiliation{University of Illinois at Chicago, Chicago, Illinois 60607, USA}
\author{L.~S.~Barnby}\affiliation{University of Birmingham, Birmingham, United Kingdom}
% ...

\collaboration{STAR Collaboration}\noaffiliation